\documentclass[12pt]{article}

\usepackage{color}
\usepackage{amsmath,amssymb}
\usepackage{amscd}
\usepackage{graphicx} 
\usepackage{epstopdf}
\usepackage{latexsym}
\usepackage{cite}
\usepackage{ascmac}
\usepackage{cases}
\usepackage[all]{xy}

\allowdisplaybreaks[1]

\newcommand{\eq}[1]{(\ref{#1})}
\newcommand{\nn}{\nonumber}
\newcommand{\ds}{\displaystyle}

\newcommand{\vev}[1]{\left\langle #1 \right\rangle}
\newcommand{\ket}[1]{\bigl|#1\bigr>}
\newcommand{\bra}[1]{\bigl<#1\bigr|}
\newcommand{\bracket}[2]{\left.\left\langle #1\right|#2\right\rangle}
\newcommand{\del}{\partial}

\newcommand{\asymeq}{\underset{asym}{\simeq}}

\newcommand{\bP}{\boldsymbol{P}}
\newcommand{\bQ}{\boldsymbol{Q}}

\DeclareMathOperator{\tr}{tr}

\DeclareMathOperator{\diag}{diag}
\DeclareMathOperator{\sgn}{sgn}

\DeclareMathOperator{\Stokes}{\mathbf S}
\DeclareMathOperator{\Cut}{\mathbf C}

\newtheorem{Theorem}{Theorem}[section]
\newtheorem{Definition}{Definition}[section]
\newtheorem{Proposition}{Proposition}[section]
\newtheorem{Conjecture}{Conjecture}[section]
\newtheorem{Lemma}{Lemma}[section]
\newtheorem{Corollary}{Corollary}[section]

\newtheorem{Remark}{Remark}[section]

\renewcommand{\thefootnote}{\fnsymbol{footnote}}

\makeatletter
\@addtoreset{equation}{section}

\begin{document}


\begin{titlepage}
\thispagestyle{empty} 
\begin{flushright}
  August 2013\\
    \vspace{0.2cm} 
  arXiv:1308.6603 \\
  \vspace{0.2cm} 
  YITP-13-64 
\end{flushright}

\vspace{2.3cm}

\begin{center}
\noindent{\Large \textbf{
Duality Constraints on String Theory \\
 \vspace{0.5cm}
 \large - Instantons and spectral networks -
}}
\end{center}

\vspace{1cm}

\begin{center}
\noindent{Chuan-Tsung Chan\footnote{ctchan@thu.edu.tw}$^{,a,b}$, 
Hirotaka Irie\footnote{irie@yukawa.kyoto-u.ac.jp}$^{,c}$
and Chi-Hsien Yeh\footnote{d95222008@ntu.edu.tw}$^{,b,d}$}
\end{center}
\vspace{0.5cm}
\begin{center}
{\em
$^{a}$Department of Physics, Tunghai University, Taichung 40704, Taiwan \\
\vspace{.3cm}
$^{b}$National Center for Theoretical Sciences, \\
National Tsing-Hua University, Hsinchu 30013, Taiwan \\
\vspace{.3cm}
$^{c}$Yukawa Institute for Theoretical Physics, \\
Kyoto University, Kyoto 606-8502, Japan \\
\vspace{.3cm}
$^{d}$Department of Physics, National Tsing-Hua University, \\
Hsinchu 30013, Taiwan 
}
\end{center}

\vspace{0.2cm}

\begin{abstract}
We study an implication of $p-q$ duality (spectral duality or T-duality) on non-perturbative completion of $(p,q)$ minimal string theory. According to the Eynard-Orantin topological recursion, spectral $p-q$ duality was already checked for all-order perturbative analysis including instanton/soliton amplitudes. Non-perturbative realization of this duality, on the other hand, causes a new fundamental issue. In fact, we find that not all the non-perturbative completions are consistent with non-perturbative $p-q$ duality; Non-perturbative duality rather provides a constraint on non-perturbative contour ambiguity (equivalently, of D-instanton fugacity) in matrix models. In particular, it prohibits some of meta-stability caused by ghost D-instantons, since there is no non-perturbative realization on the dual side in the matrix-model description. Our result is the first quantitative observation that a missing piece of our understanding in non-perturbative string theory is provided by {\em the principle of non-perturbative string duality}. To this end, we study Stokes phenomena of $(p,q)$ minimal strings with spectral networks and improve the Deift-Zhou's method to describe meta-stable vacua. By analyzing the instanton profile on spectral networks, we argue the duality constraints on string theory.
\end{abstract}

\end{titlepage}

\newpage

\renewcommand{\thefootnote}{\arabic{footnote}}
\setcounter{footnote}{0}


  \tableofcontents


\section{Introduction}

Duality means that two systems describe the same physics although they look different at the first sight. It is accompanied with a dictionary to translate observables and amplitudes among the two systems. With utilizing the dictionary, it becomes possible to study different physical regimes of the system. So far, this is the typical use of duality which has been producing a number of successes in various studies. 

String theory is a zoo of dualities, and therefore it connects a large number of theories by various means: S-, T- and U-duality, and so on \cite{PolchinskiStringTheory}. By use of duality, string theories with different appearances are related to each other. Importantly, {\em duality is believed to act non-perturbatively}, and it binds the different perturbative pictures to be a single unified theory of our universe. In string theory, therefore, duality is not only a tool for connecting different pictures. It should also be an intrinsic and fundamental ingredient which constitutes non-perturbative definition of string theory. We believe that this consideration is particularly important in non-perturbative formulation of string theory and it leads us to the following investigation. 

In this paper, we would like to put forward this idea to a quantitative level in full non-perturbative regimes, i.e.~non-perturbative completion, of string theory. In particular, we argue that {\em duality realized in non-perturbative completions imposes constraints on the system rather than a dictionary}. This new fundamental feature of non-perturbative duality is referred to as {\em duality constraints} on the system. Here we elaborate a quantitative evidence of duality constraints within a solvable framework of string theory \cite{Polyakov,DSL,BIPZ,KazakovSeries,Kostov1,Kostov2,Kostov3,BDSS,BMPNonP,TwoMatString,GrossMigdal2,DouglasGeneralizedKdV,TadaYamaguchiDouglas,Moore,FIK,GinspargZinnJustin,Shenker,fkn,fkn3,DVV,MultiCut,David0,MSS,David,EynardZinnJustin,DKK,fy1,fy2,fy3,MultiCutUniversality,McGreevyVerlinde,Martinec,KMS,AKK,KazakovKostov,MMSS,SeSh2,HHIKKMT,SatoTsuchiya,IshibashiKurokiYamaguchi,fis,fim,fi1,EynardOrantin,EynardOrantinXYsym,EynardMarino,irie2,CISY1,CIY1,CIY2,CIY3,CIY4}. As the first example of duality, we focus on spectral $p-q$ duality \cite{fkn3} in lower dimensional (non-critical) string theory \cite{Polyakov}. In particular, we argue validity of the duality constraints by analyzing the instanton spectrums on possible variety of spectral networks, which characterize non-perturbative completions of the string theory. 

A major motivation of this project is from the issue about {\em non-perturbative ambiguity in definition of string theory}. String theory is so far defined perturbatively. For instance, free-energy of string theory is given by an asymptotic expansion in string coupling $g$ \cite{Shenker}: 
\begin{align} 
\mathcal F (g) \asymeq \sum_{n=0}^\infty g^{2n-2} \mathcal F_n + \sum_{I\in \mathfrak I} \theta_I\, g^{\gamma_I} \exp\Bigl[\sum_{n=0}^\infty g^{n-1} \mathcal F_n^{(I)}\Bigr] + O(\theta^2),\qquad g\to 0, \label{EqIntroExpansionOfFreeEnergy}
\end{align}
and the perturbative definition (like worldsheet formulation) allows us to obtain all the expansion coefficients $\bigl\{\mathcal F_n\bigr\}_{n\in \mathbb Z_+}$ and $\bigr\{\mathcal F_n^{(I)}\bigr\}_{n\in \mathbb Z_+}^{I\in \mathfrak J}$ in principle. However, such a perturbative calculation does not provide any information about the relative weights for instanton sectors $\bigl\{ \theta_I \bigr\}_{I\in \mathfrak J}$, called {\em D-instanton fugacities} \cite{David}. As we will also discuss (also see \cite{CIY4}), D-instanton fugacities are directly related to {\em vacuum structure of string theory}, which is one of the most important non-perturbative physics in string theory. Despite of this importance, there is no worldsheet concept describing the D-instanton fugacity. By this fact, there appear a number of undermined free parameters in non-perturbative regimes, especially in the vacuum structure of string theory. In this sense, only after solving this issue, it is possible to capture the whole picture of non-perturbative string theory. What we argue in this paper is that this issues can be partially resolved by spelling out {\em non-perturbative duality} in string theory. 

An intrinsic reason for the non-perturbative ambiguity stems from the ambiguity appearing in Borel resummation, which we refer to as {\em Stokes ambiguity}. Note that we should not be confused among non-perturbative ambiguity and Stokes ambiguity. As is reviewed in Appendix \ref{AppendixNonperturbativeVSStokesAmbiguity}, these two concepts are different from each other. 
Stokes ambiguity takes place when the expansion is asymptotic one. In the expansion, some of the exponentially small instanton corrections are overwhelmed by relatively large instanton corrections. In this occasion, the associated D-instanton fugacity covertly changes its value discretely, i.e.~Stokes phenomenon occurs. According to the resurgent analysis \cite{Ecalle}, it is known that perturbation theory can determine all the jump information of Stokes phenomenon. However, since D-instanton fugacity plays a role of absorbent of such a jump phenomenon, perturbative description cannot assign its primary value to the D-instanton fugacity. Therefore, perturbative theory should be tolerant to variety of D-instanton fugacity and therefore non-perturbative ambiguity.  

This does not mean that D-instanton fugacity is not calculable. For example, if {\em ``a''} matrix model (or large $N$ gauge theory) is given, one can evaluate D-instanton fugacity \cite{HHIKKMT} up to the Stokes ambiguity of exponentially small corrections. One can even reproduce such a result from non-perturbative reconstruction from the perturbative string theory \cite{CIY4}. However, it is just one of the choices of matrix models. From this point of views, non-perturbative ambiguity means that there are a number of different matrix models (i.e.~with different D-instanton fugacity) which commonly include perturbative string theory as its asymptotic expansion. For example, one can find a number of matrix models with {\em a different choice of contours} which still commonly include the same perturbative string theory in its large $N$ expansion. These matrix models with different choices of contour integrals result in non-perturbatively different string theories with different vacuum structures. In this sense, {\em non-perturbative ambiguity is the ambiguity in non-perturbative physics of string theory}, and one cannot decide which non-perturbative completions are suitable for ``our'' string theory. Therefore, as theoretical physics, in order to obtain a sufficiently unique string theory, it is important to find out an additional non-perturbative principle to provide a restriction on non-perturbative formulation of string theory. It is then natural to expect that these ambiguities would be convicted by imposing non-perturbative string duality. 

\begin{figure}[htbp]
\begin{center}
\begin{align}
\begin{array}{c}
\begin{xy}
(15,20)*{\quad \text{Non-perturbative String Theory}\quad \rule[-1.4ex]{0pt}{2.8ex} }="X", 
(15,8)*{\quad\,\, \text{\footnotesize (describe the same physics) }\quad \rule[-1.4ex]{0pt}{2ex} }="Z", 
(-25,0) *{\quad \text{ A-Matrix Model } \quad \rule[-1.3ex]{0pt}{3.6ex} }="A", 
(55,0) *{\quad \text{ B-Matrix Model } \quad \rule[-1.3ex]{0pt}{3.6ex} }="B", 
(-25,-20) *{\text{ Perturbative String Theory A } \ \  \rule[-1.3ex]{0pt}{3.6ex}  }="C", 
(55,-20) *{\ \  \text{ Perturbative String Theory B }  \rule[-1.3ex]{0pt}{3.6ex}   }="D", 
\ar  @{=>} "A";"X"^{\text{  } \rule[2ex]{0pt}{0ex}}
\ar  @{=>} "B";"X"^{\text{  } \rule[2ex]{0pt}{0ex}}
\ar  @{<->} "A";"B"^{\text{ Non-Pert.~Duality }}_{\text{ (constraint) }}
\ar @<1mm> "A";"C"^{\text{ Large $N$ saddles }\rule[-0.5ex]{0pt}{2ex}}
\ar @<-1mm> "B";"D"_{\text{ Large $N$ saddles } \rule[-0.5ex]{0pt}{2ex}}
\ar @<1mm> "C";"A"^{\text{ Ensemble of saddles }\rule[-0.5ex]{0pt}{2ex}}
\ar @<-1mm> "D";"B"_{\text{ Ensemble of saddles } \rule[-0.5ex]{0pt}{2ex}}
\ar  @{<->} "C";"D"^{\text{ Pert.~Duality }}_{\text{ (dictionary) }}
\end{xy}
\end{array} \nn
\end{align}
\end{center}
\caption{\footnotesize Non-perturbative v.s.~perturbative duality }
\label{FigureIntroductionDualityConstraintonMatrices}
\end{figure}
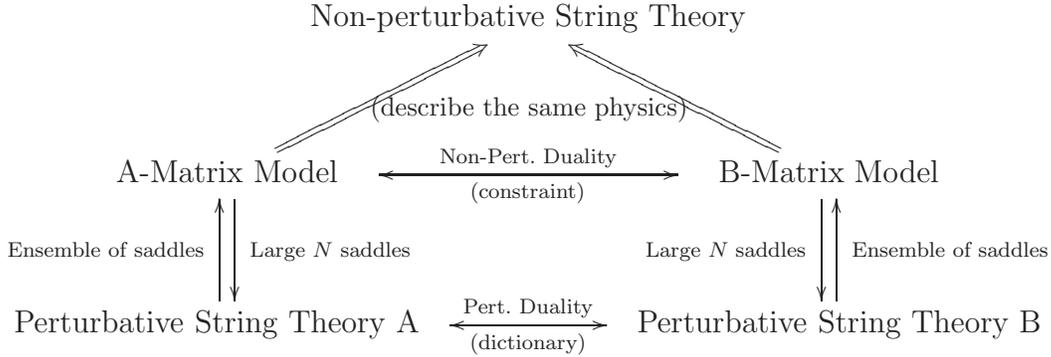

We should note how non-perturbative duality works in general, and how it is different from the perturbative one. Assume that two perturbative string theories (A and B) are dual to each other, and that there exist matrix models as non-perturbative completions of each perturbative string theory (Fig.~\ref{FigureIntroductionDualityConstraintonMatrices}). In this occasion, we will see 
\begin{itemize}
\item {\em Perturbative duality} among perturbative string theories (A and B) means the equivalence of {\em perturbative saddles} constituted in the A- and B-matrix model. 
\item {\em Non-perturbative duality} among A- and B-matrix model means the equivalence of {\em ensembles of perturbative saddles} realized by the A- and B-matrix model. 
\end{itemize}
In the usual means, string duality is only checked in a perturbative sense (by using D-branes and so on) and has rarely been checked non-perturbatively. 
The point is that {\em the equivalence of perturbative string theories does not guarantee the equivalence of the two dual matrix-model descriptions because not all the non-perturbative completions are realized by matrix models of both sides.} If there appears discrepancy, then it means that we need to restrict the non-perturbative completions, otherwise the non-perturbative string theory is described only by one side of duality. This is the key idea of duality constraints on string theory. In this paper, we will describe this new feature of non-perturbative v.s.~perturbative string duality in details, with focusing on {\em spectral $p-q$ duality} in non-perturbative completions of 
$(p,q)$ minimal string theory. 

As is reviewed in Section \ref{SectionSpectralPQDualityPreliminary}, 
from the worldsheet point of views, spectral $p-q$ duality is a kind of T-duality in $(p,q)$ minimal string theory. The worldsheet CFT of $(p,q)$ minimal string theory consists of Liouville field theory \cite{Polyakov,KPZ,DDK,DOZZ,Teschner,FZZT,ZZ,SeSh,KOPSS} and $(p,q)$ minimal CFT. If one writes the worldsheet Liouville theory as 
\begin{align}
S_{\rm Liou} = \frac{1}{4\pi } \int d^2\sigma \sqrt{g}\Bigl(g^{ab}\del_a \phi  \del_b \phi + QR\phi + 4\pi \mu e^{2b\phi}\Bigr),\qquad Q= b+\frac{1}{b}, 
\end{align}
then the $p-q$ duality is the duality under replacing the Liouville coupling $b$ by $b^{-1}$:
\begin{align}
b\leftrightarrow  \frac{1}{b}. 
\end{align}
This duality is one of the guiding principle for obtaining three-point functions \cite{DOZZ,Teschner} and is believed to exist in Liouville theory. In $(p,q)$ minimal CFT, this coupling $b$ appears in the similar place as the compactification radius of $c=1$ boson field.
 Therefore, it is also understood as a kind of T-duality in minimal string theory.%
\footnote{Spectral $p-q$ duality is a kind of T-duality but is different from the usual T-duality, in the sense that $(p,q)$ minimal models are defined by diagonal modular invariant theory, in which there is no naive spacetime picture of exchanging momentum and winding modes along the compact direction. See also \cite{FMS-CFT}.}

From the viewpoint of spectral curves, it is the duality caused by exchanging the roles of coordinates $x$ and $y$ in spectral curve: 
\begin{align}
F(x,y)=0,\qquad x\leftrightarrow y, 
\end{align}
which is generally called {\em spectral duality}. 
The best way to describe this duality in all-order large $N$ expansions is given by the Eynard-Orantin topological recursion of matrix models \cite{EynardOrantin}. In this formation, it was generally shown that this duality does hold for all-order perturbative analysis and is called $x-y$ symmetry of symplectic invariants. 

From the non-perturbative point of views, on the other hand, this duality can be understood as the equivalence of two systems encoded in two-matrix models \cite{Mehta}: 
\begin{align}
\mathcal Z = \int dX dY \, e^{-N\tr [V_1(X)+ V_2(Y) -XY]}. 
\end{align}
In two-matrix models, there are two matrices $X$ and $Y$ which describe two space-times dual to each other. In this sense, it resembles double field theory \cite{DoubleFieldTheory} since the field theory includes dual (spacetime) degree of freedom simultaneously. The spectral $p-q$ duality then means the equivalence of the system with exchanging the role of matrices $X$ and $Y$. 

Non-perturbative $p-q$ duality of two-matrix models can be understood as the non-perturbative equivalence between the following two systems (Fig.~\ref{FigureDualConst}): 
\begin{itemize}
\item \underline{$X$-system}: described by the matrix $X$ with first integrating the matrix $Y$ 
\item \underline{$Y$-system}: described by the matrix $Y$ with first integrating the matrix $X$  
\end{itemize}
Naively, one may think that the equivalence holds in general; but our result suggests that these two systems are generally different in two-matrix models. The analysis of non-perturbative comparison is shown in Section \ref{SectionStokesSpecific} with focusing on $(2,5)\leftrightarrow (5,2)$ minimal string theory. 

\begin{figure}[htbp]
\begin{center}
\includegraphics[scale=0.6]{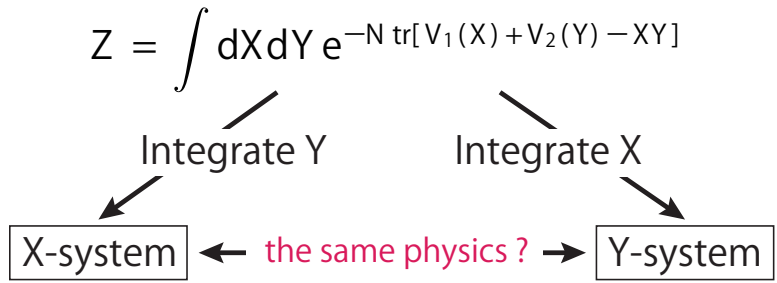}
\end{center}
\caption{\footnotesize Non-perturbative requirement of spectral $p-q$ duality is that the system is equally described as dynamics of $X$-eigenvalues and also as dynamics of $Y$-eigenvalues. }
\label{FigureDualConstIntroduction}
\end{figure}

As we have discussed in this introduction, this result of discrepancy is particularly important for fundamental understanding of non-perturbative string theory. Of course, each side of description (i.e.~$X$-system or $Y$-system) is a consistent theory and there is no intrinsic reason to be prohibited by itself. However, these descriptions are not on an equal footing in non-perturbative regimes. Therefore, we should adopt the non-perturbative duality {\em as a new non-perturbative principle of string theory} in order to retain the string duality non-perturbatively. This provides a constraint on non-perturbative ambiguity of string theory. This is our basic proposal on {\em the principle of non-perturbative string duality} or {\em duality constraints on string theory}. This consideration should supply a missing piece in the definition of non-perturbative string theory. 

It should also be noted that this is the first legitimate result on how to fix the non-perturbative ambiguity in non-critical string theory. Clearly, this consideration should be applied to various kinds of string dualities in other systems, including critical (higher-dimensional) string theory. 

\subsection{Organization: A guide for contexts}

Organization of this paper is following: 
\begin{itemize}
\item [1. ] Section 1 is here for Introduction. 
\item [2. ] Section 2 is for a review/summary of preliminary facts on spectral $p-q$ duality. 
\begin{itemize}
\item In Section \ref{SectionTwoMatrixModels}, the spectral $p-q$ duality is reviewed from the viewpoints of two-matrix models.
\begin{itemize}
\item all-order equivalence in perturbation theory, based on topological recursion, is in Section \ref{SubsubsectionAllOrderPerturbativeAnalysisAndSymplecticInvariants}. 
\item equivalence including all-order instanton corrections, based on non-perturbative partition function, is in Section \ref{SubsubsectionEquivalenceWithInstantonsForAllOrderPT}. 
\item evaluation of D-instanton fugacity, based on mean-field dynamics of eigenvalues, is in Section \ref{SubsubsectionMeanFieldApproximation}. 
\end{itemize}
\item In Section \ref{SubsubsectionBAfunctions}, the spectral $p-q$ duality is reviewed from the viewpoints of Baker-Akhiezer systems, based on KP hierarchy. 
\item In Section \ref{SectionPQdualLiouvilleTheory}, the spectral $p-q$ duality is reviewed from the viewpoints of worldsheet description, i.e.~Liouville theory. 
\end{itemize} 
\end{itemize}
In this paper, the non-perturbative completions are described by isomonodromy systems, which are associated with spectral curves. 
\begin{itemize}
\item [3. ] Section 3 is for spectral curves and related isomonodromy systems: 
\begin{itemize}
\item In Section \ref{SubsectionFurtherCommentsOnSpectralCurves}, some basics on spectral curves are summarized
\begin{itemize}
\item In Section \ref{SubsubsectionThreeDefinitionsOfSpectralCurves}, three viewpoints of spectral curves and their relations are summarized, especially with focusing on $\varphi$-functions. 
\item In Section \ref{SubsubsectionSpectralCurveForChebyshevSolutions}, the basic properties of the spectral curve for Chebyshev solutions are reviewed, since they are extensively used in this paper. 
\item In Section \ref{SubsubsectionRoleOfSpectralCurvesAndLandscapes}, the concept of universal string theory landscape $\mathcal L_{\rm str}^{\rm (univ.)}$ is introduced as a universal set of perturbative string vacua, associated with the isomonodromy systems. 
\end{itemize}
\item In Section \ref{SectionIsomonodromyEmbedding}, the associated isomonodromy systems and their symmetry properties are summarized. 
\item In Section \ref{SubsectionClassicalBAFunction}, the concept of classical BA function $\Psi_{\rm cl}(\lambda)$ is introduced and its classical monodromy matrices are obtained for general $(p,q)$-systems of Chebyshev solutions (the evaluation is shown in Appendix \ref{AppendixMonodromyCalculusOfCBAF}). 
\end{itemize}
\item [4. ] Section 4 is for Stokes phenomena and Stokes geometry in isomonodromy systems. 
\begin{itemize}
\item In Section \ref{SectionStokesMatrixIsomonodromySystems}, the basic facts/definitions on Stokes matrices and the profile of dominant exponents are reviewed/summarized. 
\item In Section \ref{SubsubsectionGeometryOfAntiStokesLines}, the geometric structure of anti-Stokes lines are studied in general $(p,q)$-systems with Chebyshev solutions. 
\item In Section \ref{RoleOfSpectralNetworks}, the concept of Deift-Zhou's spectral networks is reviewed. 
\item In Section \ref{SubsubsectionStokesGeometryProfileInstantons}, the profile of instantons on spectral networks is studied. In particular, the uniform signature property (Theorem \ref{TheoremUniformSignatureProperty}) is proved, which is one of our main results. 
\end{itemize}
\item [5. ] Section 5 is for multi-cut boundary condition in the general $(p,q)$-systems as a condition for realization in two-matrix models. 
\begin{itemize}
\item In Section \ref{SubsectionDefinitionOfMultiCutBoundaryConditionWhyNecessary}, the definition of multi-cut boundary condition is reviewed. 
\item In Section \ref{SubsectionMulticutBCforPQminimalstringtheory}, the boundary condition are explicitly evaluated. 
\begin{itemize}
\item In Section \ref{SubsubsectionExistenceOfMCBC}, existence of the boundary condition is discussed. 
\item In Section \ref{SubsubsectionMCBCequation}, equations of the boundary conditions are shown. 
\end{itemize}
\end{itemize}
\end{itemize}
In order to obtain ``non-perturbative completions'' of $(p,q)$ minimal string theory, we consider the Riemann-Hilbert problem associated with the spectral curves of Chebyshev solutions and spectral networks on them. 

\begin{itemize}
\item [6. ] Section 6 is for the Riemann-Hilbert problems of the isomonodromy systems. 
\begin{itemize}
\item In Section \ref{SubsubsectionRHPandImprovedDZmethod}, the Riemann-Hilbert problem which is associated with isomonodromy systems is reviewed. 
\item In Section \ref{SubsectionProperSpectralNetworks}, the concept of proper spectral networks is introduced. 
\item In Section \ref{SectionWeavingSpectralNetworks}, we construct the proper spectral networks from the classical monodromy matrices which are obtained in Section \ref{SubsubsectionClassicalMonodromyMatrices}. 
\item In Section \ref{SectionLocalRHP}, we evaluate the Riemann-Hilbert problem associated with Chebyshev solutions, which justifies the introduction of proper spectral networks. 
\begin{itemize}
\item In Section \ref{SubsubsectionDirectEvaluationBranchPoints}, the Riemann-Hilbert integrals around saddle points are evaluated directly. 
\item In Section \ref{SubsubsectionContractiveSolvabilityAndLRH}, the local Riemann-Hilbert problems are solved {\em by a direct sum of Airy functions} for all the branch points in general $(p,q)$-systems with Chebyshev solutions. This eventually justifies the introduction of proper spectral networks (Theorem \ref{TheoremCutJumpCancellation1}). This is also one of our main results in this paper. 
\item In Section \ref{SubsubsectionTransseriesMetastablevacua}, the concept of ``perturbative string theory landscape $\mathcal L_{\rm str}^{(\hat {\mathcal K})}$'' is introduced, which is the set of vacuum which is realized in the non-perturbative completions of $(p,q)$ minimal string theory given by a spectral network $\hat {\mathcal K}$. By this we define a set of non-perturbative completions associated with spectral curves of Chebyshev solutions. 
\end{itemize}
\item In Section \ref{SubsubsectionGeneralizedAirySystems}, the Stokes matrices of generalized Airy systems (i.e.~$(p,1)$-systems) are obtained by using the discussion given in Section 6. 
\begin{itemize}
\item In Section \ref{SubsubsectionWeavingMethodInGeneralizedAirySystems}, spectral networks are obtained by weaving procedure. 
\item In Section \ref{SubsubsectionMCBCinGeneralizedAirySystems}, Stokes matrices are obtained by using multi-cut boundary condition. 
\end{itemize}
\item In Section \ref{SubsectionAnExampleDualKazakovSeries}, the Stokes matrices of dual Kazakov series (i.e.~$(p,2)$-systems) are obtained by using weaving procedure. 
\end{itemize}
\end{itemize}
\begin{itemize}
\item [7. ] Section 7 is for non-perturbative study on spectral $p-q$ duality. 
\begin{itemize}
\item In Section \ref{SubsectionNonperturbativeAmbiguityOfMinimalStringTheory}, the concept of {\em non-perturbative ambiguity} in matrix models is reviewed. 
\item In Section \ref{SubsectionComparisonOfInstantonsInDualityConstraints}, non-perturbative comparison of spectral $p-q$ duality is presented, with focusing on profiles of instantons on spectral networks. 
\begin{itemize}
\item In Section \ref{SubsubsectionGeneralDictionaryOfInstantonsDualityConstraints}, the general dictionary of instantons is shown in the $(p,2) \leftrightarrow (2,p)$ cases.  
\item In Section \ref{SubsubsectionPureGravityDualityComparisonOfInstantons}, the comparison of proper spectral networks is given in the $(3,2) \leftrightarrow (2,3)$ cases: Pure-gravity. 
\item In Section \ref{SubsubsectionYLedgeDualityComparisonOfInstantons}, the comparison of proper spectral networks is given in the $(5,2) \leftrightarrow (2,5)$ cases: Yang-Lee edge singularity. 
\end{itemize}
\item In Section \ref{SubsectionResolventsAndDynamicsOfEigenvaluesDualityConstraints}, the eigenvalue cuts on the resolvents for $X$ and $Y$ are discussed, with condensations of eigenvalues. 
\item In Section \ref{SubsubsectionDualityConstraintsOnStringTheory}, we argue duality constraint caused by spectral $p-q$ duality and discuss further issues on it. 
\end{itemize}
\item [8. ] Section 8 is devoted to conclusion and discussions with possible applications or appearances of the duality proposal. 
\end{itemize}

\section{Spectral $p-q$ duality: Preliminary \label{SectionSpectralPQDualityPreliminary}}

In this section, preliminary results on spectral $p-q$ duality are summarized. It also clarifies what have (or have not) been known in literature. The $p-q$ duality is first considered in non-critical string theory\cite{fkn3} as a natural duality appearing in the Lax-operator formalism. Some early discussions are found in \cite{KharchevMarshakov,BEH}. From the contemporary viewpoints, spectral $p-q$ duality can be considered in any models which are described by spectral curve, $F(x,y)=0$, and duality is caused by exchanging the role of $x$ and $y$. Therefore, it is also called $x-y$ symmetry and has been formulated and proved in Eynard-Orantin topological recursion \cite{EynardOrantin,EynardOrantinXYsym}, i.e.~for all-order perturbation theory of weak string coupling $g$ or of large $N$. Therefore, the discussion in this paper should be also applied (or generalized) to such general systems. 

On the other hand, on the worldsheet formulation, the $p-q$ duality in non-critical string theory can be understand as the Kramers-Wannier duality \cite{KWduality} of $(p,q)$ minimal models which is a counterpart of T-duality in $c=1$ bosonic theory. This also provides worldsheet intuition and is reviewed in the following. We should note that there is also an attempt to define such T- (or Kramers-Wannier) duality in matrix models and to study its implication to D-instanton fugacity \cite{AKOSY,KurokiSugino}.

\subsection{The two-matrix models \label{SectionTwoMatrixModels}}

From the viewpoints of two-matrix models \cite{Mehta}, 
\begin{align}
\mathcal Z = \int_{\mathcal C_X\times \mathcal C_Y} dX dY e^{-N \tr w(X,Y)},\qquad w(X,Y)= V_1(X)+V_2(Y) - XY, 
\end{align}
spectral $p-q$ duality means that this matrix integral is described equally by matrices $X$ and $Y$. 
For later convenience, the system based on the matrix $X$ is referred to as $X$-system; and the system based on the matrix $Y$ is referred to as $Y$-system. 

\subsubsection{All-order perturbative equivalence \label{SubsubsectionAllOrderPerturbativeAnalysisAndSymplecticInvariants}}
According to topological recursion \cite{EynardOrantin}, all the perturbative information of the systems is encoded in spectral curves \cite{BIPZ,KazakovSeries,Kostov1,Kostov2,Kostov3,BDSS} which are drawn by the resolvents of matrices $X$ and $Y$: 
\begin{align}
W_1^{(0)}(x) \equiv \lim_{N\to \infty}\vev{\frac{1}{N}\tr\frac{1}{x-X}},\qquad \widetilde W_1^{(0)}(y) \equiv \lim_{N\to \infty} \vev{\frac{1}{N}\tr\frac{1}{y-Y}}. \label{EquationForResolventOpeartorsInEOrecursionSection}
\end{align}
The key of spectral $p-q$ duality is represented by the fact that the resolvents of $X$-system and $Y$-systems are both given by the same algebraic equation $F(x,y)=0$: 
\begin{align}
\left\{
\begin{array}{c}
\ds 
F_{(\text{x-sys})}\bigl(x,W_1^{(0)}(x)\bigr) =F\bigl(x,W_1^{(0)}(x)-V_1'(x)\bigr)=0 \cr
\ds
F_{(\text{y-sys})}\bigl(y,\widetilde W_1^{(0)}(y)\bigr) = F\bigl(\widetilde W_1^{(0)}(y)-V_2'(y),y\bigr)=0
\end{array}
\right.. 
\label{EqSpectralCurveDualityFiniteN}
\end{align}
In general, if a system is described by a spectral curve, $F(x,y)=0$, the invariance under changing the role of $x$ and $y$ is called {\em spectral duality} \cite{BEH}. 

Perturbative equivalence of $X$- and $Y$-systems is then studied by topological recursion (i.e.~loop equations) and was proved for arbitrary spectral curves by \cite{EynardOrantin,EynardOrantinXYsym}. In this context, $p-q$ duality (or spectral duality) is called  or $x-y$ symmetry of symplectic invariance. Their results can be summarized as follows: For each system, one can control the following perturbative amplitudes for all-order: 
\begin{align}
\text{$X$-system: }
&
\left\{
\begin{array}{l}
\ds 
\vev{\prod_{j=1}^n \frac{1}{N} \tr \frac{1}{x_j-X}}_{\rm c} \asymeq  \sum_{h=0}^\infty N^{-2h} \, W_n^{(h)}(x_1,\cdots,x_n) \cr
F\bigl(x,W_1^{(0)}(x)-V_1'(x)\bigr)=0,\qquad \mathcal F_h \equiv W_0^{(h)} 
\end{array} 
\right., \\
\text{$Y$-system: }
&
\left\{
\begin{array}{l}
\ds 
\vev{\prod_{l=1}^m \frac{1}{N} \tr \frac{1}{y_l-Y}}_{\rm c} \asymeq  \sum_{h=0}^\infty N^{-2h} \, \widetilde W_m^{(h)}(y_1,\cdots,y_m) \cr
F\bigl(\widetilde W_1^{(0)}(y)-V_2'(y),y\bigr)=0, \qquad \widetilde {\mathcal F}_h \equiv \widetilde W_0^{(h)}
\end{array}
\right.,
\end{align}
and free-energys of these systems coincide and gives the same perturbative free-energy \cite{EynardOrantinXYsym}: 
\begin{align}
\mathcal F_{\rm pert} (g) = \sum_{n=0}^\infty N^{2-2n} \mathcal F_n \qquad \bigl(\mathcal F_n = \widetilde {\mathcal F}_n, \, n=0,1,2,\cdots \bigr).
\end{align}
Therefore, spectral $p-q$ duality perfectly work for all-order perturbation theory. 

In this paper, we focus on the cases of {\em 
$(p,q)$ minimal string theory} which appear as a result of the double scaling limit of matrix models \cite{DSL}. The scaling limit of parameters are taken as 
\begin{align}
N^{-1} = a^{p+q} g,\qquad x = a^{p} \zeta,\qquad y = a^{q} \bigl[(-1)^{q+1}\beta_{p,q} \eta\bigr]\qquad \bigl(a\to 0\bigr), \label{EqDSLofNZetaEta}
\end{align}
and 
the resolvent functions ($W^{(0)}_1(x)$ and $\widetilde W^{(0)}_1(y)$) are scaled as
\begin{align}
W_1^{(0)}(x)-V_1'(x)= a^q Q(\zeta),\qquad \widetilde W_1^{(0)}(y)-V_2'(y)= a^p P(\eta) \qquad \bigl(a\to 0\bigr). 
\end{align}
The algebraic equation, Eq.~\eq{EqSpectralCurveDualityFiniteN}, is then scaled as 
\begin{align}
F(x,y)= a^{pq} F_{\rm scaled}(\zeta,\eta)=0\qquad \bigl(a\to 0\bigr). 
\end{align}
An important result of topological recursion \cite{EynardOrantin} is that the topological recursion itself does not change before/after double scaling limit. All the differences are encoded in a difference of spectral curves. It is not important whether our systems are given by large $N$ expansion or expansion of scaled variable, like string coupling $g$. In the following, the algebraic equations are represented as $F(x,y)=0$ or $F(\zeta,\eta)=0$ without any care about their differences,%
\footnote{That is, $F_{\rm scaled}(\zeta,Q)=0$ is simply written as $F(\zeta,Q)=0$. } and we do not care whether the systems are before/after scaling limit. Algebraic equations are also written as $F(\zeta,\eta)=0$, $F(\zeta,Q)=0$ or $F(P,Q)=0$, depending on its convenience. 

The most general spectral curves of minimal string theory are derived in \cite{fim} by using free-fermion formulation and $W_{1+\infty}$-constraints. The results are very simple and worth mentioning. In fact, it is relevant to isomonodromy descriptions discussed later. In general, algebraic equations of $(p,q)$ minimal string theory behave as 
\begin{align}
F(\zeta,Q) \simeq Q^p -c_{p,q} \zeta^q + \cdots \qquad \bigl(\zeta\to \infty\bigr), 
\end{align}
and are expressed as 
\begin{align}
F(\zeta,Q)=\det\bigl(Q \, I_p- \del_\zeta \varphi(\zeta)\bigr)=\sum_{n=0}^p a_{n}(\zeta) Q^{p-n} = 0: \quad \text{a polynomial in $(\zeta,Q)$} \label{FinMSTofFIM}
\end{align}
where 
\begin{align}
& \varphi(\zeta) = \underset{1\leq j\leq p}{\diag}\bigl( \varphi^{(j)}(\zeta)\bigr),\qquad 
\varphi^{(j)}(\zeta) = \varphi^{(1)}(e^{-2\pi i (j-1)}\zeta),\nn\\
& \varphi^{(1)}(\zeta) = \beta_{p,q}\Bigl[\sum_{n=1}^{p+q} t_n \zeta^{\frac{n}{p}} + \sum_{n=1}^\infty v_n \zeta^{-\frac{n}{p}}\Bigr] \qquad (\zeta\to \infty\bigr). \label{FIMformulationOfSP}
\end{align}
The requirement of loop equation (equivalently of $W_{1+\infty}$-constraints) is that the algebraic equation, Eq.~\eq{FinMSTofFIM}, is a polynomial in $\zeta$ and $Q$. The parameters $\bigl\{ t_n \bigr\}_{n=1}^{p+q}$ are given by KP hierarchy \cite{fy1,fy2,fy3}, and therefore, are also given by isomonodromy deformations of the corresponding isomonodromy description. The number of the remaining parameters are counted in \cite{fim} and given by $\mathfrak g\equiv \dfrac{(p-1)(q-1)}{2}$ which are the number of possible instantons appearing in two-matrix models, or the number of ZZ-branes in Liouville theory \cite{SeSh}, or the number of possible $A$-cycles of the spectral curve. 

Therefore, systems described by the spectral curves of polynomial-type are generally referred to as {\em minimal string theory},%
\footnote{From the viewpoint of isomonodromy descriptions, if one chooses $(p,q)$ not to be coprime, then it corresponds to multi-cut matrix models \cite{irie2,CISY1,CIY1}. For example, $(p,q)=(2,4)$ is pure-supergravity, i.e.~Painlev\'e II, and is also minimal (super)string theory. Similarly, $(p,q)=(3,6)$ is minimal ($3$rd fractional super)string theory \cite{irie2,CIY1}. } and spectral duality of this system is formally understood as a flip of the indices of $(p,q)$ in minimal string theory: 
\begin{align}
(p,q)\quad \to \quad (q,p). \label{EqPQdualPQexchange}
\end{align}
Therefore, this duality is originally called $p-q$ duality \cite{fkn3}. In this sense, we also express in the following way: 
\begin{align}
\text{$X$-system} \quad \to \quad \text{$(p,q)$-system};\qquad \text{$Y$-system} \quad \to \quad \text{$(q,p)$-system}, \label{DefPQsystemLabeling}
\end{align}
especially in the cases of minimal string theory. 

\subsubsection{Equivalence with instantons for all order perturbation theory \label{SubsubsectionEquivalenceWithInstantonsForAllOrderPT}}

Further instanton corrections are also proposed in \cite{EynardMarino} for all order in string coupling $g$, which are called {\em non-perturbative partition function}. For each $A$-cycle of spectral curve, one can consider a cycle integral:%
\footnote{Here $A_a^{(1)} \equiv A_a$. }
\begin{align}
\frac{1}{2\pi i}\oint_{A_a^{(j)}} d \varphi^{(j)} \Bigl(\equiv \frac{1}{2\pi i}\oint_{A_a} d \varphi^{(1)}\Bigr)= \epsilon_a, \qquad \bigl(a=1,2,\cdots, \mathfrak g=\frac{(p-1)(q-1)}{2}\bigr), 
\end{align}
which are called {\em filling fractions}. In particular, the following combination with $g$ is integer (even after double scaling limit): 
\begin{align}
n_a \equiv \frac{\epsilon_a}{g} \in \mathbb Z_{\geq 0},\qquad \bigl(a=1,2,\cdots, \mathfrak g=\frac{(p-1)(q-1)}{2}\bigr), 
\end{align}
which expresses a filling number of matrix-model eigenvalues. Non-perturbative partition function is then defined by summing over all the possible filling number \cite{EynardMarino}: 
\begin{align}
\mathcal Z_{\rm NP}(g;t|\theta) = \sum_{\{n_a\}_{a=1}^{\mathfrak g}\in \mathbb Z_{\geq 0}^{\mathfrak g}} \theta_1^{n_1}\times \cdots\times \theta_{\mathfrak g}^{n_{\mathfrak g}} \, \mathcal Z_{\rm pert}\bigl(g;t\big|\bigl\{n_a\bigr\}_{a=1}^{\mathfrak g}\bigr), \label{MultiInstantonNPPT}
\end{align}
where
\begin{align}
\mathcal Z_{\rm pert}\bigl(g;t\big|\bigl\{n_a\bigr\}_{a=1}^{\mathfrak g}\bigr) &\equiv \exp\Bigl[{\mathcal F_{\rm pert}\bigl(g;t\big|\bigl\{n_a\bigr\}_{a=1}^{\mathfrak g}\bigr)}\Bigr],\nn\\
\mathcal F_{\rm pert}\bigl(g;t\big|\bigl\{n_a\bigr\}_{a=1}^{\mathfrak g}\bigr) &= \sum_{n=0}^\infty g^{2n-2}\, \mathcal F_{n}\bigl(t\big|\bigl\{n_a\bigr\}_{a=1}^{\mathfrak g}\bigr). 
\end{align}
In fact, this expression around $\mathcal F_{\rm pert} \bigl(g;t\bigr) \equiv \mathcal F_{\rm pert}\bigl(g;t\big|\bigl\{0\bigr\}_{a=1}^{\mathfrak g}\bigr)$ is equivalent to the trans-series expansion of free-energy: 
\begin{align}
\mathcal F(g;t) &= \ln \mathcal Z \asymeq \ln \mathcal Z_{\rm NP} = \underbrace{\mathcal F_{\rm pert}(g;t)}_{\text{perturbative}} + \underbrace{\sum_{a=1}^{\mathfrak g} \theta_a \times \exp\Bigl[\mathcal F_{\rm inst}^{(a)}(g;t) \Bigr]}_{\text{1-instanton}} + \underbrace{O(\theta^2)}_{\text{multi-instantons}}, \label{EqFreeLargeNexpansion} 
\end{align}
where the amplitudes for 1-instanton sector are given as 
\begin{align}
\mathcal F_{\rm inst}^{(a)}\bigl(t;g\big|\bigl\{n_b\bigr\}_{b=1}^{\mathfrak g}\bigr) &= \mathcal F_{\rm pert}\bigl(g;t\big|\bigl\{n_1,n_2,\cdots,n_a+1,\cdots,n_{\mathfrak g}\bigr\}\bigr)- \mathcal F_{\rm pert}\bigl(g;t\big|\bigl\{n_b\bigr\}_{b=1}^{\mathfrak g}\bigr)\nn\\
&= \Bigl[e^{\del_{n_a}} -1\Bigr]\mathcal F_{\rm pert}\bigl(g;t\big|\bigl\{n_b\bigr\}_{b=1}^{\mathfrak g}\bigr) = \sum_{m=1}^\infty \frac{1}{m!} \del_{n_a}^m\mathcal F_{\rm pert}\bigl(g;t\big|\bigl\{n_b\bigr\}_{b=1}^{\mathfrak g}\bigr) \nn\\
&= g^{-1}\del_{\epsilon_a} \mathcal F_{0}+ \frac{1}{2}\del_{\epsilon_a}^2 \mathcal F_{0} + g\bigl(\del_{\epsilon_a} \mathcal F_{1}+ \frac{1}{3!}\del_{\epsilon_a}^3 \mathcal F_{0}\bigr) + \cdots \nn\\
&\equiv g^{-1}\sum_{m=0}^\infty g^m \mathcal F_m^{(a)}\bigl(t;g\big|\bigl\{n_b\bigr\}_{b=1}^{\mathfrak g}\bigr)
\end{align}
That is, instanton amplitudes $\mathcal F_{\rm inst}^{(a)}$ are given by deformations (of filling numbers) of the spectral curve. It is also worth noting that $m$-point instanton amplitudes are given as 
\begin{align}
\frac{\partial^m\mathcal F_{\rm pert}\bigl(g;t\big|\bigl\{n_a\bigr\}_{a=1}^{\mathfrak g}\bigr)}{\del n_{a_1}\cdots \del n_{a_m}} &= \sum_{h=0}^\infty g^{2h-2+m} \frac{\partial^m\mathcal F_h\bigl(t\big|\bigl\{n_a\bigr\}_{a=1}^{\mathfrak g}\bigr)}{\del \epsilon_{a_1}\cdots \del \epsilon_{a_m}} \nn\\
&=\sum_{h=0}^\infty g^{2h-2+m} \oint_{B_{a_1}}d\zeta_1 \cdots \oint_{B_{a_m}}d\zeta_m W_m^{(h)}(\zeta_1,\cdots,\zeta_m). \label{InstantonByBcycleContour}
\end{align}
Therefore, all-order instanton amplitudes are again obtained from perturbative amplitudes, which are uniquely fixed by spectral curve. 

For a given reference background (given by a filling fraction $\bigl\{\epsilon_a^*\bigr\}_{a=1}^{\mathfrak g}$) which are away from the boundary of moduli space (given by $n_a =0 \,\,(a=1,2,\cdots)$), this summation can be expressed by the $\Theta$-function in $g\to 0$:%
\footnote{Note that $n_a^*\to \infty$ and $n_a-n_a^* \in \mathbb Z_{\geq -n_a^*}$ becomes $n_a-n_a^* \in \mathbb Z$. } 
\begin{align}
\mathcal Z_{\rm NP}(g;t|\theta) = \mathcal Z_{\rm pert}\bigl(g;t\big|\bigl\{n_a^*\bigr\}_{a=1}^{\mathfrak g}\bigr) \Theta_{0,\nu}(\hat u, \hat \tau),
\end{align}
where 
\begin{align}
\Theta_{\mu,\nu}(u,\tau) = \sum_{n \in \mathbb Z^{\mathfrak g}} e^{(n + \mu)u} e^{\pi i (n+\mu)\tau (n+\mu)} e^{2\pi i (n+n^*) \nu},
\end{align}
and
\begin{align}
 \theta = e^{2\pi i \nu}, \qquad 
 \hat u = g^{-1}\mathcal F_0'+ O(g^1),\qquad 
\hat \tau = \frac{1}{2\pi i} \mathcal F''_0 + O(g^2). 
\end{align}
Such a choice of background is then irrelevant since we already sum over all the configurations of the filling numbers. This is the statement of background independence in matrix models \cite{EynardMarino}. 

The powerful statement of non-perturbative partition function is that, if one knows spectral curve $F(x,y)=0$, then all the coefficients of perturbation theory (including all-order instanton amplitudes) are obtained from spectral curve. However, non-perturbative partition function also depends on the choice of $A$-cycles $\bigl\{A_a\bigr\}_{a=1}^{\mathfrak g}$ and the related D-instanton fugacities $\bigl\{\theta_a\bigr\}_{a=1}^{\mathfrak g}$, which are additional inputs and originally absent from the spectral curve itself. This is the problem of non-perturbative ambiguity discussed in Introduction. 

Since the perturbative free-energy is already shown to be $p-q$ dual (as $x-y$ symmetry), this fact also concludes that all-order instanton corrections are also $p-q$ dual. Then non-perturbative differences of $X$- and $Y$-systems are then encoded in D-instanton fugacity which are differently described in each system: 
\begin{align}
\text{$X$-system: }
\Bigl\{A_a; \theta_a\Bigr\} \qquad \leftrightarrow \qquad \text{$Y$-system: }\Bigl\{\widetilde A_a; \widetilde \theta_a\Bigr\}. \label{NPdualityDef1}
\end{align}
Therefore, check of non-perturbative duality is achieved if one understands {\em equivalence with D-instanton fugacity of both sides} by comparing the behavior of partition functions: 
\begin{align}
\text{$X$-system: } \, \mathcal Z_{\rm NP}(g;t|\theta) \qquad \leftrightarrow \qquad \text{$Y$-system: } \, \widetilde {\mathcal Z}_{\rm NP}(g;t|\widetilde \theta), \label{NPdualityDef2}
\end{align}
equivalently by comparing one-instanton sectors of free-energy (i.e.~Eq.~\eq{EqFreeLargeNexpansion}). 
This objective naturally leads to a study of isomonodromy description and its Stokes phenomena.

\subsubsection{Mean-field method and D-instanton fugacity \label{SubsubsectionMeanFieldApproximation}}

In matrix models, the information of D-instanton fugacity is encoded in the mean-field approximation of matrix models \cite{David,KazakovKostov} (See also \cite{BEH} and \cite{EynardMarino}). This is obtained by focusing on a pair of eigenvalue, $(x_N,y_N)\equiv (x,y)$, and by re-expressing the two-matrix integral as follows:
\begin{align}
\mathcal Z &= \int_{\mathcal C_X\times \mathcal C_Y} dX dY e^{-N\tr w(X,Y)} = \int_{\mathcal C_x^N \times \mathcal C_y^N} \prod_{i=1}^N \Bigl[d x_i  d y_i\, e^{-N w(x_i,y_i)}\Bigl] \, \prod_{i<j}(x_i-x_j)(y_i-y_j) \nn\\
&= \int_{\mathcal C_x \times \mathcal C_y} dx dy  e^{-Nw(x,y)} \vev{\det (x-X)\det (y-Y)}_{(N-1)\times (N-1)} \nn\\
&\simeq \int_{\mathcal C_x \times \mathcal C_y} dx dy \,e^{Nxy} \,\psi_{N-1}(x)\chi_{N-1}(y)\qquad \bigl(N\to \infty\bigr) \nn\\
&\equiv \int_{\mathcal C_x \times \mathcal C_y} dx dy \,e^{-Nw_{\rm eff}(x,y)}, \label{EqEffectiveMM1}
\end{align}
where the functions,  $\bigl\{\psi_n(x);\chi_n(y)\bigr\}_{n=0}^\infty$, are bi-orthogonal (monic) pseudo-polynomials of the two-matrix models: 
\begin{align}
&\psi_n(x) = \Bigl[x^n +\cdots \Bigr] e^{-NV_1(x)},\qquad \chi_n(y) =\Bigl[y^n +\cdots \Bigr] e^{-NV_2(y)}, \nn\\
& \qquad \text{with}\qquad \int_{\mathcal C_x \times \mathcal C_y} dx dy\, e^{Nxy}\, \psi_n(x) \chi_m(y) = h_n \delta_{n,m}, \label{EqOrthgonalPoly11}
\end{align}
where $w_{\rm eff}(x,y)$ is given as
\begin{align}
w_{\rm eff}(x,y) = \Phi_1(x) + \Phi_2(y) - xy,
\end{align}
with
\begin{align}
\Phi_1(x) = \lim_{N\to \infty}\frac{-1}{N}\ln \psi_{N-1}(x),\qquad \Phi_2(y) = \lim_{N\to \infty}\frac{-1}{N}\ln \chi_{N-1}(y).
\end{align}
Here the large $N$ limit is taken for fixed values of $x$ and $y$. 

Calculation of D-instanton fugacity from the mean-field integral is performed in \cite{HHIKKMT,SatoTsuchiya,IshibashiKurokiYamaguchi}, there are two procedures: 
\begin{itemize}
\item One first integrates $y$ (of Eq.~\eq{EqEffectiveMM1}) in order to obtain the effective potential of eigenvalue $x$: 
\begin{align}
\mathcal Z &\simeq  \int_{\mathcal C_x \times \mathcal C_y} dxdy\, e^{-Nw_{\rm eff}(x,y)}  \simeq   \int_{\mathcal C_x} dx \,e^{-NV_{\rm eff}(x)}.  \label{EqTwoEffectiveXY}
\end{align}
This is usually evaluated by solving the saddle point equation, 
\begin{align}
\del_y w_{\rm eff}(x,y)=\del_y \Phi_2(y) - x = 0 \quad \Leftrightarrow \quad \bigl\{y_j^*(x)\bigr\}_{j=1,2,\cdots}
\end{align}
and substitute solutions of the saddle point equation to the original integral. This gives the effective potential of $x$: 
\begin{align}
e^{-NV_{\rm eff}(x)} \simeq  \sum_{j=1,2,\cdots} c_j e^{-Nw_{\rm eff}(x,y_j^*(x))}, 
\end{align}
where $c_j$ is some multiplier of the saddles. That is the meaning of Eq.~\eq{EqTwoEffectiveXY}. 
\item Since this is the effective theory for a single eigenvalue, it contains one-instanton sector of Eq.~\eq{MultiInstantonNPPT}. This enables us to obtain D-instanton fugacity of Eq.~\eq{EqFreeLargeNexpansion} by evaluating saddle point approximation: 
\begin{align}
\mathcal Z  \quad \to \quad \mathcal Z_{\rm NP}(g|\theta)\simeq   \int_{\mathcal C_x} dx \,e^{-NV_{\rm eff}(x)}. \label{MeanFieldPathIntegral123}
\end{align}
\end{itemize}
In this analysis, it is relatively easier to study all the possible saddles in matrix models. However, as can be seen, say in Airy functions, {\em not all the saddles contribute to the integral.} Therefore, it is much more non-trivial to know which saddle does (or does not) contribute to the system. This issue on "relevant saddles" is quite difficult from this mean-field approach, especially in the cases of two-matrix models and also of the dual-side of one-matrix models.%
\footnote{In the cases of one-matrix models, i.e.~the potential of $V_2(y)$ is gaussian (say $V_2(y)=\frac{1}{2}y^2$)), such a problem is much easier to solve.} 

If one can resolve this issue on relevant saddles, one can obtain the D-instanton fugacity on the both sides of the duality, and one can make the following comparison: 
\begin{align}
\mathcal Z_{\rm NP}(g|\theta)\simeq   \int_{\mathcal C_x} dx \,e^{-NV_{\rm eff}(x)} \qquad \leftrightarrow \qquad \widetilde {\mathcal Z}_{\rm NP}(g|\widetilde \theta) \simeq   \int_{\mathcal C_y} dy \,e^{-N\widetilde V_{\rm eff}(y)}. 
\end{align}
In this paper, we study this issue from the approach of {\em Stokes phenomena of isomonodromy systems}. 
We also come back to the relation to the mean-field approach in Section \ref{SectionDiscussionEffectivePotentials}, as discussion.

\subsection{The Baker-Akhiezer systems \label{SubsubsectionBAfunctions}}

In order to describe the two-matrix models, it is also useful to consider the bi-orthogonal pseudo-polynomials of two-matrix models, Eq.~\eq{EqOrthgonalPoly11}, which reflect integrable structure of the matrix models. In particular, it can be shown that these polynomials satisfy the following recursion equations \cite{DouglasGeneralizedKdV}: 
\begin{align}
&
\left\{
\begin{array}{cl}
\ds
x \psi_n(x) = \hat A_n(Z)\, \psi_n(x), \qquad N^{-1} \frac{\del}{\del x} \psi_n(x) = \hat B_n(Z)\, \psi_n(x) &:\quad \text{$X$-system} \cr
\ds 
y \chi_n(y) = -\hat B_n(Z)^{\rm T}\, \chi_n(y), \qquad N^{-1} \frac{\del}{\del y} \chi_n(y) = -\hat A_n(Z)^{\rm T}\, \chi_n(y) &:\quad \text{$Y$-system}
\end{array}
\right., \nn\\
&\qquad \qquad\qquad \qquad\qquad\qquad\qquad \qquad\qquad \quad \text{with \quad $\bigl[\hat A_n(Z),\hat B_n(Z)\bigr] = N^{-1}I_\infty$. } 
\end{align}
Here $Z$ is an index-shift matrix, $Z=e^{-\del_n}$, i.e.~$Z \cdot f_n = f_{n-1} \cdot Z$, and $I_\infty$ is the semi-infinite identity matrix. Since these bi-orthogonal pseudo-polynomials are given by determinant operators of two-matrix models \cite{GrossMigdal2}: 
\begin{align}
\psi_n(x) = \vev{\det \bigl(x-X\bigr)}_{n\times n} e^{-NV_1(x)},\qquad \chi_n(y) = \vev{\det \bigl(y-Y\bigr)}_{n\times n} e^{-NV_2(y)},  \label{EquationGrossMigdalFormula}
\end{align}
these systems are identified with observables in $X$-system and $Y$-system, respectively. Therefore, the $p-q$ duality is about equivalence among the two equation systems of $\psi_n(x)$ and $\chi_n(y)$. 

In the double-scaling limit, these recursion equations become differential equation systems referred to as the Baker-Akhiezer systems for $(p,q)$ minimal strings \cite{DouglasGeneralizedKdV,TadaYamaguchiDouglas}. By taking into account a change of a normalization in $\zeta$ and $\eta$, and the scaling limit of $t$ and $\del_n$, the BA systems are generally given as 
\begin{align}
\left\{
\begin{array}{cl}
\ds 
\zeta \psi(t;\zeta) = \bP(t;\del)\psi(t;\zeta),\quad
g\frac{\del \psi(t;\zeta)}{\del \zeta} = \bQ(t;\del) \psi(t;\zeta)&:\quad\text{$X$-system} \cr
\ds \eta \chi (t;\eta) = \widetilde \bP(t;\del) \chi(t;\eta),\quad
g\frac{\del \chi(t;\eta)}{\del \eta} = \widetilde \bQ(t;\del) \chi(t;\eta) &:\quad\text{$Y$-system}
\end{array}
\right. \label{EqBAfunctionPQsystem} 
\end{align}
where 
\begin{align}
\bP(t;\del) = \Bigl[2^{p-1} \del^p + \sum_{n=2}^p u_n(t) \del^{p-n}\Bigr],&\quad 
\bQ(t;\del) =\beta_{p,q} \Bigl[ 2^{q-1} \del^q + \sum_{n=2}^q v_n(t) \del^{q-n}\Bigr] \label{EqLaxOpPQ1}\\
\widetilde \bP(t;\del) = \Bigl[2^{q-1} \del^q + \sum_{n=2}^q \widetilde u_n(t) \del^{q-n}\Bigr],&\quad 
\widetilde \bQ(t;\del) = \widetilde \beta_{q,p} \Bigl[ 2^{p-1} \del^p + \sum_{n=2}^p \widetilde v_n(t) \del^{p-n}\Bigr] \label{EqLaxOpPQ2}
\end{align}
satisfying the Douglas equation \cite{DouglasGeneralizedKdV}, $\bigl[\bP(t;\del),\bQ(t;\del)\bigr] = \bigl[\widetilde \bP(t;\del),\widetilde \bQ(t;\del)\bigr] = g$. Here $\del \equiv g\del_t \equiv g\del_{t_1}$.%
\footnote{Here, $t$ means the correction of KP flows, $t= \bigl\{ t_n \bigr\}_{n=1}^r$, but if we write $\del_t$ then it means the derivative by the most relevant operator, $\del_t \equiv \del_{t_1}$. We often use this sloppy notation in the following discussions. }
 Importantly, these operators are related by {\em transpose} ${(*)}^{\rm T}$ as
\begin{align}
\widetilde \bP (t;\del) = (-1)^q \beta^{-1}_{p,q}\bQ^{\rm T} (t;\del),\qquad 
\widetilde \bQ (t;\del) = (-1)^q \beta_{p,q}  \,\bP^{\rm T} (t;\del), \label{EqPQdualPQLaxOperators}
\end{align}
with $\bigl(f(t)\circ \del^n\bigr)^{\rm T} = (-\del)^n \circ f(t)$. As a result, the following relation follows: 
\begin{align}
\widetilde \beta_{q,p} = (-1)^r \beta_{p,q},\qquad r=p+q.  \label{EqBetaRelationPQdual1}
\end{align}
Later in Theorem \ref{ThmBetaValues}, we obtain the value (or sign) of $\beta_{p,q}$ which is consistent with multi-cut boundary condition and spectral $p-q$ duality. 

Note that the two systems related by $p-q$ duality satisfy the same string (Douglas) equation, 
\begin{align}
\bigl[\widetilde \bP(t;\del),\widetilde \bQ(t;\del)\bigr]  &=
\bigl[\bQ^{\rm T}(t;\del),\bP^{\rm T}(t;\del)\bigr] \nn\\
&=\Bigl(\bigl[\bP(t;\del),\bQ(t;\del)\bigr]\Bigr)^{\rm T}  = g1. 
\end{align}
This is the original meaning of duality introduced in \cite{fkn3}. Note that this fact does not guarantee the equivalence of the two systems at the level of D-instanton fugacity (given in Eq.~\eq{NPdualityDef1} and Eq.~\eq{NPdualityDef2}). This is because information of D-instanton fugacity is encoded in integration constants of the string equation, which are again separately described by $X$- and $Y$-systems.

In order to simplify the notation, we will drop ``$\,\,\widetilde{\,}\,\,$'' in the Lax operators. This can be justified in the following way: 
For a given $(p,q)$, there are two pairs of operators $\bigl(\widetilde \bP,\widetilde \bQ\bigr)$ and $\bigl(\bP,\bQ\bigr)$. But if one considers ``$(q,p)$ case'', it is consistent to make the following identification: 
\begin{align}
\text{$\bigl(\widetilde \bP,\widetilde \bQ\bigr)$ of the $(p,q)$ case} \quad =\quad  \text{$\bigl(\bP,\bQ\bigr)$ of the $(q,p)$ case},\label{EqIdentificationOfPQandQPoperatorsInDual}
\end{align}
with 
\begin{align}
\widetilde \beta_{q,p} \equiv \beta_{q,p},\qquad \text{i.e.} \quad \beta_{q,p}= (-1)^r \beta_{p,q} \label{EqBetaRelationPQdual}
\end{align}
Therefore, we only consider a pair of operators $\bigl(\bP,\bQ\bigr)$ and we understand spectral $p-q$ duality as an exchange of the indices $(p,q) \to (q,p)$ (Also note the relation of the coefficients $\beta_{p,q}$ of Eq.~\eq{EqBetaRelationPQdual}). 
This is then consistent with ``$(p,q)$-system and $(q,p)$-system'' introduced in Eq.~\eq{DefPQsystemLabeling}. Spectral $p-q$ duality among these Lax operators can be also checked at the level of dispersion-less limit, i.e.~by using the DKK prescription \cite{DKK}, which is shown in Appendix \ref{SectionDualityDKKprescription}.

\subsection{The worldsheet description: Liouville theory \label{SectionPQdualLiouvilleTheory}}

If one chooses the KP parameters (or isomonodromy parameters) $\bigl\{ t_n \bigr\}_{n=1}^{r}$ in Eq.~\eq{FIMformulationOfSP}, one obtains a spectral curve, 
\begin{align}
F(\zeta,Q)= \frac{\bigl(\mu^{\frac{q}{2p}} \beta_{p,q}\bigr)^p}{2^{p-1}} \Bigl[T_p(Q/\mu^{q/2p} \beta_{p,q})-T_q(\zeta/\sqrt{\mu})\Bigr]=0, \label{SpectralCurveFromLiouvilleTheory}
\end{align}
which is given by Chebyshev polynomials of the first kind, $T_n(\cos\tau)=\cos(n\tau)$. This solution of loop equation is first found in matrix models by \cite{Kostov1}, and the corresponding worldsheet formulation (of non-critical string theory) is given by Liouville theory and is calculable \cite{DKK,FZZT,SeSh}. This spectral curve is referred to as {\em Chebyshev solution}. This specific choice of $t=\bigl\{ t_n \bigr\}_{n=1}^{r}$ (parametrized by $\mu$) is referred to as {\em conformal background} \cite{MSS}.  
See also the end of Section \ref{SubsubsectionRoleOfSpectralCurvesAndLandscapes} for this terminology. 

The worldsheet action (i.e.~that of Liouville theory) is given as 
\begin{align}
S_{\rm WS} &= S_{\rm Liou.}[\phi_L]+S_{\rm FF}[X_F] + S_{\rm ghost}, \nn\\
S_{\rm Liou.}\bigl[\phi_L\bigr] &= \frac{1}{4\pi }\int d^2\sigma \sqrt{g} \Bigl[g^{ab}\del_a \phi_L \del_b \phi_L + Q_L \mathcal R \phi_L + 4\pi \mu e^{2b\phi_L}\Bigr]\qquad \Bigl( Q_L = b + \frac{1}{b}\Bigr), \nn\\
S_{\rm FF}\bigl[X_F\bigr] &= \frac{1}{4\pi } \int d^2 \sigma \sqrt{g}\Bigl[g^{ab}\del_a X_F \del_b X_F + i \widetilde Q_X \mathcal R X_F\Bigr] \qquad \Bigl(\widetilde Q_X = b - \frac{1}{b}\Bigr). 
\end{align}
The $p-q$ duality in the Liouville field theory is given by a flip of the coupling $b$ and redefinition of (worldsheet) cosmological constant $\mu$ (by its dual cosmological constant $\widetilde \mu$): 
\begin{align}
b\to b^{-1}\qquad \text{and}\qquad \mu\to \widetilde \mu \qquad \Bigl( \mu^b = \widetilde \mu^{1/b},\, b = \sqrt{\frac{p}{q}} \,\Bigr). 
\end{align}
With this change of parameters, amplitudes of Liouville theory are invariant \cite{DOZZ,Teschner}. In $(p,q)$ minimal conformal field theory (given by Feign-Fuchs action), this change can be absorbed by the sign of Feign-Fuchs field $X_F \to -X_F$, and therefore minimal CFT also allows this duality. Consequently, within the worldsheet description, the spectrum of physical operators is mapped to each other and perturbative amplitudes are also equivalent (See e.g.~\cite{SeSh}). 

The spectrum of D-branes (which are called FZZT-branes \cite{FZZT} and ZZ-branes \cite{ZZ}) are mapped to D-branes on the dual side. FZZT-branes and their dual FZZT-branes are related to the resolvent operators of $X$ and $Y$ matrices \cite{FZZT,SeSh}. A simplest relation is given by disk amplitudes of FZZT-branes. If one denote $D(\zeta)$ and $\widetilde D(\eta)$ as their disk amplitudes of FZZT-branes and of dual FZZT-branes, they satisfy the same algebraic equation \cite{SeSh}: 
\begin{align}
f(x,y) = T_p(y)-T_q(x)=0. \label{EqMinimalSpectralCurve1}
\end{align}
with 
\begin{align}
\left\{
\begin{array}{ll}
\ds 
x= \frac{\zeta}{\sqrt{\mu}},\qquad y = \frac{\beta^{-1}_{p,q}}{(\sqrt{\mu})^{\frac{q}{p}}} \frac{\del D(\zeta)}{\del \zeta}&\quad :\,\text{FZZT-brane} \cr
\ds 
x=\frac{\beta^{-1}_{q,p}}{(\sqrt{\widetilde \mu})^{\frac{p}{q}}} \frac{\del \widetilde D(\eta)}{\del \eta} ,\qquad y =\frac{\eta}{\sqrt{\widetilde \mu}} &\quad :\,\text{dual FZZT-brane}
\end{array}
\right., \label{EqMinimalSpectralCurve2}
\end{align}
where $\zeta$ is boundary cosmological constant of FZZT-branes; and $\eta$ is that of dual FZZT-brane. 
Note that $\beta_{p,q}$ and $\beta_{q,p}$ are normalization factors, which are usually free parameters in worldsheet formulation. 

There are $2 \times \frac{(p-1)(q-1)}{2}$ ZZ-branes on the both sides \cite{SeSh}, labeled by $(m,n)$, 
\begin{align}
1\leq m\leq p-1,\qquad 1\leq n\leq q-1, \qquad qm-pn>0.
\end{align}
In fact, their instanton actions are manifestly $p-q$ invariant form \cite{SeSh}: 
\begin{align}
\mathcal S_{\rm ZZ_\pm}^{(m,n)} = \pm \Bigl|\beta_{p,q} \frac{2 p q }{q^2-p^2}\mu^{\frac{p+q}{2p}} \sin \Bigl(\frac{p+q}{q}n\pi\Bigr)\sin \Bigl(\frac{p+q}{p}m\pi\Bigr)\Bigr|. \label{EqZZbraneAction}
\end{align}
The sign $\pm$ represents ZZ branes of positive action (normal type $(+)$) and of negative action (ghost type $(-)$). The boundary state of the ghost type is defined by that of the normal type by putting a minus in front of the state \cite{OkudaTakayanagi}:
\begin{align}
\ket{(m,n)}_{\rm ZZ_-} = (-1)\ket{(m,n)}_{\rm ZZ_+}. \label{GhostZZBoundaryStates}
\end{align} 
It is shown that all of these $2\times \frac{(p-1)(q-1)}{2}$ ZZ-branes appear as a pinched point of their spectral curve \cite{SeSh}. 
Therefore, $p-q$ duality in the perturbative string description perfectly works for all the observables in perturbative calculus. 

One should note that all of these $2\times \frac{(p-1)(q-1)}{2}$ ZZ-branes are shown to appear as the saddle points of matrix models \cite{KazakovKostov} and also as the saddle points of the free-fermion descriptions \cite{fy1,fy2,fy3}\cite{fis,fim}. Although the ZZ branes of ghost type were not seriously taken into account, its importance only gets noticed later \cite{MarinoLecture}. One may wonder whether these ZZ branes with a negative action are really identified with ghost ZZ-branes. In the free fermion analysis, one- and two-point functions of ghost ZZ branes are already calculated in \cite{fim} and the result is consistent with the relation of ghost ZZ-branes. It is because the boundary-state string field of (ghost) ZZ-branes in the free-fermion formulation satisfies Eq.~\eq{GhostZZBoundaryStates} by definition. Furthermore, by their construction, it is obvious that these relations hold for all-order multi-point functions. Therefore, ghost ZZ branes are realized by instantons in matrix models for all order perturbation theory.%
\footnote{One can also understand in the following way: ghost-type or normal-type are flipped by changing the direction of $B$-cycle in correlators in Eq.~\eq{InstantonByBcycleContour}. This allows us to check the relation of Eq.~\eq{GhostZZBoundaryStates} for all-order perturbation theory. }

\section{Spectral curves and isomonodromy systems \label{SectionIMSandSpectralCurve}}

\subsection{Notes on spectral curves \label{SubsectionFurtherCommentsOnSpectralCurves}}

\subsubsection{Three definitions of spectral curves \label{SubsubsectionThreeDefinitionsOfSpectralCurves}}

We first define a concept of closed loop (denoted by $\gamma$): 
\begin{Definition} [Closed loop] For a given $\zeta$, consider a map $\gamma(\zeta;t)\in \mathbb C$ which is continuous (and sufficiently differentiable) in $t\in [0,1]$. If it satisfies 
\begin{align}
\gamma(\zeta;0)=\gamma(\zeta;1)=\zeta, 
\end{align}
it is referred to as a closed loop $\gamma\, \bigl(\equiv \gamma(\zeta,[0,1]) \subset \mathbb C\bigr)$. In particular, we write $\gamma(\zeta;1)\equiv \gamma(\zeta)$. If $f(\zeta)$ is holomorphic along the loop $\gamma \subset \mathbb C$, this map $\gamma(\zeta)$ can be locally extended to an open neighborhood $U (\ni \zeta)$ and $\gamma^*f(\zeta)$ is also holomorphic at $\zeta$. $\quad \square$
\end{Definition}
We then introduce three definitions of spectral curves: 
\begin{Definition} [Spectral curves] \label{DefinitionOfSpectralCurve}
Spectral curves are characterized by the following three ways, which are equivalent to each other: 
\begin{itemize}
\item [1. ] An algebraic equation $F(\zeta,Q)=0$ for a pair of variables $(\zeta,Q)$. 
\item [2. ] Three objects, $\bigl(\mathcal S;\zeta,Q\bigr)$, where $\mathcal S$ is a non-singular Riemann surface, and $(\zeta,Q)$ is a pair of  meromorphic functions on the Riemann surface: $(\zeta,Q): \mathcal S \to \mathbb C^2$. 
\item [3. ] A set of $p$ algebraic differentials $d\varphi(\zeta)= \underset{1\leq j\leq p}{\diag}\bigl(d\varphi^{(j)}(\zeta)\bigr)$ which transforms as a permutation, 
\begin{align}
&\gamma^*d\varphi(\zeta)\equiv d\varphi(\gamma(\zeta))=  \underset{1\leq j\leq p}{\diag}\bigl(d\varphi^{(\sigma_\gamma(j))}(\zeta)\bigr),\qquad \text{with} \quad {}^\exists\sigma_\gamma \in \mathfrak S_p, 
\end{align}
for any analytic continuation of $\zeta$ by a closed loop $\gamma \subset \mathbb C$.  
\end{itemize}
$\square$
\end{Definition}
{\em Explanation of equivalence} $\quad$
Equivalence of the 1st and 2nd definition (``1 $\Leftrightarrow$ 2'') is often discussed in literature (See e.g.~\cite{EynardOrantin}). Here we discuss the relation with the 3rd definition. \\
\underline{\em Relation of ``3 $\Rightarrow$ 1''}\quad We assume that $\varphi(\zeta)$ is holomorphic at $\zeta \in \mathbb Z$. For any closed loop starting at $\zeta$, the monodromy is given by a permutation of eigenvalues, which is expressed as 
\begin{align}
\gamma^* d\varphi(\zeta)=  \underset{1\leq j\leq p}{\diag}\bigl(d\varphi^{(\sigma_\gamma(j))}(\zeta)\bigr) = M_{\sigma_\gamma}\, d\varphi(\zeta) \, M_{\sigma_\gamma}^{-1}\quad \text{with} \quad  M_{\sigma_\gamma}\in GL(p,\mathbb C),\quad {}^\exists\sigma_\gamma \in \mathfrak S_p. 
\end{align}
Therefore, $F(\zeta,Q)\equiv \det\bigl(Q\, I_p - \del_\zeta \varphi(\zeta)\bigr)$ is a rational function of $\zeta$. That is, $F(\zeta,Q)=0$ is an algebraic equation. \\
\underline{\em Relation of ``2 $\Rightarrow$ 3''}\quad For any $\zeta \in \mathbb C$, there exist $p$ points $\{z^{(j)}\}_{j=1}^p \subset \mathcal S$ satisfying $\zeta(z_j)=\zeta$. Then we define 
\begin{align}
d\varphi(\zeta)\equiv \underset{1\leq j\leq p}{\diag}\bigl(d\varphi^{(j)}(\zeta)\bigr),\qquad 
d\varphi^{(j)}(\zeta) \equiv d\zeta\, Q^{(j)}(\zeta),\qquad \bigl(Q^{(j)}(\zeta) \equiv Q(z^{(j)}(\zeta))\bigr). \label{DifferentialInTermsOfPQ}
\end{align}
If we assume that $\zeta_*$ is a branch point of $\{z^{(j)}(\zeta)\}_{j=1}^p \subset \mathcal S$, then the monodromy around the branch point is given by a permutation of the roots $\{z^{(j)}(\zeta)\}_{j=1}^p \in \mathcal S$: 
\begin{align}
&z^{(j)}(\zeta_* + e^{2\pi i} (\zeta-\zeta_*))=  z^{(\sigma(j))}(\zeta),\qquad \text{with} \qquad  {}^\exists\sigma \in \mathfrak S_p. 
\end{align}
Therefore, since $Q(z)$ is a meromorphic function on $\mathcal S$, the differential $d\varphi(\zeta)$ satisfies the condition of the 3rd definition. $\quad \blacksquare$

In particular, the 3rd definition is natural and useful in isomonodromy systems because the corresponding integral $\varphi(\zeta)$, 
\begin{align}
\varphi(\zeta) \equiv \int^{\zeta}_o d\varphi = \int^{\zeta}_o d\zeta' \, Q^{(j)}(\zeta') = \int_{z^{(j)}(o)}^{z^{(j)}(\zeta)} d\zeta(z)\, Q(z), 
\end{align}
appears as the leading exponents of BA functions of Eq.~\eq{EqBAfunctionPQsystem}, and plays a central role in Riemann-Hilbert approach (See e.g.~\cite{ItsBook}). Therefore, in the following, spectral curves is also identified by $\varphi(\zeta)$. The function $\varphi(\zeta)$ itself is called {\em $\varphi$-function}%
\footnote{It is often called {\em $g$-function} and denoted by $g(\zeta)$ in literature (e.g.~\cite{ItsBook}). Since $g$ represents string coupling in this paper, it is now referred to as $\varphi$-function. } 
in this paper. 

\subsubsection{Spectral curve for Chebyshev solutions \label{SubsubsectionSpectralCurveForChebyshevSolutions}}

Here we review and summarize basics of spectral curve of Chebyshev solutions, Eq.~\eq{SpectralCurveFromLiouvilleTheory}, which is used throughout this paper. 

\paragraph{Three coordinates of spectral curves}
The following changes of coordinate in the spectral curves are also used in this paper: 
\begin{align}
1)& 
\quad \zeta = \sqrt{\mu}\cosh(p\tau) &
2)&\quad\zeta = 2^{p-1}\lambda^p  &
3)&\quad z = \mu^{\frac{1}{2p}} \cosh(\tau). 
\label{Eqtaucoordinate}
\end{align}
The coordinate $\tau$ is introduced in \cite{Kostov1,FZZT,KazakovKostov}; the coordinate $\lambda$ is introduced in \cite{fy1,CIY4}; and the coordinate $z$ is introduced in \cite{DKK,SeSh}. 
If one uses the coordinate $\tau$ and $z$, then a solution of the algebraic equation, $F(\zeta,Q)=0$, (i.e.~Eq.~\eq{SpectralCurveFromLiouvilleTheory}) is solved as 
\begin{align}
\zeta = \sqrt{\mu}\cosh(p\tau),\qquad 
Q = \beta_{p,q}\,\mu^{\frac{q}{2p}} \cosh(q\tau), 
\end{align}
and further by using $z$, they are expressed as 
\begin{align}
\zeta = \sqrt{\mu}\, T_p\bigl(z/\mu^{1/2p}\bigr),\qquad Q = \beta_{p,q}\,\mu^{\frac{q}{2p}} T_q\bigl(z/\mu^{1/2p}\bigr). 
\end{align}
Here $T_n(x)=T_n(\cos\theta)=\cos(n\theta)$ is Chebyshev polynomials of the first kind. 
Therefore, this shows that the spectral curve, $F(\zeta,Q)=0$, of Eq.~\eq{SpectralCurveFromLiouvilleTheory} is given by the two polynomial functions $(\zeta(z),Q(z))$ on $\mathcal S=\mathbb CP^1$. 

\begin{figure}[htbp]
\begin{center}
\includegraphics[scale=0.9]{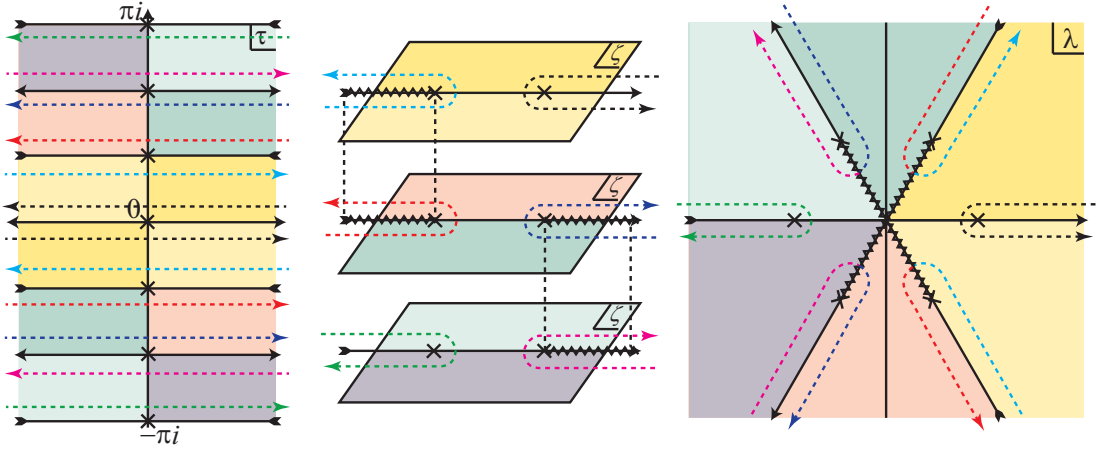}
\end{center}
\caption{\footnotesize The relation among $\tau$, $\zeta$ and $\lambda$, $\zeta  = 2^{p-1}\lambda^p =\sqrt{\mu}\cosh p\tau$. This also reflects the cut structure of $\varphi^{(1)}(\lambda)$. Here is an example of $p=3$ but structure is mostly the same for the general $p$ cases. }
\label{TauLambda}
\end{figure}

The geometrical relation among the three coordinates $\zeta$, $\lambda$ and $\tau$ are shown in Fig.~\ref{TauLambda}. 
The coordinate $\lambda \in \mathbb C$ is covered once by $\tau$ with the following fundamental region: 
\begin{align}
 -\pi \leq {\rm Im}\,\tau  \leq \pi,\qquad {\rm Re}\,\tau \geq 0.  \label{FundamentalDomainOfTau}
\end{align}
Similarly, it can cover the whole plane of the coordinate $z\in \mathbb C$ also only once.

\paragraph{$\varphi$-functions}
Here we consider $\varphi$-function of Chebyshev solutions, $\varphi(\mu;\zeta)$: 
\begin{align}
\varphi(\mu;\zeta) = \diag_{1\leq j\leq p} \Bigl( \varphi^{(j)}(\mu;\zeta)\Bigr),\qquad \varphi^{(j)}(\mu;\zeta)=\varphi^{(1)}(\mu;e^{-2\pi i (j-1)}\zeta). 
\label{PhiFunctionConformalBackground}
\end{align}
The explicit form is then given as 
\begin{align}
\varphi^{(j)}(\mu;\zeta) & = \beta_{p,q}\, \mu^{\frac{q+p}{2p}} \int^{\frac{e^{-2\pi i (j-1)}\zeta}{\sqrt{\mu}}} dx\, T_{\frac{q}{p}}(x) \qquad \bigl(1\leq j\leq p\bigr)\nn\\
&= \frac{p\beta_{p,q}\, \mu^{\frac{q+p}{2p}}}{2}\Bigl[\frac{T_{\frac{q+p}{p}}(\frac{e^{-2\pi i (j-1)}\zeta}{\sqrt{\mu}})}{q+p} - \frac{T_{\frac{q-p}{p}}(\frac{e^{-2\pi i (j-1)}\zeta}{\sqrt{\mu}})}{q-p} \Bigr] 
\label{EqSpectralCurvePhiZetaSec2} \\
&= \frac{p\beta_{p,q} \,\mu^{\frac{q+p}{2p}}}{2} 
\Bigl[
\frac{\cosh\bigl((p+q)(\tau -2\pi i \frac{(j-1)}{p})\bigr)}{p+q}-
\frac{\cosh \bigl((q-p)(\tau -2\pi i \frac{(j-1)}{p}) \bigr)}{q-p}
\Bigr]. \label{EqSpectralCurvePhiInTauSpace}
\end{align}
Note that, since $\varphi(\zeta)$ appears as the leading exponents of BA functions, the normalization factors $\beta_{p,q}$ and $\beta_{q,p}$ are fixed by the definition of BA function, Eq.~\eq{EqBAfunctionPQsystem}. Here $\mu$ (which is a KP flow parameter or an isomonodromic deformation parameter) is explicitly shown as $\varphi(\mu;\zeta)$, but is also often omitted as $\varphi(\zeta)$ in the following. We often use the following sloppy notations: 
\begin{align}
\varphi(\zeta)=\varphi(\zeta(\tau)))= \varphi(\tau) = \varphi(\zeta(\lambda))=\varphi(\lambda). 
\end{align}
In addition, if we express $\varphi(t;\zeta)$ then it means that the KP parameters $t =\bigl\{t_n\bigr\}_{n=1}^{p+q}$ are general; if we express $\varphi(\mu;\zeta)$ then it means that we choose the conformal background parametrized by a single variable $\mu$. The explicit expression of $t=\bigl\{ t_n \bigr\}_{n=1}^{p+q}$ with the variable $\mu$ can be found in \cite{fim}. 

Here we also note on the starting point of the integral: 
\begin{align}
\varphi^{(j)}(\zeta) = \int^\zeta_o d\zeta' \,Q^{(j)}(\zeta'),\qquad F(\zeta,Q)= \prod_{j=1}^p \bigl(Q-Q^{(j)}(\zeta)\bigr). 
\end{align}
Since the index $j$ is related as $\varphi^{(j)}(\zeta) = \varphi^{(1)}(e^{-2\pi i (j-1)}\zeta)$, the following combination does not depend on the starting point $o$: 
\begin{align}
\varphi^{(j,l)} (\zeta) = \varphi^{(j)}(\zeta) - \varphi^{(l)}(\zeta)  &= \int^{\zeta}_{o} d\zeta' \, \Bigl(Q^{(j)}(\zeta')-Q^{(l)}(\zeta') \Bigr)\\
&=  \int^{e^{-2\pi i (j-1)} \zeta}_{e^{-2\pi i (l-1)} \zeta} d\zeta' \, Q^{(1)}(\zeta'). 
\end{align}
In fact, various calculation is performed with this combination $\varphi^{(j,l)}(\lambda)$ and physical results do not depend on the starting point $o$. 

\subsubsection{The space of spectral curves and the landscapes \label{SubsubsectionRoleOfSpectralCurvesAndLandscapes}}

Here we consider ``string theory landscape'' in matrix models. In literature, the landscape is introduced as a collection of all the meta-stable (perturbative) vacua of string theory, and is used to discuss statistical and anthropic aspects of a huge number of string vacua \cite{CiteOfStringLandscape}. As we see in various quantum systems, however, not all the meta-stable vacua (or saddles) contribute to the path integrals. Such a set of {\em relevant meta-stable vacua} is generally smaller than the set of all the possible saddle points in the path-integral. We emphasize this fact, because it is not globally acknowledged in the community.

If one sees the non-perturbative partition function, Eq.~\eq{MultiInstantonNPPT}, of matrix models, the set of relevant saddles is determined by a choice of $A$-cycles and their non-trivial (i.e.~non-vanishing) D-instanton fugacity: 
\begin{align}
\Bigl\{A_a; \theta_a\Bigr\}_{a=1}^{\mathfrak g}\qquad 
\rightarrow \qquad \mathcal Z(t;g)&\asymeq \mathcal Z_{\rm NP}(t;g|\theta)  \nn\\
&\,\,\,\simeq  \sum_{\{n_a\}_{a=1}^{\mathfrak g}} \theta_1^{n_1} \times \cdots \times \theta_{\mathfrak g}^{n_{\mathfrak g}}e^{\mathcal F_{\rm pert}(t;g|\{ n_a\}_{a=1}^{\mathfrak g})}. 
\end{align}
Such a set of relevant saddles is non-trivial but has not been studied in much detail. Therefore, it should deserve to study in future investigation. In this paper, it is discussed later in Section \ref{SubsubsectionTransseriesMetastablevacua} but it is still convenient to define a set of all the possible vacua (or saddles), which is referred to as {\em universal landscape}. Any landscapes are subsets of this universal landscape. 

According to topological recursion \cite{EynardOrantin}, each vacuum can be identified by a spectral curve, which can be characterized by a $\varphi$-function. In terms of $\varphi$-function, solutions of loop equation (equivalently of $W_{1+\infty}$-constraints) are given by Eq.~\eq{FinMSTofFIM} \cite{fim}. That is,%
\footnote{Since there is an ambiguity by a integration constant in $\varphi(t;\zeta)$, we choose it to fix the constant part of the expansion. }
\begin{align}
&\mathcal F_{\rm pert}(t;g|\{ n_a\}_{a=1}^{\mathfrak g}) \qquad \Leftrightarrow \qquad d \varphi\bigl(t;\zeta|\{ n_a\}_{a=1}^{\mathfrak g}\bigr) \nn\\
&\qquad \Leftrightarrow \qquad \varphi\bigl(t;\zeta\bigr) \simeq \beta_{p,q}\Bigl[\sum_{n=1}^{p+q} t_n\Omega^{-n}\, \zeta^{\frac{n}{p}} \Bigr]+ O(\zeta^{-\frac{1}{p}}) \qquad (\zeta \to \infty\bigr) \nn\\
&\qquad \qquad \qquad \text{  s.t.}\quad  F(\zeta,Q)= \det\bigl(Q \, I_p - \del_\zeta \varphi(\zeta)\bigr) \quad \text{a polynomial in $\zeta$}
\label{ConditionForUnivLandscape}
\end{align}
Therefore, we define the following set of differential: 
\begin{Definition} [Universal landscape] \label{DefinitionUnivsersalLandscapes}
For $(p,q)$ minimal string theory with given KP parameters $t \equiv \bigl\{ t_n \bigr\}_{n=1}^{p+q}$, 
the following set of differentials $d\varphi(\zeta)$,   
\begin{align}
\mathcal L_{\rm str}^{(\rm univ.)}(t) = \bigl\{ \varphi(t;\zeta) \big| \text{Eq.~\eq{ConditionForUnivLandscape}}\bigr\}, 
\end{align}
is referred to as universal landscape of $(p,q)$ minimal string theory. 
$\quad \square$
\end{Definition}
Note that, in this definition, there is no quantization condition for the filling numbers, i.e.~$\bigl\{n_a\bigr\}_{a=1}^{\mathfrak g} \in \mathbb C^{\mathfrak g}$, and also there is no particular choice of $A$-cycles. It is because such a condition should not be imposed by hand (especially in two-matrix models). It should be a derived concept from the first principle, say by Riemann-Hilbert analysis. 

Here are two notes: 
\begin{itemize}
\item [1. ] The condition of Eq.~\eq{ConditionForUnivLandscape} appears around poles of the $\varphi$-function, which place at $\zeta \to \infty$ in $(p,q)$ minimal string theory. Therefore, the definition of universal landscapes can be straightforwardly extended to more general systems {\em with multiple singularities}. These systems also appear in gauge theory (or in CFT) \cite{AGT}, and extension of the formalism (developed in non-critical strings/matrix models) can also be applied to these systems, straightforwardly. 
\item [2. ] In order to have consistent terminology, we distinguish the words, {\em background} and {\em vacuum}. The specific choice of KP parameters, $t = \bigl\{ t_n \bigr\}_{n=1}^{p+q}$, is now called {\em background}; and an element $\varphi(t;\zeta) \in \mathcal L_{\rm str}^{(\rm univ.)}(t)$ is called {\em vacuum}. Note that the explicit $t$-dependence of $\varphi(t;\zeta)$ is obtained as a solution of loop equation. Therefore, $\varphi(t;\zeta)$ is also called {\em solution} (to the loop equation). See also the beginning of Section \ref{SectionPQdualLiouvilleTheory}. 
\end{itemize}

\subsection{Embedding into the isomonodromy systems \label{SectionIsomonodromyEmbedding}}

In order to capture ``the solution space'' of the BA systems (Eq.~\eq{EqBAfunctionPQsystem} and Eq.~\eq{EqLaxOpPQ1}), isomonodromy systems are known to be an efficient way to describe it. It maps the solution space to an algebraic variety of their Stokes matrices. This is a key idea of {\em inverse monodromy problem} (See e.g.~\cite{ItsBook}). Isomonodromy description is first considered in matrix models in \cite{Moore,FIK}. Roughly speaking, an isomonodromy system is an non-perturbative extension of spectral curve to integrable differential equation systems with rational-function coefficients \cite{JimboMiwaUeno}: 
\begin{align}
g\frac{\del \Psi(t;\zeta)}{\del \zeta} = \mathcal Q(t;\zeta) \Psi(t;\zeta),\qquad g\frac{\del \Psi(t;\zeta)}{\del t_n} = \mathcal B_n(t;\zeta) \Psi(t;\zeta), 
\quad \bigl(n=1,2,\cdots,r\bigr), 
\label{EqIsomonodromySystemGeneral}
\end{align}
such that the associated matrix-valued coefficient $\mathcal Q(t;\zeta)$ in weak coupling limit $g\to 0$ is given by a spectral curve of the corresponding universal landscape: 
\begin{align}
\det\bigl(Q \, I_p-\mathcal Q(t;\zeta)\bigr) \underset{g\to 0 }{\propto} \det \bigl(Q\, I_p - \del_\zeta \varphi(t; \zeta)\bigr)\qquad \bigl({}^\exists \varphi(t;\zeta) \in \mathcal L_{\rm str}^{\rm (univ.)}(t)\bigr). \label{IsomonodromyAndUnivLandscapes}
\end{align} 
Here, $t\equiv \bigl\{ t_n \bigr\}_{n=1}^r$ are isomonodromic deformation parameters and also are KP flows in $(p,q)$ minimal string theory. Due to integrability, the condition, Eq.~\eq{IsomonodromyAndUnivLandscapes}, should be satisfied for ${}^\forall t \in \mathbb C^{r}$. 
From the BA systems, there are several ways to obtain an isomonodromy system which describes $(p,q)$ minimal string theory. For example, in the Painlev\'e I case (i.e.~$(2,3)$ system), there are known two kinds of isomonodromy systems \cite{JimboMiwaUeno}\cite{Moore,KapaevPI} and often used in literature. 
For the general $(p,q)$ cases, especially in investigating spectral $p-q$ duality, it turns out to be useful to consider the $\mathbb Z_p$-symmetric $p\times p$ isomonodromy system with an essential singularity of Poincar\'e index $r=p+q$ \cite{CIY4}, which is a cousin of the $p$-cut two-matrix models \cite{fi1,irie2,CISY1,CIY1}. 

The corresponding isomonodromy system of $(p,q)$ minimal strings can be constructed from the BA systems (Eq.~\eq{EqBAfunctionPQsystem} and Eq.~\eq{EqLaxOpPQ1}). The detail construction is shown explicitly in Appendix \ref{SectionIMSLaxBQ} and here we only show the form which is necessary in the main text of this paper:%
\footnote{Sometimes, it is important to consider both {\em matrix-model basis} and {\em diagonal basis} \cite{CIY2,CIY3}, but we practically use the diagonal basis in this paper. Therefore, we here skip to introduce the matrix-model basis. }
\begin{align}
g \frac{\del \Psi(t;\lambda)}{\del t} = \mathcal B(t;\lambda) \Psi(t;\lambda),\qquad 
g \frac{\del \Psi(t;\lambda)}{\del \lambda} = \mathcal Q(t;\lambda) \Psi(t;\lambda),
\label{EqIsomonodromyMinimalStrings1}
\end{align}
where the isomonodromy Lax operators $(\mathcal B, \mathcal Q)$ are given as 
\begin{align}
\left\{
\begin{array}{l}
\ds 
\mathcal B (t;\lambda) = \Omega^{-1} \lambda - \frac{1}{2^{p-1}}\sum_{n=1}^p \frac{u_n(t) }{\lambda^{n-1}} \widetilde E_{p, p-n+1}, \cr
\ds
\mathcal Q(t;\lambda) = \frac{p\beta_{p,q}}{2}\biggl[\Omega^{-r} (2\lambda)^{r-1} + \sum_{n=1}^{r-1} 2^n\,\widetilde {\mathcal Q}_n(t)\, (2\lambda)^{r-n-1}\biggr] - \frac{\widetilde {\rm P}_1}{\lambda} 
\end{array}
\right.. \label{EqIsomonodromyMinimalStrings2}
\end{align}
Here, the matrices $\Omega$ and $U$ are defined as follows ($\omega=e^{\frac{2\pi i}{p}}$): 
\begin{align}
\Omega = 
\begin{pmatrix}
1 \cr
 & \omega \cr
 & & \omega^2 \cr
 & & & \ddots \cr
 & & & & \omega^{p-1}
\end{pmatrix},\qquad 
U=
\begin{pmatrix}
\ds \frac{1}{\sqrt{p}}\omega^{(i-1)(j-1)}
\end{pmatrix}_{1\leq i,j\leq p},  \label{DefinitionOfOmegaAndUmatrices}
\end{align}
and the matrices $\widetilde E_{j,l}$ and $\widetilde {\rm P}_1$ are defined as 
\begin{align}
\widetilde E_{j,l} \equiv U E_{j,l} U^\dagger,\qquad \widetilde {\rm P}_1 \equiv U {\rm P}_1 U^\dagger,
\end{align}
where ${\rm P}_1 = \diag \bigl(0,1,2,\cdots,p-1\bigr)$ and $E_{j,l}$ is the matrix unit, $(E_{j,l})_{ab} = \delta_{j,a} \delta_{l,b}$. 
The new spectral parameter $\lambda$ is related to the original parameter $\zeta$ as Eq.~\eq{Eqtaucoordinate}: 
\begin{align}
\zeta =  \,2^{p-1}\lambda^p. 
\end{align}
Since the Douglas equation is an integrability condition of two variables $\zeta$ and $t$, it is equivalently expressed as the isomonodromy Douglas equation: 
\begin{align}
\bigl[\bP(t;\del),\bQ(t;\del)\bigr] = g 1\qquad \Leftrightarrow \qquad 
\bigl[g \del_t - \mathcal B(t;\lambda), g \del_\lambda - \mathcal Q(t;\lambda)\bigr] = 0.  \label{EqIMSDouglasEq}
\end{align}
As related to the introduction of the new spectral parameter, $\lambda \, \bigl(=\dfrac{(2\zeta)^{1/p}}{2}\bigr)$,  
the isomonodromy systems possess the $\mathbb Z_p$-symmetry \cite{CIY2} represented as%
\footnote{$\mathbb Z_p$-symmetry in the matrix-model basis is expressed as 
\begin{align}
\mathcal B_{\rm matrix}(t;\omega^{-1}\lambda) = \Omega \,\mathcal B_{\rm matrix}(t;\lambda)\,\Omega^{-1},\qquad \omega^{-1} \mathcal Q_{\rm matrix}(t;\omega^{-1} \lambda) = \Omega \,\mathcal Q_{\rm matrix}(t;\lambda)\,\Omega^{-1}.  
\end{align}
}
\begin{align}
\mathcal B(t;\omega^{-1}\lambda) = \Gamma \,\mathcal B(t;\lambda)\,\Gamma^{-1},\qquad \omega^{-1} \mathcal Q(t;\omega^{-1} \lambda) = \Gamma \,\mathcal Q(t;\lambda)\,\Gamma^{-1},  \label{EqZpSymmetryInIsomonodromySystems}
\end{align}
where the matrix $\Gamma$ is defined as 
\begin{align}
\Gamma = \begin{pmatrix}
0& 1\cr
0& &1 \cr
\vdots & & & \ddots \cr
0 & & & &1 \cr
1 &0&0 & \cdots & 0
\end{pmatrix}. 
\end{align}
Note that the matrices $\Gamma$, $\Omega$ and $U$ are related as 
\begin{align}
\Gamma U = U \Omega,\qquad \Gamma U^\dagger = U^\dagger \Omega^{-1}. 
\end{align}

It is important to note that, because of the form of the operator $\mathcal B(t;\lambda)$, 
the BA system is just a part of more general $p\times p$ isomonodromy systems: 
\begin{align}
\mathcal B_{\rm general}(t;\lambda) = \Omega^{-1} \lambda - \frac{1}{2^{p-1}} \sum_{n=2}^p \frac{\widetilde B_n(t)}{\lambda^{n-1}}. \label{EqGeneralBoperator}
\end{align}
Therefore, there is a proper reduction of isomonodromy systems to the KP systems. This reduction is referred to as {\em KP reduction}. Because of this fact, if one constructs the most general Stokes matrices generated by the leading behavior of Eq.~\eq{EqIsomonodromyMinimalStrings2} (which is given in Section \ref{SectionStokesMatrixIsomonodromySystems}), most of such Stokes phenomena are not related to $(p,q)$ minimal string theory.  We come back to this issues in Section \ref{SectionDZmethodAndRHP}. 

\subsection{Classical Baker-Akhiezer function \label{SubsectionClassicalBAFunction}}

In this subsection, we evaluate weak coupling expansion $g\to 0$ of the isomonodromy systems (Eq.~\eq{EqIsomonodromyMinimalStrings1} and Eq.~\eq{EqIsomonodromyMinimalStrings2}), namely ``the one-loop correction to the spectral curve:''
\begin{align}
\Psi(t;\lambda) \underset{g\to 0}{\simeq} \Bigl[I_p + O(g)\Bigr]\Psi_{\rm cl}(t;g;\lambda),\qquad \Psi_{\rm cl}(t;g;\lambda) \equiv  Z_{\rm cl}(t;\lambda) e^{\frac{1}{g}\varphi(t;\lambda)} \lambda^{-\frac{p-1}{2}} , 
\end{align}
which is referred to as {\em classical BA function} of $(p,q)$ minimal-string theory.  The function $Z_{\rm cl}(t;\lambda)$ is referred to as classical $Z$ function. 
Importance of this function is understood as Theorem \ref{TheoremCutJumpCancellation1}. Here we obtain its (classical) monodromy matrices associated to branch cuts of the spectral curve. 

\subsubsection{Isomonodromy systems and solutions of the loop equation}

First of all, we discuss how the isomonodromy system admits the spectral curve of Chebyshev solutions (Eqs.~\eq{PhiFunctionConformalBackground}, \eq{EqSpectralCurvePhiZetaSec2} and \eq{EqSpectralCurvePhiInTauSpace}). This fact is essentially followed by the dispersion-less analysis of Douglas equation (discussed by Daul-Kazakov-Kostov \cite{DKK}) but it is also instructive to see how they are incorporated in the isomonodromy system. 
By expanding $\mathcal Q(t;\lambda)$ and $\mathcal B(t;\lambda)$, 
\begin{align}
\mathcal Q(t;\lambda)&= \mathcal Q^{(0)}(t;\lambda) + g \mathcal Q^{(1)}(t;\lambda) + \cdots \equiv \sum_{n=0}^\infty g^n \mathcal Q^{(n)}(t;\lambda), \nn\\
\mathcal B(t;\lambda)&= \mathcal B^{(0)}(t;\lambda) + g \mathcal B^{(1)}(t;\lambda) + \cdots \equiv \sum_{n=0}^\infty g^n \mathcal B^{(n)}(t;\lambda), \label{QBexpansionInGClassicalBAFunction}
\end{align}
one obtains the following equations: 
\begin{align}
\mathcal Q^{(0)}(t;\lambda) Z_{\rm cl}(t;\lambda) = Z_{\rm cl}(t;\lambda) \del_\lambda \varphi(t;\lambda), \qquad  \mathcal B^{(0)}(t;\lambda) Z_{\rm cl}(t;\lambda) = Z_{\rm cl}(t;\lambda) \del_t \varphi(t;\lambda) \label{FirstOrderForClassicalBAFunctions}
\end{align}
The first equation guarantees the condition of Eq.~\eq{IsomonodromyAndUnivLandscapes}. In particular, if one uses the coordinate $\zeta$, one obtains (see also e.g.~Lemma \ref{QoperatorLemmaAppendix}), 
\begin{align}
\det\bigl(Q\,I_p-\del_\zeta \varphi(t;\zeta)\bigr) = &\det\bigl(Q\,I_p-\frac{ \mathcal Q^{(0)}(t;\zeta)}{p 2^{p-1}\lambda^{p-1}}\bigr):\quad\text{a polynomial in $(\zeta,Q)$}\nn\\
&\qquad\qquad \qquad \qquad \qquad  \text{i.e.} \quad \varphi(t;\zeta) \in \mathcal L_{\rm str}^{\rm (univ.)}(t), 
\end{align}
for each value of KP parameter $t$. In this sense, for a given spectral curve $\varphi(t;\zeta) \in \mathcal L_{\rm str}^{\rm (univ.)}(t)$, there is a consistent weak-coupling expansion (Eq.~\eq{QBexpansionInGClassicalBAFunction}) with the (essentially) unique classical BA function $\Psi_{\rm cl}(t;\zeta)$. The second equation of Eq.~\eq{FirstOrderForClassicalBAFunctions} is then the equation to determine the $t$-dependence of $\varphi(t;\zeta)$ in order to satisfy loop equation of the system. 

This means that most of spectral curves $\varphi(t;\zeta) \in \mathcal L_{\rm str}^{\rm (univ.)}(t)$ are allowed to be a consistent perturbative (meta-stable) saddle of string theory. Among them, there is a special class called {\em a dispersion-less solution} where {\em the $t$-dependence of $\varphi(t;\zeta)$ is given by a power-function of $\bigl\{ t_n\bigr\}_{n=1}^{p+q}$}. This is the solution which are studied by dispersion-less analysis of Douglas equation \cite{DKK}. By construction, it is also characterized by a vanishing $A$-cycle condition \cite{DKK,fim}: 
\begin{align}
\oint_{A_a} d \varphi^{(j)}(\zeta) = 0 \qquad \bigl(a=1,2,\cdots, \mathfrak g\bigr), 
\end{align}
for some choice of $A$-cycles, $\bigl\{A_a\bigr\}_{a=1}^{\mathfrak g}$; or it is characterized by the three objects, $(\mathcal S; \zeta,Q)$ with $\mathcal S=\mathbb CP^1$. 
On the other hand, in the cases of non-dispersion-less solutions, the $t$-dependence of $\varphi(t;\zeta)$ is given by a non-trivial function, like elliptic functions (See e.g.~\cite{CIY4}). Therefore, such a solution is also called {\em non-perturbative vacua}. 

\subsubsection{Classical BA function of Chebyshev solutions}

In the cases of the Chebyshev solutions (Eqs.~\eq{PhiFunctionConformalBackground}, \eq{EqSpectralCurvePhiZetaSec2} and \eq{EqSpectralCurvePhiInTauSpace}), the second equation of Eq.~\eq{FirstOrderForClassicalBAFunctions} is expressed as 
\begin{align}
\mathcal B^{(0)}(\mu;\lambda)\, Z_{\rm cl}(\mu;\lambda) = Z_{\rm cl}(\mu;\lambda)
\begin{pmatrix}
z^{(1)}(\mu;\lambda) \cr
& z^{(2)}(\mu;\lambda) \cr
 & & \ddots \cr
 & & & z^{(p)}(\mu;\lambda)
\end{pmatrix}, 
\end{align}
with 
\begin{align}
& z^{(j)}(\mu;\lambda) \equiv \mu^{\frac{1}{2p}}\,T_{1/p}^{(-\frac{j-1}{p})}(\zeta/\sqrt{\mu})\qquad (j=1,2,\cdots,p),  \\
& \mathcal B^{(0)}(\mu;\lambda) =  \Omega^{-1} \lambda - \frac{1}{2^{p-1}}\sum_{n=1}^p \frac{T_{p,n} \,\mu^{\frac{n}{2p}}}{\lambda^{n-1}} \widetilde E_{p, p-n+1},\qquad T_p(z) = \sum_{n=0}^p T_{p,n} z^{p-n}.
\end{align}
Here we define deformed Chebyshev function $T_n^{(\nu)}(z)$ as $T_n^{(\nu)}(\cosh\theta)=\cosh(n\theta+2\pi i \nu)$. 
This relation follows since we have
\begin{align}
\det \bigl(z\, I_p - \mathcal B^{(0)}(\mu;\lambda) \bigr) = \frac{\sqrt{\mu}}{2^{p-1}} \Bigl( T_p(z/\mu^{1/2p}) - \zeta/\sqrt{\mu} \Bigr) = \prod_{j=1}^p \bigl(z- z^{(j)}(\mu;\lambda)\bigr),
\end{align}
by replacing the derivative $\del$ by a number $z$.%
\footnote{See also e.g.~\cite{DKK,fim}. }
By integrability, the classical $Z$ function also diagonalizes the $\mathcal Q$-operator: 
\begin{align}
\mathcal Q^{(0)}(\mu;\lambda) \, Z_{\rm cl}(\mu;\lambda) = Z_{\rm cl}(\mu;\lambda)
\begin{pmatrix}
\del_\lambda \varphi^{(1)}(\mu;\lambda) \cr
& \del_\lambda \varphi^{(2)}(\mu;\lambda) \cr
 & & \ddots \cr
 & & & \del_\lambda \varphi^{(p)}(\mu;\lambda)
\end{pmatrix},
\end{align}
where $\bigl\{\varphi^{(j)}(\mu;\lambda)\bigr\}_{j=1}^p$ is the spectral curve of minimal strings. We then solve these equations as follows: 
\begin{Theorem} [Classical BA functions] \label{TheoremClassicalBAFunctions}
Consider $(p,q)$-system and its Chebyshev solution $\varphi(\mu,\zeta)$ (Eqs.~\eq{PhiFunctionConformalBackground}, \eq{EqSpectralCurvePhiZetaSec2} and \eq{EqSpectralCurvePhiInTauSpace}), then the corresponding classical $Z$ function, $Z_{\rm cl}(\mu;\lambda)$, is solved as%
\footnote{One may wonder how is an ambiguity in the expression: $Z_{\rm cl}(\zeta) \to Z_{\rm cl}'(\zeta) = Z_{\rm cl}(\zeta) f(\zeta)$, where $f(\zeta) \to I_p$ ($\zeta\to \infty$). This can be killed if one requires that monodromy matrices (of Theorem \ref{TheoremClassicalMonodromyMatrices}) are independent from $\zeta$, which is also called isomonodromy property of classical BA function. }
\begin{align}
Z_{\rm cl}(\mu;\lambda) &=  \frac{1}{p} 
\begin{pmatrix}
\ds \frac{1 }{\ds \sqrt{\Delta^{(j)}(\lambda)}}
\sum_{n=1}^p \Bigl(\omega^{i-1}\gamma^{(j)}(\lambda)\Bigr)^{n-1}
\end{pmatrix}_{1\leq i,j\leq p} \nn\\
&=  U \Biggl[
\begin{pmatrix}
\ds \frac{\ds  \Bigl(\gamma^{(j)}(\lambda)\Bigr)^{i-1} }{\ds \sqrt{p \Delta^{(j)}(\lambda)}}
\end{pmatrix}_{1\leq i,j\leq p}\Biggr], \qquad 
U=
\begin{pmatrix}
\ds \frac{1}{\sqrt{p}}\omega^{(i-1)(j-1)}
\end{pmatrix}_{1\leq i,j\leq p}. 
\end{align}
with the following normalization conditions: 
\begin{align}
 \det Z_{\rm cl}(\mu;\lambda) = 1,\qquad 
Z_{\rm cl}(\mu;\lambda) = I_p + O(\lambda^{-1})\qquad\bigl(\lambda\to \infty\bigr). 
\end{align}
Here the functions $\gamma^{(j)}(\lambda)$ and $\Delta^{(j)}(\lambda)$ are defined as 
\begin{align}
\gamma^{(j)}(\lambda) \equiv \frac{z^{(j)}(\lambda)}{\lambda} \equiv \frac{\mu^{\frac{1}{2p}}\,T_{1/p}^{(-\frac{j-1}{p})}(\zeta/\sqrt{\mu})}{\lambda},\qquad \zeta = 2^{p-1}\lambda^p, \label{DefOfsmallGamma}
\end{align}
and 
\begin{align}
\sqrt{\Delta^{(j)}(\lambda) } \equiv \prod_{1\leq a \, (\neq j)\leq p} h^{(j|a)}(\lambda)\equiv \prod_{1\leq a\,(\neq j)\leq p} \sqrt{
\frac{\ds \gamma^{(j)}(\lambda)-\gamma^{(a)}(\lambda)}{\ds \omega^{-(j-1)}-\omega^{-(a-1)}}}. 
\end{align}
The branches of these functions are fixed as follows: 
\begin{align}
\gamma^{(j)}(\lambda) = \omega^{-(j-1)} + O(\lambda^{-1}),
\qquad h^{(j|a)}(\lambda)  = 1 + O(\lambda^{-1}),\qquad \lambda \to \infty. \label{EqInPropositionOfClassicalBAfunctionGammaAsympt}
\end{align}
$\square$
\end{Theorem}
Here we define deformed Chebyshev functions, $T_n^{(\nu)}(\cosh\theta)=\cosh(n\theta+2\pi i \nu)$, and therefore 
$z^{(j)}(\lambda) = \omega^{-(j-1)}\lambda +O(\lambda^{-1})$ which means Eq.\eq{EqInPropositionOfClassicalBAfunctionGammaAsympt}. 

This expression can be easily generalized to any dispersion-less solutions obtained in the KP systems (Eqs.~\eq{EqBAfunctionPQsystem} and \eq{EqLaxOpPQ1}). For example, consider the 3rd definition of a spectral curve, $(\mathcal S;\zeta,Q)$ (in Definition \ref{DefinitionOfSpectralCurve}) with $\mathcal S=\mathbb CP^1$. The coordinate of $\mathcal S = \mathbb CP^1$ is given by the dispersion-less limit of $\del$, (i.e.~$\ds z \equiv \lim_{g\to 0} \del$). Then $z^{(j)}(\lambda)$ in Eq.~\eq{DefOfsmallGamma} is just $z^{(j)}(\zeta)$ which is the inverse map of $\zeta: \mathcal S \to \mathbb C$. 

\subsubsection{Classical monodromy matrices of Chebyshev solutions \label{SubsubsectionClassicalMonodromyMatrices}}

Next we calculate the monodromy of the classical BA function. 
The spectral curves, $\varphi(\mu;\lambda)$, and the classical $Z$ function, $Z_{\rm cl}(\mu,\lambda)$, have branch points in the expression. As one can see, there are $2p$ branch points in the $\lambda$ space $(\lambda \in \mathbb C)$ and are given as 
\begin{align}
2 \lambda_{n} \equiv (2\sqrt \mu)^{1/p} \omega^{\frac{n}{2}},\qquad \tau_n \equiv \frac{n\pi }{p} i\qquad \bigl(n=0,1,\cdots,2p-1\bigr),
\end{align}
where $2 \zeta = (2\lambda)^p = 2 \sqrt{\mu} \cosh(p\tau)$. We evaluate the characteristic monodromy around a branch point $\lambda_n$. The corresponding closed loop $\gamma_n$ is defined as follows: 
\begin{align}
\gamma_n (\lambda;t) \equiv \lambda_n + e^{2\pi i t}( \lambda- \lambda_n),\qquad \bigl(n=0,1,\cdots,2p-1; \, t \in [0,1]\bigr). \label{DefClosedLoopAroundBranchPointsLambdaN}
\end{align}
The explicit evaluation is shown in Appendix \ref{AppendixMonodromyCalculusOfCBAF}, and the result is following:
\begin{Theorem} [Classical monodromy matrices] \label{TheoremClassicalMonodromyMatrices}
The monodromy of the classical BA function is given as 
\begin{align}
\gamma_n^* \, \varphi(\lambda) = C_n^{-1}\, \varphi(\lambda)\, C_n,\qquad \gamma_n^* \, Z_{\rm cl}(\lambda) = Z_{\rm cl}(\lambda) C_n, \label{GeneralDefMonodromyCutJump}
\end{align}
where the monodromy matrices $\bigl\{ C_n\bigr\}_{n=0}^{2p-1}$ are expressed as 
\begin{align}
C_n = \prod_{l>j,j+l\equiv n+2 \,\, ({\rm mod} \,p)} {\Cut}_{l,j} \bigl(\omega^{-\frac{l-j}{2}} \bigr), \label{EqCutJumpMinimalStrings}
\end{align}
where $\Cut_{j,l}(\alpha)$ is a matrix of Cut-type (it will appear again in Eq.~\eq{EqDefinitionStokesCutMatrices}), 
\begin{align}
\Cut_{l,j}(\alpha) = 
\left(
\begin{array}{c|c|c|c|c}
I_{j-1} &  &   &  & \cr\hline
 & 0 & & -\frac{1}{\alpha} &  \cr\hline
 & & I_{l-j-1} & & \cr\hline
 &\alpha  & & 0 & \cr \hline
 & & & & I_{p-l}
\end{array}
\right) \qquad \bigl(l>j\bigr). 
\end{align}
The matrices $\bigl\{ C_n\bigr\}_{n=0}^{2p-1}$ are also called cut-jump matrices of the spectral curve.  $\quad \square$
\end{Theorem}
Note that the matrix product in Eq.~\eq{EqCutJumpMinimalStrings} does not depend on the ordering of multiplication. 
Also note that the monodromy relation of the spectral curve $\varphi(t;\lambda)$ (given in Eq.~\eq{GeneralDefMonodromyCutJump}) is also written as
\begin{align}
\gamma_n^*
\varphi^{(j)}(\lambda ) = \varphi^{(n+2-j)}(\lambda),\qquad (j=1,2,\cdots,p), 
\end{align}
and the monodromy relation is summed up to the relation of classical BA function: 
\begin{align}
\gamma_n^*\Psi_{\rm cl}(g;\lambda) = \Psi_{\rm cl}(g;\lambda) \,C_n. 
\end{align}
In addition, the following remark is important in the following discussions: 
\begin{Remark} [Classical cyclic equation] \label{RemarkClassicalCyclicEquation} 
Multiplication of all the monodromy matrices $\{C_n\}_{n=0}^{2p-1}$ is given as
\begin{align}
C_0C_1\cdots C_{2p-1} = (-1)^{p-1}I_p. \label{EqClassicalCyclicEquation}
\end{align}
This is the classical counterpart of the monodromy equation of Proposition \ref{PropositionMonodromyEquationStokesMatrices}. 
$\quad \square$
\end{Remark}

\section{Stokes phenomena and Stokes geometry \label{SectionStokesGeneral}}

In this section, we discuss Stokes phenomena and the corresponding Stokes geometry in $(p,q)$ minimal string theory. 
In order to make this paper self-contained, some basic facts on Stokes matrices are reviewed and summarized in Section \ref{SectionStokesMatrixIsomonodromySystems}. Geometry of anti-Stokes lines are then studied for the cases of Chebyshev solutions (i.e.~$\varphi(\mu,\zeta)$ is given by Eqs.~\eq{PhiFunctionConformalBackground}, \eq{EqSpectralCurvePhiZetaSec2} and \eq{EqSpectralCurvePhiInTauSpace}) in Section \ref{SubsubsectionGeometryOfAntiStokesLines}. In this paper, Stokes geometry \cite{ExactWKB} itself is discussed by using {\em spectral networks}. Definition and basic facts on Deift-Zhou's spectral networks are then reviewed in Section \ref{RoleOfSpectralNetworks}. Finally, in section \ref{SubsubsectionStokesGeometryProfileInstantons}, we study {\em profile of instantons on spectral networks}. Note that the materials discussed/reviewed in this section are just frameworks to describe Stokes phenomena. The concrete solutions are obtained in Section \ref{SectionDZmethodAndRHP} with utilizing the framework discussed in this section. 

It is important to note that our convention of terminology, {\em (anti-)Stokes sectors/lines}, follows the book on isomonodromy theory \cite{ItsBook}. The convention of terminology is {\em completely opposite} to the convention which is used in context of exact WKB and resurgent analysis in the following way: 
\begin{align}
\begin{array}{ccc}
\text{Isomonodromy $\cdot$ This Paper} & & \text{Exact WKB $\cdot$ Resurgent} \cr \hline
\text{Stokes line/sector} & \leftrightarrow & \text{anti-Stokes line/sector} \cr
\text{anti-Stokes line/sector} & \leftrightarrow & \text{Stokes line/sector}
\end{array} \label{EqConventionAntiStokes}
\end{align}
Other terminology about Stokes matrices/multipliers (coefficients or constants, depending on literature) is common with other literature.

\subsection{Stokes matrices around an essential singularity \label{SectionStokesMatrixIsomonodromySystems}}
First we review and summarize some basic facts on {\em Stokes matrices around an essential singularity} in the isomonodromy systems. For more detail, see \cite{ItsBook} and \cite{CIY2}. 
All the structures are determined by the leading behavior of spectral curve $\varphi(\lambda)$ (given by Eq.~\eq{EqSpectralCurvePhiZetaSec2} with Eq.~\eq{Eqtaucoordinate}) around the singularity, which is now only at $\lambda \to \infty$: 
\begin{align}
\varphi(t;\lambda) \simeq \frac{p\beta_{p,q} }{4r} \Omega^{-r} (2\lambda)^r+ \cdots\qquad \bigl(\lambda\to \infty; \,r=p+q\bigr). \label{EqSec3First}
\end{align}
Here the power of $\lambda$ (i.e.~$r$) is called Poincar\'e index of this singularity. Therefore, it is relevant to all the spectral curve inside the universal landscape, $\varphi(t;\zeta)\in\mathcal L_{\rm str}^{\rm (univ.)}(t)$. 
As is discussed in \cite{CIY2}, for $p\times p$ isomonodromy systems with Poincar\'e index $r$, there appear $2rp$ (fine) Stokes matrices denoted by $\bigl\{S_n\bigr\}_{n=0}^{2rp-1}$, and the $\mathbb Z_p$-symmetry is also imposed (because of Eq.~\eq{EqZpSymmetryInIsomonodromySystems}), which reduces the number of independent Stokes matrices to $2r$, (i.e.~$\bigl\{S_n\bigr\}_{n=0}^{2r-1}$). The coefficient $\beta_{p,q}$ is considered to be a real number, but its sign ($\sgn(\beta)= \pm1$) is relevant to the following discussion.

\subsubsection{Stokes matrices}

There are known several ways to define Stokes matrices. Because of its technical simplicity, we define the Stokes matrices by using anti-Stokes sectors $\bigl\{\widetilde D_n\bigr\}_{n=0}^{2rp-1}$. These concepts can be defined at each pole of the spectral curve $\varphi(\lambda)$, which now places only at $\lambda\to \infty$. 
\begin{Definition} [Stokes matrices] 
\label{StokesDef}
At the essential singularity $\lambda\to \infty$ (i.e.~a pole of the $\varphi$-function $\varphi(\lambda)$), the Stokes matrices $\{S_n\}_{n=0}^{2rp-1}$ are defined as $p\times p$ matrices which satisfy 
\begin{align}
e^{\varphi(\lambda)} S_n e^{-\varphi(\lambda)} = I_p + O(\frac{1}{\lambda^{\infty}}),\qquad \lambda \to \infty \in \widetilde D_n\qquad \bigl(n=0,1,\cdots,2rp-1\bigr), 
\label{definitionStokesMatrices}
\end{align}
where $\{\widetilde D_n \}_{n=0}^{2rp-1}$ are open angular domains%
\footnote{We here use the notation, $D(a,b)=\bigl\{\lambda\in \mathbb C \,\big|\, a<\arg(\lambda)<b\bigr\}$. }  (around the essential singularity) called anti-Stokes sectors. $\quad\square$
\end{Definition}

\begin{Proposition} [Anti-Stokes sectors]
For the spectral curve of Eq.~\eq{EqSec3First}, the anti-Stokes sectors are given as 
\begin{align}
\widetilde D_n \equiv D(\chi_n, \chi_n+\frac{\pi}{r}),  \qquad \bigl(n=0,1,\cdots,2rp-1;\, r = p+q\bigr),
\end{align}
where $r$ is the Poincar\'e index of the spectral curve, $\varphi(\lambda)$, at $\lambda\to \infty$. The angles $\{\chi_n\}_{n=0}^{2rp-1}$ are those of Stokes lines around the essential singularity satisfying
\begin{align}
{\rm Re} \bigl[ \varphi^{(j)}(\lambda) - \varphi^{(l)}(\lambda)\bigr] = 0,\qquad \lambda \to \infty \times e^{i\chi_n} \qquad \bigl(1\leq {}^\exists j, {}^\exists l\leq p\bigr), 
\end{align}
and given as $\ds \chi_n = \frac{n \pi}{pr}  \,\, \bigl(n=0,1,\cdots\bigr)$. 
$\quad \square$
\end{Proposition}
Stokes matrices (defined in this way) describe Stokes phenomena of the isomonodromy systems (See \cite{ItsBook}): 
\begin{Proposition} \label{PropositionStokesMatricesCanonicalSolutions}
For a $p\times p$ isomonodromy system associated with the leading behavior of Eq.~\eq{EqSec3First} (i.e.~Poincar\'e index $r$, which includes the cases of Eq.~\eq{EqIsomonodromyMinimalStrings1} and Eq.~\eq{EqIsomonodromyMinimalStrings2}), the canonical solutions $\bigl\{\Psi_n(t;\lambda)\bigr\}_{n=0}^{2rp-1}$ and their corresponding Stokes sectors $\bigl\{D_n\bigr\}_{n=0}^{2rp-1}$ are defined as 
\begin{align}
\Psi_n(t;\lambda)\asymeq \Psi_{\rm asym}(t;\lambda)\qquad &\lambda \to \infty \in 
D_n = D(\chi_{n-1},\chi_n+ \frac{\pi}{r})\nn\\
&\qquad \qquad \quad \bigl(n=0,1,\cdots,2rp-1\bigr), \label{EqStokesSector}
\end{align}
where the asymptotic solution $\Psi_{\rm asym}(t;\lambda)$ around $\lambda\to \infty$ is given as 
\begin{align}
\Psi_{\rm asym}(t;\lambda) = \Bigl[ I_p + \frac{Y_1(t)}{\lambda} + O(\lambda^{-2})\Bigr] \exp\Bigl[\frac{p\beta_{p,q} }{4rg} \Omega^{-r} (2\lambda)^r+ O(\lambda^{r-1}) \Bigr] \lambda^\nu. \label{CanonicalAsymExpansionFormulaInPropositionStokesCanonicalSolutions}
\end{align}
Stokes phenomena of the isomonodromy systems are then generally given by Stokes matrices $\bigl\{S_n\bigr\}_{n=0}^{2rp-1}$ of Definition \ref{StokesDef} as
\begin{align}
\Psi_{n+1}(t;\lambda) = \Psi_n(t;\lambda) S_n\qquad \bigl(n=0,1,2,\cdots,2rp-1 \bigr). 
\end{align}
$\square$
\end{Proposition}
The formal monodromy $\nu$ of the isomonodromy systems in $(p,q)$ minimal string theory is discussed in \cite{CIY4} and is given as $\ds \nu= - \frac{p-1}{2}$. This results in the following (See also \cite{ItsBook}): 
\begin{Proposition} [Monodromy cyclic equation]
\label{PropositionMonodromyEquationStokesMatrices}
For the isomonodromy system of $(p,q)$ minimal strings (i.e.~Eq.~\eq{EqIsomonodromyMinimalStrings1} and Eq.~\eq{EqIsomonodromyMinimalStrings2}), 
the formal monodromy around the essential singularity ($\lambda\to \infty$) is given as%
\footnote{More generally, the monodromy equation is given as 
\begin{align}
S_0 S_1 \cdots S_{2rp-1} = e^{-2\pi i \nu} I_p,\qquad \nu = \underset{1\leq i\leq p}{\diag}\bigl(\nu^{(i)}\bigr). 
\end{align}
The diagonal matrix $\nu$ is called formal monodromy, which are related to flux in the background \cite{SeSh2}.}
\begin{align}
S_0 S_1 \cdots S_{2rp-1} = (-1)^{p-1} I_p,  \label{MonodromyEquation}
\end{align}
which is referred to as monodromy (or cyclic) equation.$\quad \square$
\end{Proposition}

\subsubsection{Profiles of dominant exponents}

{\em The profile of dominant exponents} are also important in various discussions of Stokes phenomena \cite{CIY2}: 
\begin{Definition} [The profile of dominant exponents] \label{DefProfile}
The profile of dominant exponents, $\mathcal J^{(p,r)}$, is associated with each essential singularity of Poincar\'e index $r$ in $p\times p$ isomonodromy systems. A profile is defined as a table consisting of $p$ integers, 
\begin{align}
\bigl\{j_{n,1},j_{n,2},\cdots,j_{n,p}\bigr\}=\bigl\{1,2,\cdots,p\bigr\}, 
\end{align}
for each row $(n=0,1,2,\cdots,2rp-1)$: 
\begin{align}
\mathcal J^{(p,r)} = 
\begin{tabular}{|c|c|c|c|c|c|}
\hline
$j_{2rp-1,1}$ &  $j_{2rp-1,2}$ &$\cdots$ & $j_{2rp-1,i}$ & $\cdots$ & $j_{2rp-1,p}$\cr
\hline
$\vdots$ & $\vdots$ & &  $\vdots$ & & $\vdots$ \cr
\hline
$j_{1,1}$ & $j_{1,2}$ & $\cdots$ & $j_{1,i}$ & $\cdots$ & $j_{1,p}$\cr
\hline
$j_{0,1}$  & $j_{0,2}$ & $\cdots$ & $j_{0,i}$ & $\cdots$ & $j_{0,p}$\cr
\hline
\end{tabular} 
\,\,. 
\end{align}
The $n$-th row of the profile is denoted as
\begin{align}
\mathcal J_n^{(p,r)} = 
\begin{tabular}{|c|c|c|c|}
\hline
$j_{n,1}$  & $j_{n,2}$  & $\cdots$ & $j_{n,p}$\cr
\hline
\end{tabular}\,\,, \qquad \mathcal J_{n+2rp}^{(p,r)} = \mathcal J_n^{(p,r)} \qquad \bigl(n\in\mathbb Z\bigr), 
\end{align}
and a subtable of the profile is represented as their ``union'': 
\begin{align}
\mathcal J^{(p,r)} \supset \bigcup_{n=a}^b \mathcal J_n^{(p,r)} = 
\begin{tabular}{|c|c|c|c|}
\hline
$j_{b,1}$ &  $j_{b,2}$ &$\cdots$ & $j_{b,p}$\cr
\hline
$\vdots$ & $\vdots$ &  & $\vdots$ \cr
\hline
$j_{a+1,1}$ & $j_{a+1,2}$ &  $\cdots$ & $j_{a+1,p}$\cr
\hline
$j_{a,1}$  & $j_{a,2}$ &  $\cdots$ & $j_{a,p}$\cr
\hline
\end{tabular} 
\,\, \qquad \bigl(b>a\bigr). 
\end{align}
The components of the profile, $\bigl\{j_{n,i}\bigr\}_{1\leq i\leq p}^{0\leq n\leq 2p-1}$, then represent the relative magnitudes of the exponents $\bigl\{ \varphi^{(j)}(\lambda)\bigr\}_{j=1}^p$ around the singularity (now it is only at $\lambda\to\infty$): 
\begin{itemize}
\item [1. ] A sub-profile, $\mathcal J_n^{(p,r)}$, represents the relative magnitudes in the segment, $\delta D_n$, which is defined by 
\begin{align}
\delta D_n \equiv D(\chi_n-\frac{\pi}{rp},\chi_n),\qquad \chi_n = \frac{n \pi}{pr}  \qquad \bigl(n=0,1,\cdots,2rp-1\bigr). 
\end{align}
The angles, $\bigl\{\chi_n\bigr\}_{n=0}^{2rp-1}$, are those of Stokes lines (given in Definition \ref{StokesDef}). 
\item [2. ] The integers in the profile, $\mathcal J_n^{(p,r)} = 
\begin{tabular}{|c|c|c|c|}
\hline
$j_{n,1}$  & $j_{n,2}$  & $\cdots$ & $j_{n,p}$\cr
\hline
\end{tabular}\,\,$, are then related as 
\begin{align}
{\rm Re}[\varphi^{(j_{n,1})}(\lambda)] < {\rm Re}[\varphi^{(j_{n,2})}(\lambda)] < \cdots < {\rm Re}[\varphi^{(j_{n,p})}(\lambda)] \qquad 
\bigl(\lambda \to \infty \in \delta D_n \bigr). 
\end{align}
\end{itemize}
$\square$
\end{Definition}

\begin{Definition}[Parenthesis] \label{DefParenthesis}
One can add a parenthesis structure to the profile in the following way: One writes a parenthesis for a neighboring pair $(j_{n,i},j_{n,i+1})$ in the profile $\mathcal J_n^{(p,r)}$ as 
\begin{align}
&\mathcal J_n^{(p,r)} = 
\begin{tabular}{|c|c|c|c|}
\hline
$\cdots$  & $(j_{n,i}$  & $j_{n,i+1})$ & $\cdots$\cr
\hline
\end{tabular}\,\, \quad \leftrightarrow \quad 
\delta D_n \equiv D(\chi_n-\frac{\pi}{rp},\chi_n)
\end{align}
if the pair satisfies 
\begin{align}
 {\rm Re}\bigl[ \varphi^{(j_{n,i})}(\lambda)-\varphi^{(j_{n,i+1})}(\lambda)\bigr] = 0,\qquad \lambda \to \infty \times e^{i\chi_n}. 
\end{align}
That is, the upper boundary (i.e.~given by the angle of $\chi_n$) of the corresponding segment $\delta D_n$ is the Stokes lines of $(j_{n,i},j_{n,i+1})$-kind. In particular, existence of a parenthesis is denoted by 
\begin{align}
{}^\exists (j|i) \in \mathcal J_n^{(p,r)} \quad \bigl( \text{or  ${}^\exists (j|i)_n \in \mathcal J_n^{(p,r)}$}\bigr)\quad \Leftrightarrow \quad \mathcal J_n^{(p,r)} = 
\begin{tabular}{|c|c|c|c|}
\hline
$\cdots$  & $(j$  & $i)$ & $\cdots$\cr
\hline
\end{tabular}\,\, . 
\end{align}
$\square$
\end{Definition}
The following proposition is a basic application of the profile \cite{CIY2}: 
\begin{Proposition} [Stokes matrices by a profile] \label{PropositionStokesProfileFormula}
The Stokes matrices $\{S_n\}_{n=0}^{2rp-1}$ of an essential singularity characterized by the profile are generally expressed as 
\begin{align}
S_n = I_p + \sum_{(j|i) \in \mathcal J_n^{(p,r)}} s_{n,i,j} E_{i,j}. \label{StokesProfileFormula}
\end{align}
where $n=0,1,2,\cdots, 2rp-1$. $\quad \square$
\end{Proposition}
{\em Proof}\quad By Definition \ref{StokesDef}, the associated anti-Stokes sector of the Stokes matrix $S_n$ is $\widetilde D_n = D(\chi_n, \chi_n+\frac{\pi}{r})$. By Definition \ref{DefParenthesis}, if ${}^\exists (j|i) \in \mathcal J_n^{(p,r)}$, then $\chi_n$ is the angle of Stokes line of $(j,i)$-kind. This results in 
\begin{align}
{\rm Re}[\varphi^{(i)}(\lambda)-\varphi^{(j)}(\lambda)] <0\qquad \text{i.e.} \qquad \Bigl|e^{\varphi^{(i,j)}(\lambda)}\Bigr| = O(\frac{1}{\lambda^{\infty}})\qquad \bigl(\lambda\to \infty \in \widetilde D_n\bigr), 
\end{align}
which means that the matrix $S_n$ of Eq.~\eq{StokesProfileFormula} satisfies Eq.~\eq{definitionStokesMatrices}. This proves the proposition. $\quad \blacksquare$

Here we discuss relevance of the sign of $\beta_{p,q}$ in the profile. In order to represent $\sgn(\beta_{p,q})$ dependence of the profile, we introduce the following formal notation: 
\begin{align}
\mathcal J^{(p,r)} =
\left\{
\begin{array}{rl}
|\mathcal J^{(p,r)} |& \quad \bigl(\sgn (\beta_{p,q}) =+1\bigr) \cr
-|{\mathcal J}^{(p,r)} |& \quad \bigl( \sgn(\beta_{p,q}) = -1\bigr)
\end{array}
\right.
\end{align}
The first one is called {\em positive profile} and the second one is {\em negative profile}. 
Since the spectral curve of $\sgn(\beta_{p,q})=-1$ (temporary denoted by $\bigl\{\varphi^{(j)}_{\beta_{p,q}<0}(t;\lambda) \bigr\}_{j=1}^p$) and the spectral curve of $\sgn(\beta_{p,q})=+1$ (temporary denoted by $\bigl\{\varphi^{(j)}_{|\beta_{p,q}|}(t;\lambda) \bigr\}_{j=1}^p$) are related by flipping the relative magnitudes: 
\begin{align}
\varphi^{(j)}_{\beta_{p,q}<0}(t;\lambda)  = - \varphi^{(j)}_{|\beta_{p,q}|}(t;\lambda), 
\end{align}
the components of the profiles $\pm \bigl|{\mathcal J}^{(p,r)}\bigr|$ are related by a reflection of indices in the horizontal direction: 
\begin{align}
{\mathcal J}_n^{(p,r)} = 
\begin{tabular}{|c|c|c|c|}
\hline
$j_{n,1}$  & $j_{n,2}$  & $\cdots$ & $j_{n,p}$\cr
\hline
\end{tabular}\,\,\quad \Leftrightarrow \quad 
-{\mathcal J}_n^{(p,r)} = 
\begin{tabular}{|c|c|c|c|}
\hline
$j_{n,p}$  & $j_{n,p-1}$  & $\cdots$ & $j_{n,1}$\cr
\hline
\end{tabular}\,\,.
\end{align}
This gives the following proposition: 
\begin{Proposition} [Stokes matrices by a negative profile]
In the case of ${\mathcal J}_n^{(p,r)} = -\bigl|{\mathcal J}_n^{(p,r)}\bigr|$ (i.e.~$\beta_{p,q}<0$), Proposition \ref{PropositionStokesProfileFormula} is re-expressed by using the positive profile $\bigl|{\mathcal J}_n^{(p,r)}\bigr|$ as 
\begin{align}
S_n = I_p + \sum_{(i|j) \in |{\mathcal J}_n^{(p,r)}|} s_{n,i,j} E_{i,j}\qquad \bigl(\beta_{p,q}<0\bigr), 
\end{align}
where $n=0,1,2,\cdots, 2rp-1$. $\quad \square$
\end{Proposition}
{\em Proof} \quad This can be seen by noting the following relation: 
\begin{align}
{}^\exists (j|i) \in \mathcal J_n^{(p,r)} \qquad \Leftrightarrow\qquad {}^\exists (i|j) \in |\mathcal J_n^{(p,r)}| \qquad \bigl(\beta_{p,q}<0 \bigr), 
\end{align}
and by applying it to Eq.~\eq{StokesProfileFormula} in Proposition \ref{PropositionStokesProfileFormula}. 
$\quad \blacksquare$

\begin{figure}[htbp]
\begin{center}
\includegraphics[scale=0.7]{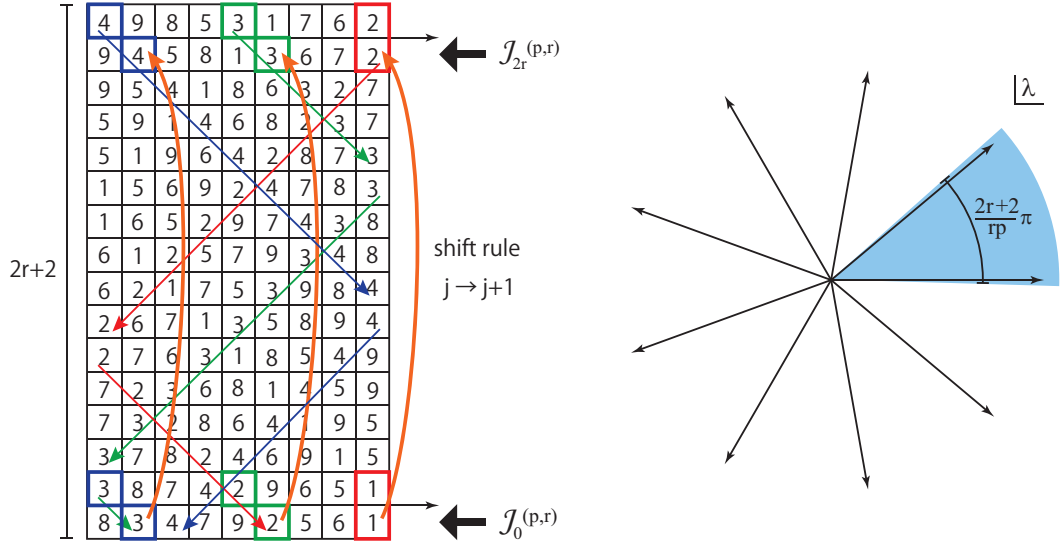}
\end{center}
\caption{\footnotesize  The left figure is the profile we draw. The right figure is the corresponding angular area in the $\lambda$ plane. A simple way to draw the profile is following: 1) $j_{0,p}=j_{1,p}=1$. 2) use shift rule, $j_{2r,p}=j_{2r+1,p}=2$. 3) fill the number in the directions of arrows. 4) use again the shift rule, $j_{n+2r,i} = j_{n,i}+1$. 5) continuing these procedure, one can fill all the boxes.}
\label{figureProfileWay}
\end{figure}

Therefore, regardless of $\sgn (\beta)$, it is convenient to consider the positive profile $|{\mathcal J}^{(p,r)}|$. For the positive profile $|{\mathcal J}^{(p,r)}|$, the explicit expression of the components is obtained in \cite{CIY2}: 
\begin{Theorem} [Explicit form of profile components] \label{TheoremExplicitFormOfProfileComponents}
The components of the profile, $j_{n,l}\in \bigl|\mathcal J^{(p,r)}\bigr|$, are given as 
\begin{align}
j_{n,l} \equiv 1 + \Bigl(\Bigl\lfloor \frac{n}{2} \Bigr\rfloor + (-1)^{p+n+l} \Bigl\lfloor \frac{p-l+1}{2} \Bigr\rfloor\Bigr)m_1 \qquad \text{mod. } p,
\end{align}
where $m_1$ is obtained by the Euclidean algorithm of $pn_1 + rm_1=1$. $\quad \square$
\end{Theorem}
The structure of parenthesis is also obtained in \cite{CIY2} as follows: 
\begin{Proposition} [Explicit form of parenthesis]
The parentheses of profiles are generally given as follows: 
\begin{align}
&\underline{\text{$p\in 2\mathbb Z+1$}} \nn\\
&\qquad |\mathcal J_{n}^{(p,r)}| = 
\left\{
\begin{array}{cl}
\ds 
\begin{tabular}{|c|c|c|c|c|c|}
\hline
$j_{n,1}$ & $(j_{n,2}$ & $j_{n,3})$ & $\,\,\,\cdots\,\,\,$ & $(j_{n,p-1}$ & $j_{n,p})$  \cr
\hline
\end{tabular}
& \quad(n\in 2\mathbb Z+1) \cr
\ds 
\begin{tabular}{|c|c|c|c|c|c|}
\hline
$(j_{n,1}$ & $j_{n,2})$ & $\cdots$ & $(j_{n,p-2}$ & $j_{n,p-1})$ & $j_{n,p}$ \cr
\hline
\end{tabular}
& \quad(n\in 2\mathbb Z)  \rule{0pt}{15pt}
\end{array}
\right., \nn\\
&\underline{\text{$p\in 2\mathbb Z$}} \nn\\
&\qquad |\mathcal J_{n}^{(p,r)} |= 
\left\{
\begin{array}{cl}
\ds 
\begin{tabular}{|c|c|c|c|c|}
\hline
$(j_{n,1}$ & $j_{n,2})$ &  $\quad\qquad \cdots \quad\qquad$ & $(j_{n,p-1}$ & $j_{n,p})$  \cr
\hline
\end{tabular}
& \quad(n\in 2\mathbb Z+1)
\cr
\ds 
\begin{tabular}{|c|c|c|c|c|c|c|}
\hline
$j_{n,1}$ & $(j_{n,2}$ & $j_{n,3})$ &$\cdots$ & $(j_{n,p-2}$ & $j_{n,p-1})$ & $j_{n,p}$ \cr
\hline
\end{tabular}
& \quad(n\in 2\mathbb Z) 
 \rule{0pt}{15pt}
\end{array}
\right.. 
\end{align}
In particular, these parentheses satisfy 
\begin{align}
(j|l) \in \mathcal J_{n}^{(p,r)} \qquad \Rightarrow\qquad j+l -2 \equiv n m_1 \mod p\qquad  \bigl(pn_1+rm_1 = 1\bigr). \label{EquationOfJLindicesInProfileStokesLinesExchanges}
\end{align}
$\square$
\end{Proposition}
Finally, we note the following conventional way to draw the profile \cite{CIY2}: 
\begin{Remark} [Drawing the profiles]
In the basis given by Eq.~\eq{EqSec3First} (i.e.~$\gamma = r$ in the notation of \cite{CIY2}), the positive profile $\bigl|{\mathcal J}_n^{(p,r)}\bigr|$ satisfies the following shift rule \cite{CIY2,CIY3}: 
\begin{align}
j_{n+2r,a} = j_{n,a}+1. \label{EqShiftRule}
\end{align}
With use of this rule, one can draw the profile as follows (also shown in Fig.~\ref{figureProfileWay}):
\begin{itemize}
\item [0. ] Consider a profile of the following range (i.e.~a table of $p\times (2r+2)$ boxes): 
\begin{align}
\bigcup_{n=0}^{2r+1} |\mathcal J_n^{(p,r)}|. 
\end{align}
It is the fundamental domain of $\mathbb Z_p$-symmetry ($+2$ segments). 
\item [1. ] Put $j_{0,p}=j_{1,p}=1$ ($\because$ Eq.~\eq{EqSec3First} (with positive $\beta>0$)). 
\item [2. ] Put $j_{2r,p}=j_{2r+1,p}=2$ ($\because$ the shift rule, Eq.~\eq{EqShiftRule}). 
\item [3. ] Fill the number ``\,$2$'' (i.e. that of $j_{2r,p}=j_{2r+1,p}=2$) along the directions (shown as tilting arrows in Fig.~\ref{figureProfile}) till one arrives at $\bigl|\mathcal J_0^{(p,r)}\bigr|$.
\item [4. ] By using the shift rule, $j_{2r,i} = j_{0,i}+1$, put the number ``\,$3$'' in $\bigl|\mathcal J_{2r}^{(p,r)}\bigr| \cup \bigl|\mathcal J_{2r+1}^{(p,r)}\bigr|$. 
\item [5. ] By continuing the procedure 3. and 4., one can fill all the boxes in the table. 
\end{itemize}
$\square$
\end{Remark}

\begin{Remark} [Profile in the $p=2$ cases] \label{RemarkOnProfileOfPeq2}
The behavior of the profiles in the $p=2$ cases are a little bit different from others, since $\bigl\{\chi_{2n}\bigr\}_{n\in\mathbb Z}$ are not Stokes lines. Therefore, we also write in the following way: 
\begin{align}
\bigl|\mathcal J^{(2,r)}\bigr| = 
\begin{tabular}{|c|c|}
\hline
 $\vdots$ & $\vdots$ \cr
\hline
$(2$ & $1)_5$ \cr
\hline
$2$ & $1$ \cr
\hline
$(1$ & $2)_3$ \cr
\hline
$1$ & $2$ \cr
\hline
$(2$ & $1)_1$ \cr
\hline
$2$ & $1$ \cr
\hline
\end{tabular}
\qquad\to \qquad 
\bigcup_{m=0,1,2,\cdots} \bigl|\mathcal J^{(2,r)}_{2m+1} \bigr|= 
\begin{tabular}{|c|c|}
\hline
 $\vdots$ & $\vdots$ \cr
\hline
$(2$ & $1)_5$ \cr
\hline
$(1$ & $2)_3$ \cr
\hline
$(2$ & $1)_1$ \cr
\hline
\end{tabular}
\end{align}
$\square$
\end{Remark}

\subsubsection{$\theta$-parameterization}

\begin{figure}[htbp]
\begin{center}
\includegraphics[scale=0.7]{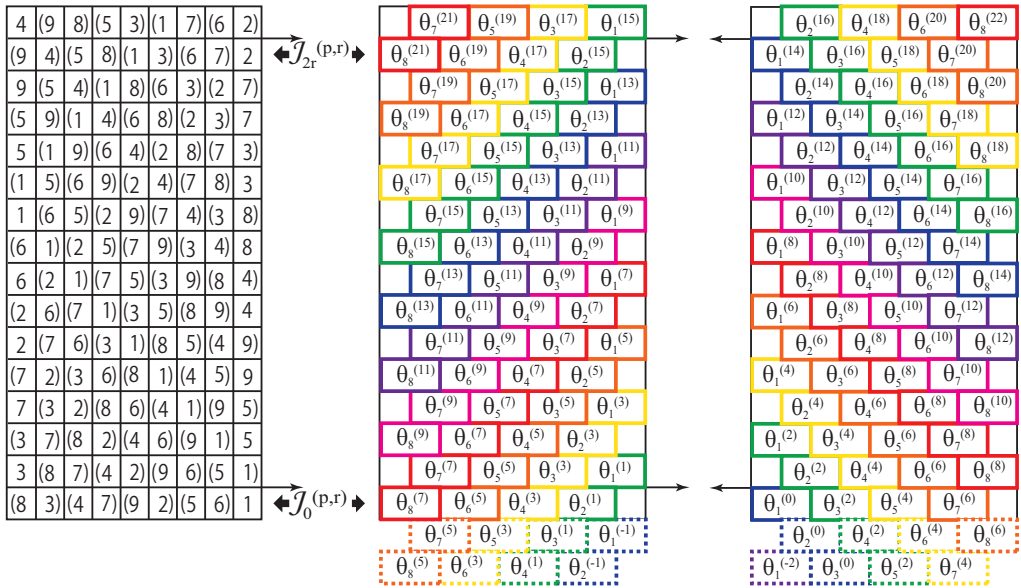}
\end{center}
\caption{\footnotesize The profile of dominant exponent of $9\times 9$ system with $r=7$. The parentheses and $\theta$-parametrization of the Stokes multipliers are shown.  }
\label{figureProfile}
\end{figure}

It is also convenient to introduce the following $\theta$-parameterization of Stokes multipliers: 
\begin{Definition} [$\theta$-parameterization] \label{DefinitionOfThetaParametrization}
The $\theta$-parametrization of Stokes multipliers \cite{CIY2} is introduced as follows: 
\begin{align}
\text{1) If $\beta>0$: }&\qquad  \theta_n^{(a)} \equiv  s[a-n+1,j_{a-n+1,p-n+1},j_{a-n+1,p-n}], \nn\\
&\qquad\qquad \qquad \qquad \qquad \bigl(a \in 2 \mathbb Z+1,\quad 1\leq n\leq p-1\bigr), \nn\\
\text{2) If $\beta<0$: }&\qquad \theta_n^{(a)} \equiv s[a-n+1,j_{a-n+1,n},j_{a-n+1,n+1}], \nn\\
&\qquad\qquad \qquad \qquad \qquad \bigl(
\left\{
\begin{array}{ll}
a \in 2 \mathbb Z  & (\text{p: odd})\cr
a \in 2 \mathbb Z +1 & (\text{p: even})
\end{array}
\right\},\quad 1\leq n\leq p-1\bigr),
\end{align}
where $s_{n,i,j} = s[n,i,j]$, which is shown in Fig.~\ref{figureProfile}. It is also convenient to extend the index $n$ as 
\begin{align}
\theta_0^{(a)} = -1,\qquad \theta_p^{(a)} = 1\qquad \bigl(a\in \mathbb Z + r\bigr). 
\end{align}
There is also the following periodicity: 
\begin{align}
\theta_n^{(a+2rp)} = \theta_n^{(a)}\qquad \bigl(a\in \mathbb Z + r,0\leq n\leq p\bigr). 
\end{align}
$\square$
\end{Definition}
With use of the $\theta$-parameters, the following symmetry relations are simply expressed \cite{CIY2}: 
\begin{Proposition} [Symmetry for Stokes multipliers]
In terms of the $\theta$-parameter, the following constraints are expressed as follows: 
\begin{align}
\left\{
\begin{array}{rcl}
1) & \theta_n^{(a)} = \theta_n^{(a+2r)} & \quad \text{$\mathbb Z_p$-symmetric condition} \cr
2) & (\theta_n^{(a)})^* = -\theta_{p-n}^{(2r-a-2)} & \quad \text{Hermiticity condition}
\end{array}
\right.. \label{StokesSymmetryEquationsZpAndHermiticity}
\end{align}
$\square$
\end{Proposition}
As an example, the parentheses of the profile and its $\theta$-parameters are shown in the case of $9\times 9$ system with $r=7$ in Fig.~\ref{figureProfile}. 

\subsection{Geometry of anti-Stokes lines \label{SubsubsectionGeometryOfAntiStokesLines}}

From this subsection, we start discussing so-called Stokes geometry \cite{ExactWKB} for the spectral curve of Chebyshev solutions (i.e.~$\varphi(\mu,\zeta)$ of Eqs.~\eq{PhiFunctionConformalBackground}, \eq{EqSpectralCurvePhiZetaSec2} and \eq{EqSpectralCurvePhiInTauSpace}) in $(p,q)$ minimal string theory. Stokes geometry consists of a collection of anti-Stokes lines which are studied in this subsection. 
We first start with the definition of {\em saddle points}: 
\begin{Definition} [Saddle points of spectral curve] \label{DefinitionGeneralSaddlePoints}
For a given spectral curve, characterized by $\varphi(t;\zeta)$, the following points $\zeta_*$ are called saddle points: 
\begin{align}
\zeta_*\in \mathbb C\qquad \text{s.t.}\qquad d \varphi^{(j,l)}(\zeta_*) = 0. \label{SaddleAntiStokes}
\end{align}
By expressing this condition with using $(\mathcal S;\zeta,Q)$ (i.e.~Eq.~\eq{DifferentialInTermsOfPQ}), 
\begin{align}
0=d \varphi^{(j,l)}(\zeta) = \bigl(Q^{(j)}(\zeta)- Q^{(l)}(\zeta)\bigr) d\zeta = \bigl(Q(z^{(j)}(\zeta))- Q(z^{(l)}(\zeta))\bigr) d\zeta, 
\end{align}
one can categorize the saddle points into two classes: 
\begin{itemize}
\item [1. ] A point $\zeta_*$ is called branch points if 
\begin{align}
z^{(j)}(\zeta_*) = z^{(l)}(\zeta_*). 
\end{align}
That is, $z_*\equiv z^{(j)}(\zeta_*)=z^{(l)}(\zeta_*)$ is a double point of $\zeta(z)$. 
\item [2. ] A point $\zeta_*$ is called singular points if 
\begin{align}
Q(z^{(j)}(\zeta))- Q(z^{(l)}(\zeta))=0\qquad \bigl( z^{(j)}(\zeta)\neq z^{(l)}(\zeta) \bigr). 
\end{align}
That is, the point $(\zeta_*,Q(\zeta_*))$ on spectral curve is a singular point of the algebraic curve, $F(\zeta,Q)=0$. 
\end{itemize}
$\square$
\end{Definition}
The reason why we call them ``saddle points'' is clear if one sees free-fermion analysis \cite{fy1,fy2,fy3,fis,fim,fi1} and also Riemann-Hilbert analysis (e.g.~\cite{ItsBook}). 
Note that if one changes the coordinate to $\lambda$ (i.e.~$\zeta=2^{p-1} \lambda^p$), then a new branch point appear at $\lambda_*=0$. We however do not pay much attention to this branch point since it does not contribute to the results since it disappears in $\zeta$-plane, i.e.~physical space-time of the system. The singular points will not change since they are a coordinate invariant concept on spectral curve. 

\begin{figure}[htbp]
\begin{center}
\includegraphics[scale=0.9]{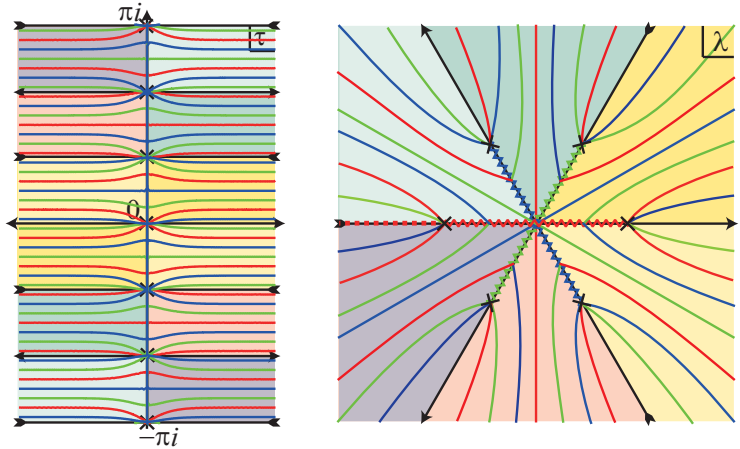}
\end{center}
\caption{\footnotesize Stokes geometry of the $(3,4)$ case is shown in the $\tau$ and $\lambda$ space. The $\tau$-space is plotted by Mathematica. Different colors represent anti-Stokes lines of different types ${\rm ASL}^{(j,l)}$ of $(j,l)=(1,2),(2,3),(3,1)$. ${\rm ASL}^{(1,2)}$ is green; ${\rm ASL}^{(2,3)}$ is red; ${\rm ASL}^{(3,1)}$ is blue. This can be checked by the shift rule, Eq.~\eq{EqShiftRule}. }
\label{FigStokesGeometry34}
\end{figure}

The definition of anti-Stokes lines is as follows (See also \cite{ItsBook}): 
\begin{Definition}[anti-Stokes lines] \label{DefAntiStokesLines}
For a given spectral curve $\varphi(\lambda)$, anti-Stokes lines of $(j,l)$-type are (real) one-dimensional lines in $\lambda \in \mathbb C$ which are drawn from each saddle point $\lambda_*$ of $\varphi^{(j,l)}(\lambda)$ (i.e.~$d\varphi^{(j,l)}(\lambda_*)=0$), satisfying 
\begin{align}
{\rm Im}\Bigl[\varphi^{(j,l)}(\lambda)\Bigr] = {\rm Im}\Bigl[\varphi^{(j,l)}(\lambda_*)\Bigr]. 
\end{align}
The set of anti-Stokes lines of $(j,l)$-type is denoted by ${\rm ASL}^{(j,l)}$. Therefore, $d\varphi^{(j,l)}(\lambda)$ is a real differential along the lines ${\rm ASL}^{(j,l)}$:  
\begin{align}
{\rm Im}\Bigl[f^*d \varphi^{(j,l)}(t)\Bigr] = 0 \qquad \Bigl(f: [0,1] \ni t \ \mapsto \ \lambda(t) \in {\rm ASL}^{(j,l)} \subset \mathbb C\Bigr). 
\end{align}
Here $f(t)$ represents a line-segment in ${\rm ASL}^{(j,l)}$. 
$\quad \square$
\end{Definition}
It is relatively easier to draw these lines in the $\tau$ space (defined by Eq.~\eq{Eqtaucoordinate}) by using Mathematica. For an instruction, we show anti-Stokes lines of the $(p,q)=(3,4)$ case in both $\tau$ and $\lambda$ planes, in Fig.~\ref{FigStokesGeometry34}. ${\rm ASL}^{(1,2)}$ is drawn in green; ${\rm ASL}^{(2,3)}$ is in red; ${\rm ASL}^{(3,1)}$ is in blue. 

We note {\em branch cuts} in Fig.~\ref{FigStokesGeometry34} with associated colors. From Definition \ref{DefinitionOfSpectralCurve}, the differentials $d\varphi^{(j)}(\lambda)$ are mapped by a permutation for any analytic continuation of a closed loop $\gamma$: 
\begin{align}
&\gamma^*d\varphi(\zeta)=  \underset{1\leq j\leq p}{\diag}\bigl(d\varphi^{(\sigma_\gamma(j))}(\zeta)\bigr),\qquad \text{with} \quad {}^\exists\sigma_\gamma \in \mathfrak S_p. 
\end{align}
This means that the branch cuts of the differential $d\varphi(\lambda)$ (or $\varphi$-function) are understood as a permutation of indices (given by $\sigma\in \mathfrak S_p$). Therefore, if a line of ${\rm ASL}^{(j,l)}$ crosses a branch cut of $\sigma$-type, then the anti-Stokes line will be connected to a line of ${\rm ASL}^{(\sigma(j),\sigma(l))}$. In Fig.~\ref{FigStokesGeometry34}, therefore, the branch cut of $\sigma={(1,2)}$ is drawn in green; $\sigma={(2,3)}$ is in red; $\sigma={(3,1)}$ is in blue. The branch cuts are movable (by definition), and therefore they are usually put in $\mathbb Z_p$-symmetric manner in the $\lambda$-plane.

\subsubsection{Symmetry and degeneracy of anti-Stokes lines}

Although the coordinate $\lambda$ is suitable in describing the isomonodromy systems, it is often useful to use the coordinate $\zeta$, since it is the spectral parameter naturally defined in matrix models. This procedure is guaranteed by the $\mathbb Z_p$-symmetry of anti-Stokes lines: 
\begin{Proposition} [$\mathbb Z_p$-symmetry of anti-Stokes lines]
In the $(p,q)$-system, anti-Stokes lines of Chebyshev solutions possess the $\mathbb Z_p$-symmetry by the following shifts:
\begin{align}
\lambda  \to \omega^n \lambda,\qquad j \to j-n,\qquad \bigl(n =0,1,\cdots,p-1\bigr). 
\end{align}
That is, ${\rm ASL}^{(j,l)} = \omega^n \times {\rm ASL}^{(j-n,l-n)}$. 
 $\quad \square$
\end{Proposition}
{\em Proof} \quad By the $\mathbb Z_p$ rotation of the $\lambda$ coordinate $\lambda \to \omega^{n} \lambda$, the functions $\bigl\{\varphi^{(j)}(\lambda)\bigr\}_{j=1}^p$ are mapped as 
\begin{align}
\varphi^{(j)}(\lambda) \to \varphi^{(j)}(\omega^{n} \lambda) = \varphi^{(j-n)}(\lambda). 
\end{align}
This gives the $\mathbb Z_p$-symmetry relation: ${\rm ASL}^{(j,l)} = \omega^n \times {\rm ASL}^{(j-n,l-n)}$. 
 $\quad \blacksquare$ 

Therefore, the following domain is the fundamental domain of $\mathbb Z_p$-symmetry: 
\begin{align}
-\frac{\pi}{p} \leq {\rm lm}\, \tau \leq \frac{\pi}{p},\qquad {\rm Re}\,\tau \geq 0, 
\end{align}
which covers once the domain of $\zeta\in \mathbb C$. In this sense, anti-Stokes lines in the $\lambda$-plane can be recovered from anti-Stokes lines drawn only in $\zeta$-plane. 
In the $\zeta$-coordinate, it is useful to move the branch cuts to those stretching to $\zeta \to \pm \infty$ (like in the $\zeta$ coordinate of Fig.~\ref{TauLambda}). It is because eigenvalue cuts in matrix models always place along such a semi-infinite direction, and also because one can see the anti-Stokes lines which are drawn behind the $\mathbb Z_p$-symmetric branch cuts. 

Also note the following degeneracy of anti-Stokes lines: 
\begin{Proposition} [Degeneracy of anti-Stokes lines] 
In the $(p,q)$-system, anti-Stokes lines of Chebyshev solutions generally degenerate as 
\begin{align}
{\rm ASL}^{(l,j)} = {\rm ASL}^{(j,l)} = {\rm ASL}^{(j',l')}\qquad j+l -2 \equiv j'+l'-2 \quad \mod p. 
\end{align}
Therefore, there are $p$ classes of anti-Stokes lines ${\rm ASL^{[n]}}$ ($n=0,1,\cdots,p-1$): 
\begin{align}
{\rm ASL}^{[n]} = {\rm ASL}^{(j,l)} \qquad \text{for ${}^\forall (j,l)$ s.t.}\quad j+l -2 \equiv n \quad \mod p, \label{EqPairOfDegenerateAntiStokesLines}
\end{align}
which does not degenerate if $n\not \equiv n'$ (mod.~$p$). 
This class of anti-Stokes lines ${\rm ASL}^{[n]}$ (labelled by this $n$) is referred to as anti-Stokes lines of class-$n$.  $\quad \square$
\end{Proposition}
{\em Proof}\quad With using Eq.~\eq{EqSpectralCurvePhiInTauSpace}, the spectral curve $\varphi^{(j,l)}(\tau)$ is given as 
\begin{align}
\varphi^{(j,l)}(\tau) &=  p \beta_{p,q} \mu^{\frac{r}{2p}}\, i\, \sin \bigl[(p+q) \frac{l-j}{p} \pi \bigr]\times \nn\\
&\times \Bigl[
\frac{\sinh\bigl((p+q)(\tau -\pi i \frac{(j+l-2)}{p})\bigr)}{p+q}-
\frac{\sinh \bigl((q-p)(\tau -\pi i \frac{(j+l-2)}{p}) \bigr)}{q-p}
\Bigr]. \label{EqSpectralCurvePhiOfTauInProofOfProposition}
\end{align}
Therefore, the $\tau$ dependence of ${\rm Im}\bigl[\varphi^{(j,l)}(\tau)\bigr]$ only comes from 
\begin{align}
{\rm Im}\bigl[\varphi^{(j,l)}(\tau)\bigr] = &{\rm Re}\Bigl[
\frac{\sinh\bigl((p+q)(\tau -\frac{\pi i n}{p})\bigr)}{p+q}-
\frac{\sinh \bigl((q-p)(\tau -\frac{\pi i n}{p}) \bigr)}{q-p}
\Bigr]  \times {\rm Cost}_{j,l}, 
\end{align}
where ${\rm Cost}_{j,l} \, (\in \mathbb R)$ is a real constant depending only on $(j,l)$. Therefore, anti-Stokes lines ${\rm ASL}^{(j,l)}$ of Eq.~\eq{EqPairOfDegenerateAntiStokesLines} degenerate in the whole plane.  $\quad \blacksquare$

\subsubsection{Branch points and singular points}

As can be seen in Definition \ref{DefAntiStokesLines}, it is important to know the saddle points on the spectral curve. Here we review the results of saddle points which have been studied in minimal string theory: 
\begin{Theorem} [Branch points] \label{TheoremBranchPointsMinimalStringTheory}
In the $(p,q)$-system, spectral curve of Chebyshev solutions has $2p$ branch points $\bigl\{\lambda_n\bigr\}_{n=0}^{2p-1}$ in the $\lambda$ plane $(\lambda \in \mathbb C)$, which are given as 
\begin{align}
\lambda_{n} \equiv \frac{ (2\sqrt \mu)^{1/p} \omega^{\frac{n}{2}}}{2},\qquad \tau_n \equiv \frac{n\pi }{p} i\qquad \bigl(n=0,1,\cdots,2p-1\bigr),
\end{align}
where $2 \zeta = (2\lambda)^p = 2 \sqrt{\mu}\cosh(p\tau)$. In $\zeta$-plane, they are given as
\begin{align}
\zeta_n = e^{\pi i n} \sqrt{\mu} \qquad (n=0,1,\cdots,2p-1). 
\end{align}
By analytic continuing around the branch point $\lambda_n$ by a cycle $\gamma_n$ (i.e. Eq.~\eq{DefClosedLoopAroundBranchPointsLambdaN}), the $\varphi$-functions $\bigl\{ \varphi^{(j)}(\lambda)\bigr\}_{j=1}^p$ are reshuffled as
\begin{align}
\gamma_n^*\varphi^{(j)}(\lambda) = \varphi^{(l)}(\lambda) \qquad j+l-2 \equiv n \quad \text{mod}\quad p.  \label{EqCutReshuffle}
\end{align}
Therefore, the combination $\varphi^{(j,l)}(\lambda)$ has a simple square-root branch cut at $\lambda=\lambda_n$: 
\begin{align}
\varphi^{(j,l)} (\lambda) =& \frac{8\sqrt{2}}{3} i c_{j,l} \Bigl(\frac{2(\lambda-\lambda_n)}{p\lambda_n}\Bigr)^{\frac{3}{2}} \bigl(1+O(\frac{\lambda-\lambda_n}{\lambda_n})\bigr), \label{BranchCutAroundBranchPointsInTheorem} 
\end{align}
with 
\begin{align}
c_{j,l} \equiv & \frac{3p \beta_{p,q} \mu^{\frac{q+p}{2p}}}{8\sqrt{2}} \frac{2pq}{3} (-1)^{\frac{r(j+l-2-n)}{p}}\sin\bigl(\frac{(q+p)(l-j)}{p} \pi \bigr) \in \mathbb R. 
\label{ProofOfPropositionDirectEvaluationVanishingCriterionCoefficientCJL}
\end{align}
where $\lambda$ and $\zeta$ are related as $\ds \frac{(\zeta-\zeta_n)}{\zeta_n} \simeq \frac{p(\lambda-\lambda_n)}{\lambda_n}$ around the branch point. 
$\quad \square$
\end{Theorem}
It is worth noting the types of branch cuts: 
\begin{Corollary} \label{CorollaryForTheTypeOfBranchCuts}
The type of branch cuts generated from the branch point $\lambda_n$ ($n\in \mathbb Z/ 2p \mathbb Z$) is given by a permutation $\sigma_n \in\mathfrak S_p$, 
\begin{align}
\sigma_n \equiv  \prod_{j<l, j+l-2\equiv n} (j,l), 
\end{align}
such that $\gamma_n^* \varphi^{(j)} (\lambda) = \varphi^{(\sigma_n(j))}(\lambda)$. 
The branch cuts of $\sigma_n$-type are also referred to as the branch cuts of class-$n$. $\quad \square$
\end{Corollary}

Also the following fact has been obtained since the Liouville theory calculation \cite{SeSh} and also the matrix-model one \cite{KazakovKostov} and those of the free-fermion \cite{fis,fim}: 

\begin{Theorem} [Singular points] \label{TheoremSingularPointsMinimalStringTheory}
In the $(p,q)$-system, spectral curve of Chebyshev solutions has singular points for the function $\varphi^{(j,l)}(\lambda)$ which are given in the $\tau$ space as 
\begin{align}
\tau_n^{(j,l)} =\frac{pn+q(j+l-2)}{pq} \pi i \qquad \bigl(n\in \mathbb Z\bigr),\label{EqInTheoremTauForInstantons}
\end{align}
where $(2\lambda)^p = 2\sqrt{\mu} \cosh p\tau$. They are related to $(m,n)$ ZZ-branes 
with the following identification: 
\begin{align}
(m,n) = (l-j,n),\qquad m,n\in \mathbb Z. \label{EqZZbraneLabelingInFreeFermion}
\end{align}
In fact, $\varphi^{(j,l)}(\tau)$ around the singular points are given by 
\begin{align}
\varphi^{(j,l)}(\tau) = \varphi^{(j,l)}(\tau_n^{(j,l)}) + \frac{1}{2} \frac{\del^2\varphi^{(j,l)}(\tau_{n}^{(j,l)})}{\del \tau^2} \bigl(\tau-\tau_n^{(j,l)}\bigr)^2 + \cdots \label{SingularPointsBehaviorForPhiInTheorem}
\end{align}
with 
\begin{align}
\varphi^{(j,l)}(\tau_n^{(j,l)}) &= \varphi^{(1)}(\tau_{m,n}^{(+)})-\varphi^{(1)}(\tau_{m,n}^{(-)})\qquad \Bigl(\tau_{m,n}^{(\pm)}= \frac{pn\pm qm}{pq}\pi i\Bigr) \nn\\
 &=-\beta_{p,q} \frac{2 p q }{q^2-p^2}\mu^{\frac{r}{2p}} \sin \Bigl(\frac{rn\pi}{q}\Bigr)\sin \Bigl(\frac{rm\pi}{p}\Bigr),  \label{SaddleValue} \\
 \frac{\del^2\varphi^{(j,l)}(\tau_{n}^{(j,l)})}{\del \tau^2} &= -\beta_{p,q} \bigl(2 p q \bigr) \mu^{\frac{r}{2p}} \sin \Bigl(\frac{rn\pi}{q}\Bigr)\sin \Bigl(\frac{rm\pi}{p}\Bigr).
\end{align}
In particular, Eq.~\eq{SaddleValue} is identified as the ZZ-brane action (Eq.~\eq{EqZZbraneAction}). 

Furthermore, the instanton actions (Eq.~\eq{SaddleValue}) are invariant under the following shifts: 
\begin{align}
(m,n) \qquad \to \qquad  (m+p,n\pm q)\qquad \to \qquad (p-m,q-n),
\end{align}
and $(m,n)\to (m,-n)$ changes the sign of the action. Therefore, the saddle points of the $(p,q)$ minimal-string spectral curve realize all the $\frac{(p-1)(q-1)}{2}$ distinct ZZ and ghost-ZZ branes in the following range of indices: 
\begin{align}
1\leq m\leq p-1,\qquad 1\leq n\leq q-1. 
\end{align}
$\square$
\end{Theorem}

\subsubsection{Global structure of anti-Stokes lines}

With these facts, anti-Stokes lines around branch points and around singular points are simply given by 
\begin{align}
{\rm Im}\bigl[\varphi^{(j,l)}(\lambda)\bigr] = 
{\rm Im}\bigl[\varphi^{(j,l)}(\lambda_*)\bigr] = 0.   \label{EqAntiStokesLinesImEq0}
\end{align}
It is then convenient to define the following concept of {\em instanton interval}: 
\begin{Definition} [Instanton interval] The following regions (in $\zeta$-plane; in $\lambda$-plane; and in $\tau$-plane) are referred to as instanton intervals: 
\begin{align}
\bigl(\text{Instanton Interval for $\zeta$}\bigr) & \equiv \Bigl\{\zeta\Big| -\sqrt{\mu} \leq \zeta\leq \sqrt{\mu}  \Bigr\} \subset \mathbb R,  \label{InstantonIntervalL} \\
\bigl(\text{Instanton Interval for $\lambda$}\bigr)  & \equiv \bigcup_{n=0}^{p-1}  \omega^n \times \Bigl\{\lambda\Big| - \Bigl(\frac{\sqrt{\mu}}{2^{p-1}}\Bigr)^{\frac{1}{p}} \leq \lambda \leq \Bigl(\frac{\sqrt{\mu}}{2^{p-1}}\Bigr)^{\frac{1}{p}}  \Bigr\}  \subset  \bigcup_{n=0}^{p-1}  \omega^n \times \mathbb R, \label{InstantonIntervalZ} \\
\bigl(\text{Instanton Interval for $\tau$}\bigr) & \equiv \Bigl\{\tau\Big| -\pi  \leq (-i\tau) \leq +\pi \Bigr\} \subset i \mathbb R
\end{align}
and 
\begin{itemize}
\item [a. ] All the saddle points are located in the intervals. 
\item [b. ] All the branch points are located in the end points of the intervals for $\zeta$- and $\lambda$-plane. 
\item [c. ] All the anti-Stokes lines degenerate along the intervals. 
\end{itemize}
$\square$
\end{Definition}
{\em Proof}\quad The statements of {\em a.}~and {\em b.}~are obvious by Theorem \ref{TheoremBranchPointsMinimalStringTheory} and \ref{TheoremSingularPointsMinimalStringTheory}.  From Eq.~\eq{EqSpectralCurvePhiInTauSpace}, one can see that ${\rm Im}\bigl[ \varphi^{(j)}(\tau)\bigr]=0$ ($j=1,2,\cdots,p$) in the instanton interval. This means that all the anti-Stokes lines degenerate along the intervals. $\quad \blacksquare$

Since all the saddle points are located along the instanton intervals, all the local behaviors of anti-Stokes lines (in $\zeta$-plane) are then categorized into the following three cases: 
\begin{Proposition} [Three situations for saddle points] \label{PropositionThreeCasesOfSaddlePointsInInstantonIntervals}
All the local behaviors of anti-Stokes lines (in $\zeta$-plane) are categorized into the following three cases (the part around $\zeta=\zeta_{2n+1}$ is just reflection of them, $\zeta\to-\zeta$): 
\begin{align}
\begin{array}{c}
\includegraphics[scale = 1]{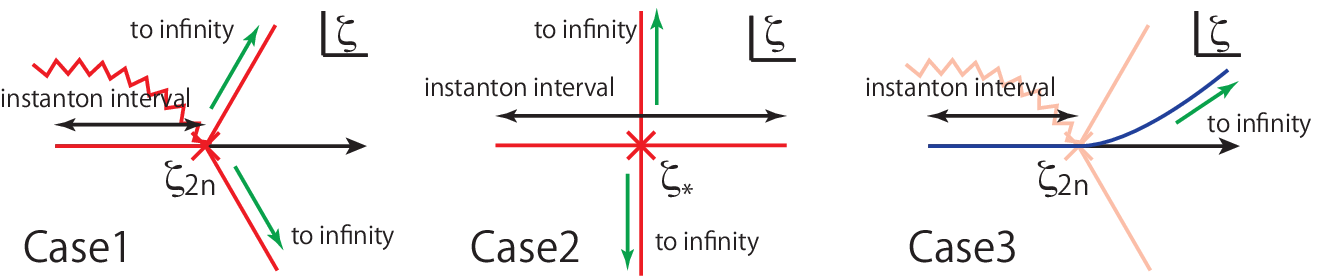}
\end{array} \label{EqFigureAiryAntiStokesLocalBehaviorOfDZnetworks}
\end{align}
\begin{itemize}
\item [1. ] Around the branch point $\zeta_n = e^{\pi i n}\sqrt{\mu}\, \, (n\in \mathbb Z/2p \mathbb Z)$, there are three anti-Stokes lines of class-$n$ (mod.~$p$): Two of them go to infinity without hitting any saddle points; and one of them goes along the instanton interval, Eq.~\eq{InstantonIntervalL}. 
\item [2. ] Around the singular points of $\varphi^{(j,l)}(\zeta)$, there appear four anti-Stokes lines of $(j,l)$-type. In the $\zeta$ plane, two of them go to vertical directions and to infinity without hitting any saddle points; and the other two of them go to horizontal directions, i.e.~along the instanton interval, Eq.~\eq{InstantonIntervalL}. 
\item [3. ] For anti-Stokes lines of class-$n'$ ($n'\not \equiv n \mod p$), the end point of instanton interval $\zeta_n$ are neither branch points nor singular points. Therefore, such lines just pass through the end point and go to infinity, without hitting any saddle points. 
\end{itemize}
$\square$
\end{Proposition}
{\em Proof} \quad \underline{Case {\em 1.}}\quad  Anti-Stokes lines around the branch points are given by Eq.~\eq{BranchCutAroundBranchPointsInTheorem} as 
\begin{align}
0={\rm Im}\Bigl[i \Bigl(\frac{2(\zeta-\zeta_n)}{\zeta_n}\Bigr)^{\frac{3}{2}}\Bigr] \qquad \Leftrightarrow \qquad \arg(\zeta-\zeta_n) = n \pi + \frac{2m+1}{3} \pi\qquad \bigl(m\in \mathbb Z/3\mathbb Z\bigr).   
\end{align}
Therefore, the lines are locally drawn as in Eq.~\eq{EqFigureAiryAntiStokesLocalBehaviorOfDZnetworks} around $\zeta = \sqrt{\mu}$ (i.e.~$n\in 2\mathbb Z$), and the cases around $\zeta = -\sqrt{\mu}$ (i.e.~$n\in 2\mathbb Z+1$) are also the same (but left-right reflected). \\
\underline{Case {\em 2.}}\quad  Anti-Stokes lines around singular points are given by Eq.~\eq{SingularPointsBehaviorForPhiInTheorem} as 
\begin{align}
0={\rm Im}\Bigl[(\zeta-\zeta_*)^2\Bigr] \qquad \Leftrightarrow \qquad \arg(\zeta-\zeta_*) = \frac{m}{2} \pi\qquad \bigl(m\in \mathbb Z/4\mathbb Z\bigr).   
\end{align}
Therefore, the lines are locally drawn as in Eq.~\eq{EqFigureAiryAntiStokesLocalBehaviorOfDZnetworks}. \\
\underline{Case {\em 3.}}\quad Note that $\varphi^{(j,l)}(\tau)$ is a real function along instanton interval. Therefore, around the ending point $\zeta_n$ of instanton interval, anti-Stokes lines of class-$n'$ ($n'\not\equiv n \mod p$) gives 
\begin{align}
0={\rm Im}\Bigl[(\zeta-\zeta_n)\Bigr] \qquad \Leftrightarrow \qquad \arg(\zeta-\zeta_n) = m \pi\qquad \bigl(m\in \mathbb Z/2\mathbb Z\bigr),   
\end{align}
and the lines are locally drawn as in Eq.~\eq{EqFigureAiryAntiStokesLocalBehaviorOfDZnetworks}. 
$\quad \blacksquare$

\begin{figure}[htbp]
\begin{center}
\includegraphics[scale=0.9]{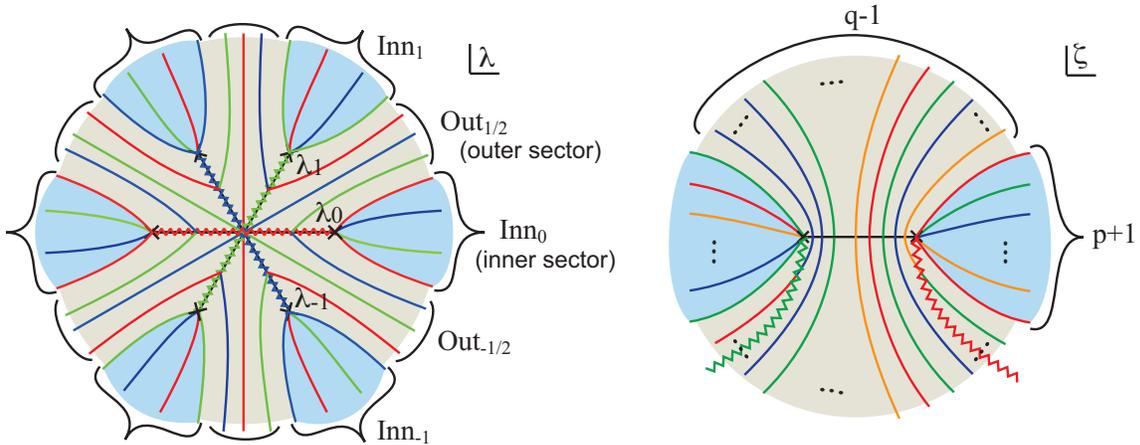}
\end{center}
\caption{\footnotesize The left figure shows inner and outer sectors in the $(p,q)=(3,4)$ case. The right figure is the general $(p,q)$ cases in the $\zeta$-plane: In an inner sector, there are $(p+1)$ anti-Stokes lines (mod.~degeneracy) generated from the associated branch point; and in an outer sector, there are $(q-1)$ anti-Stokes lines (mod.~degeneracy) generated from singular points. }
\label{FigInnerOuter}
\end{figure}

For later convenience, we further introduce {\em inner and outer sectors}. It would be easier to understand by looking at Fig.~\ref{FigInnerOuter}.

\begin{Definition} [Inner/outer sector] \label{DefInnerOuterSectors} 
The $\lambda$-plane can be divided into $2p$ inner sectors $\bigl\{{\rm Inn}_{n}\bigr\}_{n\in \mathbb Z/2p\mathbb Z}$ and $2p$ outer sectors $\bigl\{{\rm Out}_{n+\frac{1}{2}}\bigr\}_{n\in \mathbb Z/2p \mathbb Z}$. 
\begin{itemize}
\item At the branch point of $\lambda_n$, there are the two anti-Stokes lines of class-$n$ (going to infinity) from the branch point. The sector in between the two lines are referred to inner sectors and denoted by ${\rm Inn}_n$. Inner sector ${\rm Inn}_n$ is defined to include the two anti-Stokes lines of class-$n$ from the branch points 
\item There are still some sectors in between two different inner sectors (and also divided by instanton intervals). These sectors are referred to as outer sector. The outer sector in between ${\rm Inn}_n$ and ${\rm Inn}_{n+1}$ is denoted by ${\rm Out}_{n+ \frac{1}{2}}$. 
\end{itemize}
An example of the $(p,q)=(3,4)$ case is shown in Fig.~\ref{FigInnerOuter}. 
$\quad \square$
\end{Definition}

\begin{Proposition} \label{PropositionOfAnti-StokesLinesBehaviors1}
For general $(p,q)$ cases, the counting of lines (mod.~degeneracy) is following: 
\begin{itemize}
\item [1. ] In between the lines of class-$n$, there are $(p-1)$ lines (mod.~degeneracy).
\item [2. ] In an inner sector, there are $(p+1)$ anti-Stokes lines (mod.~degeneracy) generated from the associated branch point. 
\item [3. ] In an outer sector, there are $(q-1)$ anti-Stokes lines (mod.~degeneracy) generated from singular points.  
\end{itemize}
 $\square$
\end{Proposition}
{\em Proof}\quad If one sees $\lambda\to \infty$, the equation of the anti-Stokes lines (Eq.~\eq{EqAntiStokesLinesImEq0}) becomes 
\begin{align}
0&={\rm Im}\bigl[\bigl(\omega^{-(j-1)r}-\omega^{-(l-1)r}\bigr)\lambda^r\bigr] =2\big| \lambda\bigr|^r\cos\bigl(r\theta- \frac{\pi r(j+l-2)}{p}\bigr)\sin\bigl(\frac{\pi r(l-j)}{p}\bigr) \nn\\
&\qquad \Leftrightarrow\qquad p \theta = \frac{rn+p(m+\frac{1}{2})}{r}\pi \qquad \bigl(j+l-2\equiv n \mod p\,;\, m\in \mathbb Z;\, \lambda = |\lambda|e^{i\theta}\bigr) \label{Eq1inProofOfAntiStokesTheorem}\\
& \therefore\quad  \zeta = e^{ip\theta} |\zeta|,\qquad p \theta = \frac{\widetilde n + p (\widetilde m+\frac{1}{2})}{r} \pi \qquad \bigl(rn = \widetilde n + p m_n;\, \widetilde m = m + m_n \in\mathbb Z\bigr). \label{Eq2inProofOfAntiStokesTheorem}
\end{align}
for anti-Stokes lines of class-$n$. Since $(p,r)$ are coprime, Eq.~\eq{Eq1inProofOfAntiStokesTheorem} tells us that there are $2(p+q)$ anti-Stokes lines (mod.~degeneracy) in $\zeta$-plane. Also noting that $\widetilde n \in \mathbb Z/p\mathbb Z$ has 1-1 correspondence with $n \in \mathbb Z/p\mathbb Z$ (in Eq.~\eq{Eq1inProofOfAntiStokesTheorem}), one can see that in between the lines of class-$n$, there are $(p-1)$ lines (mod.~degeneracy). This is statement {\em 1.} This explains that there are $(p+1)$ lines (mod.~degeneracy) inside an inner sector (statement {\em 2.}). Therefore, one also concludes that there are $(q-1)$ lines (mod.~degeneracy) in an outer sector (statement {\em 3.}). 
$\quad \blacksquare$

We are then ready to describe the whole geometry of anti-Stokes lines: 
\begin{Theorem} [Geometry of anti-Stokes lines] \label{TheoremGeometryOfAnti-StokesLines}
Consider $(p,q)$-system. There are $p$ different classes of anti-Stokes lines (i.e.~of class-$n$ labeled by $n \in \mathbb Z/p\mathbb Z$). We divide $r$ by $p$ as 
\begin{align}
r =p+q = mp + l\qquad (m=1,2,\cdots;1\leq l \leq p-1),  \label{RintermsofPinTheoremAntiStokesLines}
\end{align}
and draw these anti-Stokes lines (in $\zeta$-plane including $\zeta_0$ and $\zeta_1$) as follows: 
\begin{itemize}
\item [1.] The lines of class-$0$ (red lines on the left-hand-side) and the lines of class-$1$ (green lines on the right-hand-side) are drawn as 
\begin{align}
\begin{array}{c}
\includegraphics[scale = 0.8]{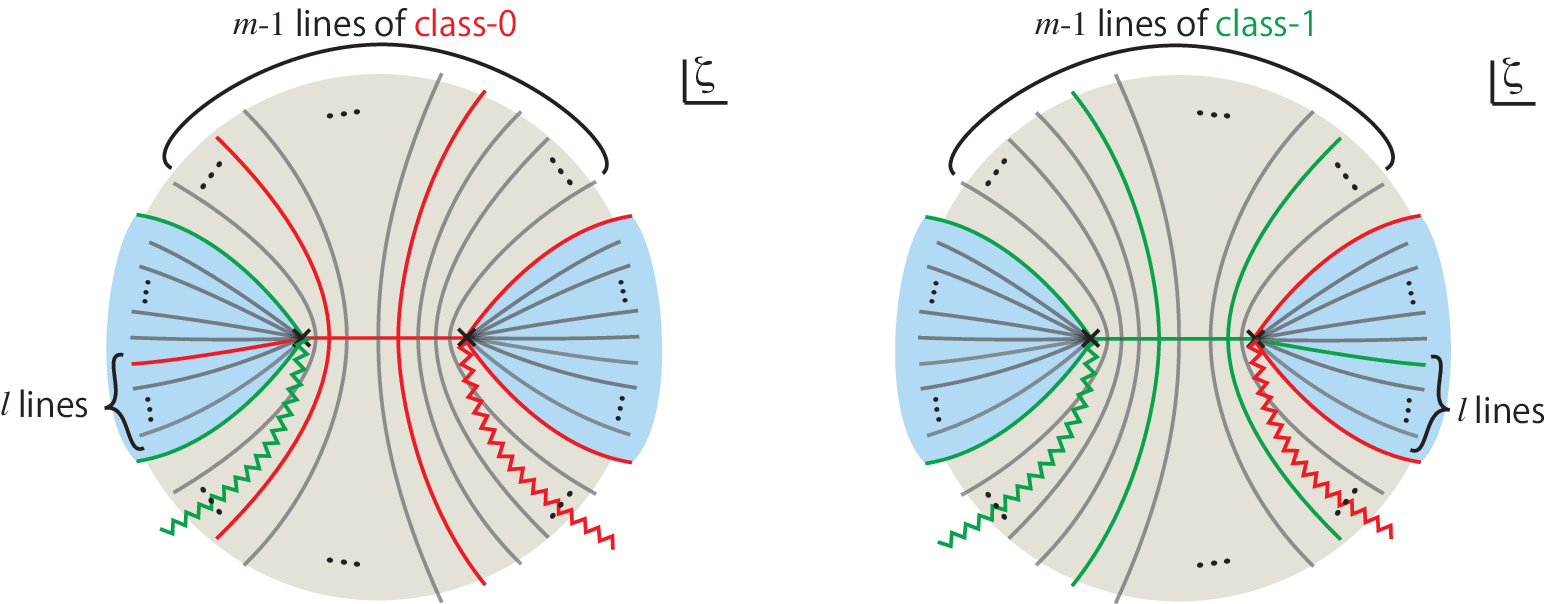}
\end{array}. 
\end{align}
The lines of class-$0$ and class-$1$ are left-right symmetric to each other in the $\zeta$ plane. 
\begin{itemize}
\item [a. ] The lines of class-$0$ has a branch point at $\zeta (= \zeta_0)=+\sqrt{\mu}$ and forms an inner sector. In an outer sector, there are $(m-1)$ lines of class-$0$ going to infinity. 
\item [c. ] The lines of class-$1$ has the same story, but is left-right reflected. 
\item [b. ] In the inner sector of class-$1$, there is an anti-Stokes line of class-$0$. The reverse is also true. 
\end{itemize}
\item [2.] The lines of class-$n$ ($n\neq 0,1$) are generally drawn (by blue lines) as 
\begin{align}
\begin{array}{c}
\includegraphics[scale = 0.8]{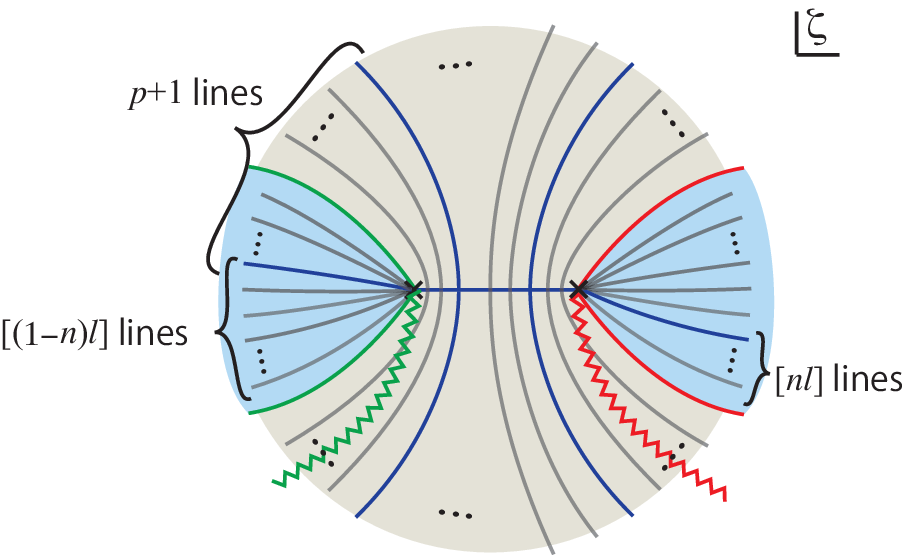}
\end{array}, \label{EqFigureAiryAntiStokesNthEPS}
\end{align}
where $[nl]$ and $[(1-n)l]$ are integers modulo $p$. 
Note that, in between the anti-Stokes lines of class-$n$, there are $(p-1)$ lines (mod.~degeneracy). 
\end{itemize}
$\square$
\end{Theorem}
{\rm Proof}\quad \underline{Statement {\em 1.}} \quad Around the branch point, $\zeta (=\zeta_0)= \sqrt{\mu}$, the lines of class-$0$ have its branch point and behaves as the case {\em 1.} of Proposition \ref{PropositionThreeCasesOfSaddlePointsInInstantonIntervals}. According to Proposition \ref{PropositionOfAnti-StokesLinesBehaviors1}, around $\zeta\to \infty$, a line of class-$0$ appears once every $p$ lines (mod.~degeneracy). By taking into account the decomposition of Eq.~\eq{RintermsofPinTheoremAntiStokesLines}, one can see that there are $(m-1)$ lines of class-$0$ in ${\rm Out}_{1/2}$ (statement {\em a.}), and there is a single line of class-$0$ in ${\rm Inn}_1$ (statement {\em c.}). It is clear that the same story can apply to the lines of class-$1$ whose branch point is at $\zeta (=\zeta_1)=-\sqrt{\mu}$ (statement {\em b.}). \\
\underline{Statement {\em 2.}} \quad The non-trivial statement is about the position of lines in inner sectors. This can be seen by focusing on Eq.~\eq{Eq2inProofOfAntiStokesTheorem}. Noting that the angle of ${\rm Inn}_0$ is given by 
\begin{align}
{\rm Inn}_0 \qquad \Leftrightarrow \qquad -\frac{p\pi}{2r} \leq \arg(\zeta) \leq  \frac{p\pi}{2r}, 
\end{align}
a line of class-$n$ is located at $\arg(\zeta) = \frac{\widetilde n}{r} -\frac{p\pi}{2r}$ (See Eq.~\eq{Eq2inProofOfAntiStokesTheorem}), which is $[nl]$-th line (as shown in Eq.~\eq{EqFigureAiryAntiStokesNthEPS}). Similarly, the angle of ${\rm Inn}_1$ is given by 
\begin{align}
{\rm Inn}_1 \qquad \Leftrightarrow \qquad \frac{(r-\frac{p}{2})\pi}{r} \leq \arg(\zeta) \leq \frac{(r+\frac{p}{2})\pi}{r}. 
\end{align}
Therefore, a line of class-$n$ is located at $\arg(\zeta) = \frac{[r(n-1)]}{r} +\frac{(r-\frac{p}{2})\pi}{r} = - \frac{[r(1-n)]}{r} +\frac{(r+\frac{p}{2})\pi}{r}$, which is the position shown in Eq.~\eq{EqFigureAiryAntiStokesNthEPS}.  $\quad\blacksquare$

\subsection{Spectral networks \label{RoleOfSpectralNetworks}}

Here we review the concept of {\em spectral networks} to describe Stokes geometry. The concept of spectral network itself is introduced and developed in the context of Riemann-Hilbert problem for isomonodromy systems \cite{DZmethod,ItsKapaev} (See \cite{ItsBook} for historical constructions). It is also called {\em Riemann-Hilbert graph} in literature. The name, ``spectral networks,'' itself is introduced in \cite{SpectralNetworks}, which discuss implications of spectral networks in gauge theory. Therefore, it is also called {\em Deift-Zhou (DZ) networks}. 

Roughly speaking, spectral networks provide an algebraic way to describe {\rm Stokes data} along anti-Stokes lines, which formulate Stokes geometry on spectral curve. This construction is possible because the system is described by isomonodromy systems. The definition is given as follows: 

\begin{figure}[htbp]
\begin{center}
\includegraphics[scale=1.2]{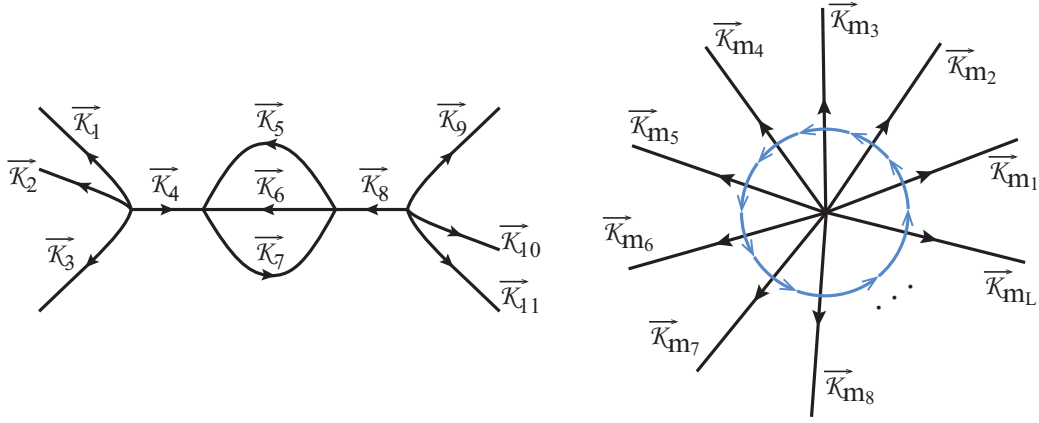}
\end{center}
\caption{\footnotesize Typical examples of DZ networks. }
\label{FigureDZnetworks}
\end{figure}

\begin{Definition} [Deift-Zhou's spectral network] Consider a $p\times p$ isomonodromy system and an associated spectral curve $\varphi(t;\lambda) \in \mathcal L_{\rm str}^{\rm (univ.)}(t)$. 
Deift-Zhou spectral network $\hat {\mathcal K}$ consists of two objects: A graph $\mathcal K$ and its associated matrices given by $G$. 
\begin{itemize}
\item [I) ] The graph $\mathcal K$ is a collection of oriented smooth lines in $\mathbb CP^1$ (or $\mathbb C$): 
\begin{align}
\mathcal K \equiv \bigcup_{m=1}^{\# \mathcal K} \vec{\mathcal K}_m \subset \mathbb CP^1,
\end{align}
where $\bigl\{\vec{\mathcal K}_m\bigr\}_{m=1}^{\#\mathcal K}$ are real 1-dimensional line-elements and 
 the ``arrow'' on the top of $\vec{\mathcal K}_ m$ indicates that each line-element possesses an orientation. 
 A flip of the orientation is denoted by ``$-1\times $'': 
 \begin{align}
 \vec{\mathcal K}_m \to \vec{ \mathcal K}'_m \equiv  -\vec{\mathcal K}_m
 \end{align}
 A typical graph is shown in Fig.~\ref{FigureDZnetworks}. 
\item [II) ]  $G$ is a map from a graph $\mathcal K$ to $p\times p$ non-singular matrices in ${\rm GL}(p;\mathbb C)$: 
\begin{align}
\begin{array}{rllcc}
G: & \mathcal K & & \to& {\rm GL}(p;\mathbb C) \cr
 & \,\rotatebox{90}{$\subset$} & & & \rotatebox{90}{$\in$} \cr
 & \vec{\mathcal K}_m & \ni \lambda & \mapsto & G_m
\end{array}
\end{align}
The matrix $G_m (= G(\vec{\mathcal K}_m))$ is called the matrix associated with a line-element $\vec{\mathcal K}_m$. A flip of the orientation gives the inverted matrix: 
\begin{align}
G(-\vec{\mathcal K}_m) = G(\vec{\mathcal K}_m)^{-1}. \label{EqInvertedNetwork}
\end{align}
\end{itemize}
A pair of these two objects, $\hat {\mathcal K} \equiv \bigl(\mathcal K,G\bigr)$, is called Deift-Zhou spectral network (or shortly, DZ network), if it satisfies the following conditions: 
\begin{itemize}
\item [1. ] Assume that there is a junction of lines (say by $\bigl\{\vec{\mathcal K}_{m_i}\bigr\}_{i=1}^L$). One can choose their orientation so that all the directions are directed to outside from the junction point (as in Fig.~\ref{FigureDZnetworks}). If they are ordered in counter-clockwise as 
\begin{align}
\vec{\mathcal K}_{m_1} \to \vec{\mathcal K}_{m_2} \to \cdots \to \vec{\mathcal K}_{m_L}\to \vec{\mathcal K}_{m_1}
\end{align}
(as is shown in Fig.~\ref{FigureDZnetworks}), then the associated matrices $\bigl\{G_{m_i}\bigr\}_{i=1}^L$ satisfy the following matrix conservation condition: 
\begin{align}
G_{m_1}G_{m_2}\cdots G_{m_L} = I_p. \label{MatrixConservationLawInDefinition}
\end{align}
\item [2. ] The essential singularity (now only at $\lambda \to \infty$) forms a junction point (say by $\bigl\{\vec{\mathcal K}_{m_{i}}^{(\infty)}\bigr\}_{i=1}^{L^{(\infty)}}$). In particular, the associated matrices $\bigl\{G_{m_{i}}^{(\infty)}\bigr\}_{i=1}^{L^{(\infty)}}$ satisfy 
\begin{align}
e^{\varphi(\lambda)} G_{m_i}^{(\infty)} e^{-\varphi(\lambda)} \asymeq I_p + O(\lambda^{-\infty})\qquad \lambda \to \infty \in \vec{\mathcal K}_{m_i}^{(\infty)},  \label{EqStokesInDZnetworkCondition}
\end{align}
for $i=1,2,\cdots,L$. This means that these associated matrices are Stokes matrices ($\because$ Definition \ref{StokesDef}).  
\end{itemize}
$\square$
\end{Definition}
The precise relation between the associated matrices $\bigl\{G_{m_{i}}^{(\infty)}\bigr\}_{i=1}^{L^{(\infty)}}$ around the essential singularity and the Stokes matrices $\bigl\{S_n\bigr\}_{n=0}^{2rp-1}$ is discussed later. Before that, we first define {\em equivalence deformations} of DZ network. This deformation is an essential part of Deift-Zhou's method \cite{DZmethod} (See also \cite{ItsBook}): 
\begin{Definition}[Equivalence deformations] \label{DefDeformEquiv}
One can continuously deform a DZ network $\hat {\mathcal K}$ to another DZ network $\hat {\mathcal K}'$: 
\begin{align}
\hat {\mathcal K} = \bigl(\mathcal K,G\bigr) \qquad\leadsto \qquad 
\hat {\mathcal K}'= \bigl(\mathcal K',G'\bigr),
\end{align}
by combining the following three procedures:  
\begin{itemize}
\item [1. ] [Continuous deformations] One can continuously deform the graph $\mathcal K$  by a continuous automorphism $f_t: \mathbb CP^1\to \mathbb CP^1$, parametrized by $t \in [0,1]$, as 
\begin{align}
f_t:\, \mathcal K\quad \to \quad \mathcal K^{[t]} = \bigcup_{m=1}^{\#\mathcal K} f_t(\vec{\mathcal K}_m),\qquad \text{with \quad $G^{[t]}(f_t(\vec{\mathcal K}_m)) = G(\vec{\mathcal K}_m)$},
\end{align}
where $f_0$ is the identity map on $\mathbb CP^1$ ($f_0(\lambda) = \lambda \in \mathbb CP^1$) and $f_t$ keeps the essential singularities (i.e.~$f_t(\infty) = \infty$). This induces a continuous map of DZ networks: $f_t: \hat {\mathcal K} \to \hat{\mathcal K}^{[t]} = \bigl(\mathcal K^{[t]},G^{[t]}\bigr)$. 
These deformations are locally drawn as 
\begin{align}
\begin{array}{c}
\includegraphics{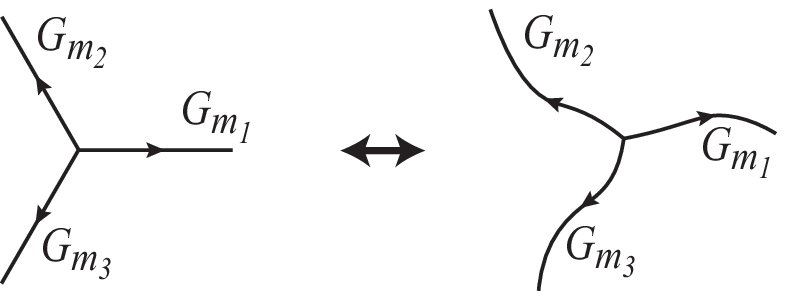}
\end{array}
\end{align}
\item [2. ] [Split/Merge of junctions] One can split/merge a junction of line-element as 
\begin{align}
\begin{array}{c}
\includegraphics[scale=1]{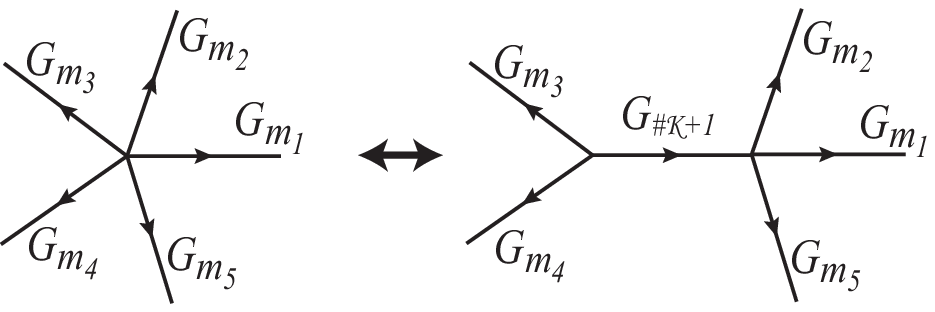}
\end{array},
\end{align}
by adding/deleting a new line-element (i.e.~$\vec{\mathcal K}_{\#\mathcal K+1}$ in the figure). In that case, the associated matrix of the additional line-element is given as 
\begin{align}
G_{\#\mathcal K+1} = G_{m_5} G_{m_1} G_{m_2} = G_{m_4}^{-1} G_{m_3}^{-1} \qquad \bigl(\text{i.e.} \quad  G_{m_1} G_{m_2}  G_{m_3} G_{m_4}  G_{m_5} = I_p\bigr). 
\end{align}
\item [3. ] [Split/Merge of lines] One can split/merge a line-element as 
\begin{align}
\begin{array}{c}
\includegraphics[scale=1]{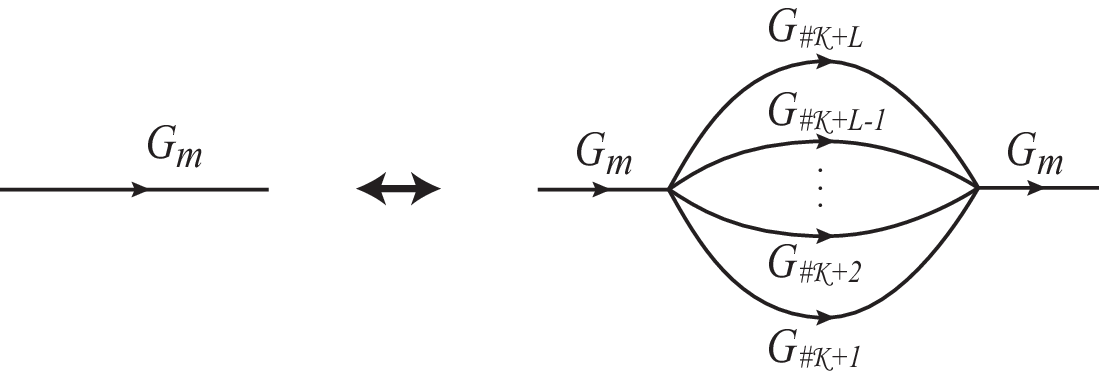}
\end{array},
\end{align}
with satisfying the split of the associated matrices: 
\begin{align}
G_m = G_{\#\mathcal K+1}G_{\#\mathcal K+2}\cdots G_{\#\mathcal K+L-1} G_{\#\mathcal K+L}. 
\end{align}
\end{itemize}
In particular, two DZ networks are said to be equivalent if there exists an equivalence deformation given by the above. In that case, they are written as $\hat {\mathcal K} \sim \hat {\mathcal K}'$. 
$\quad \square$
\end{Definition}
With use of equivalence deformations, one obtains the relation to Stokes matrices (See e.g.~\cite{ItsBook}): 
\begin{Remark}[Stokes matrices and DZ networks] \label{RemarkStokesDZnetworks}
Consider $(p,q)$-system, and its corresponding $p\times p$ isomonodromy system with an essential singularity (only at $\lambda\to \infty$) of Poincar\'e index $r (=p+q)$. The Stokes matrices are given by $\bigl\{S_n\bigr\}_{n=0}^{2rp-1}$. One can define a corresponding DZ network $\hat {\mathcal K}^{(\epsilon)} = \bigl(\mathcal K^{(\epsilon)},G\bigr)$ $\bigl(0<\epsilon < \frac{\pi}{r}\bigr)$ as follows: 
\begin{itemize}
\item [I) ] The graph $\mathcal K^{(\epsilon)}$ is given by 
\begin{align}
 \mathcal K^{(\epsilon)} = \vec{\mathcal K}_\phi \cup \bigcup_{n=0}^{2rp-1} \vec{\mathcal K}_n^{(\epsilon)},\qquad 
\left\{
\begin{array}{ll}
|\vec{\mathcal K}_n^{(\epsilon)}| = \mathbb R_{\geq 0} \times e^{i(\chi_n + \epsilon)} &\quad \bigl(\chi_n = \frac{n \pi}{pr}\bigr) \cr
|\vec{\mathcal K}_\phi| = \mathbb R_{\geq 0}
\end{array}
\right., 
\end{align}
where $|\vec{\mathcal K}_m^{(\epsilon)}|$ indicates a simple line-element without direction, and the direction of $\vec{\mathcal K}_m^{(\epsilon)}$ is directed to $\lambda= \infty$ (the graph is like the right-hand figure of Fig.~\ref{FigureDZnetworks}). 
\item [II) ] The associated matrices of $\mathcal K^{(\epsilon)}$ are given by the Stokes matrices $\bigl\{S_n\bigr\}_{n=0}^{2rp-1}$ as 
\begin{align}
G(\vec{\mathcal K}_n^{(\epsilon)}) = S_n\qquad \bigl(n=0,1,\cdots,2rp-1\bigr),\qquad G(\vec{\mathcal K}_\phi) = (-1)^{p-1} I_p. 
\end{align}
The monodromy cyclic equation (Eq.~\eq{MonodromyEquation}) guarantees the matrix conservation law, Eq.~\eq{MatrixConservationLawInDefinition}. 
\end{itemize}
Here $\hat {\mathcal K}^{(\epsilon)}$ is parametrized by $\epsilon$ $(0<\epsilon < \frac{\pi}{r})$ but they are equivalent: $\hat {\mathcal K}^{(\epsilon)}\sim \hat {\mathcal K}^{(\epsilon')}$. 

By using equivalence deformations, one obtains various other (equivalent) DZ networks $\hat{\mathcal K}' (\sim \mathcal K^{(\epsilon)})$. Conversely, since any graph of DZ networks can be continuously deformed to the above network $\hat {\mathcal K}^{(\epsilon)}$. 
$\quad \square$
\end{Remark}
This means that the both concepts of DZ networks and of Stokes matrices have the same information. 
The DZ networks further become like ``networks'' if one re-expresses them by the following two kinds of matrices:
\begin{Definition} [Stokes/Cut-type matrices] \label{DefinitionStokesCutMatrices}
Consider $p\times p$ isomonodromy systems. The following $p\times p$ matrices, $\Stokes_{l,j}(\alpha)$ and $\Cut_{l,j}(\alpha)$, are referred to as matrices of Stokes-type and matrices of Cut-type, respectively: 
\begin{align}
\Stokes_{l,j}(\alpha) \equiv I_p + \alpha E_{l,j},\qquad \Cut_{l,j}(\alpha) \equiv I_p +\alpha E_{l,j} + (-1/\alpha) E_{j,l} - E_{l,l} - E_{j,j}
\end{align} 
If we assume $l>j$, then these matrices are expressed as 
\begin{align}
\left\{
\begin{array}{c}
\Stokes_{l,j}(\alpha) = 
\left(
\begin{array}{c|c|c|c|c}
I_{j-1} &  &   &  & \cr\hline
 & 1 & &0 &  \cr\hline
 & & I_{l-j-1} & & \cr\hline
 &\alpha  & & 1 & \cr \hline
 & & & & I_{p-l}
\end{array}
\right)
\cr 
\Cut_{l,j}(\alpha) = 
\left(
\begin{array}{c|c|c|c|c}
I_{j-1} &  &   &  & \cr\hline
 & 0 & & -\frac{1}{\alpha} &  \cr\hline
 & & I_{l-j-1} & & \cr\hline
 &\alpha  & & 0 & \cr \hline
 & & & & I_{p-l}
\end{array}
\right) 
\end{array}
\right.\qquad (l>j),\label{EqDefinitionStokesCutMatrices}
\end{align}
In DZ networks, line-elements associated with these matrices are drawn in a special manner:%
\footnote{Here $<i,j>$ means a Stokes tail, and the associated indices $(i,j)$ cannot be reversed: $<i,j> \not \to <j,i>$. Therefore, the Stokes tail of $<i,j>$ kind are denoted as ``$i<j$'' \cite{ExactWKB}.  $[i,j]$ means a cut line, and the associated indices $(i,j)$ can also be reversed (see Eq.~\eq{EqNetworkRuleBreflection}). }
\begin{align}
\begin{array}{c}
\includegraphics[scale=1]{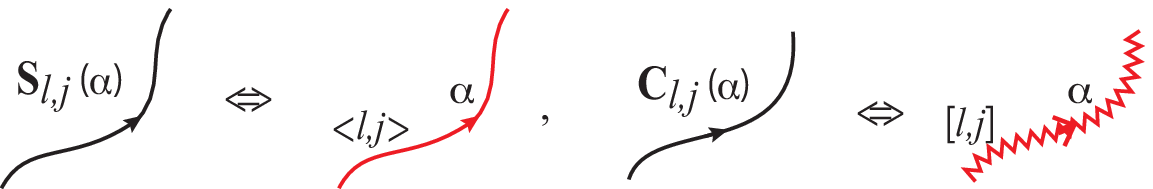}
\end{array}.
\end{align}
Along each line, there is a number associated, $\alpha$. 
\begin{itemize}
\item [1. ] The lines of Stokes-type are called Stokes tails (the number $\alpha$ is called its weight) 
\item [2. ] The lines of Cut-type are called cut lines (the number $\alpha$ is called its phase). 
\end{itemize}
Some proper colors are chosen to distinguish the associated indices $(i,j)$.
$\quad \square$
\end{Definition}
The following relations are basic in the DZ method (See also \cite{ItsBook}): 
\begin{Proposition} [Network rules]
Matrices of Stokes/Cut-types satisfy the following algebraic relations: 
\begin{itemize}
\item [(a)] Inverse: \qquad $\Stokes_{i,j}(\alpha)^{-1} = \Stokes_{i,j}(-\alpha),\qquad \Cut_{i,j}^{-1}(\alpha) = \Cut_{i,j}(-\alpha)$ 
\begin{align}
\begin{array}{c}
\includegraphics[scale = 0.9]{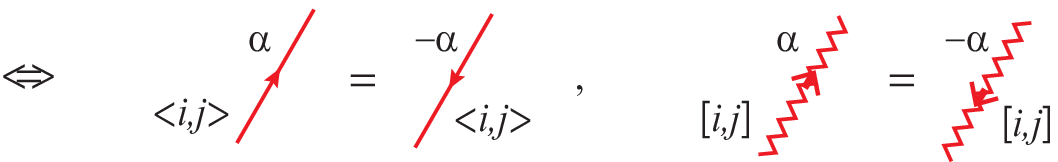}
\end{array} 
\end{align}
\item [(b)] Reflection: \qquad $\Cut_{i,j}(\alpha) = \Cut_{j,i}(-1/\alpha)$
\begin{align}
\begin{array}{c}
\includegraphics[scale = 0.9]{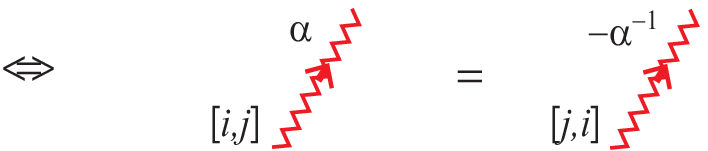}
\end{array}  \label{EqNetworkRuleBreflection}
\end{align}
\item [(c)] Three junction: \qquad $\Stokes_{i,j}(\alpha) \Stokes_{i,j}(-1/\alpha)  \Stokes_{i,j}(\alpha) =\Cut_{i,j}(\alpha)$
\begin{align}
\begin{array}{c}
\includegraphics[scale = 0.9]{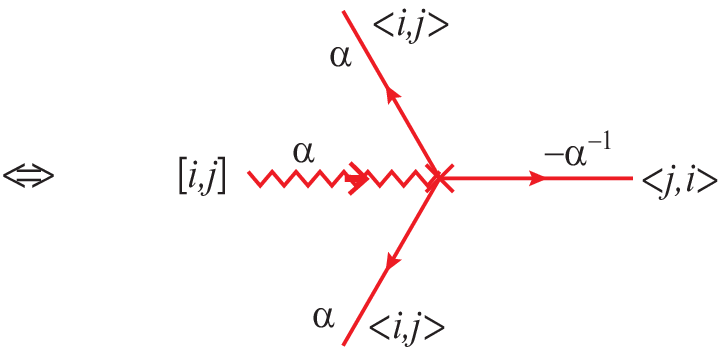}
\end{array}  \label{EqNetworkRulesThreeJunctions}
\end{align}
\item [(d)]  Crossing cuts: \qquad $ \ds \left\{
\begin{array}{l}
\Cut_{i,j}(\nu) \Stokes_{j,k}(\alpha) \Cut_{i,j}^{-1}(\nu) = \Stokes_{i,k}(\nu\alpha)\cr
 \Cut_{j,k}(\nu) \Stokes_{i,j}(\alpha) \Cut_{j,k}^{-1}(\nu) = \Stokes_{i,k}(-\nu\alpha) \cr
 \Cut_{i,j}(\nu) \Stokes_{j,i}(\alpha) \Cut_{i,j}^{-1}(\nu) = \Stokes_{i,j}(-\nu^2\alpha)
\end{array}
\right.$
\begin{align}
\begin{array}{c}
\includegraphics[scale = 0.9]{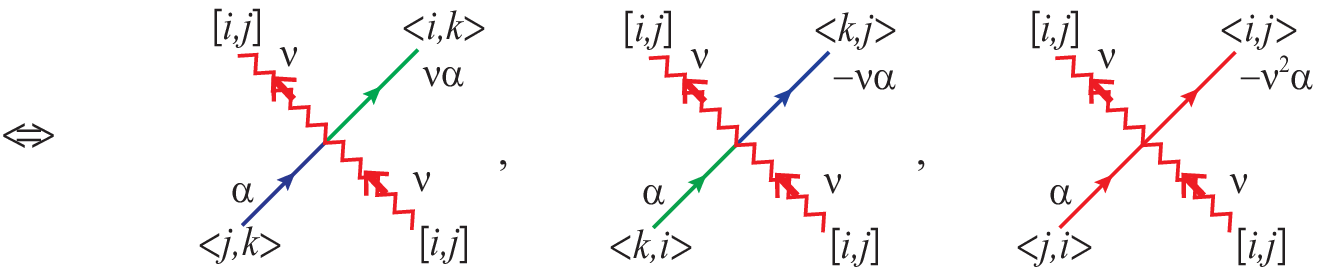}
\end{array} 
\end{align}
\item [(e)] Crossing tails: \qquad $\left\{
\begin{array}{l}
\Stokes_{i,j}(\alpha) \Stokes_{j,k}(\beta) = \Stokes_{i,k}(\alpha\beta) \Stokes_{j,k}(\beta)\Stokes_{i,j}(\alpha)\cr
\Stokes_{j,k}(\beta) \Stokes_{i,j}(\alpha)  = \Stokes_{i,k}(-\alpha\beta) \Stokes_{i,j}(\alpha)\Stokes_{j,k}(\beta)
\end{array}
\right.$
\begin{align}
\begin{array}{c}
\includegraphics[scale = 0.9]{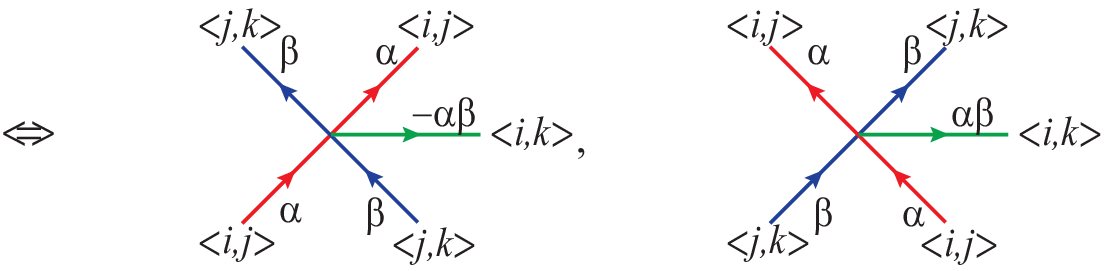}
\end{array} \label{EqNetworkRulesE}
\end{align}
\end{itemize}
$\square$ \label{PropositionNetworkRules}
\end{Proposition}
\begin{Proposition}
With use of equivalence deformations (Definition \ref{DefDeformEquiv}), any DZ networks can be expressed by lines of Stokes/Cut-type. $\quad \square$
\end{Proposition}
{\em Proof}\quad Any DZ networks can be deformed to the DZ network of Remark \ref{RemarkStokesDZnetworks}. This DZ network only includes lines of Stokes-type. $\quad \blacksquare$

\subsection{Stokes geometry, the profile and instantons \label{SubsubsectionStokesGeometryProfileInstantons}}

In this subsection, we consider the following problem: In anti-Stokes lines ${\rm ASL}^{(j,l)}$, there are many singular points of $\varphi^{(j,l)}(\zeta)$, which are instantons of the system. The formula for the positions of instantons (i.e.~Eq.~\eq{EqInTheoremTauForInstantons}) is not very useful to specify the positions in ${\rm ASL}^{(j,l)}$ and also which kinds of instantons are located. We here show a new treatment to deal with this problem, by using the profile of dominant exponents. 

\subsubsection{The location of instantons via the profile}

We first make a connection between the profile of dominant exponents $\mathcal J^{(p,r)}$ and anti-Stokes lines $\bigcup_{j,l}{\rm ASL}^{(j,l)}$ around $\lambda\to \infty$. The key is spectral networks. {\em A spectral network is said to describe Stokes geometry of $\varphi(t;\lambda)$ if its Stokes tails can be adjusted to anti-Stokes lines of spectral curve.} It is a non-trivial problem whether a spectral network can be adjustable to anti-Stokes lines in the whole plane, $\lambda\in\mathbb C$. This issue is studied later; but at least around $\lambda\to \infty$, it is always possible to consider the adjustment. This justifies the following relation: 

\begin{Proposition} 
[Profile and anti-Stokes lines ($\lambda\to\infty$)] \label{PropositionProfileAntiStokesLines}
The $n$-th inner sector, ${\rm Inn}_n$, and the $\bigl(n+ \frac{1}{2}\bigr)$-th outer sector, ${\rm Out}_{n+ \frac{1}{2}}$, are related to the profile of dominant exponents as 
\begin{align}
 {\rm Inn}_n \quad \leftrightarrow \quad\bigcup_{m=0}^p \mathcal J_{nr -p + m}^{(p,r)},\qquad {\rm Out}_{n+\frac{1}{2}} \quad \leftrightarrow \quad \bigcup_{m=1}^{q-1} \mathcal J_{nr + m}^{(p,r)}. 
\end{align}
Graphically this relation is drawn as follows: 
\begin{align}
\begin{array}{c}
\includegraphics[scale = 1]{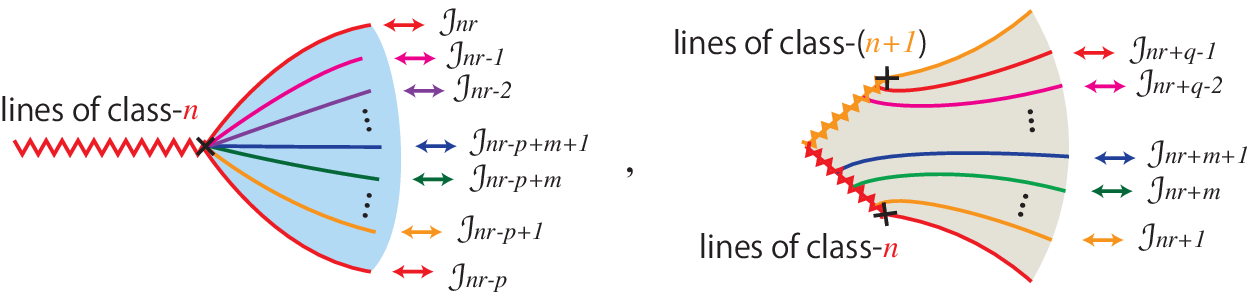}
\end{array}. \label{EqStokesGeoToProfile}
\end{align}
This relation means that 
\begin{align}
\text{a line in ${\rm ASL}^{(l,j)}$} \quad \underset{\text{adjust}}{\leftarrow} \quad \text{a Stokes tail $<\!l,j\!>_\alpha$} \quad \underset{\text{Remark~\ref{RemarkStokesDZnetworks}}}{\Leftrightarrow} \quad (j|l)_n \in \mathcal J^{(p,r)}_n, 
\end{align}
around $\lambda\to \infty$.  $\quad \square$
\end{Proposition}
{\em Proof}\quad \underline{Adjustability}\quad Adjustability of Stokes tails to anti-Stokes lines can be seen from Remark \ref{RemarkStokesDZnetworks}. In fact, Stokes tails in the DZ network $\hat {\mathcal K}^{(\epsilon)}$ (of Remark \ref{RemarkStokesDZnetworks}) are asymptotically approached to anti-Stokes lines (around $\lambda\to \infty$) if one chooses $\epsilon = \frac{\pi}{2r}$. \\
\underline{Correspondence}\quad The correspondence is seen as follows: An inner sector ${\rm Inn}_n$ is formed by the lines of class-$n$ (Proposition \ref{PropositionThreeCasesOfSaddlePointsInInstantonIntervals} and Definition \ref{DefInnerOuterSectors}). Angles of the inner sector (around $\lambda\to \infty$) are then given (by Eq.~\eq{Eq1inProofOfAntiStokesTheorem}) as
\begin{align}
{\rm Inn}_n \qquad \Leftrightarrow \qquad \frac{(nr-\frac{p}{2})\pi}{rp} \leq \arg(\lambda) \leq \frac{(nr+\frac{p}{2})\pi}{rp}. 
\end{align}
This can be compared with the Stokes tails of $\hat {\mathcal K}^{(\epsilon)}$ (of Remark \ref{RemarkStokesDZnetworks}), and one obtains 
\begin{align}
{\rm Inn}_n \qquad \Leftrightarrow \qquad \chi_{nr-p} + \epsilon\leq \arg(\lambda) \leq \chi_{nr} + \epsilon \qquad \bigl(\chi_m= \frac{m \pi}{rp},\,\epsilon = \frac{\pi}{2r}\bigr). 
\end{align}
This is the correspondence stated in the proposition. The correspondence in outer sectors also follows from this result. 
 $\quad \blacksquare$
 
It is then useful to represent the inner sector in the profile. Some of the examples are shown in Fig.~\ref{FigureMCBCs}. By this proposition, we conclude the following: 
\begin{Corollary} [The profile of instantons] \label{CorollaryProfileOfInstantons} The profile of dominant exponents $\mathcal J^{(p,r)}$ corresponding to outer sectors $\bigl\{{\rm Out}_{n+\frac{1}{2}}\bigr\}_{n\in \mathbb Z/2p \mathbb Z}$ represents a table of instantons located in Anti-Stokes lines. $\quad \square$
\end{Corollary}
{\em Proof}\quad Each anti-Stokes line around $\lambda\to \infty$ are connected to instantons in the instanton interval. The correspondence of Proposition \ref{PropositionProfileAntiStokesLines} can be translated to the correspondence between the profile and instantons on anti-Stokes lines.  $\quad \blacksquare$

\subsubsection{Identification of instantons/ZZ-branes}

We further discuss how to specify instantons (or ZZ-branes) from the components of profile. 
Since the $\tau$ space itself nicely parametrizes the coordinate of the instantons (i.e.~Eq.~\eq{EqInTheoremTauForInstantons}), we consider the following labeling $L$ of the instanton: 
\begin{align}
\tau_n^{(j,l)}  = \Bigl(\frac{n}{q} + \frac{j+l-2}{p}\Bigr)\pi i \equiv \frac{L}{pq} \pi i. 
\end{align}
Note the following range is related to the instanton interval of the $\zeta$ and $\lambda$ space: 
\begin{align}
0\leq L \leq q\qquad &  \Leftrightarrow \qquad \zeta \in  \bigl[-\sqrt{\mu},\sqrt{\mu}\bigr], \label{EqInstantonIntervalLspaceZeta} \\
0\leq L \leq 2pq\qquad & \Leftrightarrow \qquad \lambda \in \bigcup_{n=0}^{p-1}  \bigl[-\lambda_{0},\lambda_0\bigr] \times \omega^{n}, \label{EqInstantonIntervalLspaceLambda}
\end{align}
and therefore is also related to outer sectors and the profile as
\begin{align}
L = a q + b \qquad\leftrightarrow \qquad \mathcal J_{ar + b}^{(p,r)} \subset {\rm Out}_{a+\frac{1}{2}}\qquad \bigl(0 \leq a < 2p,\,\, 0 < b<q\bigr). 
\end{align}
This leads us to the following result: 
\begin{Proposition} [ZZ-branes on the profile]
For a given $L$ $($i.e.~$\tau(L) \equiv \dfrac{L}{pq} \pi i)$, there is the corresponding profile $\mathcal J_{ar + b}^{(p,r)}$ ($L=aq+b$). For a parenthesis $(j|l)\in \mathcal J_{ar + b}^{(p,r)}$, the following holds: 
\begin{itemize}
\item [1. ] The point $\tau = \tau(L)$ is a saddle (i.e.~an instanton) of $\varphi^{(j,l)}(\tau)$. 
\item [2. ] There exists an integer $n \in \mathbb Z$, satisfying
\begin{align}
q(j+l-2) + pn = L \qquad \bigl(j-l \not \equiv 0  \mod p;\, n\in \mathbb Z\bigr).  \label{EqSaddelChaEq}
\end{align}
\item [3. ] The relation to the ZZ-brane labeling $(m,n)$ is given by Eq.~\eq{EqZZbraneLabelingInFreeFermion}, i.e. 
\begin{align}
(m,n) = (l-j,n),\qquad m,n\in \mathbb Z.
\end{align}
\end{itemize}
$\square$
\end{Proposition}
{\em Proof}\quad Statement {\em 1.}~follow from Corollary \ref{CorollaryProfileOfInstantons}. This implies Statement {\em 2.}. Statement {\em 3.}~is from Theorem \ref{TheoremSingularPointsMinimalStringTheory}. $\quad \blacksquare$

\subsubsection{Uniform signature property}

We further discuss how to see {\em distributions of large/small instantons} on the spectral networks. We first re-express Eq.~\eq{SaddleValue} with use of $L$ as 
\begin{align}
\varphi^{(j,l)}(\tau(L)) = -\frac{2pq}{q^2-p^2} \underbrace{\Bigl[\beta_{p,q} (-1)^n\sin \Bigl(\frac{q}{p}(l-j)\pi \Bigr)\Bigr]}_{(*1)} \underbrace{\sin\Bigl(\frac{L}{q}\pi \Bigr)}_{(*2)}, \label{EqInstantonL}
\end{align}
with an integer $n$ defined by Eq.~\eq{EqSaddelChaEq}:
\begin{align}
n = \frac{L-q(j+l-2)}{p} \in \mathbb Z. 
\end{align}
We first consider ${\rm Out}_{\frac{1}{2}}$ (i.e.~$0<L<q$), then the part $(*2)$ in Eq.~\eq{EqInstantonL} is always positive: 
\begin{align}
\sin\Bigl(\frac{L}{q}\pi \Bigr)>0 \qquad \bigl(1\leq L\leq q-1\bigr). 
\end{align}
On the other hand, the corresponding anti-Stokes lines (around $\lambda\to \infty$) are given as
\begin{align}
\tau(L)=\frac{L}{pq}\pi i \qquad \leftrightarrow \qquad \lambda\to \infty \times e^{i\theta_L}\qquad \theta_L = \frac{L}{rp} \pi + \frac{\pi}{2r}. 
\end{align}
If we assume that $(l|j) \in \mathcal J_{L}^{(p,r)}$, then ${\rm Re}\bigl[\varphi^{(j,l)}(\lambda)\bigr] < 0$ along the anti-Stokes lines (around $\lambda\to \infty$). Therefore, we obtain 
\begin{align}
{\rm Re}\bigl[\varphi^{(j,l)}(\lambda)\bigr] &= \frac{p\beta_{p,q} \mu^{\frac{r}{2p}} |2\lambda|^r}{4(p+q)} {\rm Re} \Bigl[ e^{ri \theta_L }\Bigl( e^{-2\pi i \frac{r(j-1)}{p}} - e^{-2\pi i \frac{r(l-1)}{p}}\Bigr)\Bigr] + O(\lambda^{r-1}) \nn\\
&= -  \frac{p \mu^{\frac{r}{2p}}|2\lambda|^r}{2(p+q)} \Bigl[\beta_{p,q} (-1)^n \sin \Bigl( \frac{q}{p}(l-j) \pi \Bigr)\Bigr]  + O(\lambda^{r-1})< 0, 
\end{align}
around $\lambda\to\infty$. 
This means that the part $(*1)$ in Eq.~\eq{EqInstantonL} is also positive: 
\begin{align}
\beta_{p,q} (-1)^n \sin \Bigl( \frac{q}{p}(l-j) \pi \Bigr) > 0 \qquad \bigl(1\leq L \leq q-1\bigr). 
\end{align}
With taking into account the $\mathbb Z_p$ symmetry, these results lead to the following uniform signature property of instanton on the profile: 
\begin{Theorem}[Uniform signature property 1] \label{TheoremUniformSignatureProperty}
Consider $(p,q)$-systems with Chebyshev solutions. The sign of instanton actions on the Chebyshev solution are given as follows: 
\begin{itemize}
\item If $q>p$ then 
\begin{align}
{}^\exists (j|l) \in \mathcal J_{ar + b} \quad \Leftrightarrow \quad \del_\tau \varphi^{(l,j)}(\tau(aq+b)) = 0\qquad \varphi^{(l,j)} (\tau(aq+b)) <0. 
\end{align}
\item If $p>q$ then 
\begin{align}
{}^\exists (j|l) \in \mathcal J_{ar + b} \quad \Leftrightarrow \quad \del_\tau \varphi^{(l,j)}(\tau(aq+b)) = 0\qquad \varphi^{(l,j)} (\tau(aq+b)) >0. 
\end{align}
\end{itemize}
Here $\tau(L) = \dfrac{L}{pq} \pi i$. The outer sectors are labelled as $\bigl\{ {\rm Out}_{a+ \frac{1}{2}} \bigr\}_{a\in \mathbb Z/2p \mathbb Z}$ and the integer $b$ labels anti-Stokes lines inside of them $\bigl(0<b<q\bigr)$. $\quad \square$
\end{Theorem}

Similarly, we also show the uniform signature property on branch points: 
\begin{Theorem} [Uniform signature property 2] \label{TheoremUniformSignaturePropertyOnBranchPoints}
Consider $(p,q)$-systems with Chebyshev solutions. The branch points $\lambda=\lambda_n$ $(n\in \mathbb Z/2p\mathbb Z)$ are connected to anti-Stokes lines labeled by $\mathcal J_{nr}^{(p,r)}$. Then the following uniform signature property holds: 
\begin{align}
(l|j)\in \mathcal J_{nr}^{(p,r)}\qquad\Rightarrow \qquad 
\varphi^{(j,l)} (\lambda) = \frac{8\sqrt{2}}{3} i c_{j,l} \Bigl(\frac{2x}{p}\Bigr)^{\frac{3}{2}} (1+O(x)), 
\qquad c_{j,l}>0, 
\label{BranchCutAroundBranchPointsInTheorem1} 
\end{align}
where $c_{j,l}$ is given by Eq.~\eq{ProofOfPropositionDirectEvaluationVanishingCriterionCoefficientCJL} and $x=\frac{\lambda-\lambda_n}{\lambda_n}$. $\quad \square$
\end{Theorem}
{\em Proof} \quad Similarly, one considers anti-Stokes lines along $\lambda\to \infty \times \exp\bigl[i (\frac{n\pi }{p} + \frac{\pi}{2r} )\bigr]$ corresponding to $\mathcal J_{nr}^{(p,r)}$ and ${\rm Re}[\varphi^{(j,l)}(\lambda)]<0$. This gives 
\begin{align}
\beta_{p,q} \cos\bigl(\frac{r(j+l-2-n)}{p} \pi \bigr) \sin\bigl(\frac{r(l-j)}{p} \pi \bigr) >0. 
\end{align} 
This gives the statement. $\quad \blacksquare$

\section{The multi-cut boundary conditions \label{SectionMulticutBC}}

Here we consider the multi-cut boundary condition (BC) \cite{CIY2} in $(p,q)$ minimal string theory. The multi-cut BC can be generally considered in each essential singularity of Poincar\'e index $r$ in $k\times k$ isomonodromy systems. 
\begin{itemize}
\item In the cases of $k>r$, multi-cut BC becomes quantum integrability (i.e.~T-systems) of quantum integrable systems \cite{CIY3}, as an extension of ODE/IM correspondence \cite{ODEIMCorrespondence}. This quantum integrability is strong enough to derive explicit solutions of Stokes phenomena. 
\item In the cases of $r>k$ (which are the cases of minimal strings), the quantum integrability becomes {\em drastically simple} and cannot capture all the degree of freedom. This issue would be related to the fact discussed in Section \ref{SectionIsomonodromyEmbedding}, i.e.~$(p,q)$ minimal string theory (described by matrix models) is a subsystem in the most general isomonodromy descriptions. 
\end{itemize}
Beside this issue, there is a new feature in the $r>k$ cases (i.e.~minimal string cases). We find new non-perturbative constraints on the sign of the parameter $\beta_{p,q}$ caused by spectral $p-q$ duality. In particular, we will see that, in some cases of $p+q \in 2\mathbb Z$, spectral $p-q$ duality cannot close in the theories of bosonic minimal strings. 

\subsection{Definition and why it is necessary \label{SubsectionDefinitionOfMultiCutBoundaryConditionWhyNecessary}}

We first review the definition of multi-cut boundary condition and why it is necessary for isomonodromy systems to describe matrix models \cite{CIY2}: 

\begin{Definition} [Multi-cut boundary condition] \label{DefinitionMCBC}
Consider a $k\times k$ isomonodromy system, 
\begin{align}
g\frac{\del \Psi(t;\lambda)}{ \del \lambda} = \mathcal Q(t;\lambda) \Psi(t;\lambda),
\qquad g\frac{\del \Psi(t;\lambda)}{ \del t} = \mathcal B(t;\lambda) \Psi(t;\lambda), \label{IMSinDefinitionOfMCBC}
\end{align}
and its essential singularity (now only at $\lambda\to \infty$). This system is said to satisfy multi-cut boundary condition, if there exists a vector solution of Eq.~\eq{IMSinDefinitionOfMCBC}, $\vec{\psi}_{\rm orth}(t;\lambda) = \Psi(t;\lambda) \vec v$, such that its leading behavior around $\lambda\to \infty$ changes along $k$ ($\mathbb Z_k$-symmetric) rays: 
\begin{align}
\vec{\psi}_{\rm orth} (t;\lambda) &\asymeq c_a \times \vec {\chi}^{(j_a)}(t) e^{\frac{1}{g}\varphi^{(j_a)}(t;\lambda)} \lambda^{\nu^{(j_a)}}\Bigl(1 + O(\lambda^{-1}) \Bigr) \nn\\
&  \text{with}\qquad \lambda\to \infty \in D\bigl(\theta_0 + \frac{2\pi (a-1)}{p}, \theta_0 + \frac{2\pi a}{p}\bigr), \label{EqForBCInDefinitionOfMCBC}
\end{align}
where $j_{a} \not \equiv  j_{a+1}$ ${\rm mod}$ $p$ and $c_a\neq 0$ $(a\in \mathbb Z/p\mathbb Z)$. 
Here $\varphi(t;\lambda) \in \mathcal L_{\rm str}^{\rm (univ.)}(t)$,%
\footnote{Note that it is irrelevant which element $\varphi(\lambda) \in \mathcal L_{\rm str}^{\rm (univ.)}$ is chosen because $O(\lambda^{-1})$-corrections are irrelevant in Eq.~\eq{EqForBCInDefinitionOfMCBC}, and $\varphi(\lambda)-\varphi'(\lambda)=O(\lambda^{-1})$ $(\lambda\to\infty)$ for $\varphi,\, \varphi' \in \mathcal L_{\rm str}^{\rm (univ.)}$.  } 
and $\bigl\{\vec \chi^{(j)}(t)\bigr\}_{j=1}^k$ are some vector functions. $\quad \square$
\end{Definition}

\begin{figure}[htbp]
\begin{center}
\includegraphics[scale=1]{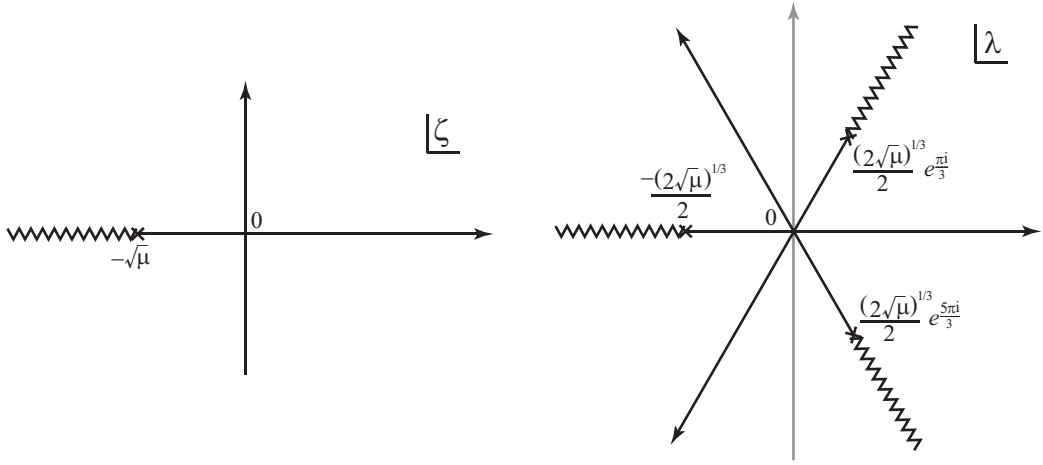}
\end{center}
\caption{\footnotesize Here is a case of $p=3$. The cuts are shown in the $\zeta$ plane and the $\lambda$ plane, which represent the condensation of eigenvalues in matrix models. }
\label{multicutBCfigure}
\end{figure}

Differential equations of isomonodromy systems (Eq.~\eq{IMSinDefinitionOfMCBC}) are originally the BA system (Eq.~\eq{EqBAfunctionPQsystem}). Therefore, one of the solutions is given by orthogonal polynomials of matrix models, $\vec\psi_{\rm orth}(t;\lambda)$. 
By Gross-Migdal formula (Eq.~\eq{EquationGrossMigdalFormula}) \cite{GrossMigdal2}, the orthogonal polynomials are given by determinant operators which are essentially the resolvent operators (Eq.~\eq{EquationForResolventOpeartorsInEOrecursionSection}). Eigenvalue cuts are defined by the determinant operator \cite{MMSS}: 
\begin{Definition} [Eigenvalue branch cuts] \label{DefinitionOfPhysicalCuts}
For a given determinant operator, 
\begin{align}
\psi_N(x) \equiv \vev{\det(x-X)}_{N\times N},\qquad x \in \mathbb C, 
\end{align}
the eigenvalue distribution of the matrix $X$ is on the support $\mathfrak B \subset \mathbb C$, along which the large $N$ behaviors of left/right sides of the support $\mathfrak B$ (i.e.~$x_\pm \equiv x \pm \epsilon$, $x\in \mathfrak B$) are different: 
\begin{align}
\psi_N(x_\pm) \underset{N\to \infty}{\simeq} e^{N \phi(x_{\pm}) +O(N^0)}, \qquad\phi(x_+)\neq \phi(x_-) \qquad \bigl(x\in \mathfrak B\bigr). 
\end{align}
In particular, its difference gives the eigenvalue distribution function,
\begin{align}
d\bigl(\phi(x_+) - \phi(x_-)\bigr) = 2\pi i \rho(x) dx \in i \mathbb R. 
\end{align}
That is, $\mathfrak B$ is Stokes lines of determinant operator $\psi_N(x)$.  $\quad \square$
\end{Definition}
The meaning of Definition \ref{DefinitionMCBC} is then a requirement that the resolvent has $k$ ($\mathbb Z_k$-symmetric) branch cuts around $\lambda\to \infty$ \cite{CIY2}, as in the right figure of Fig.~\ref{multicutBCfigure}.

In $(p,q)$ minimal string theory, this multi-cut boundary condition should be also imposed \cite{CIY4}. This boundary condition requires that there appears $p$ branch cuts around $\lambda\to\infty$. Since $\lambda$ is related to the standard spectral parameter $\zeta$ as $\zeta = 2^{p-1} \lambda^p$, there is a single cut in the resolvent around $\zeta\to\infty$ (as in the left figure of Fig.~\ref{multicutBCfigure}). In the following discussion (especially in Section \ref{SectionStokesSpecific}), we should keep in mind that, {\em by double scaling limit, we focus on an edge of eigenvalue cuts and, at least on the top of critical points, the system always satisfies the multi-cut boundary condition for both $X$-system and $Y$-system. }

\subsection{Multi-cut BC for $(p,q)$ minimal string theory \label{SubsectionMulticutBCforPQminimalstringtheory}}

\begin{figure}[htbp]
\begin{center}
\includegraphics[scale=0.6]{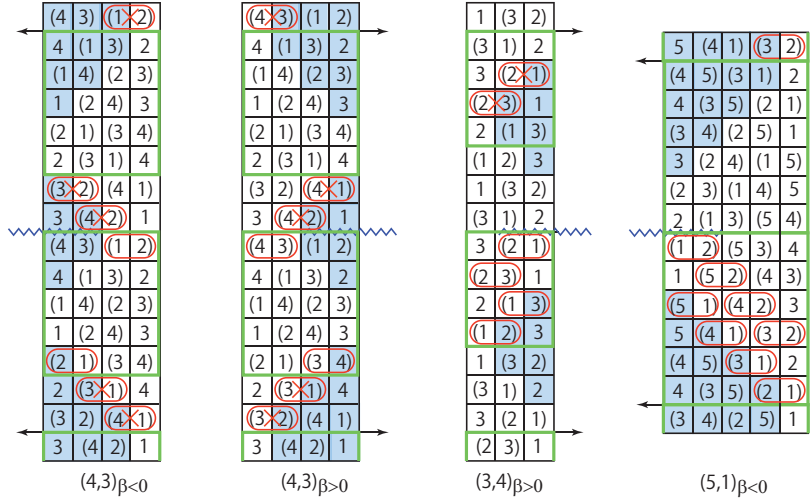}
\end{center}
\caption{\footnotesize Here is the multi-cut BC in the profile of $(4,3)$; $(3,4)$; $(5,1)$. Shaded boxes represent the vanishing coefficients $\bigl\{ v_n^{(j)} \bigr\}_{n,j}$ of the multi-cut BC. Stokes multipliers which are relevant to multi-cut BC equations are marked by circles. In particular, vanishing multipliers are marked with ``$\times$''. Green boxes are inner sectors of anti-Stokes lines defined in Definition \ref{DefInnerOuterSectors}.  }
\label{FigureMCBCs}
\end{figure}

General discussions on multi-cut BC are already given in \cite{CIY2}, and therefore, we here briefly review the procedure. 
\begin{itemize}
\item [1. ] By Proposition \ref{PropositionStokesMatricesCanonicalSolutions}, there are $2rp$ canonical solutions $\bigl\{ \Psi_n(t;\lambda)\bigr\}_{n=0}^{2rp-1}$, associated with Stokes sectors $\bigl\{ D_n \bigr\}_{n=0}^{2rp-1}$. Therefore, ${\vec \psi}_{\rm orth}(t,\lambda)$ in Definition \ref{DefinitionMCBC} is given as
\begin{align}
{\vec \psi}_{\rm orth}(t,\lambda) = \Psi_n(t;\lambda) {\vec v}_n\qquad \bigl(0 \leq n \leq 2rp-1\bigr). \label{ExpressionByCanonicalSolutionsInMCBC}
\end{align}
\item [2. ] In each Stokes sector $D_n$ ($0\leq n\leq 2rp-1$), the expression Eq.~\eq{ExpressionByCanonicalSolutionsInMCBC} has a canonical expansion: 
\begin{align}
{\vec \psi}_{\rm orth}(t,\lambda) \asymeq \Psi_{\rm asym}(t;\lambda) \vec v_n = \sum_{j=1}^p v_n^{(j)}\times {\vec \psi}_{\rm asym}^{(j)}(t,\lambda)\qquad \bigl(\lambda\to \infty \in D_n\bigr), 
\end{align}
where $\Psi_{\rm asym}(t;\lambda)$ is given in Eq.~\eq{CanonicalAsymExpansionFormulaInPropositionStokesCanonicalSolutions}, and we define 
\begin{align}
\Psi_{\rm asym}(t;\lambda) = \bigl( {\vec \psi}_{\rm asym}^{(1)}(t,\lambda),\cdots, {\vec \psi}_{\rm asym}^{(p)}(t,\lambda)\bigr),\qquad 
\vec v_n = 
\begin{pmatrix}
v_n^{(1)} \cr
\vdots \cr
v_n^{(p)}
\end{pmatrix}. \label{ExpressionByCanonicalSolutionsInMCBC22}
\end{align}
\item [3. ] Multi-cut BC is then expressed as 
\begin{align}
\vec{\psi}_{\rm orth} (t;\lambda) \asymeq v_n^{(j_a)} \times {\vec \psi}_{\rm asym}^{(j_a)}(t,\lambda) \qquad \lambda\to \infty \in D\bigl(\frac{(2a-1)\pi}{p}, \frac{(2a+1)\pi}{p} \bigr), \label{EqForBCInMCBCofMST}
\end{align}
where the index $a$ is re-defined according to the value of $\theta_0$: 
\begin{align}
a\in \mathbb Z \qquad \bigl(\text{$\theta_0 = \frac{\pi}{p}$}\bigr),\qquad a\in \mathbb Z + \frac{1}{2}\qquad \bigl(\text{$\theta_0 = 0$}\bigr). 
\label{LabelingOfAInMCBCDefinition}
\end{align}
This boundary condition gives a constraint on $\bigl\{ \vec v_n \bigr\}_{n=0}^{2rp-1}$ and, as a result, on Stokes matrices $\bigl\{ S_n \bigr\}_{n=0}^{2rp-1}$. 
\item [4. ] This boundary condition requires some of the components of the coefficient vectors $\bigl\{ \vec v_n \bigr\}_{n=0}^{2rp-1}$ should vanish. This information is drawn in the profile of dominant exponents, as in Fig.~\ref{FigureMCBCs}, where shaded boxes are those of vanishing coefficients: 
\begin{align}
{}^\exists\quad 
\begin{picture}(0,0)(5,5)
\includegraphics[scale=0.7]{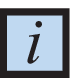} 
\end{picture}
\quad \in \mathcal J_n^{(p,r)}\qquad \Leftrightarrow \qquad v_n^{(i)} = 0.
\end{align}
\end{itemize}

\subsubsection{Existence of the boundary condition \label{SubsubsectionExistenceOfMCBC}}

We here discuss possible multi-cut BC's in $(p,q)$ minimal string theory. 
It is instructive to see some typical examples of the multi-cut BC. Three typical cases, $(p,q) =(4,3)$, $(3,4)$ and $(5,1)$, are shown in Fig.~\ref{FigureMCBCs}. Here are some notes on the boundary condition: 
\begin{itemize}
\item [1. ] Multi-cut BC should be consistent with $\mathbb Z_p$-symmetry of the system. Therefore, the boundary condition should be also symmetric in the following shift of indices: 
\begin{align}
v_n^{(i)} = 0\quad \Leftrightarrow \quad v_{n+2r}^{(i+1)}=0 \qquad \bigl(\text{or }\quad v_{n+2r}^{(i+1)}\neq 0 \quad \Leftrightarrow \quad v_n^{(i)}\neq 0\bigr). \label{ZpSymForVsinMCBC}
\end{align}
\item [2. ] If the eigenvalue branch cuts are along the angle, $\arg (\lambda) = \frac{(2a +1) \pi}{p}\,\bigl( \Leftrightarrow \, \arg (\zeta) = (2a+1)\pi\bigr)$, then the indices $\bigl\{j_a\bigr\}_{\text{$a$ of Eq.~\eq{LabelingOfAInMCBCDefinition}}}$ of Eq.~\eq{EqForBCInMCBCofMST} should satisfy  
\begin{align}
(j_{a+1}|j_{a}) \in \mathcal J_{(2a+1)r}^{(p,r)}\qquad \text{with}\qquad v_{(2a+1)r}^{(j_a)} \neq 0 \neq v_{(2a+1)r+1}^{(j_{a+1})}. \label{JaIndicesOfMCBC}
\end{align}
In particular, $v_{n}^{(j_a)} \neq 0$ for $(2a-1)r+1\leq n \leq (2a+1)r$. 
\item [3. ] The above two conditions are consistent with each other if the indices $\bigl\{j_a\bigr\}_{\text{$a$ of Eq.~\eq{LabelingOfAInMCBCDefinition}}}$ satisfy $j_{a+1} = j_a + 1$. 
Due to Eq.~\eq{EquationOfJLindicesInProfileStokesLinesExchanges}, one concludes 
\begin{align}
&(j_{a+1}|j_{a}) \in \mathcal J_{(2a+1)r}^{(p,r)} \qquad \Rightarrow \qquad 2j_a -1 \equiv 2a+1 \mod p\nn\\
&\qquad \therefore \qquad j_a = a +1 \quad\text{or}\quad j_a=\frac{2(a+1) + p}{2}. 
\end{align}
\end{itemize}
One then sees the following statement: 

\begin{Proposition} [Possible multi-cut BC] \label{PropositionPossibleMCBC}
Consider $(p,q)$-systems and $r\,(=p+q)=mp+l$ $(1\leq l\leq p-1)$. 
Multi-cut BC's exist (only) in the following cases: 
\begin{itemize}
\item [1. ] \underline{Even $p$}
\begin{itemize}
\item [a) ]   $j_a = a+1$  \qquad  with \qquad $\sgn(\beta_{p,q})=(-1)^m$,\qquad $\theta_0=\frac{\pi}{p}$. 
\item [b) ]   $j_a = a+\dfrac{p}{2}+1$ \qquad with \qquad   $\sgn(\beta_{p,q})=(-1)^{m+1}$,\qquad $\theta_0=\frac{\pi}{p}$. 
\end{itemize}
\item [2. ] \underline{Odd $p$}
\begin{itemize}
\item [c) ] $j_a = a+1$  \qquad with\qquad   $\sgn(\beta_{p,q})=(-1)^m$,\qquad $\theta_0=\frac{\pi}{p}$. 
\item [d) ] $j_a = \dfrac{2a+p}{2} + 1$\qquad with \qquad   $\sgn(\beta_{p,q})=(-1)^{m+r}$,\qquad $\theta_0=0$. 
\end{itemize}
\end{itemize}
$\square$
\end{Proposition}
{\em Proof}\quad The remaining statement is on the sign of $\beta_{p,q}$. We show it only in the case of odd $p$ with $\theta_0 = \frac{\pi}{p}$, but other cases can be also carried out similarly. In this case, components of the positive profile, $j_{n,l} \in |\mathcal J_n^{(p,r)}|$, satisfy  
\begin{align}
j_{0,p} = j_{1,p} = j_0 = 1,\qquad j_{2r,p} = j_{2r,p} = j_1 = 2. 
\end{align}
The branch cut along $\arg(\lambda)= \frac{r \pi}{rp}$ is formed by the change of dominance among $\varphi^{(1)}$ and $\varphi^{(2)}$. Then one can see that the sign of $\beta_{p,q}$ should  be $\sgn(\beta_{p,q}) = (-1)^m$, where $r=mp+l$. This is the statement. $\quad \blacksquare$

Here is also some notes: 
\begin{itemize}
\item [I. ] \underline{\em Even $p$} \quad Typical cases, $(p,q)=(4,3)$, are shown in Fig.~\ref{FigureMCBCs}. Note that the two multi-cut BC's of (a) and (b) are essentially the same since they are related by a rename of the index: 
\begin{align}
\varphi_{\beta<0}^{(j)}(\lambda) = \varphi_{|\beta|}^{(j+ \frac{p}{2})}(\lambda). 
\end{align}
Other examples of even $p$ cases are the cases of Kazakov series $(2,2l+1)$. 
\item [II. ] \underline{\em Odd $p$} \quad  
The typical cases, $(p,q)=(3,4)$ and $(5,1)$, are shown in Fig.~\ref{FigureMCBCs}. 
The two multi-cut BC's of (c) and (d) are related as follows: By shifting $\tau$ (i.e.~$\tau \to \tau + \pi i$) of 
\begin{align}
\zeta = \sqrt{\mu} \cosh(p\tau),\qquad \del_\zeta\varphi^{(0)}(t;\zeta) =(-1)^m|\beta_{p,q} |\,\mu^{\frac{q}{2p}} \cosh(q\tau). 
\end{align}
one obtains the spectral curve with negative sign in front of $\zeta$: 
\begin{align}
\zeta = -\sqrt{\mu} \cosh(p\tau),\qquad \del_\zeta\varphi^{(0)}(t;\zeta) =(-1)^{m+q}|\beta_{p,q} |\,\mu^{\frac{q}{2p}} \cosh(q\tau). 
\end{align}
This is the second type of multi-cut BC. In this sense, these boundary conditions are again essentially the same. 
\end{itemize}

We further consider consistency with the spectral $p-q$ duality. As discussed before, the spectral duality relates $\beta_{p,q}$ and $\beta_{q,p}$ in the following way:
\begin{align}
\beta_{p,q} = (-1)^r \beta_{q,p}. \label{BetaPQdualityEquationSectionMCBC}
\end{align}
Therefore, one obtains the following result: 
\begin{Theorem} [Duality-consistent sign of $\beta_{p,q}$]
\label{ThmBetaValues}
Consider $(p,q)$-systems with $p<q$, and $r \, (=p+q) = mp + l$.  Consistent multi-cut BC's are obtained on both $(p,q)$- and $(q,p)$-systems in the following cases: 
\begin{itemize}
\item [1.] \underline{Even $r\, (=p+q)$} \quad 
\begin{itemize}
\item [A) ] If $m\in 2\mathbb Z+1$ then  
\begin{align}
\sgn (\beta_{p,q}) =-1,\qquad  \sgn(\beta_{q,p}) = -1. 
\end{align}
\item [B) ] If $m\in 2\mathbb Z$ then there is no consistent BC. 
\end{itemize}
\item [2.] \underline{Odd $r\, (=p+q)$}
\begin{itemize}
\item [C) ] If $p$ is even then
\begin{align}
\sgn (\beta_{p,q}) = +\epsilon,\qquad  \sgn(\beta_{q,p}) = -\epsilon
\end{align}
\item [D) ] If $p$ is odd then
\begin{align}
\sgn (\beta_{p,q}) = (-1)^m\epsilon,\qquad  \sgn(\beta_{q,p}) = (-1)^{m+1}\epsilon, 
\end{align}
\end{itemize}
where $\epsilon$ is given by 
\begin{align}
\epsilon = 
\left\{
\begin{array}{cl}
+1 & \text{: multi-cut BC of (c)-type} \cr
-1 & \text{: multi-cut BC of (d)-type}
\end{array}
\right.. 
\end{align}
\end{itemize}
$\square$
\end{Theorem}
{\em Proof}\quad Consider $r=mp+l = \widetilde m q + \widetilde l$. Since $q>p$, one obtains $\widetilde m=1$ and $\widetilde l=p$. \\
\underline{\em 1. Even $r$}\quad  Here $(p,q)$ are both odd integers. Proposition \ref{PropositionPossibleMCBC} gives 
\begin{align}
\sgn(\beta_{p,q}) = (-1)^m,\qquad \sgn (\beta_{q,p}) = (-1)^{\widetilde m} = -1, 
\end{align}
whatever types of multi-cut BC's. By duality relation of Eq.~\eq{BetaPQdualityEquationSectionMCBC}, one obtains that $(-1)^m = (-1) \,\, \Leftrightarrow\,\, m\in 2\mathbb Z+1$, which is the statement. \\
\underline{\em 2. Odd $r$}\quad {\em C)} In this case of even $p$, we have $\sgn(\beta_{q,p}) = (-1)^{\widetilde m} \epsilon = -\epsilon$. Therefore, by the duality relation of Eq.~\eq{BetaPQdualityEquationSectionMCBC}, one obtains that $\sgn(\beta_{p,q})=+\epsilon$. {\em D)} In this case of odd $p$, we have $\sgn(\beta_{p,w}) = (-1)^{m} \epsilon$. Therefore, by the duality relation of Eq.~\eq{BetaPQdualityEquationSectionMCBC}, one obtains that $\sgn(\beta_{p,q})=(-1)^{m+1}\epsilon$.
$\quad \blacksquare$

If one only uses the multi-cut BC of (c)-type, then Kazakov series $(2,2k+1)$ are given as 
\begin{align}
\sgn(\beta_{2,2k+1}) = +1,\qquad \sgn(\beta_{2k+1,2}) = -1, 
\end{align}
and unitary series $(p,p+1)$ are given as
\begin{align}
\sgn(\beta_{p,p+1}) = +1,\qquad \sgn(\beta_{p+1,p}) = -1. 
\end{align}
It is interesting to compare this values with those of matrix models, since there are explicit constructs of the critical points for Kazakov series and unitary series in two-matrix models \cite{DKK}. 

It is further interesting that there is a restriction on $(p,q)$: For example, the (3,5) and (5, 3) models cannot be connected by $p-q$ duality with keeping a single cut in the $\zeta$ and $\eta$ plane. Note that, if $(p,q)$ are both odd, then some systems are not consistent with $p-q$ duality. However, with taking into account a suggestive relation with type 0 $(p,q)$ minimal superstring theory of odd models \cite{SeSh,fi1}, our result implies that the systems should be embedded into minimal superstring theory. In this way, one expects that the $p-q$ duality may complete the systems by higher embeddings into $(p,q)$ minimal $k$-fractional superstring theory shown in \cite{irie2}.

\subsubsection{Multi-cut BC equations for Stokes multipliers \label{SubsubsectionMCBCequation}}

As in \cite{CIY2}, we evaluate the multi-cut BC equations from 
\begin{align}
\vec v_n = S_n \vec v_{n+1} \qquad \Leftrightarrow \qquad v_n^{(j)} = v_{n+1}^{(j)} + \sum_{l \text{ of }{}^\exists (l|j) \in \mathcal J_n} s_{n,j,l}\times v_{n+1}^{(l)}, 
\end{align}
and the boundary condition given by the shaded boxes in the profiles. The multi-cut BC equations are then expressed as 

\begin{Proposition} [Multi-cut BC equations] \label{PropositionMultiCutBCequation}
Based on $\theta$-parameters defined in Definition \ref{DefinitionOfThetaParametrization}, the multi-cut BC equations are given as follows: 
\begin{align}
1) \qquad & 
\theta_{n}^{(r+l-1+ 2ra)} v_{2r(a+1)}^{(j_{a+1})} + \theta_{n-l}^{(r-1-l+2ra)} v_{2ra}^{(j_a)}=0, \qquad \bigl(1\leq n \leq p+l-1\bigr) \label{EqMCBC1stEq} \\
2)\qquad  & 
\theta_{n}^{(r+l-1 + 2pb+2ra)} = 0\qquad \bigl(1\leq n \leq p-1;\,1 \leq b \leq m-1\bigr) 
\end{align}
Here $v_{2ra}^{(j_a)}\neq 0$, and 
\begin{align}
a\in \mathbb Z \qquad \bigl(\text{$\theta_0 = \frac{\pi}{p}$}\bigr),\qquad a\in \mathbb Z + \frac{1}{2}\qquad \bigl(\text{$\theta_0 = 0$}\bigr). 
\end{align}
By the monodromy cyclic equation Eq.~\eq{MonodromyEquation} and $\mathbb Z_p$-symmetry Eq.~\eq{StokesSymmetryEquationsZpAndHermiticity}, the coefficients $\bigl\{v_{2ra}^{(j_a)}\bigr\}_{a=1}^p$ are solved as 
\begin{align}
v_{2r(a+1)}^{(j_{a+1})} = \rho^{-1} \times v_{2ra}^{(j_a)},\qquad \rho^p = (-1)^{p-1}, 
\end{align}
and Eq.~\eq{EqMCBC1stEq} is given as 
\begin{align}
\left\{
\begin{array}{rll}
1) \quad &
\theta_{n}^{(r+l-1 + 2ra)} = 0\qquad \theta_{p-n}^{(r-l-1+2ra)} = 0 & (1 \leq n \leq l-1) \cr
2) \quad &
\theta_l^{(r+l-1+2ra)} = \rho\qquad \theta_{p-l}^{(r-l-1+2ra)} = -1/\rho  \cr
3) \quad &
\theta_{l+n}^{(r+l-1+2ra)} + \rho \,\theta_{n}^{(r-l-1+2ra)}=0 & (1\leq n  \leq p-l-1) 
\end{array}
\right.,\label{EqMCBC1stEqPart2}
\end{align}
for $a \in \mathbb Z/p \mathbb Z$ or $a \in \mathbb Z/p \mathbb Z + \frac{1}{2}$. 
$\quad \square$
\end{Proposition}
{\em Proof}\quad 
The proof is essentially the same as \cite{CIY2,CIY3}. It is a little bit a tedious task (even though the resulting equations are simple), so we skip it here. $\quad \blacksquare$

The Stokes multipliers which are relevant to multi-cut BC equations are marked by red circle in the profile (as in Fig.~\ref{FigureMCBCs}). In particular, some of multipliers are required to vanish and therefore are marked with ``${\color{red}\times}$''. Green boxes are {\em inner sectors of anti-Stokes lines} which are defined in Definition \ref{DefInnerOuterSectors}. 

Note that the result of the boundary condition (Eq.~\eq{EqMCBC1stEqPart2}) is particularly simple compared with the cases of ($k>r$) \cite{CIY2,CIY3} where quantum integrability takes place. The current cases ($k<r$) are also understood as a T-system of quantum integrability \cite{CIY3}; however, the multi-cut BC does not possess ``$\mathbb Z_r$-symmetry'' which was important structure for adding complementary boundary conditions. Therefore, the situations of minimal strings ($k<r$) are completely different from the cases of $k>r$.


\section{Riemann-Hilbert analysis and weaving networks \label{SectionDZmethodAndRHP}}
In this section, we discuss the explicit solutions of Stokes matrices (i.e.~spectral networks) which describe non-perturbative completions of $(p,q)$ minimal string theory. For that purpose, we first review and summarize Deift-Zhou's method \cite{DZmethod} for the Riemann-Hilbert problem (RHP) \cite{RHcite,JimboMiwaUeno} in Section \ref{SubsubsectionRHPandImprovedDZmethod}. 

From Section \ref{SubsectionProperSpectralNetworks}, we start to discuss a sequence of spectral networks in $(p,q)$ minimal string theory, with the procedure of {\em weaving networks}. In this procedure, we define a concept of {\em proper spectral networks}, which are essentially introduced to parametrize the possible non-perturbative completions of $(p,q)$ minimal string theory. After that, we evaluate the RH integrals and discuss how such a choice of solutions is ``proper'' to describe the system. In the last two sections, we demonstrate these discussions in two basic examples of $(p,q)=(p,1)$ and $(p,2)$.  

\subsection{Deift-Zhou's steepest descent method \label{SubsubsectionRHPandImprovedDZmethod}}
Here we review the RH approach based on \cite{ItsBook}. 
Given spectral curves and spectral networks, we can formulate the RHP of isomonodromy systems. 
We first recall the definition of Riemann-Hilbert problem and its integrals: 

\begin{Definition}[Riemann-Hilbert problem] \label{DefinitionRHP}
For a given oriented graph $\mathcal K$ and a $p\times p$ matrix-valued non-singular function $\mathcal G(\lambda)$ on the graph ($\lambda \in \mathcal K$), the Riemann-Hilbert problem is a problem to find a $p\times p$ matrix-valued (sectional) holomorphic function $\chi(\lambda)$ in $\lambda \in \mathbb C\setminus \mathcal K$ which satisfies 
\begin{align}
\left\{
\begin{array}{rl}
(1) & \chi(\lambda) = I_p + O(\lambda^{-1})\qquad \lambda \to \infty \cr
(2) & \chi(\lambda_+) = \chi(\lambda_-) \,\mathcal G(\lambda) \qquad \lambda \in \mathcal K 
\end{array}
\right.. 
\end{align}
where $\lambda_+$ and $\lambda_-$ are $\lambda\in {\mathcal K}$ approaching from the left-hand side ($\lambda_+$) and the right-hand side ($\lambda_-$). Therefore, RHP is defined by the pair $\bigl(\mathcal K,\mathcal G(\lambda)\bigr)$. $\square$ 
\end{Definition}
The RHP is then formally solved by the following singular integral equation (See e.g.~\cite{ItsBook}): 
\begin{Proposition} [Riemann-Hilbert integral]
The RHP of $\chi(\lambda)$ can be formally solved by a holomorphic function $\rho(\lambda)$ on $\lambda \in \mathcal K$ as 
\begin{align}
\chi(\lambda) = I_p+ \int_{\mathcal K} \frac{d\xi}{2\pi i} \frac{\rho(\xi)(\mathcal G(\xi)-I_p)}{\xi-\lambda}, \label{RHintegralChi}
\end{align}
where $\rho(\lambda)$ is given by $\rho(\lambda) = \chi(\lambda_-)$ ($\lambda\in \mathcal K$) and therefore satisfies the following singular equation: 
\begin{align}
\rho(\lambda) = I_p+ \int_{\mathcal K} \frac{d\xi}{2\pi i} \frac{\rho(\xi)(\mathcal G(\xi)-I_p)}{\xi-\lambda_-}, \label{RHintegral}
\end{align}
where $\lambda_-$ is $\lambda \in \mathcal K$ approaching from its right-hand side. $\square$
\end{Proposition}
The RHP of isomonodromy systems (with string-coupling $g$) is then formulated as follows: 

\begin{Definition}[$M$-th order $\mathcal G$-function]
For a given spectral curve $\varphi(\lambda)\in \mathcal L_{\rm str}^{\rm (univ.)}(t)$ and spectral network, $\hat {\mathcal K}= \bigl(\mathcal K,G\bigr)$, one defines $M$-th order $\mathcal G$-function $\mathcal G^{(M)}(g;\lambda)$ on $\mathcal K$ as 
\begin{align}
\mathcal G^{(M)}(g;\lambda) = \Psi_{\rm pert}^{(M)}(g;\lambda_-)\, G(\lambda) \Psi_{\rm pert}^{(M)}(g;\lambda_+)^{-1}\qquad \bigl(\lambda \in \mathcal K\bigr), \label{DefinitionEquationMthOrderGfunction}
\end{align}
where $\Psi_{\rm pert}^{(M)}(g;\lambda)$ is the $M$-th order perturbatively corrected BA function around the spectral curve $\varphi(\lambda)$ (in string coupling $g$) given by 
\begin{align}
\Psi_{\rm pert}^{(M)}(g;\lambda)& = \underbrace{\Bigl[I_p + \sum_{n=1}^M g^n Z^{(n)}(\lambda)\Bigr]}_{\ds \equiv Z_{\rm pert}^{(M)}(g;\lambda)} \underbrace{Z_{\rm cl}(\lambda) \exp\bigl[\frac{1}{g}\varphi(\lambda)\bigr] \lambda^{-\nu}}_{\ds \equiv \Psi_{\rm cl}(g;\lambda)} \nn\\
&\equiv Z_{\rm pert}^{(M)}(g;\lambda) \Psi_{\rm cl}(g;\lambda). \label{EqDefinitionPerturbativeBAfunctions}
\end{align}
$\square$
\end{Definition}

\begin{Proposition} [RHP of isomonodromy systems]
Consider the RHP defined by the pair $\bigl(\mathcal K,\mathcal G^{(M)}\bigr)$ (as in Definition \ref{DefinitionRHP}) and the solution is denoted by $\chi^{(M)}(g;\lambda)$. Then the following sectional holomorphic function $\Psi_{\rm RH}(g;\lambda)$ on $\mathbb CP^1\setminus \mathcal K$\begin{align}
\Psi_{\rm RH}(g;\lambda) = \chi^{(M)}(g;\lambda) \Psi_{\rm pert}^{(M)}(g;\lambda)\qquad \lambda \in \mathbb CP^1\setminus \mathcal K, \label{EqSolutionOfRHinIMS}
\end{align}
solves the isomonodromy system, Eq.~\eq{EqIsomonodromyMinimalStrings1}, with general Lax operators of Eq.~\eq{EqGeneralBoperator}. $\quad \square$
\end{Proposition} 
Deift-Zhou's steepest descent method \cite{DZmethod} for evaluating the RH integral equation is based on the following consideration: The resulting BA function $\Psi_{\rm RH}(g;\lambda)$ (of Eq.~\eq{EqSolutionOfRHinIMS}) does not depend on the following two kinds of deformations in RHP: 
\begin{itemize}
\item [1. ] Equivalence deformations of DZ networks $\hat {\mathcal K}$ (Definition \ref{DefDeformEquiv}) \cite{DZmethod}: $\hat{\mathcal K} \to  \hat{\mathcal K}' \sim \hat{\mathcal K}. $
It is because such deformations are analytic continuation of integration contours.  
\item [2. ] Deformations of the spectral curve $\varphi(\lambda)$ within the universal landscape \cite{ItsKapaev}: $\varphi(\lambda) \to  \varphi'(\lambda)\in \mathcal L_{\rm str}^{\rm (univ)}(t)$. It is because such deformations are just changing the reference frame of Eq.~\eq{EqSolutionOfRHinIMS}
\end{itemize}
One therefore considers 
\begin{itemize}
\item [1. ] Adjust the DZ network $\hat {\mathcal K}$ to the steepest descent curve of $\varphi(\lambda)\in \mathcal L_{\rm str}^{\rm (univ)}(t)$ \cite{DZmethod}. 
\item [2. ] Choose $\varphi(\lambda)\in \mathcal L_{\rm str}^{\rm (univ)}(t)$ such that it can best approximate Eq.~\eq{RHintegral} \cite{ItsKapaev}. 
\end{itemize}
This searching procedure of $\varphi(t;\lambda) \in \mathcal L_{\rm str}^{\rm (univ)}(t)$ in the Deift-Zhou method corresponds to the vacuum-search in string-theory landscape. The background independence of matrix models \cite{EynardMarino} can be also seen simply in the Riemann-Hilbert approach. 

\subsection{Proper spectral networks \label{SubsectionProperSpectralNetworks}}

It is convenient to express Eq.~\eq{RHintegralChi} and Eq.~\eq{RHintegral} as
\begin{align}
\chi(g;\lambda) = I_p + K\Bigl[\rho(g;\cdot)\Bigr](\lambda),\qquad 
\rho(g;\lambda) = I_p + K_-\Bigl[\rho(g;\cdot)\Bigr](\lambda),
\end{align}
with use of kernels: 
\begin{align}
K\bigl[f\bigr] (\lambda) &= \int_{\mathcal K} \frac{d\xi}{2\pi i} f(\xi)\frac{ (\mathcal G(g;\xi)-I_p)}{\xi - \lambda},\qquad \bigl(\lambda\in \mathbb C\setminus \mathcal K\bigr)\\
K_\pm \bigl[f\bigr] (\lambda)& \equiv K\bigl[f\bigr] (\lambda_\pm),\qquad \bigl(\lambda\in \mathcal K\bigr). 
\end{align}
For further discussions, we express the the RH integral by matrices of Stokes-type and Cut-type matrices of DZ networks (given in Definition \ref{DefinitionStokesCutMatrices}) \cite{DZmethod,ItsKapaev}. 
For the sake of component representation, we introduce the following notation about decomposition of indices into Stokes/Cut types:
\begin{align}
\bigl\{1,2,\cdots,\#\mathcal K\bigr\} = \Stokes \cup \Cut. 
\end{align}
Here $\bigl\{\vec{\mathcal K}_m\bigr\}_{m\in \Stokes}$ means that the corresponding contour elements are those of Stokes-type; and $\bigl\{\vec{\mathcal K}_m\bigr\}_{m\in \Cut}$  means that those are of Cut-type. 

\begin{Lemma} [Component representation of RH kernel]
The Riemann-Hilbert kernel can be expressed as 
\begin{align}
K^{(M)}[f](\lambda) =& \sum_{m\in \Stokes} \sum_{j,l}\alpha_{m,j,l}\int_{\vec{\mathcal K}_m} \frac{d\xi}{2\pi i} f(\xi)\frac{\Xi_{j,l}^{(M)}(g;\xi) e^{\frac{1}{g}\varphi^{(j,l)}(\xi)}}{\xi-\lambda}  + \nn\\
&+ \sum_{m\in \Cut}\int_{\vec{\mathcal K}_m} \frac{d\xi}{2\pi i}f(\xi) \frac{\bigl(\Psi_{\rm pert}^{(M)}(g;\xi_-)G_m\Psi_{\rm pert}^{(M)}(g;\xi_+)^{-1} - I_p\bigr)}{\xi-\lambda},
\label{EqRHkernel}
\end{align}
where $\Xi_{j,l}^{(M)}(g;\xi)$ is given by the perturbative BA function $\Psi_{\rm pert}^{(M)}(g;\xi)$ as 
\begin{align}
\Xi_{j,l}^{(M)}(g;\xi) e^{\frac{1}{g}\varphi^{(j,l)}(\xi)}\equiv \Psi_{\rm pert}^{(M)}(g;\xi) E_{j,l} \Psi_{\rm pert}^{(M)}(g;\xi)^{-1}, 
\end{align}
and $\alpha_{m,j,l}$ is the weight of a Stokes tail $<j,l>$ along $\vec{\mathcal K}_m$ $(m\in \Stokes)$. $\quad \square$
\end{Lemma}
We consider adjusting the spectral networks to anti-Stokes lines (Definition \ref{DefAntiStokesLines}). The following concept is then important in discussion below: 

\begin{Definition}[Proper networks] \label{DefinitionProperNetworks}
For a given spectral curve $\varphi(t;\lambda) \in \mathcal L_{\rm str}^{\rm (univ.)}(t)$, if a spectral network $\hat {\mathcal K}$ satisfies the following conditions: 
\begin{itemize}
\item [1. ] All the Stokes tails and cut lines of the network, $\hat {\mathcal K}$, can be adjusted to anti-Stokes lines and branch cuts of the spectral curve $\varphi(\lambda)$. 
\item [2. ] Cut-type matrices of the networks, $\bigl\{G_m\bigr\}_{m\in \Cut}$, are given by classical monodromy matrices of the the classical BA function $\Psi_{\rm cl}(g;\lambda)$:
\begin{align}
\Psi_{\rm cl}(g;\lambda_+) = \Psi_{\rm cl}(g;\lambda_-) G_m \qquad \lambda \in \vec{\mathcal K}_m\qquad m\in \Cut. 
\end{align}
\end{itemize}
then the network is called a proper (spectral) network of the spectral curve $\varphi(t;\lambda) \in \mathcal L_{\rm str}^{\rm (univ.)}(t)$. $\quad \square$
\end{Definition}
With a proper network of a spectral curve, the kernel integral \eq{EqRHkernel} becomes an integral only along Stokes tails:  
\begin{align}
K^{(M)}[f](\lambda) = \sum_{j\neq l}\sum_{m\in \Stokes}\alpha_{m,j,l}\int_{\vec{\mathcal K}_m} \frac{d\xi}{2\pi i}f(\xi) \frac{\Xi_{j,l}^{(M)}(g;\xi) e^{\frac{1}{g}\varphi^{(j,l)}(\xi)}}{\xi-\lambda}.  \label{EquationKernelOnlyStokeTails}
\end{align}
This situation is said that the spectral curve is consistent with the DZ network. This choice of spectral networks should be justified. We will come back to this issue in Section \ref{SectionLocalRHP}. 
For later convenience, we define the following decomposition of the kernel: 
\begin{align}
K^{(M)}[f] (\lambda) = \sum_{j<l} K^{(j,l|M)}[f](\lambda), \label{ComponentDecompositionOfKernelEquationJL}
\end{align}
where $K^{(j,l|M)}[f](\lambda) = K^{(l,j|M)}[f](\lambda)$ is the kernel contributed by $\varphi^{(j,l)}$ and $\varphi^{(l,j)}$: 
\begin{align}
K^{(j,l|M)}[f](\lambda) =  \sum_{m\in \Stokes}\Bigl[& \alpha_{m,j,l}\int_{\vec{\mathcal K}_m} \frac{d\xi}{2\pi i}f(\xi) \frac{\Xi_{j,l}^{(M)}(g;\xi) e^{\frac{1}{g}\varphi^{(j,l)}(\xi)}}{\xi-\lambda} +\nn\\
&+ \alpha_{m,l,j}\int_{\vec{\mathcal K}_m} \frac{d\xi}{2\pi i}f(\xi) \frac{\Xi_{l,j}^{(M)}(g;\xi) e^{\frac{1}{g}\varphi^{(l,j)}(\xi)}}{\xi-\lambda} \Bigr]. 
\end{align}

\subsection{Weaving a proper spectral network \label{SectionWeavingSpectralNetworks}}
In this subsection, based on the classical monodromy matrices (i.e.~Eq.~\eq{EqCutJumpMinimalStrings}), we generate a sequence of proper networks consistent with Chebyshev solutions in $(p,q)$ minimal string theory. 

\subsubsection{Classical monodromy matrices revisited}

\begin{figure}[htbp]
\begin{center}
\includegraphics[scale=1.2]{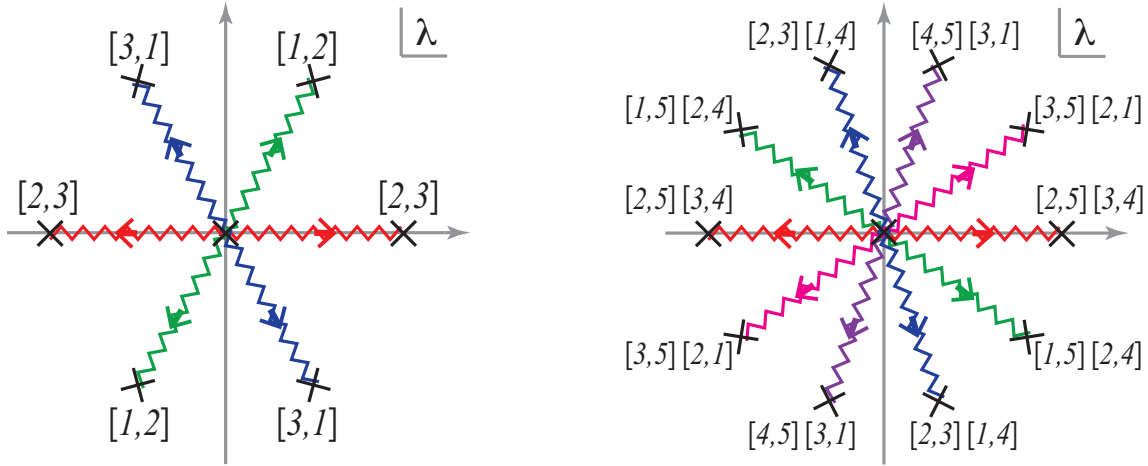}
\end{center}
\caption{\footnotesize Cut-jump structure of $p=3$ and $p=5$ classical BA function. Here we skip writing the phases since they are all canonical (See Definition \ref{DefCanoWeiPha}). }
\label{FigureCBAp3}
\end{figure}

As we have calculated in Section \ref{SubsectionClassicalBAFunction}, classical monodromy matrices of the Chebyshev solutions are obtained in Theorem \ref{TheoremClassicalMonodromyMatrices}. Since these matrices are parts of proper spectral networks (Definition \ref{DefinitionProperNetworks}), it is natural to draw as in Fig.~\ref{FigureCBAp3}. The examples are of $p=3$ and of $p=5$. The classical cyclic equation (Eq.~\eq{EqClassicalCyclicEquation}) is then naturally understood as a part of the cyclic monodromy equation (Eq.~\eq{MonodromyEquation}). 

Furthermore, in the expression of the cut-jump matrices, there appears a particular phase in its matrix of Cut-type. Here we regard it as a canonical value for spectral networks:

\begin{Definition}[Canonical weight/phase] \label{DefCanoWeiPha}
For Stokes tails of $<l,j>$ or cut lines of $[l,j]$, the following weight/phase $\omega_{l,j}$ is called the canonical weight/phase: 
\begin{align}
\omega_{l,j} \equiv 
\left\{
\begin{array}{ll}
\omega^{-\frac{|l-j|}{2}} & (l>j) \cr
-\omega^{\frac{|l-j|}{2}} & (j>l)
\end{array}
\right.. 
\end{align}
In particular, if Stokes tails or Cut lines have these canonical weight/phase, we skip writing the value of weight/phase in the DZ networks: 
\begin{align}
\begin{array}{c}
\includegraphics[scale=1]{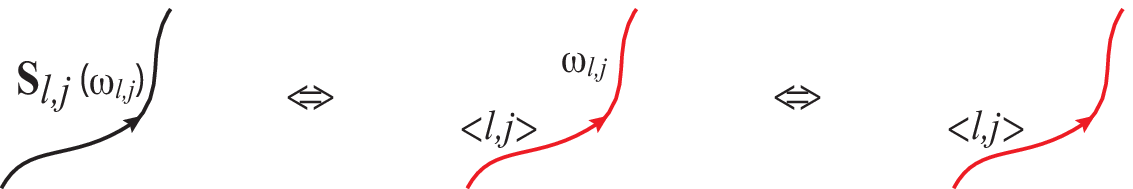}
\end{array}, \\
\begin{array}{c}
\includegraphics[scale=1]{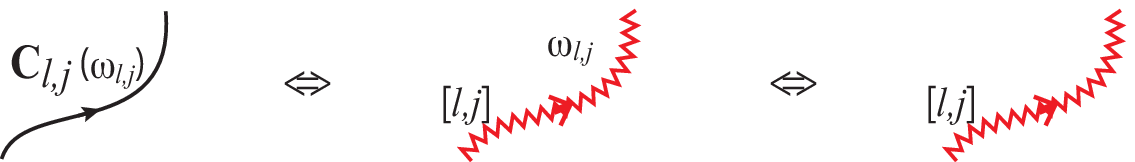}
\end{array}
\end{align}
$\square$
\end{Definition}

\subsubsection{Enhancing the classical monodromy matrices}

\begin{figure}[htbp]
\begin{center}
\includegraphics[scale = 0.9]{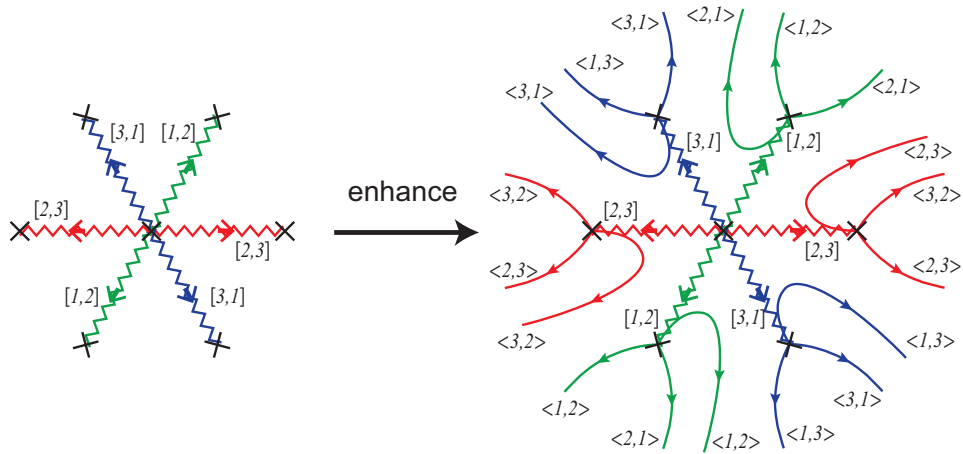}
\end{center}
\caption{\footnotesize  Enhancement of the cut-jump matrices in the $(3,4)$ case. }
\label{FigureWeaving1Example}
\end{figure}

In order to obtain a proper network (defined in Definition \ref{DefinitionProperNetworks}), it is natural to add {\em three Stokes tails} spreading from a branch point of classical monodromy lines (of Fig.~\ref{DefCanoWeiPha}):%
\footnote{Here the three-junction rule is used but more general branch points should be analyzed by the combinations of these basic three junctions. See also a similar situation for generalized topological recursions \cite{GeneralizedRecursions}}
\begin{align}
\begin{array}{c}
\includegraphics[scale = 0.7]{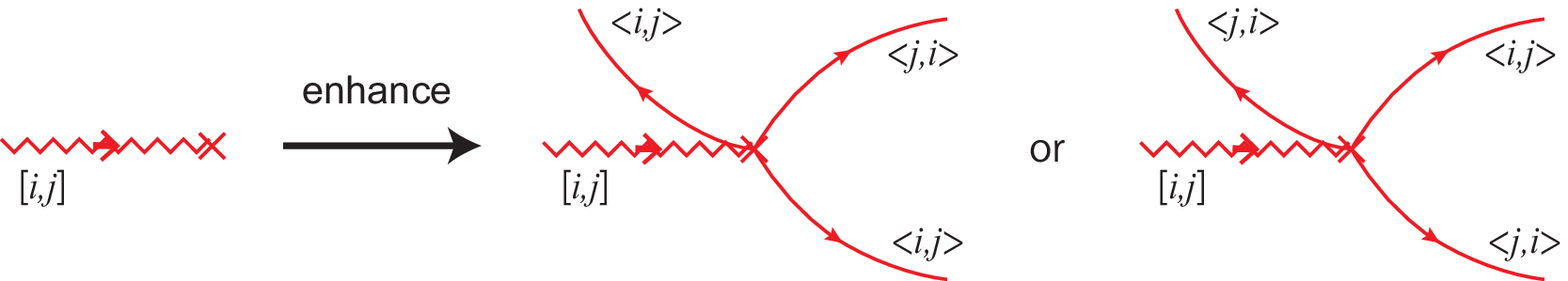}
\end{array}. \label{EquationFigureEnhancements1}
\end{align}
In fact, by this procedure, the resulting graph satisfy the matrix conservation law (of Eq.~\eq{EqNetworkRulesThreeJunctions}). 
Enhancement of the $(3,4)$ case is also shown in Fig.~\ref{FigureWeaving1Example}. 

The next problem is to adjust the additional Stokes tails to anti-Stokes lines. Since geometry of anti-Stokes lines is already studied in Section \ref{SubsubsectionGeometryOfAntiStokesLines}, it is not difficult to carry it out. Here we force on a special class of solutions, referred to as {\em primitive solution}. 

They are classified into two cases: We first show it in the cases of $p=2$. 

\begin{Definition} [Primitive solution of Kazakov series] \label{DefinitionPrimitiveSolutionPeq2}
Consider $(2,q)$-systems. The following proper network is called primitive solution: 
\begin{itemize}
\item [I. ] \underline{The case of $(2,q)=(2,1)$ } This is Airy system and is in Fig.~\ref{FigureDZAiry} (in Section \ref{SubsubsectionGeneralizedAirySystems}). 
\item [II. ] \underline{The case of $(2,q)$ $(q>2)$ } The cases of $r = p+q=m p+1$ with $m \in 2\mathbb Z+1$: 
\begin{align}
\begin{array}{c}
\includegraphics[scale = 0.8]{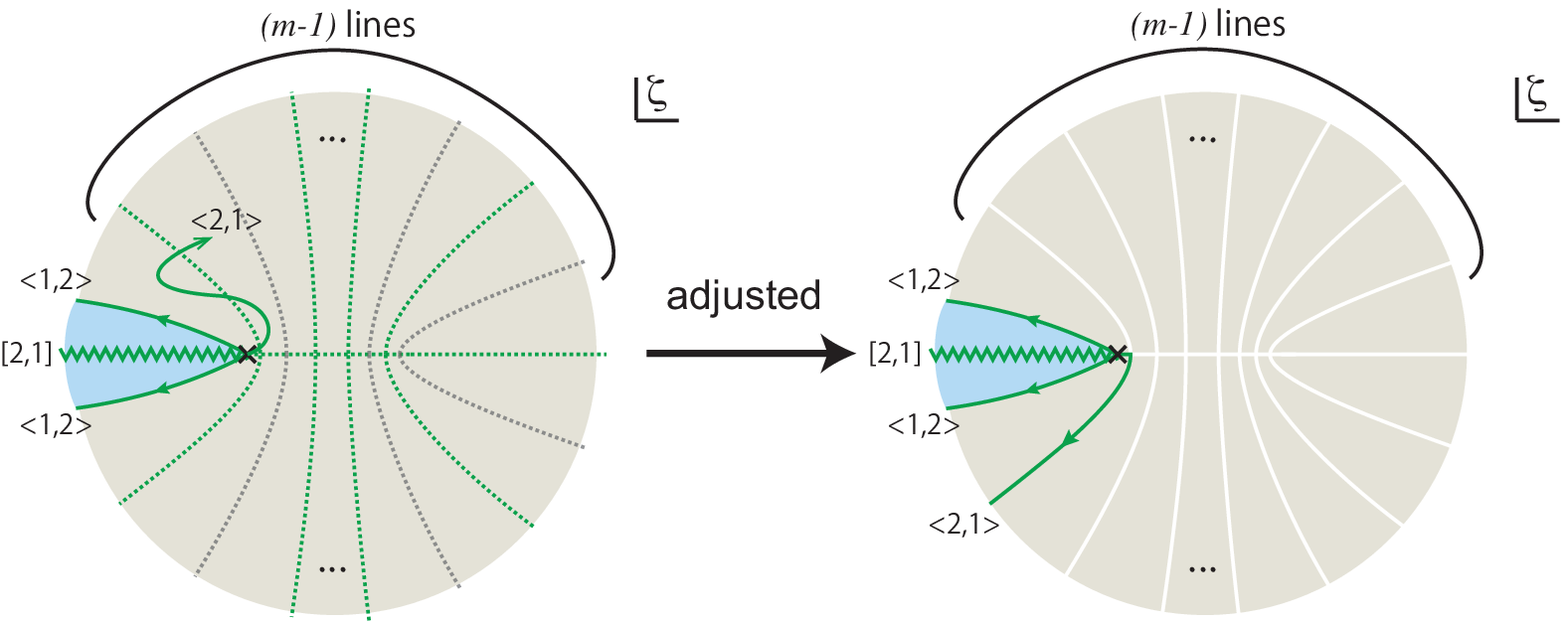}
\end{array}. 
\end{align}
Here green dashed lines are anti-Stokes lines along which the Stokes tail $<2,1>$ can be adjusted; and grey dashed lines are anti-Stokes lines along which the Stokes lines $<1,2>$ can be adjusted (See Proposition \ref{PropositionProfileAntiStokesLines}). 

The cases with $m \in 2\mathbb Z$ are also essentially the same around the branch point, except for the labeling of $<1,2>$ or $<2,1>$: 
\begin{align}
\begin{array}{c}
\includegraphics[scale = 0.8]{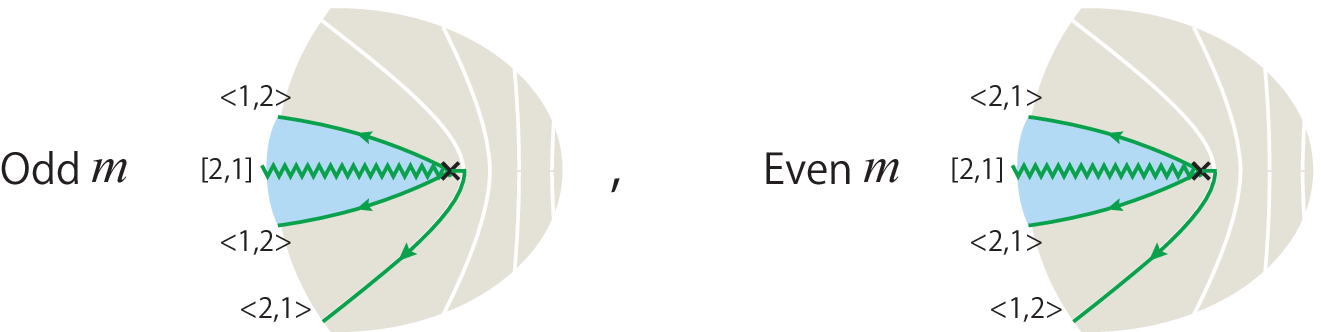}
\end{array}. 
\end{align}
\end{itemize}
$\square$
\end{Definition}
These DZ networks are obtained in \cite{CIY4} by solving multi-cut boundary condition. It is argued in \cite{CIY4} that the adjustable Stokes tail of the DZ network (say, $<2,1>$ in the case of $m \in 2\mathbb Z$) corresponds to the contours in the matrix-model integral.%
\footnote{In the sense that the RH calculus based on the above contour (evaluated in \cite{CIY4}) correctly reproduces the results obtained by the mean-field method based on a single eigenvalue contour integral (first proposed by David \cite{David} and carefully evaluated by \cite{HHIKKMT,SatoTsuchiya}).} Further discussions are given in Section \ref{SectionDiscussionEffectivePotentials}. 

Similarly, the general $(p,q)$ cases are given as follows: 

\begin{Definition} [Primitive solution] \label{DefinitionPrimitiveSolution}
Consider $(p,q)$-systems $(p>2)$. The following DZ networks are proper networks consistent with Chebyshev solutions, and are called primitive solutions: 
\begin{itemize}
\item [I. ] \underline{The cases of $p>q$:}\quad In this case, the remaining tail should across the anti-Stokes lines of the neighborhood inner sector  so that the tail is adjusted to the anti-Stokes line inside the inner sector: 
\begin{align}
\begin{array}{c}
\includegraphics[scale = 0.8]{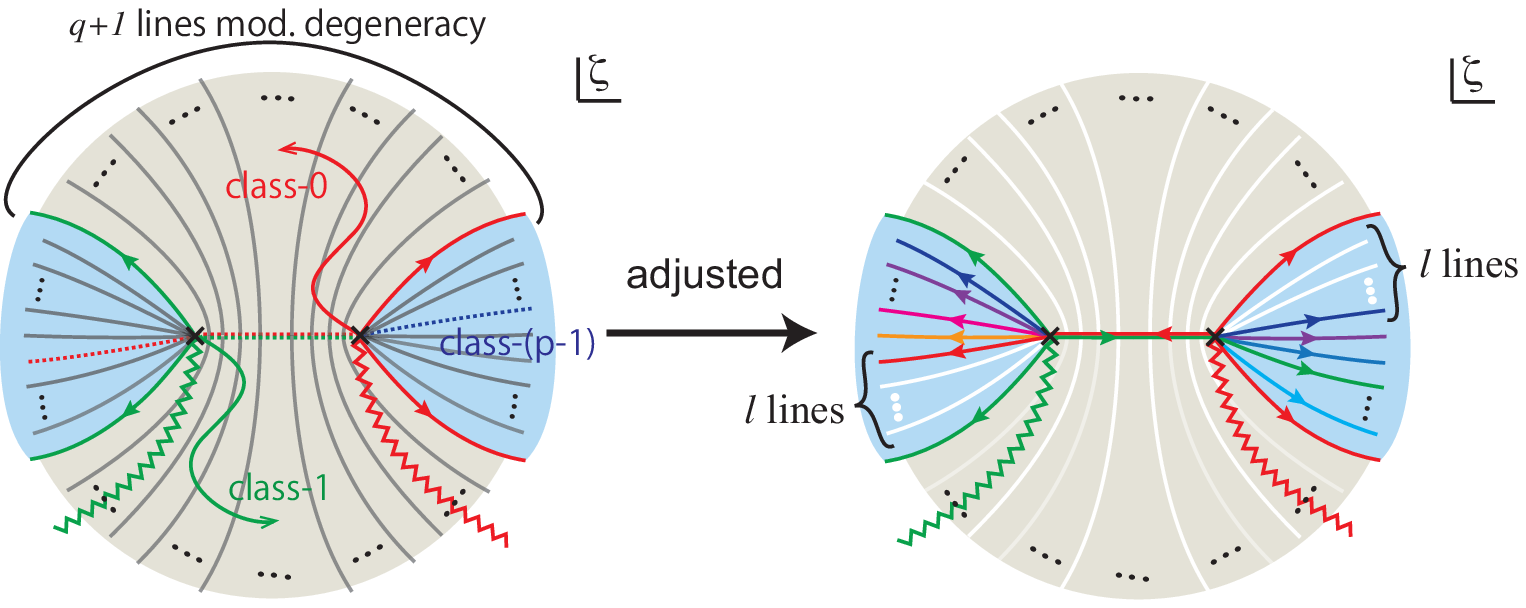}
\end{array}. \label{EqFigurePrimitive1}
\end{align}
Since the Stokes tail crosses another Stokes tails, it generates a bunch of other Stokes tails (according to the rule, Eq.~\eq{EqNetworkRulesE}). 

It is a non-trivial issue whether such extra Stokes tails can be adjusted to anti-Stokes lines. We can check it on one-by-one basis. In particular, we have checked it up to $p=16$, and all are consistently adjustable. Explicit expressions of generalized Airy systems $(q=1)$ and dual Kazakov series $(q=2)$, are explicitly shown later. 
\item [II. ] 
\underline{The cases of $p<q$:}\quad In this case, since $r (=p+q) = mp+l$ $(m\geq 2)$, the remaining tail does not need to across the anti-Stokes lines of the neighborhood inner sector. Therefore, one can adjust the Stokes tails as follows: 
\begin{align}
\begin{array}{c}
\includegraphics[scale = 0.8]{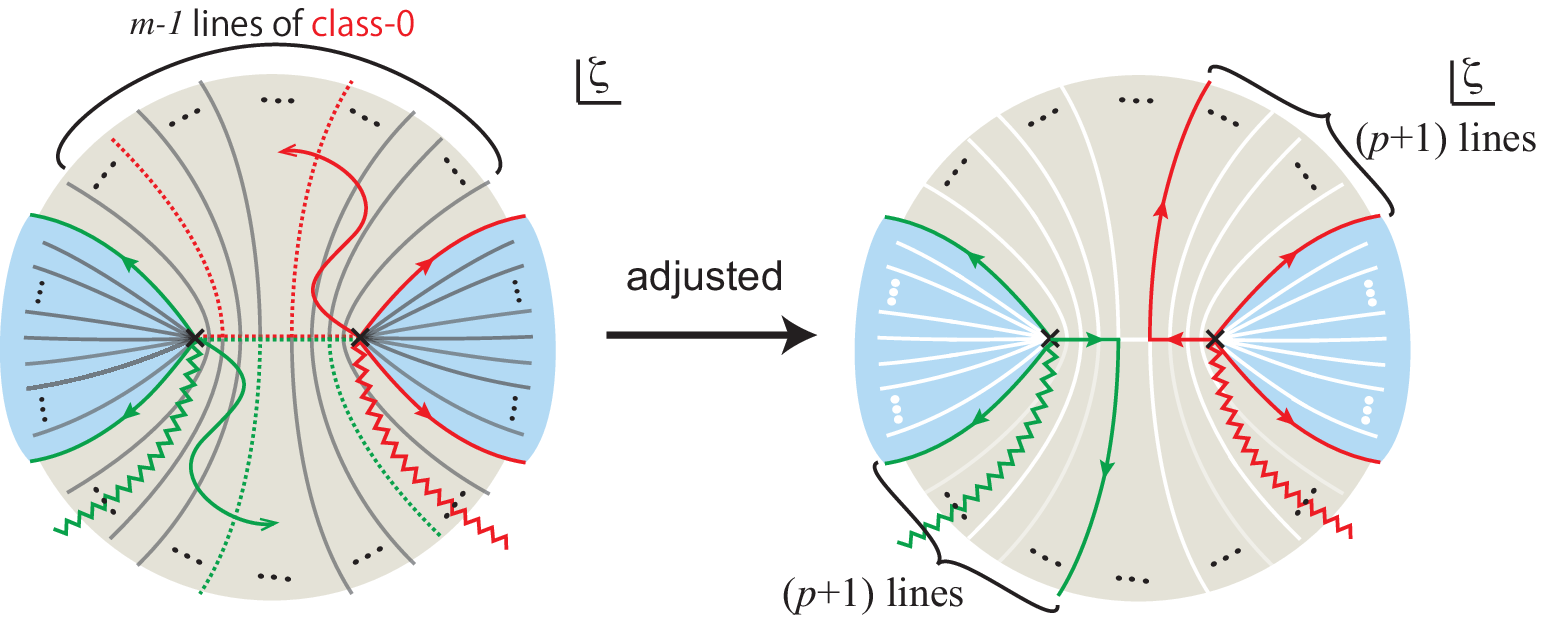}
\end{array}. 
\end{align}
By construction, this DZ network is a proper network consistent with the Chebyshev solutions. 
\end{itemize}
$\quad \square$
\end{Definition}

\subsubsection{Weaving other proper networks}
Primitive solutions are proper networks. A sequence of other proper networks are also obtained from the primitive solutions, with adding Stokes tails starting from the essential singularity (now only at $\lambda=\infty$) to the essential singularity. This is the procedure called {\em weaving a proper network}. 
Here, in order to avoid technical complexity, we show this in the cases of Kazakov series: 

\begin{itemize}
\item [A) ] One can add a Stokes tail in the way that it cancels an original tail:%
\footnote{Originally, in adjusting the Stokes tail in the primitive solutions (Definition \ref{DefinitionPrimitiveSolutionPeq2} and \ref{DefinitionPrimitiveSolution}), one could choose several choices. This procedure realizes such other choices of adjustment. In particular, these discrete deformations of primitive solutions are referred to as {\em single-line deformations of primitive solutions}. }
\begin{align}
\begin{array}{c}
\includegraphics[scale = 0.7]{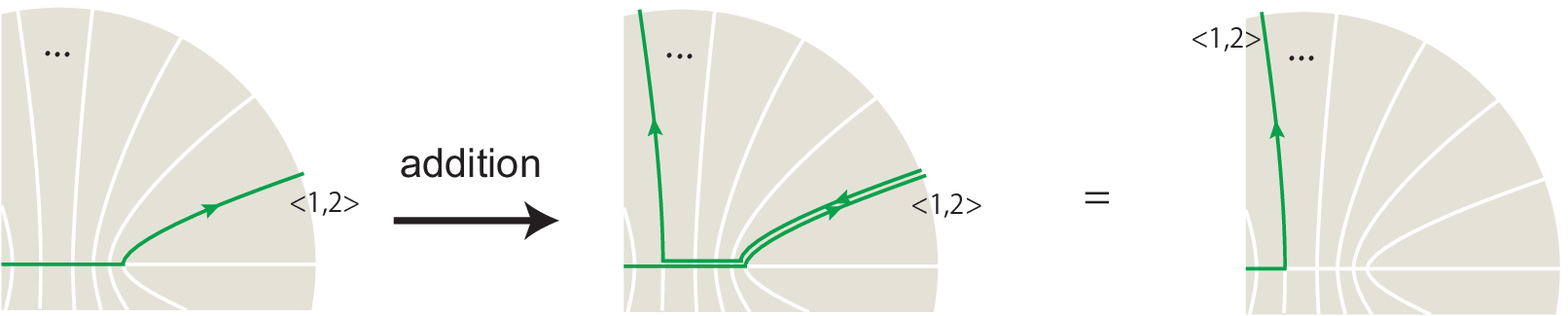}
\end{array}. \label{EqFigureWeavingAdeformSingleLine}
\end{align}
\item [B) ] One can add a new Stokes tail with an arbitrary weight $\theta$:%
\footnote{It is interesting to note that these procedures are naturally understood as an addition of a D-instanton operator of the above contour in free-fermion analysis \cite{fy1,fy2,fy3}: 
\begin{align}
\tau (x) = \bracket{x}{\Phi} \to \bra{x} e^{\theta D_{1,2}} \ket{\Phi}. 
\end{align}}
\begin{align}
\begin{array}{c}
\includegraphics[scale = 0.7]{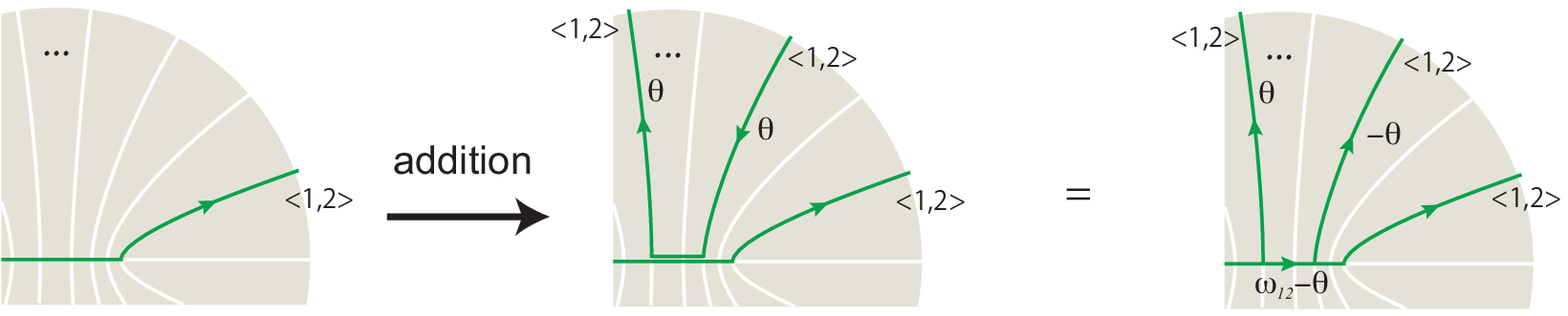}
\end{array}. 
\end{align}
\item [C) ] One can add $<j,l>$ and $<l,j>$ Stokes tails simultaneously (i.e.~$<1,2>$ and $<2,1>$ which are now colored differently to distinguish them) with assigning an arbitrary weight $\theta$: 
\begin{align}
\begin{array}{c}
\includegraphics[scale = 0.7]{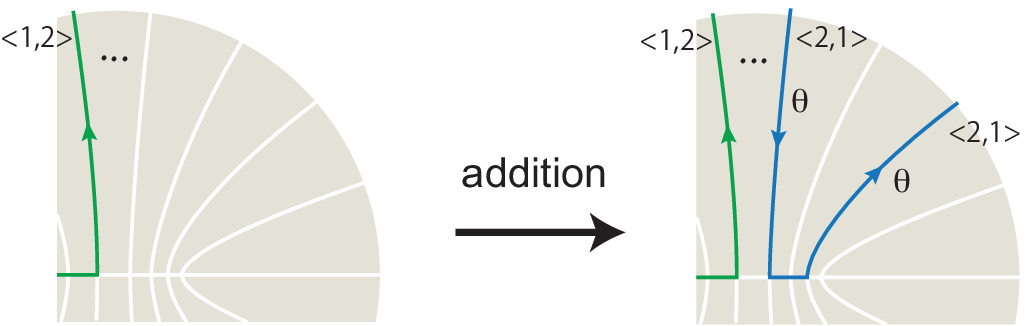} \label{EqWeavingC1}
\end{array},
\end{align}
unless these two lines cross each other.%
\footnote{Crossing of these lines are not allowed because it creates a new cut-jump matrix which does not exist in the classical BA function. Naively, it is no problem to add such Stokes tails with opposite indices in the general non-perturbative completion, but sometimes it is not realized in matrix integral. This issue will come back in Section \ref{SectionStokesSpecific}. }

\item [D) ] A different feature of the general $(p,q)$ cases ($p>2$) is given by a creation of Stokes tails in crossing other Stokes tails (i.e.~Eq.~\eq{EqNetworkRulesE}). Some examples are already shown in Eq.~\eq{EqFigurePrimitive1} and shown also in Eq.~\eq{EqFigureAiryWeaving}. 
\end{itemize}

Note that these procedures naturally correspond to drawing Stokes geometry of a given ODE system in exact WKB analysis (See e.g.~\cite{ExactWKB}).

\subsection{Evaluation of the RH integral and its solvability \label{SectionLocalRHP}}

We here justify the choice of proper networks (Definition \ref{DefinitionProperNetworks}) in Riemann-Hilbert problem. 
We consider the decomposition Eq.~\eq{ComponentDecompositionOfKernelEquationJL}: 
\begin{align}
K^{(M)}[f] (\lambda) = \sum_{j<l} K^{(j,l|M)}[f](\lambda), 
\end{align}
and we analyze it in week coupling $g\to 0$, i.e.~saddle point analysis. The contributions to the integral are then localized around saddle points (of Definition \ref{DefinitionGeneralSaddlePoints}): 
\begin{align}
K^{(j,l|M)}[f](\lambda) \asymeq  \sum_{\xi_* \in \mathfrak J^{(j,l)}_{\rm relev}[\hat {\mathcal K}]}   K^{(j,l|M)}_{\xi_*}[f](\lambda)\qquad \bigl(g\to 0\bigr), \label{EqKernelinstantonExpansionFirst} 
\end{align}
where the set $\mathfrak J_{\rm relev}^{(j,l)}[\hat{\mathcal K}]$ is a set of {\em relevant saddle points}, which are passed by DZ network: 
\begin{align}
\mathfrak J^{(j,l)}_{\rm relev}[\hat {\mathcal K}] = \Bigl\{\xi_* \in \mathbb C \, \Big| \, &\del_\xi \varphi^{(j,l)}(\xi_*)=0\,\, \bigl(\xi_*\in \mathcal K\bigr), \nn\\
& \quad \text{and}\quad  \alpha_{j,l}(\xi_*) \neq 0 \quad \text{ or } \quad \alpha_{l,j}(\xi_*) \neq 0 \Bigr\}. 
\end{align}
Since the behavior of kernel around saddle points (Eq.~\eq{EqKernelinstantonExpansionFirst}) is given as 
\begin{align}
\Bigl|K^{(j,l|M)}_{\xi_*}[f](\lambda) \Bigr|  \simeq e^{\frac{1}{g}A^{(j,l)}_{\xi_*} (1+O(g))},\qquad A^{(j,l)}_{\xi_*} 
\equiv \left\{
\begin{array}{cl}
{\rm Re}\bigl[\varphi^{(j,l)}(t;\xi_*)\bigr] & \text{if $\alpha_{j,l}(\xi_*)\neq 0$} \cr
{\rm Re}\bigl[\varphi^{(l,j)}(t;\xi_*)\bigr] & \text{if $\alpha_{l,j}(\xi_*)\neq 0$}
\end{array}
\right., \label{EquationKernelEvaluationAndLocalActionA}
\end{align}
the following condition is necessary \cite{ItsKapaev}: 
\begin{Definition} [Boutroux condition] \label{DefinitionDoutrouxCondition}
Consider a spectral network $\hat {\mathcal K}$ and a spectral curve $\varphi(t;\lambda) \in \mathcal L_{\rm str}^{\rm (univ)}(t)$. If the corresponding RH kernel satisfies 
\begin{align}
A_{\xi_*}^{(j,l)} \leq 0\qquad {}^\forall \xi_*\in \mathfrak J^{(j,l)}_{\rm relev} [\hat {\mathcal K}] \qquad \bigl(1\leq j < l\leq p\bigr), \label{EqBoutrouxConditionIK}
\end{align} 
of Eq.~\eq{EquationKernelEvaluationAndLocalActionA}, 
then $\varphi(t;\lambda) \in \mathcal L_{\rm str}^{\rm (univ)}(t)$ is called a Boutroux solution. 
$\quad\square$
\end{Definition}
If the Boutroux condition is satisfied, then all the instanton corrections are exponentially small instantons.%
\footnote{Or corresponding instanton actions are all positive.} 
The leading behavior of this integral kernel is then given by the points of 
\begin{align}
\del_\lambda\varphi^{(j,l)}(\lambda_*)=0,\qquad {\rm Re}\bigl[\varphi^{(j,l)}(\lambda_*)\bigr]=0. 
\end{align}
In our cases of minimal string theory, these points are always realized by {\em branch points} of the spectral curve. The issue of evaluating the RH integral around branch points is that the classical BA function generally diverges around the branch points. 

\subsubsection{Direct evaluation of the integral around branch points \label{SubsubsectionDirectEvaluationBranchPoints}}

Although there is a divergence at the branch points, we can show the following: 
\begin{Proposition} 
For a given branch point $\xi_n$ of class-$n$, the corresponding RH integral with $0$-th order perturbative BA function (i.e.~$M=0$) converges, and it is given as 
\begin{align}
K^{(j,l|0)}_{\xi_n}[f](\lambda) = O(g),\qquad \bigl(j+l-2\equiv n \mod p, \text{ and }\, f(\xi_n)\neq 0\bigr), \label{InPropositionKernelAroundBranchPoints}
\end{align}
if the spreading three Stokes tails (from branch points) are assigned with the canonical weight. $\quad \square$
\end{Proposition}
{\em Proof}\quad We evaluate $K^{(j,l|0)}[f](\lambda)$ by saddle point analysis around $\xi=\xi_n$. By using the associated coordinates around it, 
\begin{align}
x \equiv \frac{\xi-\xi_n}{\xi_n}, \qquad 
\delta \tau \simeq \Bigl( \frac{2x}{p} \Bigr)^{\frac{1}{2}}, 
\end{align}
one writes 
\begin{align}
\varphi^{(j,l)} (\xi) = i \frac{8\sqrt{2}}{3} c_{j,l} \Bigl(\frac{2x}{p}\Bigr)^{\frac{3}{2}} (1+O(x)) = i \frac{8\sqrt{2}}{3} c_{j,l} \, \delta \tau^{3} (1+O(\delta\tau^2)). 
\end{align}
where $c_{j,l}$ is given by Eq.~\eq{ProofOfPropositionDirectEvaluationVanishingCriterionCoefficientCJL}. 
Below, we assume $c_{j,l}>0$, since the cases of $c_{j,l}<0$ are also essentially the same. 
The integral is then given as 
\begin{align}
K^{(j,l|0)}[f](\lambda) \simeq \frac{ f(\xi_n) \xi_n}{2\pi i(\xi_n-\lambda)} &\int_{-\infty}^0 dx \Bigl[  \omega_{j,l}\Xi_{j,l}(x)     + a \omega_{l,j} \Xi_{l,j}(a x)  + a^{-1} \omega_{l,j} \Xi_{l,j}(a^{-1}x)\Bigr] \times \nn\\
&\quad \times \exp\Bigl[-\frac{8\sqrt{2}}{3g}c_{j,l} \Bigl(\frac{-2x}{p} \Bigr)^{\frac{3}{2}}\Bigr]\qquad \bigl(a= e^{\frac{2\pi}{3}i },\,\, g\to0 \bigr). 
\end{align}
According to the assumption of this proposition, weights of the spreading three Stokes tails are chosen as the canonical value (i.e.~$\alpha_{j,l}(\xi)=\omega_{j,l}$ and so on). 
Note that the leading divergence of $\Xi_{j,l}^{(0)}$ is due to $\sqrt{\Delta^{(j)} (\xi)}$,%
\footnote{It is already evaluated in Appendix \ref{AppendixMonodromyCalculusOfCBAF}.  }
and that 
\begin{align}
\Xi_{j,l}^{(0)}(\xi) = \frac{A_{-1}}{\delta \tau} + A_0 + A_1 \delta \tau + \cdots. 
\end{align}
Therefore, the integrand of $K^{(j,l|0)}[f](\lambda)$ has an integrable singularity at $\xi=\xi_n$. 
We next note the following relation: 
\begin{align}
\gamma_n^{*}\Bigl[\omega_{j,l} \Xi_{j,l}(x) \Bigr] =\Bigl[\omega_{j,l} \Xi_{j,l}(e^{2\pi i}x) \Bigr]= \omega_{l,j} \Xi_{l,j}(x) \qquad \bigl(j+l-2 \equiv n \quad \text{mod. } p\bigr),  
\end{align}
which can be expressed in the coordinate of $\delta\tau$ as
\begin{align}
\gamma_n^{*}\Bigl[\omega_{j,l} \Xi_{j,l}(\delta\tau) \Bigr] =\Bigl[\omega_{j,l} \Xi_{j,l}( e^{\pi i}\delta\tau) \Bigr] = \omega_{l,j} \Xi_{l,j}(\delta \tau) \qquad \bigl(j+l-2 \equiv n \quad \text{mod. } p\bigr). 
\end{align}
By this relation, the following combination of $\Xi_{j,l}(\xi)$ with $a= e^{\frac{2\pi}{3}i }$: 
\begin{align}
dx \Bigl[\omega_{j,l}\Xi_{j,l}^{\rm [sym]}(x) \Bigr]\equiv &dx \Bigl[ \omega_{j,l}\Xi_{j,l}(x) + a \omega_{l,j} \Xi_{l,j}(a x) + a^{-1}\omega_{l,j}\Xi_{l,j}(a^{-1} x)\Bigr] \nn\\
=& dx\Bigl[ \omega_{j,l}\Xi_{j,l}(\delta \tau) + a \omega_{l,j} \Xi_{l,j}(a^{1/2} \delta\tau ) + a^{-1}\omega_{l,j}\Xi_{l,j}(a^{-1/2} \delta\tau)\Bigr] \nn\\
=& dx\Bigl[ \omega_{j,l}\Xi_{j,l}(\delta \tau) + a \omega_{j,l} \Xi_{j,l}(a^{-1} \delta\tau ) + a^2\omega_{j,l}\Xi_{j,l}(a^{-2} \delta\tau)\Bigr], 
\end{align}
behaves as 
\begin{align}
\Bigl[\omega_{j,l}\Xi_{j,l}^{\rm [sym]}( a^n\delta \tau) \Bigr] = a^{n} \Bigl[\omega_{j,l}\Xi_{j,l}^{\rm [sym]}( \delta \tau) \Bigr]\qquad \bigl(n\in \mathbb Z /3\mathbb Z\bigr). 
\end{align}
Therefore, one obtains 
\begin{align}
K^{(j,l|0)}[f](\lambda) \simeq \frac{ f(\xi_n) \xi_n}{2\pi i(\xi_n-\lambda)} \int_{-\infty}^0 dx \Bigl[3A_1 \Bigl(\frac{2x}{p}\Bigr)^{\frac{1}{2}}\Bigr]\exp\Bigl[-\frac{8\sqrt{2}}{3g}c_{j,l} \Bigl(\frac{-2x}{p} \Bigr)^{\frac{3}{2}}\Bigr] = O(g). 
\end{align}
This is the statement. 
$\quad \blacksquare$

Here are some notes on this result: 
\begin{itemize}
\item This results implies that, if $\varphi(t;\lambda) \in \mathcal L_{\rm str}^{\rm (univ)}(t)$ can approximate the system in a good precision, weights of three spreading Stokes tails in the enhancement (Eq.~\eq{EquationFigureEnhancements1}) should be the canonical weights. 
\item Since the integrand includes an integrable singularity, this result is not enough to evaluate all the $O(g)$ corrections. For further precision, we should solve local RH problem around the branch points. 
\end{itemize}

\subsubsection{Contractive solvability and local RH problem \label{SubsubsectionContractiveSolvabilityAndLRH}}

For solving integral equation, the following Theorem is basic (See e.g.~\cite{KressLIE}): 
\begin{Theorem} [Contractive solvability] \label{TheoremContractiveSolvability}
If the kernel is contractive kernel: 
\begin{align}
\lVert K \rVert <1,
\end{align}
for some norm, one can solve the RH integral recursively: 
\begin{align}
\rho(g;\lambda) = I_p+\sum_{n=1}^\infty K_-^{n}\bigl[I_p\bigr](\lambda),\qquad 
 K_-^{n+1}\bigl[I_p\bigr](\lambda) \equiv K_-\bigl[ K_-^{n}\bigl[I_p\bigr]\bigr](\lambda). 
\end{align}
The leading behavior is obtained by a truncation: 
\begin{align}
\rho(g;\lambda) = I_p+ K_-\bigl[I_p\bigr](\lambda) + O(\lVert K \rVert^2),\quad 
\chi(g;\lambda) = I_p+ K\bigl[I_p\bigr](\lambda) + O(\lVert K \rVert^2)
\end{align}
$\square$
\end{Theorem}
This situation can be realized by solving the local RH problem \cite{ItsBook}. 

The problem is that the $M$-th order $\mathcal G$-function $\mathcal G^{(M)}(\xi)$ (Eq.~\eq{DefinitionEquationMthOrderGfunction}) has a divergence around the branch points ($\xi=\xi_n$, $n\in \mathbb Z/2p\mathbb Z$). Therefore, we cut off a ball region ${\mathfrak B}_R(\xi_n)$ around the branch point $\xi=\xi_n$, 
\begin{align}
{\mathfrak B}_R(\xi_n) \equiv \Bigl\{\xi \in \mathbb C\Big| |\xi-\xi_n| < R\Bigr\}, 
\end{align}
and find the following new BA function $\Psi_{{\rm local}}^{(n)}(g;\lambda)$ defined inside the ball region: 
\begin{itemize}
\item [1. ] $\Psi_{{\rm local}}^{(n)}(g;\lambda)$ is bounded in the ball (including the boundary): 
\begin{align}
\bigl\lVert\Psi_{{\rm local}}^{(n)}(g;\lambda)\bigr\rVert < \infty \qquad \Bigl(\lambda \in \overline{\mathfrak B_R(\lambda_n)}\Bigr). 
\end{align}
\item [2. ] On the boundary, $\Psi_{\rm cl}(g;\lambda)$ and $\Psi_{{\rm local}}^{(n)}(g;\lambda)$ are close to each other: 
\begin{align}
\Psi_{\rm cl}(g;\lambda) \simeq \Psi_{{\rm local}}^{(n)}(g;\lambda), \qquad \lambda\in \del \mathfrak B_{R}(\lambda_n)
\end{align}
if the radius $R$ is properly chosen, according to the value of $g$, as follows: 
\begin{align}
|\lambda_n|>\!\!> R >\!\!> g^{1/\gamma} \qquad \Leftrightarrow \qquad 
\ds x \equiv \frac{\lambda-\lambda_n}{\lambda_n} \to 0 \quad \text{and}\quad 
\ds \frac{x^{\gamma}}{g} \to \infty \label{ScalingLimitOfLocalBAFunctionsAndClassicalBAFunctions}
\end{align}
That is, by $|\lambda_n|>\!\!>|\lambda-\lambda_n|$ we focus on the branch point of $\Psi_{\rm cl}(g;\lambda)$ (typical distances between saddle points are almost $\sim |\lambda_n|$). By $|\lambda-\lambda_n|>\!\!> g^{1/\gamma}$, we consider a large $\lambda$ expansion of $\Psi_{{\rm local}}^{(n)}(g;\lambda)$. In our cases, we will see that $\gamma=\frac{3}{2}$. 
\end{itemize}
If one can find such a local BA function $\Psi_{{\rm Local}}^{(n)}(g;\lambda)$, then it can replace $\Psi_{\rm cl}(g;\lambda)$ inside the ball region. 
This defines a new RH problem given by $(\mathcal K^{\rm (reg.)}, \mathcal G^{\rm (reg.)})$, where 
\begin{align}
\mathcal K^{\rm (reg.)} &= \mathcal K \cup \bigcup_{n\in \mathbb Z/2p\mathbb Z} \del \, \mathfrak B_{R_n}(\lambda_n), \nn\\
\mathcal G^{\rm (reg.)}(\lambda) &= 
\left\{
\begin{array}{cc}
\ds
\Psi_{\rm cl}(\lambda_-) G(\lambda) \Psi_{\rm cl}(\lambda_+)^{-1} & \ds \left(\lambda \in \mathcal K \setminus \bigcup_{n\in \mathbb Z/2p\mathbb Z} \mathfrak B_{R_n}(\lambda_n) \right)\cr
\ds
\Psi_{\rm local}^{(n)}(\lambda_-) G(\lambda) \Psi_{\rm local}^{(n)}(\lambda_+)^{-1} & \ds \left(\lambda \in \mathcal K \cap \mathfrak B_{R_n}(\lambda_n) \quad \bigl(n\in \mathbb Z/2p\mathbb Z\bigr) \right)\cr
\Psi_{\rm cl}(\lambda_-) \Psi_{\rm local}^{(n)}(\lambda_+)^{-1}& \ds \left(\lambda \in \del\, \mathfrak B_{R_n}(\lambda_n) \quad \bigl(n\in \mathbb Z/2p\mathbb Z\bigr) \right)
\end{array}
\right. 
\end{align}
This new RH problem satisfies 
\begin{itemize}
\item [1. ] The new $\mathcal G$-function is bounded along $\mathcal K^{\rm (reg.)}$. 
\item [2. ] On the boundary of ball regions, $\mathcal G$-function is identical when $g\to 0$: 
\begin{align}
\mathcal G^{\rm (reg.)}(\lambda) \simeq I_p \qquad \lambda \in \bigcup_{n\in \mathbb Z/2p\mathbb Z} \del\, \mathfrak B_{R_n}(\lambda_n). 
\end{align}
That is, its contribution is suppressed in the RH integral. 
\end{itemize}
Therefore, remaining contributions to the RH integral are only from Stokes tails with small instanton saddles (due to the Boutroux condition, Definition \ref{DefinitionDoutrouxCondition}). This realizes the condition for the contractive solvability (of Theorem \ref{TheoremContractiveSolvability}) \cite{ItsKapaev}.%
\footnote{Here we emphasize that it is not necessary to solve all the local jump behavior inside the ball regions. }

In the cases of our $(p,q)$-systems with Chebyshev solutions, what we need is to find such a local BA function $\bigl\{\Psi_{{\rm local}}^{(n)}(g;\lambda)\bigr\}_{n\in \mathbb Z/2p\mathbb Z}$. Here we show it explicitly. Before that, we first show an old result on the solution of the RH problem in $(2,1)$-system, i.e.~Airy system: 

\begin{Theorem} [The Airy system] \label{TheoremOnAirySystem21}
Consider Airy system, i.e.~$(2,1)$-system: 
\begin{align}
g\frac{\del A(\zeta)}{\del \zeta} = \beta
\begin{pmatrix}
 & 1 \cr 
 \frac{\zeta}{2} & 
\end{pmatrix}
 A(\zeta),\qquad \beta = \sqrt{2}, \label{TheoremAirySystemEquationIsomonodromy}
\end{align}
then the system is solved by using Airy function ${\rm Ai}(\zeta)$ as follows: 
\begin{itemize}
\item [1. ] Canonical solutions $\bigl\{A_n (\zeta) \bigr\}_{n\in \mathbb Z/6\mathbb Z}$ of Eq.~\eq{TheoremAirySystemEquationIsomonodromy} and corresponding Stokes sectors $\bigl\{ D_n^{\rm (Airy)} \bigr\}_{n\in \mathbb Z/6\mathbb Z}$ are given by 
\begin{align}
&\left\{
\begin{array}{l}
A_{2n}(\zeta) \equiv 
\begin{pmatrix}
{\rm Ai}^{(2n+1)}(\zeta) & {\rm Ai}^{(2n)}(\zeta) \cr
\frac{g}{\sqrt{2}} \del_\zeta {\rm Ai}^{(2n+1)}(\zeta) & \frac{g}{\sqrt{2}} \del_\zeta {\rm Ai}^{(2n)}(\zeta)
\end{pmatrix}  \cr
A_{2n+1}(\zeta) \equiv 
\begin{pmatrix}
{\rm Ai}^{(2n+1)}(\zeta) & {\rm Ai}^{(2n+2)}(\zeta) \cr
\frac{g}{\sqrt{2}} \del_\zeta {\rm Ai}^{(2n+1)}(\zeta) & \frac{g}{\sqrt{2}} \del_\zeta {\rm Ai}^{(2n+2)}(\zeta)
\end{pmatrix}
\end{array}
\right.
, \nn\\
&\qquad D_n^{\rm (Airy)} \equiv D\bigl(\frac{2n-1}{3} \pi , \frac{2n+3}{3} \pi \bigr)
\end{align}
where ${\rm Ai}^{(n)}(\zeta) \equiv e^{-\frac{\pi i}{6} n} {\rm Ai}\bigl( e^{-\frac{2\pi i}{3} n} \zeta \bigr)$ $\bigl(n\in \mathbb Z\bigr)$ are solutions of Airy equation: 
\begin{align}
g^2 \frac{\del^2 {\rm Ai}^{(n)}(\zeta)}{\del \zeta^2} = \zeta {\rm Ai}^{(n)}(\zeta), \label{TheoremAirySystemEquationNthAirySolutions}
\end{align}
and are given by Airy function ${\rm Ai}(\zeta)$ (normalized as)
\begin{align}
{\rm Ai}(\zeta) \asymeq  \dfrac{\exp\Bigl[{-\frac{4\sqrt{2}}{3g} (\frac{\zeta}{2})^{ \frac{3}{2} }} \Bigr] 
}{ (2\zeta)^{1/4}} \Bigl[1 + O(\frac{g}{\zeta^{3/2}})\Bigr] \qquad \bigl(\zeta \to \infty \in D(-\pi , \pi)\bigr). \label{TheoremAirySystemExpansionFormulaOfAiryFunction12}
\end{align}
In particular, the asymptotic behavior of the canonical solutions is given as 
\begin{align}
A_n(\zeta)&\asymeq \frac{1}{\sqrt{2}} 
\begin{pmatrix}
\bigl(\frac{\zeta}{2}\bigr)^{-\frac{1}{4}} & \bigl(\frac{\zeta}{2}\bigr)^{-\frac{1}{4}} \cr
\bigl(\frac{\zeta}{2}\bigr)^{\frac{1}{4}} & -\bigl(\frac{\zeta}{2}\bigr)^{\frac{1}{4}} 
\end{pmatrix}
 \Bigl[I_2 + O(\frac{g}{\zeta^{3/2}})\Bigr]
\exp\Bigl[{+\frac{4\sqrt{2}}{3g} \sigma_3 \Bigl(\frac{\zeta}{2} \Bigr)^{ \frac{3}{2} }} \Bigr] \nn\\
&\qquad \qquad \qquad \qquad \qquad \qquad \qquad \qquad \bigl(\zeta \to \infty \in D_n^{\rm (Airy)}\bigr). 
\label{TheoremAirySystemEquationAsymptoticExpansionOfIMSBAfunctions}
\end{align}
\item [2. ] The sectionally holomorphic BA function $A_{\rm RH}(\zeta)$ 
\begin{align}
A_{\rm RH}(\zeta) = 
\left\{
\begin{array}{cc}
A_1(\zeta) & \quad \zeta \in D(\frac{2\pi }{3},\pi) \cr
A_0(\zeta) & \quad \zeta\in D(0, \frac{2\pi }{3}) \cr
A_{-1}(\zeta) & \quad \zeta \in D(-\frac{2\pi }{3}, 0) \cr
A_{-2}(\zeta) & \quad \zeta \in D(-\pi, -\frac{2\pi }{3}) 
\end{array}
\right., 
\end{align}
satisfies the uniform asymptotic expansion: 
\begin{align}
A_{\rm RH}(\zeta)& \simeq \frac{1}{\sqrt{2}} 
\begin{pmatrix}
\bigl(\frac{\zeta}{2}\bigr)^{-\frac{1}{4}} & \bigl(\frac{\zeta}{2}\bigr)^{-\frac{1}{4}} \cr
\bigl(\frac{\zeta}{2}\bigr)^{\frac{1}{4}} & -\bigl(\frac{\zeta}{2}\bigr)^{\frac{1}{4}} 
\end{pmatrix}
 \Bigl[I_2 + O(\frac{g}{\zeta^{3/2}})\Bigr]
\exp\Bigl[{+\frac{4\sqrt{2}}{3g} \sigma_3 \Bigl(\frac{\zeta}{2} \Bigr)^{ \frac{3}{2} }} \Bigr]\qquad \bigl(\zeta \to \infty \bigr). \label{AsymptoticExpansionOfARHInAirySystemTheorem}
\end{align}
The jump relation of this sectional function $A_{\rm RH}(\zeta)$ is expressed by a spectral network $\hat {\mathcal K}^{\rm (Airy)}=\bigl(\mathcal K^{\rm (Airy)},G^{\rm (Airy)}\bigr)$ defined by 
\begin{align}
&\mathcal K^{\rm (Airy)}= \vec{\mathcal K}_{\pi}^{\rm (Airy)} \cup \bigcup_{n=-1,0,1} \vec{\mathcal K}_n^{\rm (Airy)},\qquad 
\left\{
\begin{array}{lc}
|\vec{\mathcal K}_n^{\rm (Airy)}| = e^{\frac{2\pi i}{3} n}\times \mathbb R_{\geq 0} & \bigl(n=-1,0,1\bigr) \cr
|\vec{\mathcal K}_\pi^{\rm (Airy)}| = e^{\pi i }\times  \mathbb R_{\geq 0} &
\end{array}
\right., \\
&G^{\rm (Airy)} (\vec{\mathcal K}_{\pi})= 
\begin{pmatrix}
0 & i  \cr
i & 0
\end{pmatrix},\qquad 
G^{\rm (Airy)} (\vec{\mathcal K}_{0})= 
\begin{pmatrix}
1 & 0 \cr
-i & 1 
\end{pmatrix},\qquad 
G^{\rm (Airy)} (\vec{\mathcal K}_{\pm 1})= 
\begin{pmatrix}
1 & -i \cr
0 & 1 
\end{pmatrix},
\end{align}
and all the line-elements are directed from origin to $\zeta\to \infty$. The graph $\mathcal K$ is also constituted of anti-Stokes lines of $\varphi(\zeta) = +\frac{4\sqrt{2}}{3g} \sigma_3 \bigl(\frac{\zeta}{2} \bigr)^{ \frac{3}{2} }$. 
\item [3. ] On the other hand, consider the Riemann-Hilbert problem defined by the above spectral network $\hat {\mathcal K}^{\rm (Airy)}=(\mathcal K^{\rm (Airy)},G^{\rm (Airy)})$ and by the spectral curve $\varphi(\zeta) = +\frac{4\sqrt{2}}{3g} \sigma_3 \bigl(\frac{\zeta}{2} \bigr)^{ \frac{3}{2} }$, then the solution is given by $A_{\rm RH}(\zeta)$. 
\end{itemize}
$\square$
\end{Theorem}
{\em Proof}\quad Since these facts on Airy function are well-studied, we only note some key relations in this system. 
\begin{itemize}
\item Note that the isomonodromy system of Eq.~\eq{TheoremAirySystemEquationIsomonodromy} is obtained from the KP system, which is equivalent to Airy equation: 
\begin{align}
&\zeta \psi = 2 \del^2 \psi,\qquad 
 g\frac{\del \psi}{\del \zeta} = \beta_{2,1} \del \psi
\qquad \bigl(\beta_{2,1} = \sqrt{2}\bigr) 
\qquad \Leftrightarrow \qquad 
\zeta \psi = g^2\frac{\del^2 \psi}{\del \zeta^2} \nn\\
&  \qquad \qquad \Leftrightarrow \qquad g\frac{\del \vec \psi}{\del \zeta} = \sqrt{2}
\begin{pmatrix}
 & 1 \cr 
 \frac{\zeta}{2} & 
\end{pmatrix}
 \vec \psi,\qquad \vec \psi = 
\begin{pmatrix}
\psi \cr 
\frac{g}{\sqrt{2}} \del_\zeta \psi
\end{pmatrix}. 
\end{align}
Therefore, $\Psi(\zeta)$ of Eq.~\eq{TheoremAirySystemEquationIsomonodromy} is constructed by solutions of Airy equation. 
\item Note that $\bigl\{ {\rm Ai}^{(n)}(\zeta) \bigr\}_{n\in \mathbb Z}$ are all solutions of Airy equation (Eq.~\eq{TheoremAirySystemEquationNthAirySolutions}), since there is $\mathbb Z_3$-symmetry in Eq.~\eq{TheoremAirySystemEquationNthAirySolutions}. In particular, from Eq.~\eq{TheoremAirySystemExpansionFormulaOfAiryFunction12}, one obtains 
\begin{align}
{\rm Ai}^{(1)}(\zeta) \asymeq  \dfrac{\exp\Bigl[{+\frac{4\sqrt{2}}{3g} (\frac{\zeta}{2})^{ \frac{3}{2} }} \Bigr] 
}{ (2\zeta)^{1/4}} \Bigl[1 + O(\frac{g}{\zeta^{3/2}})\Bigr] \qquad \bigl(\zeta \to \infty \in D_0^{\rm (Airy)}\bigr). 
\end{align}
Similarly applying to ${\rm Ai}^{(n)}(\zeta)$, one obtains the expansion formula, Eq.~\eq{TheoremAirySystemEquationAsymptoticExpansionOfIMSBAfunctions}. 
\item The solutions $\bigl\{ {\rm Ai}^{(n)}(\zeta) \bigr\}_{n\in \mathbb Z}$ are not independent, and are related as 
\begin{align}
{\rm Ai}^{(n+2)}(\zeta) & = {\rm Ai}^{(n)}(\zeta) - i \, {\rm Ai}^{(n+1)}(\zeta). 
\end{align}
This leads to the Stokes matrices: 
\begin{align}
A_{n+1}(\zeta) = A_{n}(\zeta) S_n,\qquad S_{2m+1} = 
\begin{pmatrix}
1 & 0 \cr
-i & 1 
\end{pmatrix},\qquad 
S_{2m} = 
\begin{pmatrix}
1 & -i \cr
0 & 1 
\end{pmatrix}
\end{align}
and $A_{-2}(\zeta) = A_1(\zeta) 
\begin{pmatrix}
0 & i  \cr
i & 0
\end{pmatrix}$. 
This gives the spectral network $\hat{\mathcal K}^{\rm (Airy)}$ shown in the theorem. 
\item Finally note that the standard normalization of the BA function $\Psi_{\rm RH}^{\rm (Airy)}(\lambda)$ (i.e.~that of Eq.~\eq{EqSolutionOfRHinIMS}) and $A_{\rm RH}(\zeta)$ are related as 
\begin{align}
A_{\rm RH}(\zeta) = V
U^\dagger \times \Psi_{\rm RH}^{\rm (Airy)}(\lambda),\qquad 
V = \begin{pmatrix}
1 & \cr
  & \bigl(\frac{\zeta}{2}\bigr)^{\frac{1}{2}}
\end{pmatrix},\qquad U
= \frac{1}{\sqrt{2}}
\begin{pmatrix}
1 & 1 \cr
1 & -1
\end{pmatrix}. 
\end{align}
In this sense, $A_{\rm RH}(\zeta)$ is the solution of the RH problem. 
\end{itemize}
$\blacksquare$

With use of the Airy BA function, the local RH problem for Chebyshev solutions in $(p,q)$-systems is solved as follows: 

\begin{Theorem} [Local RH problem for Chebyshev solutions] 
Consider $(p,q)$-systems with Chebyshev solutions. 
Around the branch point of class-$n$ ($\lambda= \lambda_n$), the classical BA function $\Psi_{\rm cl}(g;\lambda)$ can be replaced by the following local BA function $\Psi_{\rm local}^{(n)}(g;\lambda)$ which is bounded from above: 
\begin{align}
\Psi_{\rm local}^{(n)}(g;\lambda) = U B &\prod_{(l|j)\in \mathcal J_{nr}^{(p,r)}} P_{j,l}
\left[
\begin{pmatrix}
1 & \cr
 & i\omega_{l,j}
\end{pmatrix}
A_{\rm RH} (\bigl(ic_{j,l}\bigr)^{2/3} \frac{4x}{p})
\begin{pmatrix}
1 & \cr
 & i\omega_{j,l}
\end{pmatrix}
e^{\frac{\varphi^{(j)}(\lambda)+\varphi^{(l)}(\lambda)}{2g} I_2}
\right] \times \nn\\
& \times \prod_{2(j-1)\equiv n}P_{j}\left[
e^{\frac{\varphi^{(j)}(\lambda)}{g}}
\right],
\end{align}
where $P_{j,l}$ ($j<l$) and $P_j$ are embedding maps to $p\times p$ matrices, 
\begin{align}
P_{j,l}\left[
\begin{pmatrix}
a_{11} & a_{12} \cr
a_{21} & a_{22}
\end{pmatrix}
\right] 
= 
\begin{pmatrix}
I_{j-1} &          &            &           &  \cr
         & a_{11} &            & a_{12} &  \cr
         &           & I_{l-j-1} &            & \cr
         & a_{21} &            & a_{22} & \cr
         &           &            &            & I_{p-j}
\end{pmatrix},
\qquad P_j \left[
a
\right] 
= \begin{pmatrix}
I_{j-1} &    &              \cr
         & a &              \cr
         &    & I_{p-j} 
\end{pmatrix},
\end{align}
and $P_{l,j}[A] \equiv P_{j,l}[A]$. $c_{j,l}$ is given in Eq.~\eq{ProofOfPropositionDirectEvaluationVanishingCriterionCoefficientCJL}. $B$ is a $p\times p$ matrix given as 
\begin{align}
&\left\{
\begin{array}{l}
B_{aj} = \sqrt{2} \bigl(ic_{j,l}\bigr)^{1/6} \frac{\sqrt{\omega^{\frac{j-l}{2}} \sinh\big(\frac{l-j}{p} \pi i\bigr)}}{p (\lambda_n/2^{1/p})^{\frac{p-1}{2}}} \Bigl( \frac{\mu^{\frac{1}{2p}}}{\lambda_n} \cos\bigl(\frac{l-j}{p} \pi \bigr)\Bigr)^{a-1} \cr
B_{al} = \sqrt{2} \bigl(ic_{j,l}\bigr)^{-1/6} \frac{\sqrt{\omega^{\frac{l-j}{2}} \sinh\big(\frac{j-l}{p} \pi i\bigr)}}{p (\lambda_n/2^{1/p})^{\frac{p-1}{2}}} \Bigl( \frac{\mu^{\frac{1}{2p}}}{\lambda_n} \cos\bigl(\frac{j-l}{p} \pi \bigr)\Bigr)^{a-1} \times (a-1) \tanh \bigl( \frac{j-l}{p} \pi i\bigr)
\end{array}
\right. \nn\\
&\qquad \qquad \qquad\qquad \qquad \qquad \qquad \bigl( j+l-2\equiv n,\,\, j<l\bigr) \\
&B_{aj'} =  \frac{2^{\frac{(p-1)a}{p}}}{\omega^{\frac{n}{2}(a-1)}\sqrt{U_{p-1}(1)}} \qquad \bigl(2(j'-1)\equiv n,\,\, U_n(\cos(\theta)) = \frac{\sin(n\theta)}{\sin(\theta)}\bigr). 
\end{align}
and $A_{\rm RH}(\zeta)$ is the sectionally holomorphic BA function given in Theorem \ref{TheoremOnAirySystem21}. 
$\quad \square$
\end{Theorem}
{\em Proof} \quad Consider the branch point of class-$n$, $\lambda= \lambda_n$, and we first note the following: 
\begin{itemize}
\item The jump relation of the sectionally holomorphic function 
\begin{align}
\widetilde A_{\rm RH}(\zeta) \equiv \begin{pmatrix}
1 & \cr
 & i\omega_{l,j}
\end{pmatrix}
A_{\rm RH}(\zeta) 
\begin{pmatrix}
1 & \cr
 & i\omega_{j,l}
\end{pmatrix}
\end{align}
is given by the spectral network of $\hat {\mathcal K}^{\rm (\widetilde{Airy})} = (\mathcal K^{\rm (Airy)}, \widetilde{G}^{\rm (Airy)})$ defined as 
\begin{align}
&\widetilde G^{\rm (Airy)} (\vec{\mathcal K}_{\pi})= 
\begin{pmatrix}
0 & -\omega_{j,l}  \cr
-\omega_{l,j} & 0
\end{pmatrix},\nn\\
&\widetilde G^{\rm (Airy)} (\vec{\mathcal K}_{0})= 
\begin{pmatrix}
1 & 0 \cr
\omega_{l,j} & 1 
\end{pmatrix},\qquad 
\widetilde G^{\rm (Airy)} (\vec{\mathcal K}_{\pm 1})= 
\begin{pmatrix}
1 & \omega_{j,l} \cr
0 & 1 
\end{pmatrix}. 
\end{align}
Note that this jump relation is associated with Stokes tails of type-$(j,l)$ with their canonical weights. 
\item If one focuses on the $j$-th component of $\Psi_{\rm cl}(\mu;\lambda) = \bigl(\psi^{(1)}_{\rm cl}(\lambda),\cdots,\psi^{(p)}_{\rm cl}(\lambda)\bigr)$, 
the behavior around the branch point $x= \frac{\lambda-\lambda_n}{\lambda_n}\to 0$ is given as 
\begin{align}
\psi^{(j)}_{\rm cl}(\lambda) \simeq & U \times \biggl[
\dfrac{\sqrt{\omega^{\frac{j-l}{2}} \sinh \bigl(\frac{l-j}{p} \pi i\bigr)}}{2^{-\frac{p-1}{2p} } \sqrt{p}} \Bigl(\frac{2^{\frac{p-1}{p}}}{\omega^{\frac{n}{2}}} \cos \bigl(\frac{l-j}{p} \pi \bigr)\Bigr)^{a-1} \times \nn\\
&\times 
\Bigl\{ 
\Bigl(\frac{2x}{p} \Bigr)^{-1/4} + (a-1) \tanh \bigl(\frac{l-j}{p} \pi i\bigr) \Bigl(\frac{2x}{p} \Bigr)^{+1/4} + O(x^{3/4})
\Bigr\}
\biggr] \times e^{\frac{1}{g} \varphi^{(j)}(\lambda)}, \nn\\
&\qquad \qquad \qquad \qquad \qquad \bigl(j+l-2\equiv n,\, j\not \equiv l \mod p\bigr), 
\end{align}
or 
\begin{align}
\psi^{(j)}_{\rm cl}(\lambda) \simeq  U\times  \frac{2^{\frac{(p-1)a}{p} } e^{\frac{1}{g} \varphi^{(j)}(\lambda)}}{\omega^{\frac{n}{2}(a-1)}\sqrt{U_{p-1}(1)}} \qquad \bigl(2(j-1) \equiv n \mod p \bigr). 
\end{align}
\item 
We then focus on the pair $(j,l)$ satisfying $j<l$ and $j+l-2 \equiv n \mod p$, and consider the exponents: 
\begin{align}
\diag\bigl(e^{\frac{1}{g} \varphi^{(j)}(\lambda)}, e^{\frac{1}{g} \varphi^{(l)}(\lambda)}\bigr) = \exp\Bigl[ \sigma_3{\frac{\varphi^{(j,l)}(\lambda)}{2g} }  + \frac{\varphi^{(j)}(\lambda) + \varphi^{(l)}(\lambda)}{2g} I_2\Bigr]. 
\end{align}
By this, we read a coordinate $\zeta_{j,l}(x)$ as
\begin{align}
\sigma_3{\frac{\varphi^{(j,l)}(\lambda)}{2g} } 
= & i \frac{4\sqrt{2}}{3} c_{j,l} \Bigl(\frac{2x}{p}\Bigr)^{\frac{3}{2}} (1+O(x)) \equiv + \frac{4\sqrt{2}}{3} \sigma_3 \Bigl(\frac{\zeta_{j,l}(x)}{2}\Bigr)^{\frac{3}{2}} \nn\\
&\therefore \qquad \zeta_{j,l}(x) = \bigl(ic_{j,l}\bigr)^{2/3} \frac{4x}{p}(1+O(x)). 
\end{align}
Note that, since $(l|j)\in \mathcal J_{nr}^{(p,r)}$, it is guaranteed that $c_{j,l}>0$ (i.e.~Theorem \ref{TheoremUniformSignaturePropertyOnBranchPoints}). 
With use of this coordinate $\zeta_{j,l}(x)$, one can see that the scaling limit of $\bigl(\psi^{(j)}_{\rm cl}(\lambda),\psi^{(l)}_{\rm cl}(\lambda)\bigr)$  coincides with the asymptotic expansion of $A_{\rm RH}(\zeta_{j,l}(x))$ given in Eq.~\eq{AsymptoticExpansionOfARHInAirySystemTheorem}. 
\end{itemize}
With these facts, one can conclude that the classical BA function and local BA function are close to each other in the region of Eq.~\eq{ScalingLimitOfLocalBAFunctionsAndClassicalBAFunctions}. 
$\quad \blacksquare$

By this result, one obtain the following statement: 
\begin{Theorem} [Cut-jump cancellation criterion 1] \label{TheoremCutJumpCancellation1}
Consider $(p,q)$-systems and its Chebyshev solutions $\varphi(\mu;\lambda)$. There exists a proper network $\hat{\mathcal K}$ of the spectral curve $\varphi(\mu;\lambda)$ such that $\varphi(\mu;\lambda)$ is a Boutroux solution of the system.  The correction function $\chi(\lambda)$ of Eq.~\eq{EqSolutionOfRHinIMS} is then suppressed: 
\begin{align}
\Psi_{\rm RH}(\mu;\lambda)\simeq \Bigl[1 + O(g)\Bigr] \Psi_{\rm cl}(\mu;\lambda)\qquad \bigl(g\to 0\bigr). 
\end{align}
In this case, $\varphi(\mu;\lambda)$ is called the true vacuum of the system. $\quad \square$
\end{Theorem}
{\em Proof} \quad The non-trivial statement is the existence of a spectral network satisfying the above condition. In fact, the primitive solution is one of such spectral networks. Therefore, the statement holds. $\quad \blacksquare$

This theorem is important in the sense that it guarantees that the proper networks with small instantons (i.e.~Boutroux condition) provide non-perturbative completions of Chebyshev solutions. However, in these non-perturbative completions, the Chebyshev solutions are true vacuum of the system. This is not the most general situation in string theory because perturbative string theory is at least {\em meta-stable vacuum}. This leads to the next consideration:

\subsection{Meta-stable vacua and perturbative string landscapes \label{SubsubsectionTransseriesMetastablevacua}}

As a principle of string theory, the system is assumed to possess the topological (or power-series) expansion by string coupling $g$, at least as a local saddle of path-integral in string theory. This means that observables, say $u_n(t;g)$, are given by a simple expansion in the string coupling $g \to +0 \in \mathbb R$: 
\begin{align}
u_n(t;g) \asymeq  u_{n,0}(t) +  u_{n,1}(t)\, g^2 + \cdots = \sum_{m=0}^\infty u_{n,m}(t)\,g^{2m} + (\text{Non-pert.}),
\end{align}
and these expansion coefficients are obtained, say by the worldsheet description. 
If this vacuum is stable, the non-perturbative corrections are exponentially small, compared with the power-series expansion. Therefore, we have the following asymptotic expansion: 
\begin{align}
u_n(t;g) \asymeq  \sum_{m=0}^\infty u_{n,m}(t)\,g^{2m},\qquad g\to +0,
\end{align}
in precision of neglecting exponentially small contributions. 
If this is a wrong vacuum (i.e.~meta-stable vacuum), it receives exponentially large corrections by non-perturbative effects of instantonic objects, and then we should consider the full-expansion with all the (large) multi-instanton corrections. Here, for simplicity, we consider the cases of a single kind of instantons of ghost type (i.e.~$A>0$): 
\begin{align}
u_n(t;g) \asymeq  \sum_{m=0}^\infty u_{n,m}(t)\,g^{2m} +\sum_{l=1}^\infty \, g^{1/2}\Bigl(\theta_Ae^{\frac{A}{g}}\Bigr)^l\Bigl[ \sum_{m=0}^\infty u_{n,m}^{[l]} g^m \Bigr] ,\quad g\to +0, \label{EqPainleveTransseriesExpansion}
\end{align}
which is also known as transseries in resurgent analysis \cite{Ecalle}. The parameter $\theta_A$ is called a trans-series parameter and is identified with D-instanton fugacity in the terminology of string theory. 
In asymptotic analysis of Painlev\'e equation, this behavior is known to appear when the system is in the phase of elliptic functions (See e.g.~\cite{ItsBook}). From the viewpoints of physics, we understand that this behavior is an indication of meta-stability of the reference vacuum, and the large instanton corrections are corrections from relatively stable saddles.
For more precise approximation, one should change the reference vacuum so that the large instanton does not appear as its corrections:%
\footnote{Note that $u_{n,m}^{\rm (elliptic)}(t;\kappa)$ is generally given by algebraic functions on the spectral curves, which represent the target-space geometry of the corresponding vacuum.}
\begin{align}
u_n(t;g) \asymeq  \sum_{m=0}^\infty u_{n,m}^{\rm (elliptic)}(t;\kappa)\,g^{2m},\qquad \bigl(\kappa \equiv g/(-t)^{\frac{p+q}{p+q-1}},\, g\to +0\bigr),
\label{EqPainleveElliptic}
\end{align}
which is the expansion around the true vacuum in string theory landscape.%
\footnote{Note that Eq.~\eq{EqPainleveTransseriesExpansion} and Eq.~\eq{EqPainleveElliptic} are {\em different asymptotic expansions of the same transcendental function}. Usually, string theory is given by the form of Eq.~\eq{EqPainleveTransseriesExpansion} and its non-perturbative vacuum is given by the form of Eq.~\eq{EqPainleveElliptic}. In order to achieve vacuum search in string theory, therefore, it is necessary to become able to make the connection between these two asymptotic expansions. }

In the context of RHP, Eq.~\eq{EqPainleveElliptic} rather naturally appears as a consequence of the Boutroux condition, Eq.~\eq{EqBoutrouxConditionIK} \cite{ItsKapaev}.%
\footnote{It is explained in \cite{ItsBook}.} That is, the analysis is based on the true vacuum. On the other hand, for application to physics, it is also important to analyze meta-stable vacua of the system. It is achieved by simply turning off the Boutroux condition in the RH calculus. 

\begin{Proposition} [Cut-jump cancellation criterion 2]
Consider $(p,q)$-systems. For a given spectral network $\hat{\mathcal K}$ and a solution $\varphi(t;\lambda) \in \mathcal L_{\rm str}^{\rm (univ.)}$, if the spectral network $\hat{\mathcal K}$ is a proper network of the spectral curve $\varphi(t;\lambda)$, one can extract the perturbative amplitude as zero-instanton sector: 
\begin{align}
\Psi_{\rm RH}(\mu;\lambda)\simeq \Bigl[1 + O(g)\Bigr] \Psi_{\rm cl}(\mu;\lambda) + (\text{Non-pert.})\qquad \bigl(g\to 0\bigr). 
\end{align}
The solution $\varphi(t;\lambda) \in \mathcal L_{\rm str}^{\rm (univ.)}$ is then called a meta-stable vacuum of the system. 
$\quad \square$
\end{Proposition}

Therefore, by this proposition, we say that the space of non-perturbative completions of Chebyshev solution is given by the set of proper networks consistent with the Chebyshev solution. 
Also, this consideration on meta-stable vacua naturally leads the following definition of the landscape of perturbative string vacua: 
\begin{Definition}[Perturbative string landscape] \label{DefinitionPerturbativeLandscapes}
For a given proper network $\hat {\mathcal K}$ and a spectral curve $\varphi(t;\lambda) \in \mathcal L_{\rm str}^{\rm (univ.)}(t)$, there is a sequence of spectral curves $\varphi'(t;\lambda)\in \mathcal L_{\rm str}^{\rm (univ.)}(t)$ such that the spectral curves $\varphi'(t;\lambda)\in \mathcal L_{\rm str}^{\rm (univ.)}(t)$ are consistent with a spectral network $\hat{\mathcal K}'$ which is equivalent to the spectral network $\hat {\mathcal K}$. The set of these consistent spectral curves: 
\begin{align}
\mathcal L_{{\rm str}_{\varphi}}^{(\hat {\mathcal K})}(t) \equiv \Bigl\{\varphi'(t;\lambda)\in \mathcal L_{\rm str}^{\rm (univ.)}(t)\Big| \text{$\varphi'(t;\lambda)$ is consistent with $\hat {\mathcal K}'\sim \hat {\mathcal K}$}\Bigr\} \subset \mathcal L_{\rm str}^{\rm (univ.)}(t),
\end{align}
is called a perturbative string theory landscape, based on $\bigl(\varphi,\hat {\mathcal K}\bigr)$. $\quad \square$
\end{Definition}

\subsection{An example: Generalized Airy systems, $(p,1)$ \label{SubsubsectionGeneralizedAirySystems}}
We here obtain concrete data of Stokes phenomena in $(p,1)$-systems by applying the discussion in previous sections. 
The cases of $(p,q)=(p,1)$ are known as topological minimal string theory, and the system becomes particularly simple. By the reason mentioned below, this system is referred to as {\em generalized Airy systems}. 

In these topological cases, the BA systems with general KP flows $t=\bigl\{t_n\bigr\}_{n=1}^p$ are given as 
\begin{align}
\zeta \psi(t;\zeta)  = \bP(t;\beta^{-1}_{p,1}\del_\zeta)\psi(t;\zeta),\qquad
\Bigl[ \del_\zeta- \beta_{p,1} \del_t \Bigr]\psi(t;\zeta)= 0, \label{EqBATopMinimalStrings}
\end{align}
where $\bP(t;z)$ is a $p$-th order polynomial in $z$, 
\begin{align}
\bP(t;z) = 2^{p-1} z^p + \sum_{n=1}^{p} t_n z^{p-n},
\end{align}
the coefficients of which are just given by the KP flows $t=\bigl\{t_n\bigr\}_{n=1}^p$ (as a result of string equation).%
\footnote{Note that there is also $t_{p+1}$ which is chosen as $t_{p+1} = 2^{p-1}$. }
\begin{itemize}
\item If one sends all the KP-flow parameters $t$ to zero, the system becomes {\em generalized Airy equations}: 
\begin{align}
(-1)^p\frac{\del^p f(\zeta)}{\del \zeta^p} =  \zeta f(\zeta),\qquad f(\zeta) = \psi(0;\zeta),\qquad \beta_{p,1} = (-1)^p 2^{\frac{p-1}{p}}. 
\end{align}
This choice of KP-flow parameters is referred to as {\em Airy background}. 
\item The KP-flow parameters $t$ are also isomonodromy deformations of the system. Therefore, 
 the Stokes multipliers of this system do not depend on $t$. This means that the behavior of the systems can essentially be extracted from the Airy background. This is a reason why this system is called generalized Airy systems. 
\item Because of this simplicity, there is {\em no} non-perturbative ambiguity, and Stokes phenomenon of generalized Airy systems is uniquely determined. In other words, {\em the non-perturbative completion of this topological theory is completely fixed by the perturbative string theory. } 
In fact, generalized Airy systems can be expressed by a matrix model \cite{ShihAiry}. This means that multi-cut BC in this case is not a constraint but an identity of the system. 
\item A technical specialty of generalized Airy systems is that there is the following extra $\mathbb Z_r$-symmetry, 
\begin{align}
\zeta \to a \zeta,\qquad a^r =1,\qquad r=p+q=p+1, \label{GeneralizedAirySystemZRsymmetry}
\end{align}
in the Airy background. 
\end{itemize}
We first show the result on Stokes phenomenon of generalized Airy systems: 

\begin{Theorem} [Generalized Airy systems] \label{TheoremGeneralizedAirySystem}
The Stokes phenomenon of the generalized Airy systems, i.e.~$(p,1)$-systems (we choose $\beta_{p,1} = (-1)^p 2^{\frac{p-1}{p}}$): 
\begin{align}
g \frac{\del \Psi(\lambda)}{\del \lambda } = \Bigl[\frac{p \beta_{p,1}}{2} \Bigl(\Omega^{-p-1} (2 \lambda)^{p} + \sum_{n=1}^p 2^n t_n \widetilde E_{p,p-n+1}\,(2\lambda)^{p-n}\Bigr) - \frac{\widetilde {\rm P}_1}{\lambda}\Bigr] \Psi(\lambda), \label{TheoremGeneralizedAirySystemsAiryIMSequations}
\end{align} 
is given by the Stokes matrices $\bigl\{ S_n \bigr\}_{n=0}^{2rp-1}$ of 
\begin{align}
S_n= I_p + \sum_{(j|l)\in \mathcal J_n^{(p,r)}} \omega_{l,j} E_{l,j}, \qquad 
\bigl(n=0,1,\cdots,2rp-1\bigr). \label{StokesMatricesOfGeneralizedAirySystem}
\end{align}
Stokes matrices are based on Definition \ref{StokesDef} and Proposition \ref{PropositionStokesMatricesCanonicalSolutions}. 
$\quad \square$
\end{Theorem}
A proof of this theorem is given in the following two sub-subsections. Before that, we make a comment on the more conventional definition of generalized Airy systems: 
\begin{align}
A(\zeta) = 
V
U^\dagger \times \Psi(\lambda),\qquad 
\bigl(\zeta= 2^{p-1} \lambda^p\bigr), 
\end{align}
where $V$ is given by Eq.~\eq{DefinitionOfVmatrix} and $U$ is given by Eq.~\eq{DefinitionOfOmegaAndUmatrices}. 
It satisfies 
\begin{align}
g\frac{\del A(\zeta)}{\del \zeta} =  \beta_{p,1}
\begin{pmatrix}
0 & 1 \cr
\vdots & & \ddots \cr
0 & & & 1 & 0\cr
0 & & & 0& 1 \cr
\bigl[\frac{\zeta}{2^{p-1}}- \frac{t_p}{2^{p-1}}\bigr] & - \frac{t_{p-1}}{2^{p-1}} & \cdots &  - \frac{t_{2}}{2^{p-1}} & 0
\end{pmatrix} 
 A(\zeta),  
\end{align}
with $\beta_{p,1} = (-1)^p 2^{\frac{p-1}{p}}$. 

\subsubsection{Weaving method \label{SubsubsectionWeavingMethodInGeneralizedAirySystems}}

\begin{figure}[htbp]
\begin{center}
\includegraphics[scale = 0.7]{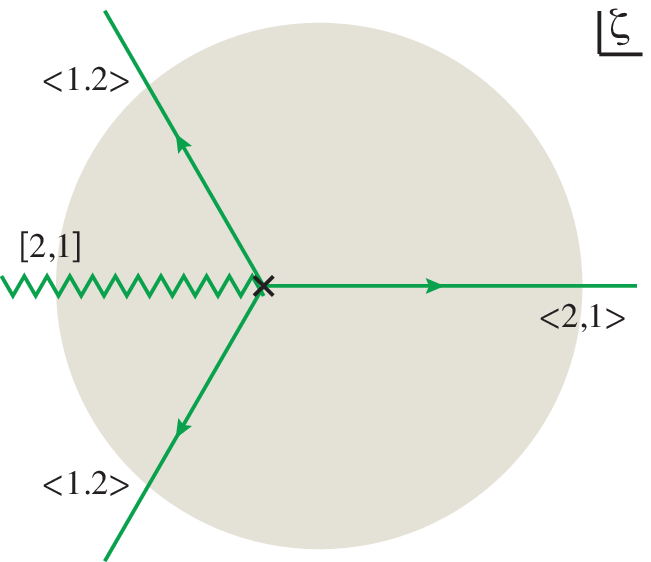}\qquad 
\includegraphics[scale = 0.7]{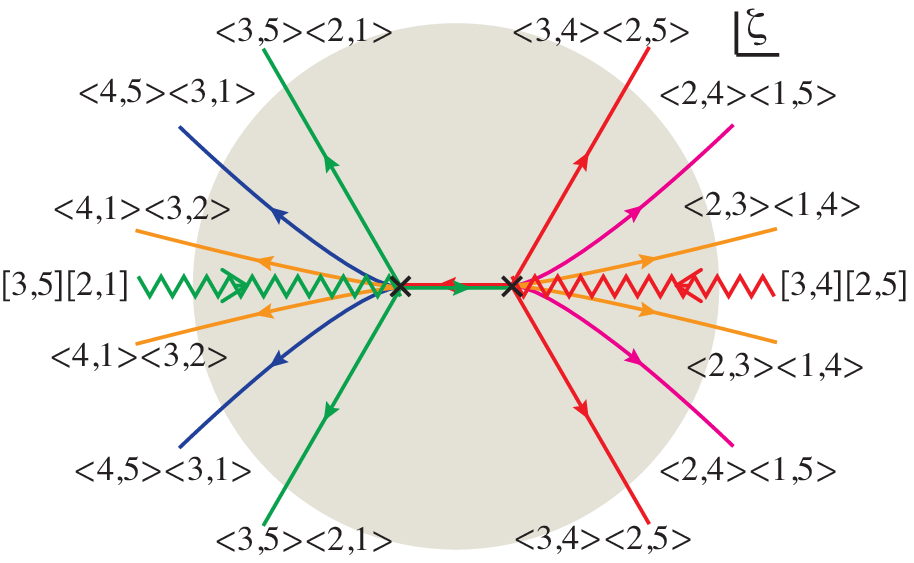}
\end{center}
\caption{\footnotesize  The DZ network of generalized Airy systems: the $(2,1)$ case and $(5,1)$ case. Note that all the Stokes tails and cut-jump lines are associated with their canonical weights/phases. }
\label{FigureDZAiry}
\end{figure}

Here we first apply the weaving method to this system. 
One can choose the KP-flow parameters as follows:  
\begin{align}
\bP(t;z) = \sqrt{\mu}\, T_p(z/\mu^{1/2p}) = 2^{p-1} z^p + \sum_{n=1}^p T_{p,n}\, \mu^{\frac{n}{2p}} z^{p-n}, \qquad \text{i.e.}\,\, t_n \equiv T_{p,n}\,\mu^{\frac{n}{2p}},
\end{align}
which is the conformal background and gives the spectral curve of Chebyshev solution. The Stokes multipliers are then constructed by the weaving procedure. For example, the primitive solutions are shown as the spectral network in the $(2,1)$ (i.e.~Airy function) case and of the $(5,1)$ case in Fig.~\ref{FigureDZAiry}. {\em Actually, these primitive solutions are identified with the Stokes multipliers of Theorem \ref{TheoremGeneralizedAirySystem}. }

Here is shown how to weave the DZ network in the $(5,1)$ case: 
\begin{align}
\begin{array}{c}
\includegraphics[scale = 0.7]{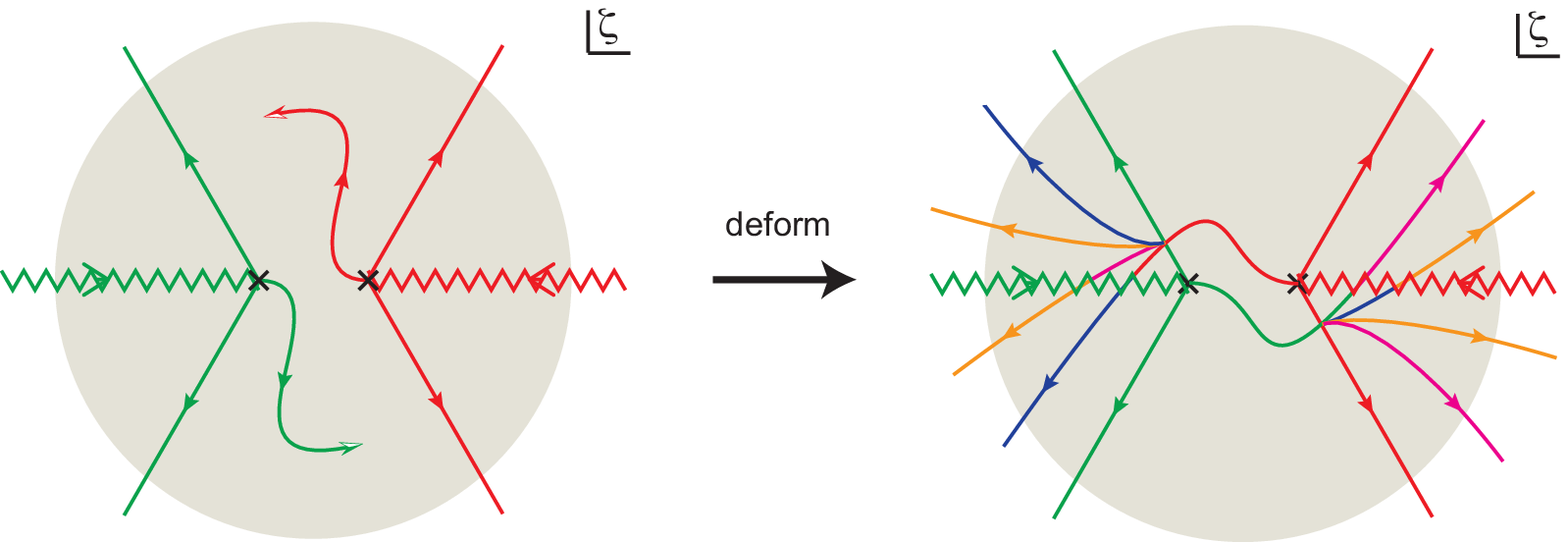} 
\end{array}. \label{EqFigureAiryWeaving}
\end{align}
When the $<4,3><5,2>$ tails cross the $<3,5><2,1>$ tails, the branching of Stokes tails occurs, which should be carefully evaluated as 
\begin{align}
\begin{array}{c}
\includegraphics[scale = 0.7]{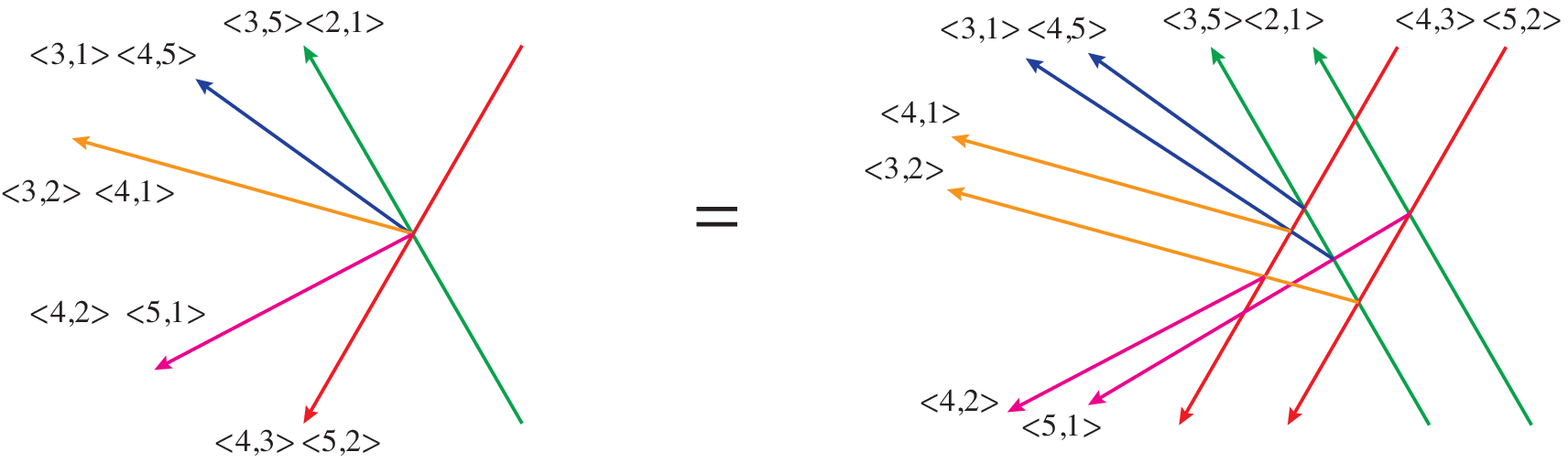} 
\end{array}. 
\end{align}
It is quite non-trivial that all the Stokes tails appearing here are associated with their canonical weights (therefore their weights are not particularly written). 
In this way, (although we show it on a case-by-case basis) one can see that, for the general $(p,1)$ cases, the primitive solution can be expressed in the simple form. 

It is important to note that, from the weaving approach, one can generate many other proper networks, by adding other Stokes tails. However, as mentioned above, there is no non-perturbative ambiguity in this system. In fact, many of such proper networks are not realized by Eq.~\eq{TheoremGeneralizedAirySystemsAiryIMSequations}, or equivalently are not realized by matrix models. This is related to the issue commented in the end of Section \ref{SectionIsomonodromyEmbedding}, and ``proper networks'' only guarantee existence of perturbative vacuum in the landscape (See Theorem \ref{TheoremCutJumpCancellation1}). In the generalized Airy system, it can be fixed by multi-cut boundary condition. 

\subsubsection{Multi-cut boundary condition \label{SubsubsectionMCBCinGeneralizedAirySystems}}

We first note the $\mathbb Z_r$-symmetry of the system (Eq.~\eq{GeneralizedAirySystemZRsymmetry}) in the Airy background. This means that, by isomonodromy property, Stokes phenomenon of this system has the $\mathbb Z_p\times \mathbb Z_r$-symmetry, {\em even in the general background} (i.e.~Eq.~\eq{TheoremGeneralizedAirySystemsAiryIMSequations}). By applying this symmetry to multi-cut BC equations (Proposition \ref{PropositionMultiCutBCequation} with $r=p+1$, i.e.~$m=l=1$, and $\sgn(\beta_{p,1}) = (-1)^p$), one obtains the following solutions given by a discrete parameter $\rho$: 
\begin{align}
 \theta_n^{(1+2a)} = (-1)^{n+1} \rho^n,\qquad \rho^p = (-1)^{p+1}\qquad \bigl(a\in \mathbb Z/rp\mathbb Z\bigr). 
\end{align}
Here we choose the multi-cut BC of $\theta_0=\frac{\pi}{p}$. By noting the relation with the Stokes multipliers, \begin{align}
s_{m,l,j} = \theta_{n}^{(*)}=(-1)^{n+1} \rho^n\qquad \bigl(n\equiv j-l \mod p,\,\, m\in \mathbb Z/2rp \mathbb Z\bigr), 
\end{align}
one obtains the Stokes matrices: 
\begin{align}
S_m = I_p + \sum_{(j|l)\in \mathcal J_m^{(p,r)}} - (-\rho)^{n_{j,l}} E_{l,j}\qquad n_{j,l} \equiv j-l \mod p. 
\end{align}
Note that multi-cut BC (and also $\mathbb Z_p$-symmetry) is not affected by the following similarity transformation: 
\begin{align}
\Psi(\lambda) &\to \Psi'(\lambda) = \Omega^{-u} \Psi(\lambda) \Omega^{u},\qquad 
g \frac{\del \Psi'(\lambda)}{\del \lambda } \sim \frac{p \beta_{p,1}}{2} \Omega^{-1} (2 \lambda)^{p} \Psi'(\lambda). 
\end{align}
Therefore, this discrete residual degree of freedom (to change the basis of the isomonodromy system) is related to the discrete parameter $\rho$:  
\begin{align}
S_m &\to S_m'=\Omega^{-u} S_m \Omega^{u} = I_p + \sum_{(j|l)\in \mathcal J_m^{(p,r)}} - (-\rho\omega^u)^{n_{j,l}} E_{l,j}. 
\end{align}
By choosing $\rho\omega^u = -\omega^{\frac{1}{2}}$ with $\omega=e^{\frac{2\pi i}{p}}$, one obtains
\begin{align}
S_m' = I_p + \sum_{(j|l)\in \mathcal J_m^{(p,r)}} - \omega^{\frac{n_{j,l}}{2}} E_{l,j}\qquad n_{j,l} \equiv j-l \mod p. \label{StokesPrime12345}
\end{align}
Noticing that the indices $n_{j,l}$ are given by 
\begin{align}
n_{j,l} = 
\left\{
\begin{array}{cc}
|l-j| & (j>l) \cr
p-|l-j| & (l>j)
\end{array}
\right.,
\qquad \text{i.e. } - \omega^{\frac{n_{j,l}}{2}} = \omega_{l,j}, 
\end{align}
one sees that Eq.~\eq{StokesPrime12345} is the Stokes matrices of Eq.~\eq{StokesMatricesOfGeneralizedAirySystem}. Since this Stokes matrices are consistent with the weaving procedure (i.e.~with the basis of classical BA function given in Theorem \ref{TheoremClassicalBAFunctions}), the choice of the basis (given above) is the basis of the Stokes multipliers chosen in Eq.~\eq{TheoremGeneralizedAirySystemsAiryIMSequations}. Therefore, the theorem follows. $\quad \blacksquare$

\subsection{An example: Dual Kazakov series, $(p,2)$ \label{SubsectionAnExampleDualKazakovSeries}}
The next simplest examples are the cases of $(p,q) = (2k+1,2)$, which are $p-q$ dual of the Kazakov series, $(p,q)=(2,2k+1)$. We first show the primitive solutions in the most important two examples, $(3,2)$ and $(5,2)$: 
\begin{align}
\begin{array}{c}
\includegraphics[scale = 0.7]{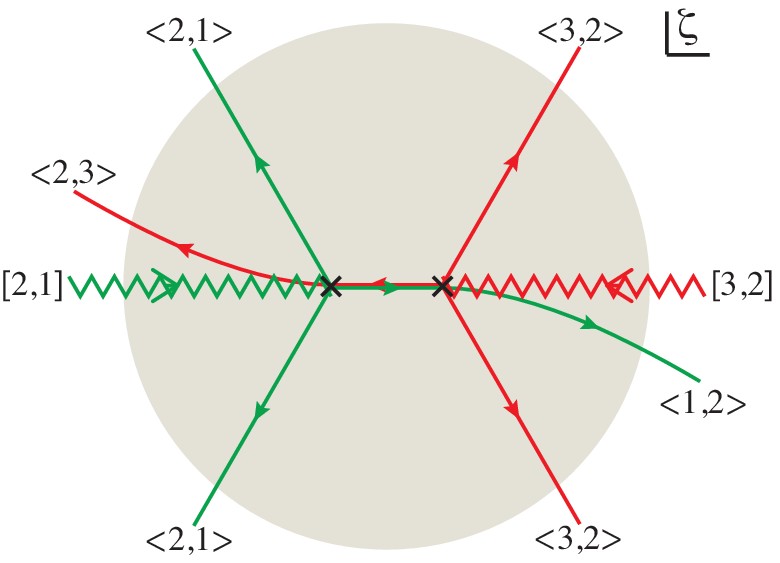}
\end{array},\qquad 
\begin{array}{c}
\includegraphics[scale = 0.7]{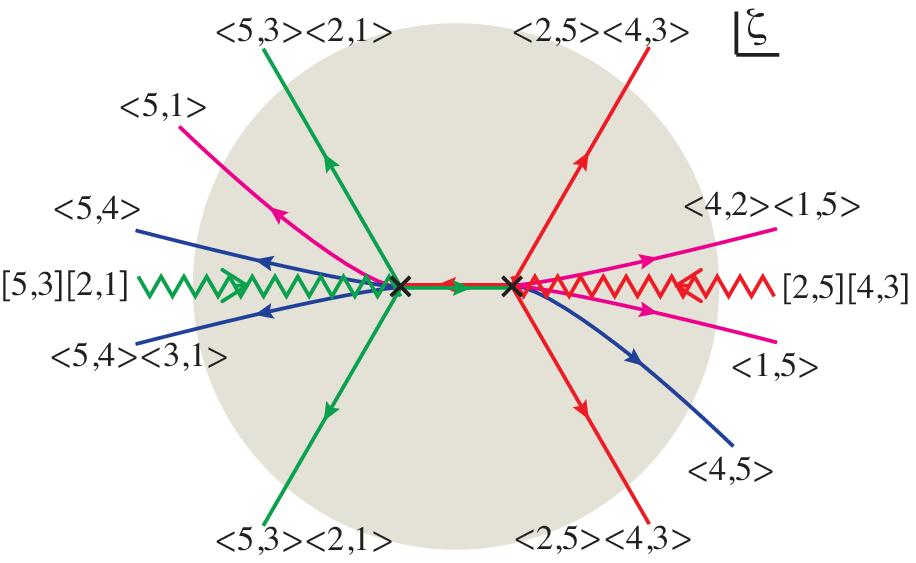}
\end{array}. \nn
\end{align}
The weaving procedure itself is essentially the same as $(p,1)$-systems as in Eq.~\eq{EqFigureAiryWeaving}. 

\begin{figure}[htbp]
\begin{center}
\includegraphics[scale=0.6]{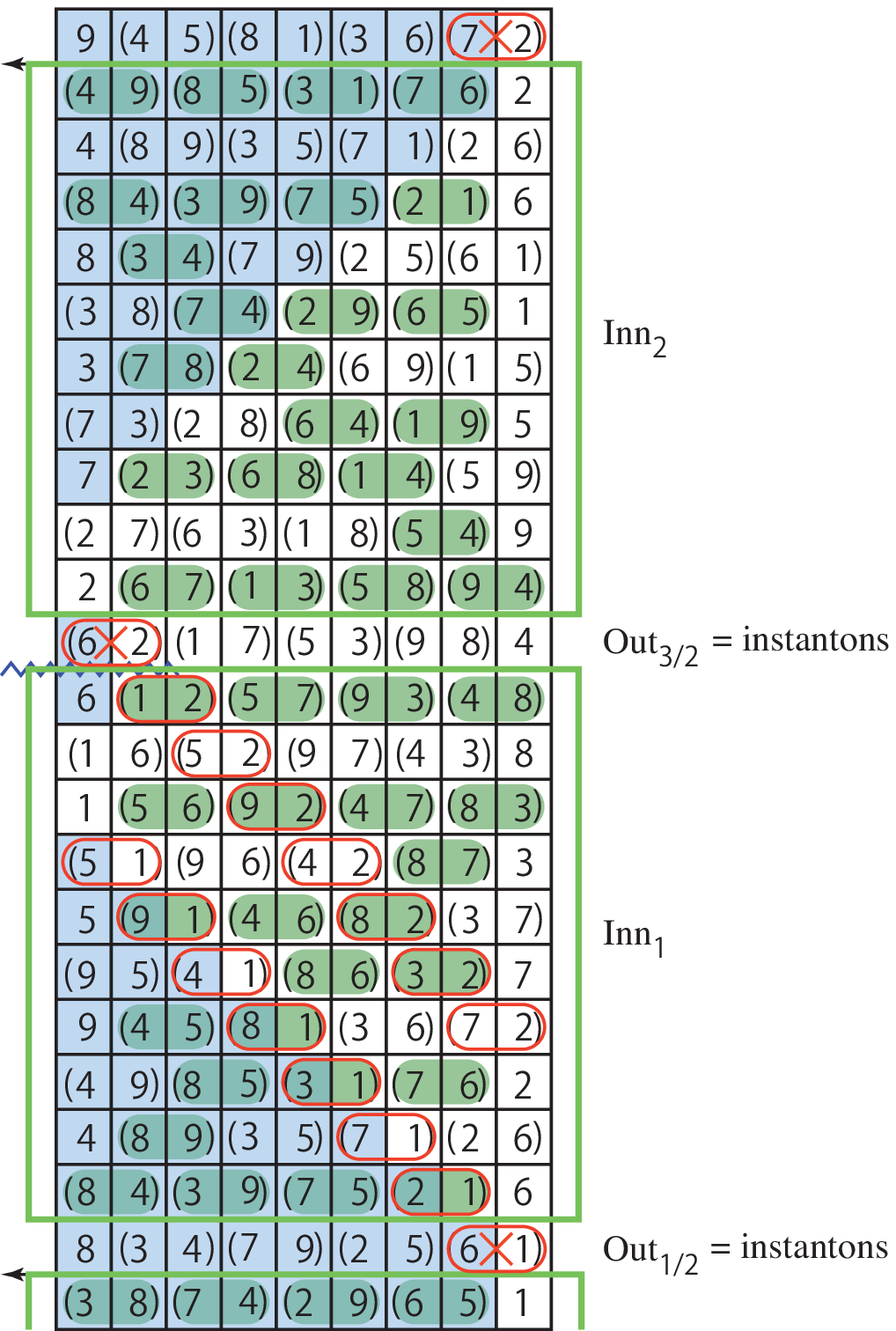}\quad 
\includegraphics[scale=0.5]{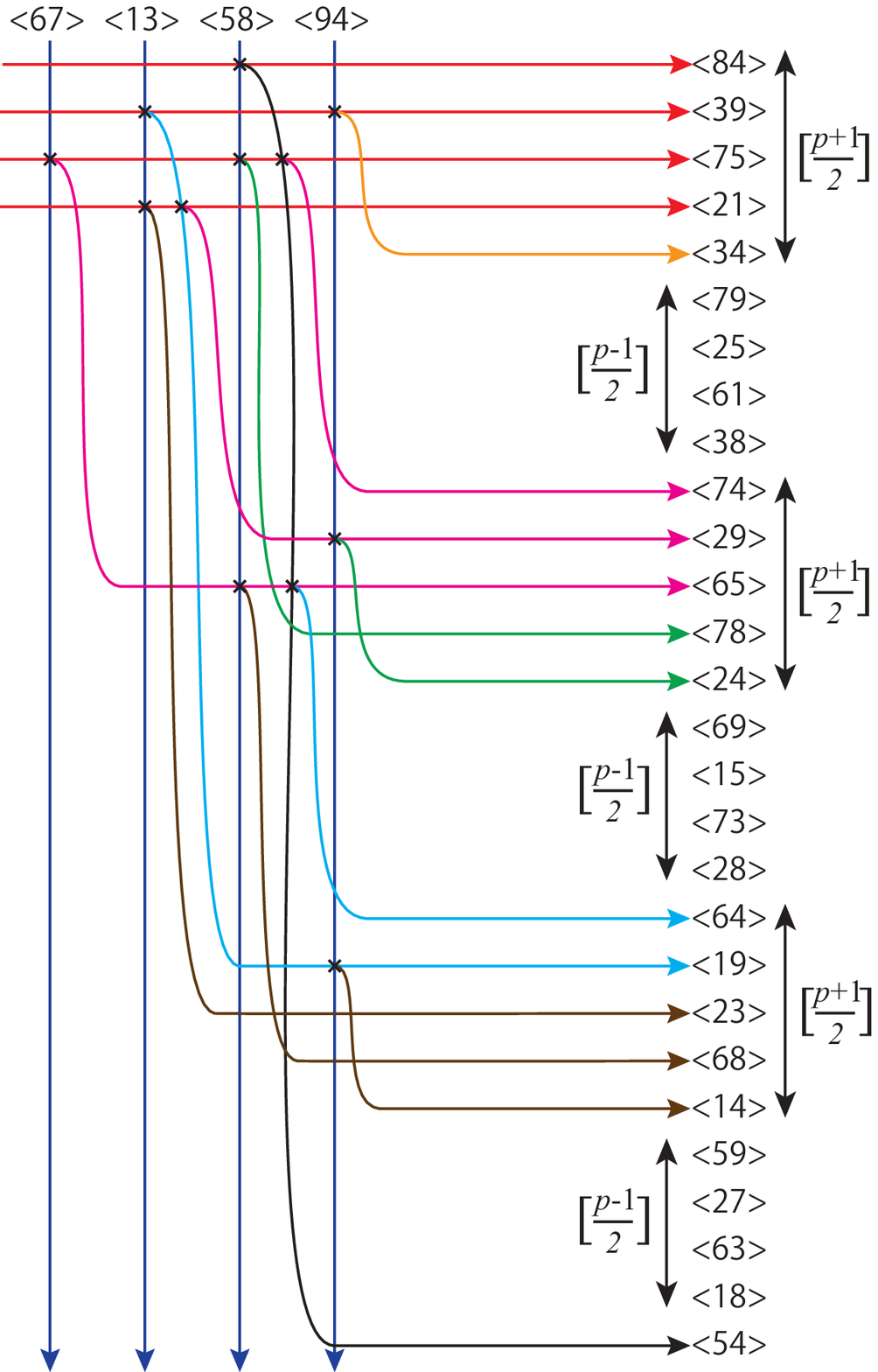}\quad 
\includegraphics[scale=0.6]{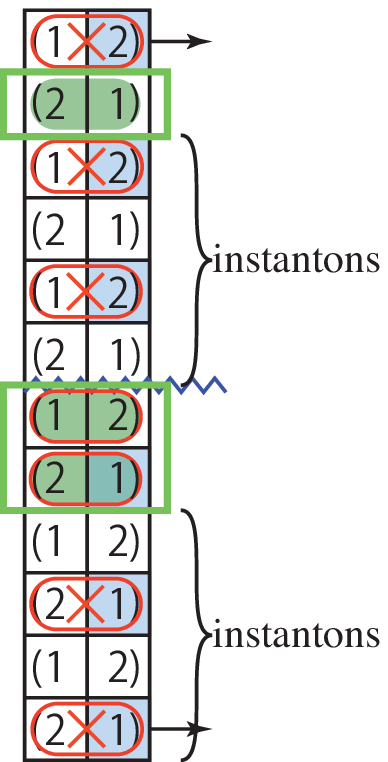}
\end{center}
\caption{\footnotesize The primitive solution of the $(p,q)=(9,2)$ case. Green shaded circles are non-zero multipliers with canonical weights. The primitive solution is consistent with the multi-cut boundary condition (red circles). General structure of $(p,2)$ cases are easily seen. The figure in the middle shows how other Stokes tails are generated (in ${\rm Inn}_2$). Remarkably, all the tails are associated with the canonical weight. The profile of the $(p,q)=(2,9)$ case is also shown with its primitive solution (See also Remark \ref{RemarkOnProfileOfPeq2}). }
\label{FigurePrimitiveQ2}
\end{figure}

The general structure of $(2k+1,2)$-systems can be also seen if one uses the profile of dominant exponents. We show it in Fig.~\ref{FigurePrimitiveQ2} (in the example of the $(p,q)=(9,2)$ case). 

Note that these weaving procedure can be performed in more general cases, say $(p,3)$ and $(p,4)$. Although it is doable, we have not found the general pattern to express them in a simple form. 

\section{Non-perturbative study on string duality \label{SectionStokesSpecific}}

So far, we have discussed proper spectral networks $\hat{\mathcal K}$ consistent with a spectral curve $\varphi(\mu;\lambda)$, which is the Chebyshev solution. The proper networks parametrize non-perturbative string theories which include the Chebyshev solution as a (meta-)stable vacuum of their string theory landscape. In this sense, we say that {\em proper networks $\hat {\mathcal K}$ (of $\varphi(\mu;\zeta)$) parametrize possible non-perturbative completions of the perturbative string theory $\varphi(\mu;\zeta)$}. 

Now we make a comparison between the non-perturbative completions constructed from the two sides of spectral $p-q$ duality. 
\begin{itemize}
\item [1. ] We introduce the pairs of $\bigl(\varphi(t;\zeta), \hat{\mathcal K}\bigr)$ and $\bigl(\widetilde \varphi(t;\eta), \widetilde{\mathcal K}\bigr)$ from both sides. 
\item [2. ] Evaluate the RH problem on both sides to obtain the BA functions $\Psi(g,t;\zeta|\theta)$ and $\widetilde \Psi(g,t;\eta|\widetilde \theta)$. Here D-instanton fugacity ($\{\theta_a\}_{a=1}^{\mathfrak g}$ and $\{\widetilde \theta_a\}_{a=1}^{\mathfrak g}$) are translated from spectral networks ($\hat{\mathcal K}$ and $\widetilde {\mathcal K}$). 
\item [3. ] From the BA functions $\Psi(g,t;\zeta|\theta)$ and $\widetilde \Psi(g,t;\eta|\widetilde \theta)$, one can construct its (isomonodromic) tau-functions ($\tau_X$ for $X$-system; $\tau_Y$ for $Y$-system), whose weak-coupling asymptotic expansions are non-perturbative partition functions $\mathcal Z_{\rm NP}(g,t|\theta)$ and $\widetilde{\mathcal Z}_{\rm NP}(g,t|\widetilde \theta)$. 
\end{itemize}
What one should ultimately compare is the non-perturbative partition functions:
\begin{align}
\mathcal Z_{\rm NP}(g,t|\theta)\qquad \leftrightarrow\qquad \widetilde{\mathcal Z}_{\rm NP}(g,t|\widetilde \theta), 
\end{align}
but conventionally, it is enough to compare the coefficient function, 
\begin{align}
 u_2(t) = g^2 \del_{t_1}^2\mathcal F(g,t) \qquad \leftrightarrow \qquad \widetilde v_2(t) = g^2 \del_{t_1}^2\widetilde{\mathcal F}(g,t)
\end{align}
especially in the $(p,2) \leftrightarrow (2,p)$ systems. Note that the functions $u_2(t)$ and $\widetilde v_2(t)$ are the same function under the duality relation (See Section \ref{SubsubsectionBAfunctions}): 
\begin{align}
\bP(t;\del) = 2 \del^2 + u_2(t) \qquad \Leftrightarrow \qquad \widetilde \bQ(t;\del) = \beta_{p,2} \bigl[2 \del^2 + \widetilde v_2(t)\bigr] \qquad \bigl(\widetilde v_2(t) = u_2(t)\bigr). 
\end{align}
Therefore, we evaluate the same observable from the different pictures. 
Schematically, it is shown in Fig.~\ref{FigureDualityCheckDiagram}. 

\begin{figure}[htbp]
\begin{center}
\begin{align}
\begin{array}{c}
\begin{xy}
(-25,7)*{\underline{\text{$X$-system}}\quad \rule[-1.4ex]{0pt}{2.8ex} }="X", 
(55,7)*{\quad \underline{\text{$Y$-system}} \rule[-1.4ex]{0pt}{2.8ex} }="Y", 
(-25,0) *{\quad \bigl(\varphi(t;\zeta), \hat{\mathcal K}\bigr)\quad \rule[-1.4ex]{0pt}{2.8ex} }="A", 
(55,0) *{\quad \bigl(\widetilde \varphi(t;\eta), \widetilde{\mathcal K}\bigr)\quad \rule[-1.4ex]{0pt}{2.8ex}}="B", 
(-25,-20) *{\Psi(g,t;\zeta|\theta)\quad \rule[-1.3ex]{0pt}{2.8ex} }="C", 
(55,-20) *{\quad \widetilde \Psi(g,t;\eta|\widetilde \theta) \rule[-1.3ex]{0pt}{2.8ex}}="D", 
(-25,-40)*{\tau_X=\mathcal Z_{\rm NP}(g,t|\theta)\quad \rule[-1.4ex]{0pt}{2.8ex}}="E", 
(55,-40)*{\quad \tau_Y=\widetilde{\mathcal Z}_{\rm NP}(g,t|\widetilde \theta)\rule[-1.4ex]{0pt}{2.8ex}}="F",
(-25,-60)*{\ds \mathcal F\simeq \mathcal F_p + \sum_I \theta_I e^{\frac{1}{g}\mathcal F_{\rm inst}^{(I)}}+\cdots\quad }="G", 
(55,-60)*{\ds \quad \widetilde {\mathcal F}\simeq \mathcal F_p + \sum_I \widetilde \theta_I e^{\frac{1}{g}\mathcal F_{\rm inst}^{(I)}}+\cdots}="H", 
(-25,-80)*{\ds \mathcal Z\simeq  \int_{\mathcal C_x} dx \, e^{-\frac{1}{g}V_{\rm eff}(x) } }="I",
(55,-80)*{\ds \widetilde {\mathcal Z} \simeq \int_{\mathcal C_y} dy \, e^{-\frac{1}{g}\widetilde V_{\rm eff}(y) } }="J",
\ar  @{<->} "A";"B"|{\text{ dual spectral curves }}
\ar @{->} "A";"C"|{\text{RH problem} \rule[-0.5ex]{0pt}{2ex}}
\ar @{->} "B";"D"|{\text{RH problem} \rule[-0.5ex]{0pt}{2ex}}
\ar @{<=>} "C";"E"|{\text{$\tau$/BA relation} \rule[-0.5ex]{0pt}{2ex}}
\ar @{<=>} "D";"F"|{\text{$\tau$/BA relation} \rule[-0.5ex]{0pt}{2ex}}
\ar @{<=>} "G";"E"|{\mathcal F = \ln \mathcal Z \rule[-0.5ex]{0pt}{2ex} }
\ar @{<=>} "H";"F"|{ \widetilde {\mathcal F} = \ln \widetilde {\mathcal Z} \rule[-0.5ex]{0pt}{2ex} }
\ar  @{<=>} "G";"H"|{\text{ comparison }}
\ar @{=>} "I";"G"|{\text{described by $X$-matrix} \rule[-0.5ex]{0pt}{2ex}}
\ar @{=>} "J";"H"|{\text{described by $Y$-matrix} \rule[-0.5ex]{0pt}{2ex}}
\end{xy}
\end{array} \nn
\end{align}
\end{center}
\caption{\footnotesize Comparison for spectral $p-q$ duality. Here we check the duality including D-instanton fugacity, which is generated by proper spectral networks. }
\label{FigureDualityCheckDiagram}
\end{figure}

\subsection{Non-perturbative ambiguity of minimal string theory \label{SubsectionNonperturbativeAmbiguityOfMinimalStringTheory}}

In this subsection, we discuss {\em non-perturbative ambiguity}. To avoid confusion in terminology, we should note the difference between non-perturbative ambiguity and Stokes ambiguity, which is discussed in Appendix \ref{AppendixNonperturbativeVSStokesAmbiguity}. Note that the ambiguity often discussed in resurgent analysis is Stokes ambiguity and is not the non-perturbative ambiguity discussed in this section. 

As is discussed in Introduction, the issue about non-perturbative ambiguity is that {\em non-perturbative string theory is not uniquely obtained from perturbative string theory}. In other words, there is a number of non-perturbative completions of a perturbative string theory, and there is no criterion/principle to specify a particular non-perturbative completion which identifies ``non-perturbative string theory.'' In this sense, our duality constraints (proposed in this paper) is one of the candidates for non-perturbative principle of string theory. 

In minimal string theory, we can say more specifically: Non-perturbative ambiguity is parametrized by proper spectral networks of ``a perturbative vacuum'' $\varphi(\lambda)$. Note that however not all the proper networks are realized by matrix models. It is clear if one sees the $(2,9)$-system, for example: 
\begin{align}
\begin{array}{c}
\includegraphics[scale=0.6]{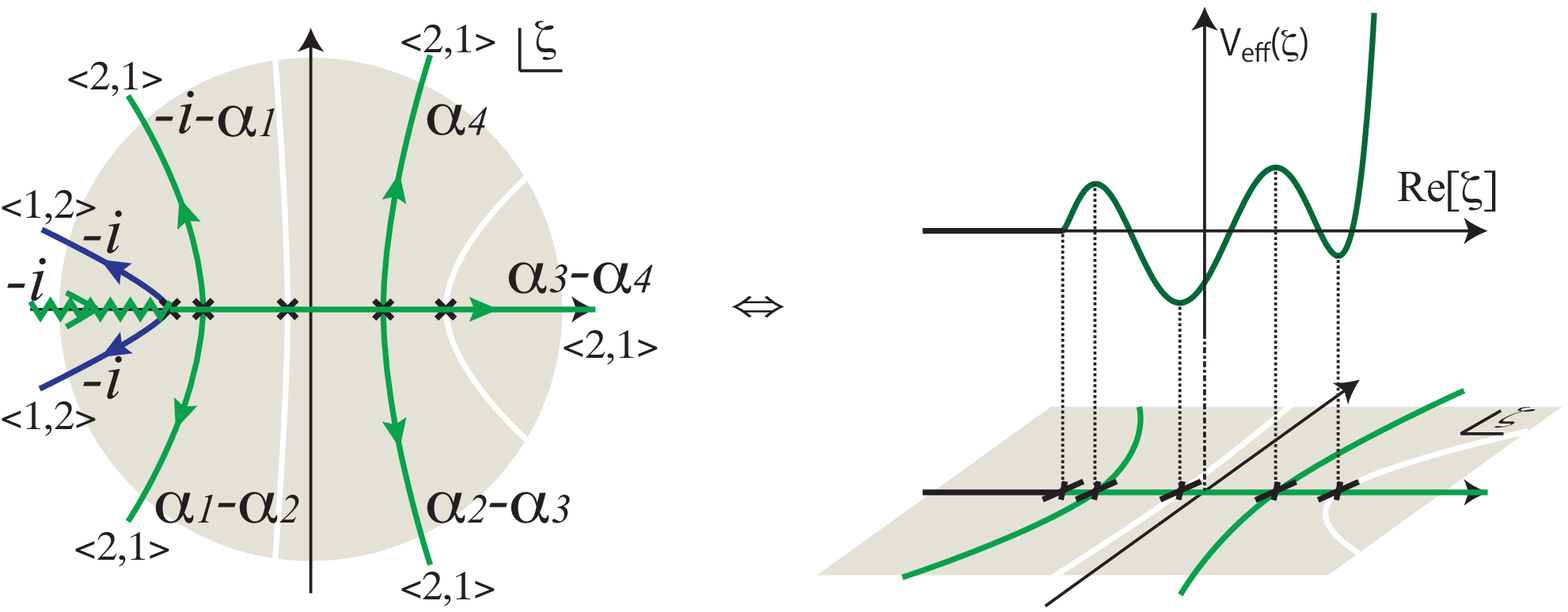}
\end{array}. \label{Equation29systemDZnetworkMM}
\end{align}
Here is shown the most general proper network realized within the ``conventional'' one-matrix models: 
\begin{align}
\mathcal Z = \int_{\mathcal C_X} dX e^{-N\tr V(X)}. \label{EquationConventionalMatrixModels}
\end{align}
Some comments are in order: 
\begin{itemize}
\item The parameters $\bigl\{\alpha_n\bigr\}_{n=1}^4$ are weights of Stokes tails which can be taken to be any complex values $\alpha_n\in \mathbb C$ ($n=1,2,3,4$). 
\item The right hand side of Eq.~\eq{Equation29systemDZnetworkMM} represents the corresponding contour of the matrix model with effective potential. 
\item This proper network is the most general solution of the multi-cut BC in one-matrix model \cite{CIY4}. 
\item According to the RH calculus, the parameters $\bigl\{\alpha_n\bigr\}_{n=1}^4$ reflects the values of D-instanton fugacity $\bigl\{\theta_I\bigr\}_I$ of free-energy. Therefore, this system has non-perturbative ambiguity parametrized by four complex variables $\bigl\{\alpha_n\bigr\}_{n=1}^4$. 
\end{itemize}
The corresponding mean-field integral (Eq.~\eq{MeanFieldPathIntegral123}) is given by the following contour $\mathcal C_x$ \cite{CIY4}: 
\begin{align}
\begin{array}{c}
\ds \mathcal Z \simeq \Bigl(\int_{\mathcal C_0}d\zeta + \sum_{n=1}^4 i \alpha_n \int_{\mathcal C_n} d\zeta \Bigr) e^{-\frac{1}{g} V_{\rm eff}(\zeta)} 
\cr \ds
\text{i.e. }\quad \mathcal C_x = \mathcal C_0 + \sum_{n=1}^4 i \alpha_n \mathcal C_n
\end{array}
\qquad \text{with}\qquad 
\begin{array}{c}
\includegraphics[scale=0.6]{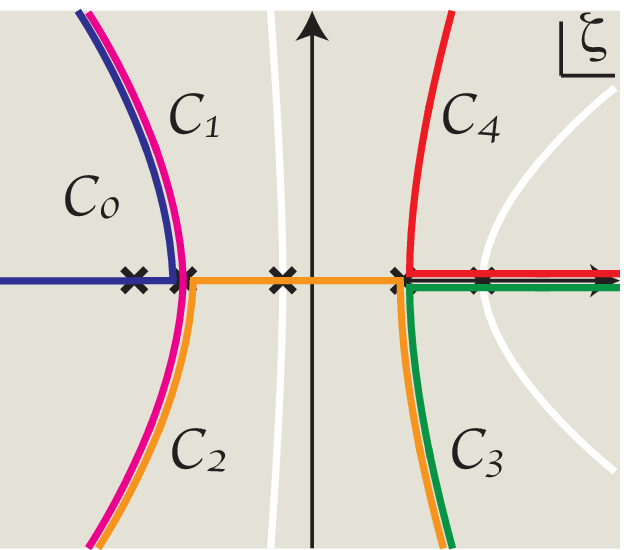}
\end{array}.
\end{align}
We put ``$i$'' in front of $\alpha_n$ because this factor is necessary to match the results from matrix models and from the Riemann-Hilbert calculus (See \cite{CIY4} for details). 

On the other hand, one can also generate (or weave) a proper network by adding an extra Stokes tail of $<1,2>$ as follows: 
\begin{align}
\begin{array}{c}
\includegraphics[scale=0.6]{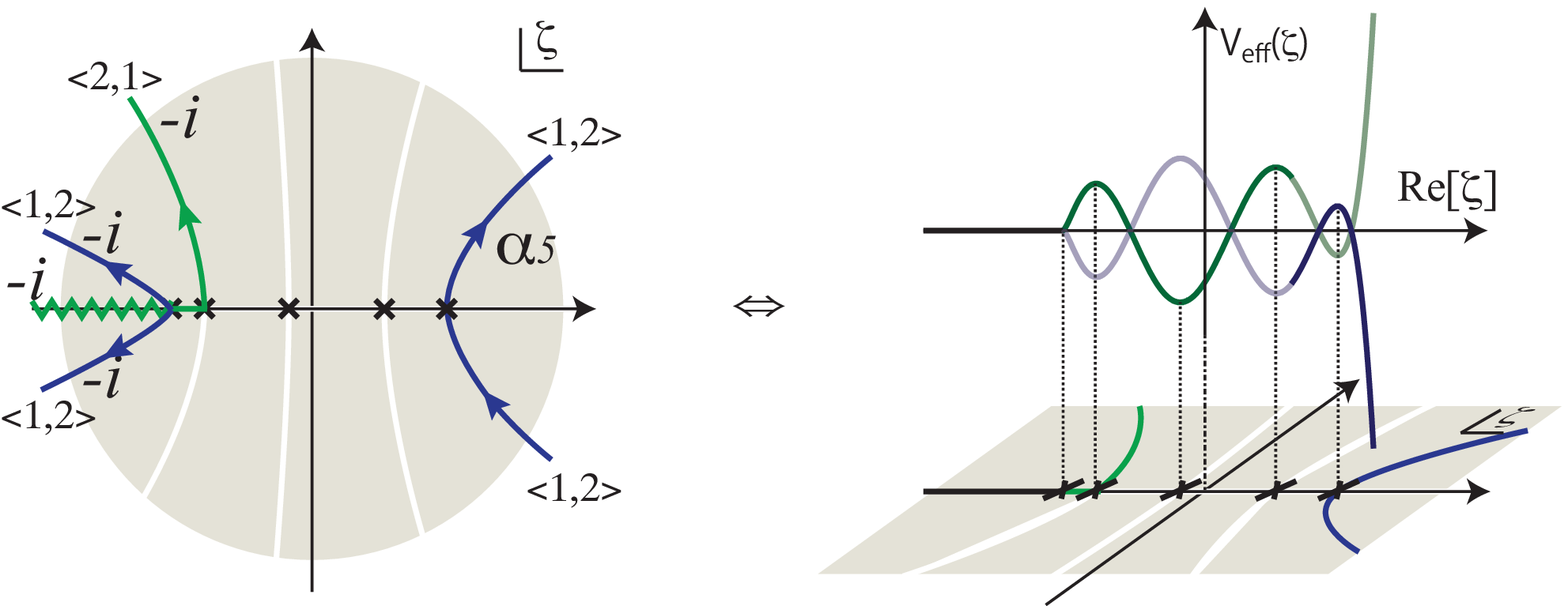}
\end{array}. \label{EquationIrregularContour}
\end{align}
This spectral network is also a proper network of the Chebyshev solution. However, {\em this network cannot be realized by the conventional matrix-model description (given by Eq.~\eq{EquationConventionalMatrixModels}).} If one pushes ahead with a formal concept of matrix models, it is also possible to say that there is ``a matrix model'' with up-side-down effective potential (as in the right-hand-side of Eq.~\eq{EquationIrregularContour}):
\begin{align}
\mathcal Z \simeq \int_{\mathcal C_0}d\zeta e^{-\frac{1}{g} V_{\rm eff}(\zeta)}  + i \alpha_5 \int_{\mathcal C_5} d\zeta e^{+\frac{1}{g} V_{\rm eff}(\zeta)}. 
\end{align}
However, it is clearly not of the conventional matrix model. Therefore, we consider that there is no matrix-model description corresponding to this proper network. 

\subsubsection{Spectral networks and effective potentials \label{SectionDiscussionEffectivePotentials}}

Here we briefly comment on the relation between spectral networks and matrix-model effective potentials. Mean field method is reviewed in Section \ref{SubsubsectionMeanFieldApproximation}, and it is not difficult to show that the mean-field effective potential is generally expressed as 
\begin{align}
e^{-\frac{1}{g} V_{\rm eff}(\zeta)} =\sum_{j\neq l} a_{j,l}(\zeta)\, e^{\frac{1}{g}\varphi^{(j,l)}(\zeta)}\qquad \zeta \in \mathcal C_x. 
\end{align}
Here $a_{j,l}(\zeta)$ is a piecewise function along the matrix-model contour, $\zeta\in \mathcal C_x$. A natural expectation is that there is a relation with spectral network $\hat {\mathcal K}$. 

In the cases of one-matrix models, as discussed above (and also in \cite{CIY4}), $\hat {\mathcal K}$ and $\mathcal C_x$ is almost identical and given as 
\begin{align}
a_{j,l}(\zeta) = i\alpha_{j,l}(\zeta)\qquad \zeta \in \mathcal C_x, 
\end{align}
where $\alpha_{j,l}(\zeta)$ is the weight of Stokes tail $<j,l>$ along $\zeta\in \mathcal K$. 
The only difference occurs around branch points, but it can be resolved by shrinking the inner sector to a line along the negative real axes: 
\begin{align}
\begin{array}{c}
\includegraphics[scale = 0.8]{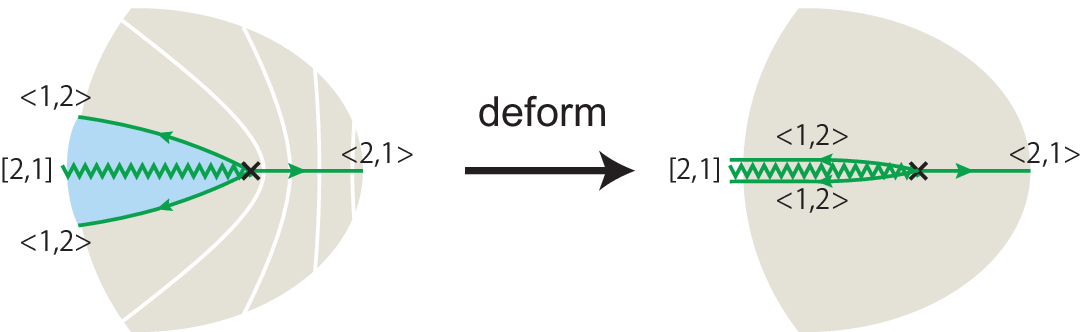}
\end{array}. 
\end{align}
Here is shown the case of $(2,p)=(2,4m+1)$, but other cases are also the same. This means that the DZ network represents the following Stokes phenomena of the effective function: 
\begin{align}
e^{-\frac{1}{g} V_{\rm eff}(\zeta)} 
=
\left\{
\begin{array}{cc}
e^{\frac{1}{g} \varphi^{(2,1)}(\zeta)} & \quad :\zeta > -\sqrt{\mu} \cr
e^{\frac{1}{g} \varphi^{(1,2)}(\zeta_+)} + e^{\frac{1}{g} \varphi^{(1,2)}(\zeta_-)}& \quad :\zeta < -\sqrt{\mu}
\end{array}
\right.\quad \bigl(g\to 0 \bigr). 
\end{align}
Since the potential along the branch cut is an oscillating function, the result of integration along the branch cuts should vanish almost everywhere in $g\to 0$. This behavior explains the flat direction of the effective potential $V_{\rm eff}(x)$ along the branch cut, which is usually understood by taking the principle value of potential.

\subsection{Comparison of instantons \label{SubsectionComparisonOfInstantonsInDualityConstraints}}

Here further comparisons are performed, based on instantons along spectral networks. 
From this subsection, we use the following notations for the $(p,2)\leftrightarrow (2,p)$ systems: 
\begin{align}
\begin{tabular}{c|c|c}
 & $(2,p)$-system (or $X$) & $(p,2)$-system (or $Y$) \cr
\hline
$\zeta$- and $\lambda$-coordinate & $\zeta = 2 \lambda^2$ & $\eta = 2^{p-1} \widetilde \lambda^{p}$ \cr
\hline
$\varphi$-function & $\varphi(\zeta)= \varphi(\lambda)= \varphi(\tau)$ & $\widetilde \varphi(\eta) =\widetilde \varphi( \widetilde \lambda)  =\widetilde \varphi( \widetilde \tau)$ \cr
\hline 
ambiguity parameters & $\alpha_1,\alpha_2,\alpha_3,\cdots$ & $s_1, s_2, s_3,\cdots$ \cr
\hline 
BA functions & $\Psi(\zeta),\, \vec \psi_{\rm orth}(\zeta),\cdots$ & $\widetilde \Psi(\eta),\, \vec \chi_{\rm orth}(\eta),\cdots$
\end{tabular}
\end{align}

\subsubsection{General dictionary in $(p,2)\leftrightarrow (2,p)$ cases \label{SubsubsectionGeneralDictionaryOfInstantonsDualityConstraints}}

As is discussed in Section \ref{SubsubsectionStokesGeometryProfileInstantons}, the location of instantons is labelled by the profile of dominant exponents. Since $(p,2)\leftrightarrow (2,p)$ systems are dual to each other, such components of the profiles are also mapped to each other. 
Here we show such a dictionary in $(p,2)\leftrightarrow (2,p)$ cases: 
\begin{Proposition} [Duality dictionary for instantons] \label{PropositionDualityDictionaryForInstantons}
Consider $(p,2)\leftrightarrow (2,p)$ systems ($p=2k+1$). The instantons on the profiles on both sides ($\mathcal J^{(p,p+2)}$ and $\mathcal J^{(2,p+2)}$) are translated into each other as follows (Note also Remark \ref{RemarkOnProfileOfPeq2}): 
\begin{align}
\bigl|\mathcal J_{1}^{(p,p+2)}\bigr| = 
\begin{tabular}{|c|c|c|c|c|c|}
\hline
{\,} & $(\,k\,)$ & $\cdots$ &$(\,3\,)$ & $(\,2\,)$ & $(\,1\,)$  \cr
\hline
\end{tabular}
\quad \leftrightarrow \quad \bigcup_{m=0}^{k-1} \mathcal J_{2m+1}^{(2,p+2)}
= \begin{tabular}{|c|}
\hline
$(\,k\,)_{2k-1}$ \cr
\hline
$\vdots$ \cr
\hline
$(\,3\,)_{5}$ \cr
\hline
$(\,2\,)_{3}$ \cr
\hline
$(\,1\,)_{1}$ \cr
\hline
\end{tabular}
\end{align}
and 
\begin{align}
\bigl|\mathcal J_{r+1}^{(p,p+2)}\bigr| = 
\begin{tabular}{|c|c|c|c|c|c|}
\hline
 $(\,1\,)$  &$(\,2\,)$ & $(\,3\,)$ & $\cdots$& $(\,k\,)$ & {\,} \cr
\hline
\end{tabular}
\quad \leftrightarrow \quad \bigcup_{m=0}^{k-1} \mathcal J_{-2m-3}^{(2,p+2)}
= \begin{tabular}{|c|}
\hline
$(\,1\,)_{-3}$ \cr
\hline
$(\,2\,)_{-5}$ \cr
\hline
$(\,3\,)_{-7}$ \cr
\hline
$\vdots$ \cr
\hline
$(\,k\,)_{-2k-1}$ \cr
\hline
\end{tabular}
\end{align}
Here $(*)$ means some parenthesis $(j|l)$ in the profile. 
In equation, it is expressed as 
\begin{align}
&\bigl|\mathcal J_1^{(p,p+2)} \bigr|\ni (j_{1,p-2m-1}| j_{1,p-2m})_1\quad \leftrightarrow \quad (1|2)_{2m+1}\text{ or } (2|1)_{2m+1} \in \mathcal J_{2m+1}^{(2,p+2)}, \label{PropositionDualityDictionaryOfInstantonsResult1}\\
&\bigl|\mathcal J_{r+1}^{(p,p+2)} \bigr| \ni (j_{r+1,2m+1}| j_{r+1,2m+2})_{r+1}\quad \leftrightarrow \quad (1|2)_{-2m-3}\text{ or } (2|1)_{-2m-3} \in \mathcal J_{-2m-3}^{(2,p+2)}\label{PropositionDualityDictionaryOfInstantonsResult2}
\end{align}
where $0 \leq m\leq k-1$. $\quad \square$
\end{Proposition}
{\em Proof}\quad Because of the uniform signature property (Theorem \ref{TheoremUniformSignatureProperty}), it is enough to check the absolute value of the instanton actions (of Eq.~\eq{EqInstantonL}). We here check Eq.~\eq{PropositionDualityDictionaryOfInstantonsResult1}. 
\\
\underline{\em 1. $(p,2)$ case} \quad The absolute value of the action is given as 
\begin{align}
\bigl|\widetilde \varphi^{(j_{1,p-2m-1},j_{1,p-2m})}(\widetilde \tau(1))\bigr| = \Bigl|\frac{4p\beta_{p,2}}{p^2-4}\Bigr|\bigl| \sin\big(\frac{2(j_{1,p-2m-1}-j_{1,p-2m})\pi}{p}\bigr)\sin \bigl(\frac{\pi}{2}\bigr)\bigr| \label{ActionOfJLinProfileCorrespondenceDictionaryInstantonDC}
\end{align}
By Theorem \ref{TheoremExplicitFormOfProfileComponents}, one obtains 
\begin{align}
j_{1,p-2m-1}-j_{1,p-2m} = (2m+1)m_1,\qquad\bigl(p(n_1+m_1) + 2m_1 = 1\bigr). 
\end{align}
and therefore 
\begin{align}
\text{Eq.~\eq{ActionOfJLinProfileCorrespondenceDictionaryInstantonDC}}
= \Bigl|\frac{4p\beta_{p,2}}{p^2-4}\Bigr|\bigl| \sin\big(\frac{(2m+1)\pi}{p}\bigr)\bigr| \qquad\bigl(0\leq m\leq k-1\bigr). \label{ActionOfJLinProfileCorrespondenceDictionaryInstantonDC2}
\end{align}\\
\underline{\em 2. $(2,p)$ case} \quad The absolute value of the action is given as 
\begin{align}
\bigl|\varphi^{(1,2)}(\tau(2m+1))\bigr| = \Bigl|\frac{4p\beta_{p,2}}{p^2-4}\Bigr|\bigl| \sin\big(\frac{p\pi}{2}\bigr)\sin \bigl(\frac{(2m+1)\pi}{p}\bigr)\bigr| = \text{Eq.~\eq{ActionOfJLinProfileCorrespondenceDictionaryInstantonDC2}}
\end{align}
This is the statement. Eq.~\eq{PropositionDualityDictionaryOfInstantonsResult2} is also the same. $\quad \blacksquare$

\begin{figure}[htbp]
\begin{center}
\begin{align}
\includegraphics[scale=0.6]{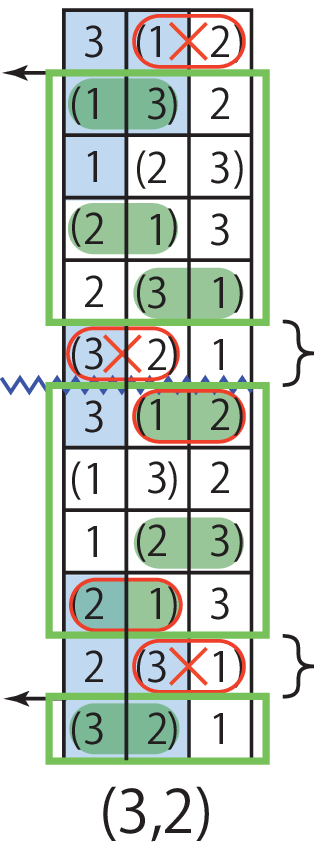}
\quad \raisebox{5ex}[1ex][0ex]{$\leftrightarrow$} \quad 
\includegraphics[scale=0.6]{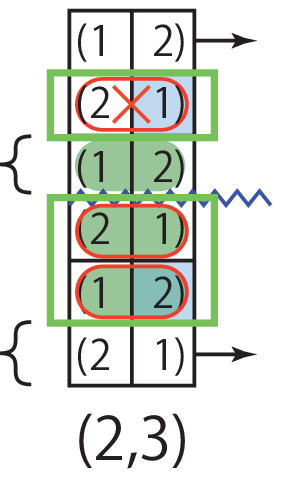}
\quad \raisebox{5ex}[1ex][0ex]{\Large ;}\quad 
\includegraphics[scale=0.6]{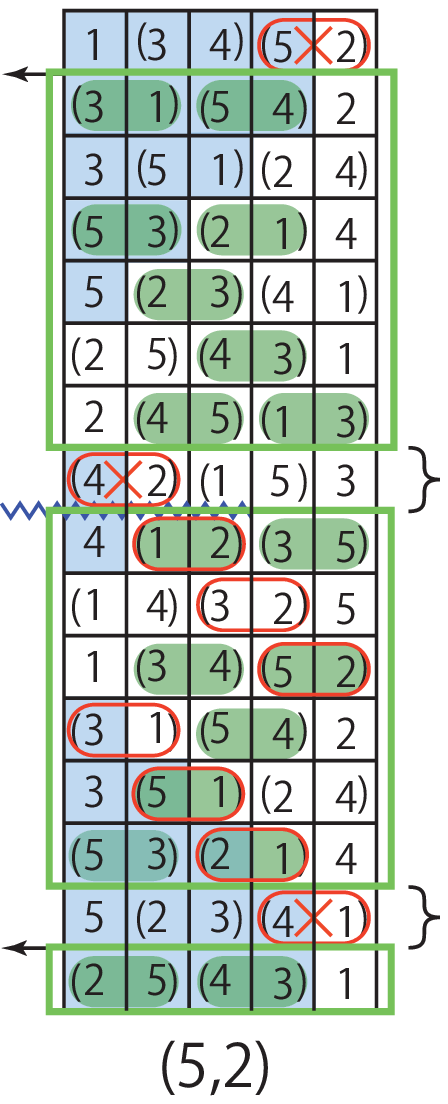}
\quad \raisebox{5ex}[1ex][0ex]{$\leftrightarrow$} \quad 
\includegraphics[scale=0.6]{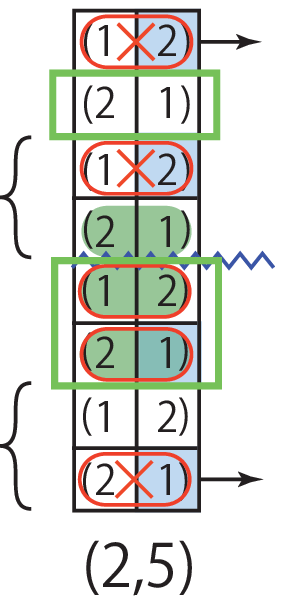}
\nn
\end{align}
\end{center}
\caption{\footnotesize Duality dictionary of $(3,2) \leftrightarrow (2,3)$ and $(5,2) \leftrightarrow (2,5)$. The parentheses ``$\bigl\}$'' or ``$\bigl\{$'' represent the part for the table of instantons. } 
\label{FigureDualityDictionary23322552}
\end{figure}

It is also instructive to see Fig.~\ref{FigurePrimitiveQ2}, which is the $(9,2)\leftrightarrow (2,9)$ systems. Four important examples (discussed in later) are shown in Fig.~\ref{FigureDualityDictionary23322552}. 

\subsubsection{Pure-gravity: $(3,2) \leftrightarrow (2,3)$ cases \label{SubsubsectionPureGravityDualityComparisonOfInstantons}}

The general matrix-model solution in $(2,3)$-system has one-complex-dimensional non-perturbative ambiguity (parametrized by $\alpha_1$) and is expressed as 
\begin{align}
\begin{array}{c}
\includegraphics[scale=0.6]{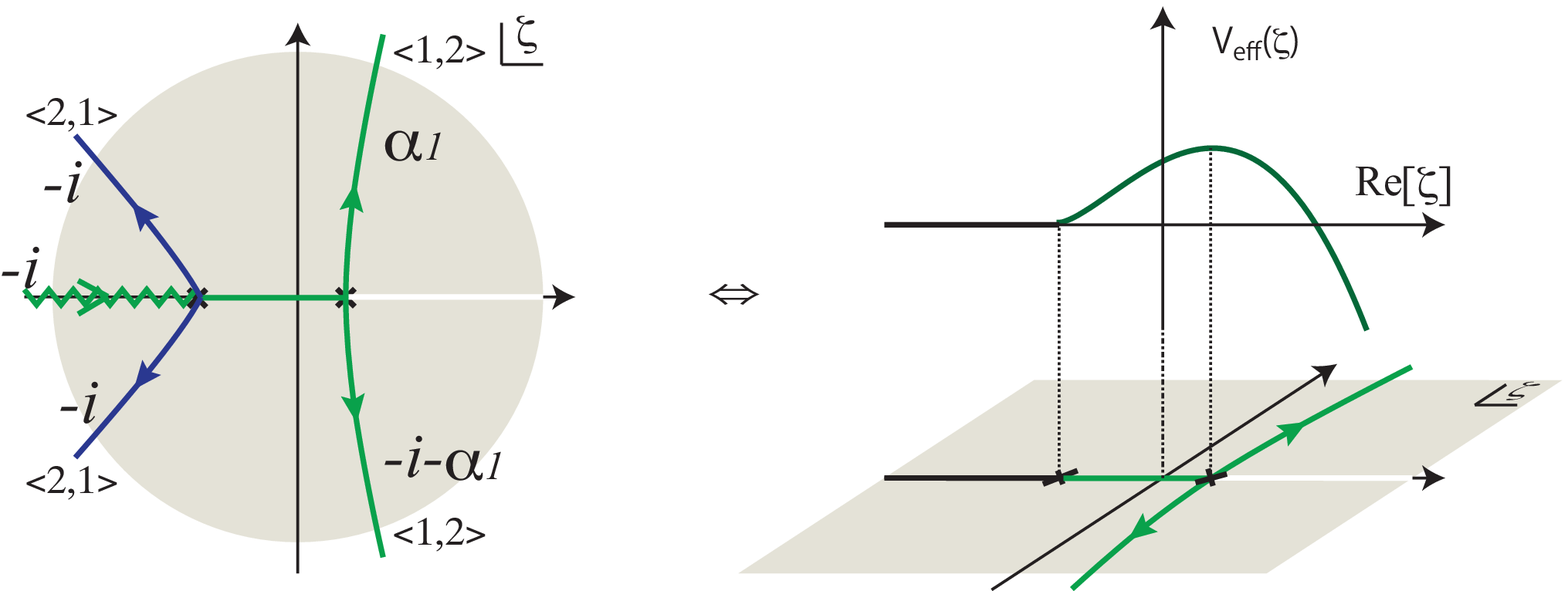}
\end{array}. \label{FigureEquationDZnetwork23systemMatrixModels}
\end{align}
It has a small instanton correction (related to $\alpha_1$) to the free-energy. 

On the other hand, the general proper network in $(3,2)$-system (which satisfies multi-cut BC) also possesses one-complex-dimensional non-perturbative ambiguity (parametrized by $s_1$) and is obtained by weaving the primitive network: 
\begin{align}
\begin{array}{c}
\includegraphics[scale=0.8]{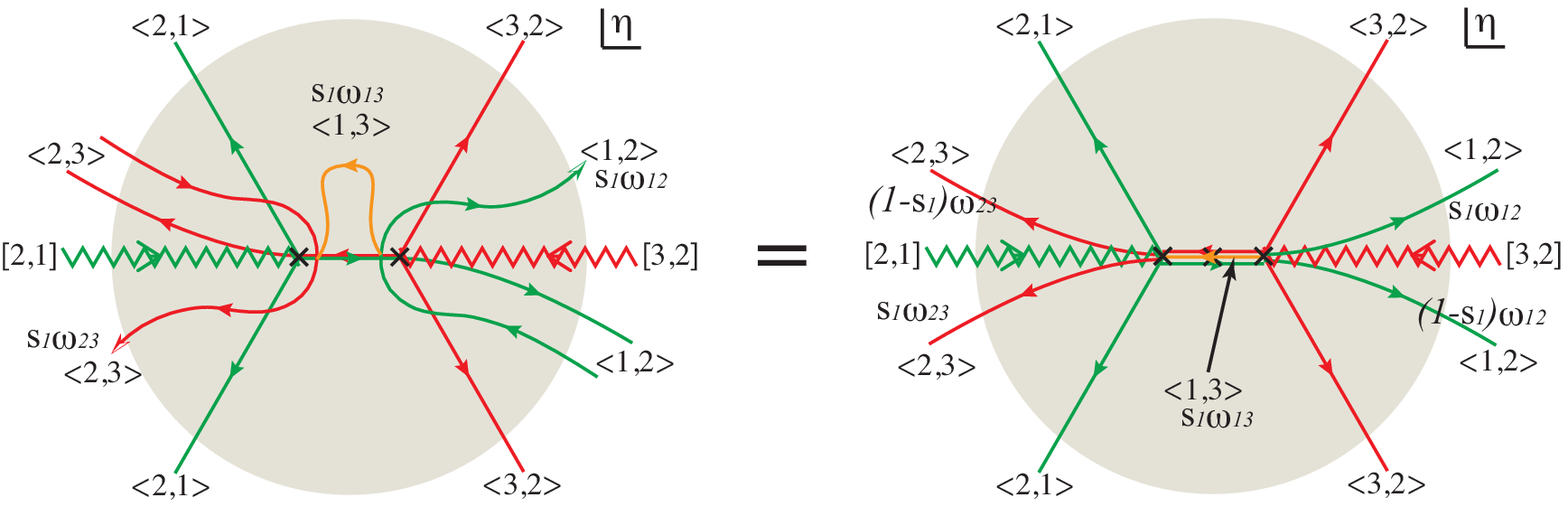}
\end{array}. 
\end{align}
This proper network also has the same small instanton correction (related to $s_1$ and given by $\widetilde \varphi^{(1,3)}(\eta_*)$) to the free-energy. 

Here we have found that the solutions on the both sides have one-complex-parameter ambiguity related to the same instanton. For further precise relationship, it is necessary to evaluate the RH problem and is remained to future investigation. At least the solutions of $i\alpha_1=s_1=0$ and $1$ have been checked to coincide \cite{SecondPaper}, since both solutions describe the same Stokes phenomenon in analytic continuation of $g$ (or $\mu$).

\subsubsection{Yang-Lee edge singularity: $(5,2) \leftrightarrow (2,5)$ cases \label{SubsubsectionYLedgeDualityComparisonOfInstantons}}

The general matrix-model solution in $(2,5)$-system has two-complex-dimensional non-perturbative ambiguity (parametrized by $\alpha_1$ and $\alpha_2$) and is expressed as 
\begin{align}
\begin{array}{c}
\includegraphics[scale=0.6]{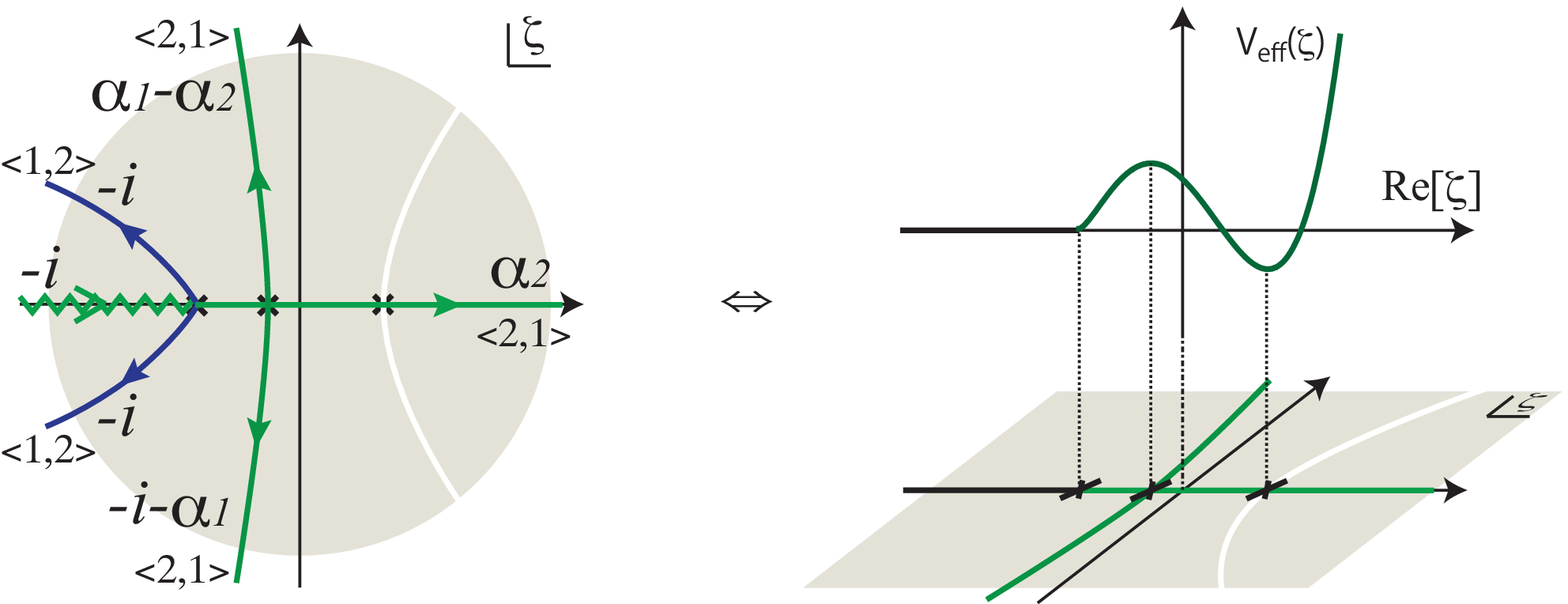}
\end{array}. \label{25systemDZnetworkMatrixModelSpectralNetworks}
\end{align}
It has a small instanton correction (related to $\alpha_1$) to the free-energy and also a large (or ghost) instanton correction (proportional to $\alpha_2$). 

If one turns off the large instanton ($\alpha_2=0$), then the situation is similar to that of pure-gravity. One can construct the similar solution in $(5,2)$-system, which has one-complex-dimensional non-perturbative ambiguity (parametrized by $s_1$), by weaving the primitive network: 
\begin{align}
\begin{array}{c}
\includegraphics[scale=0.8]{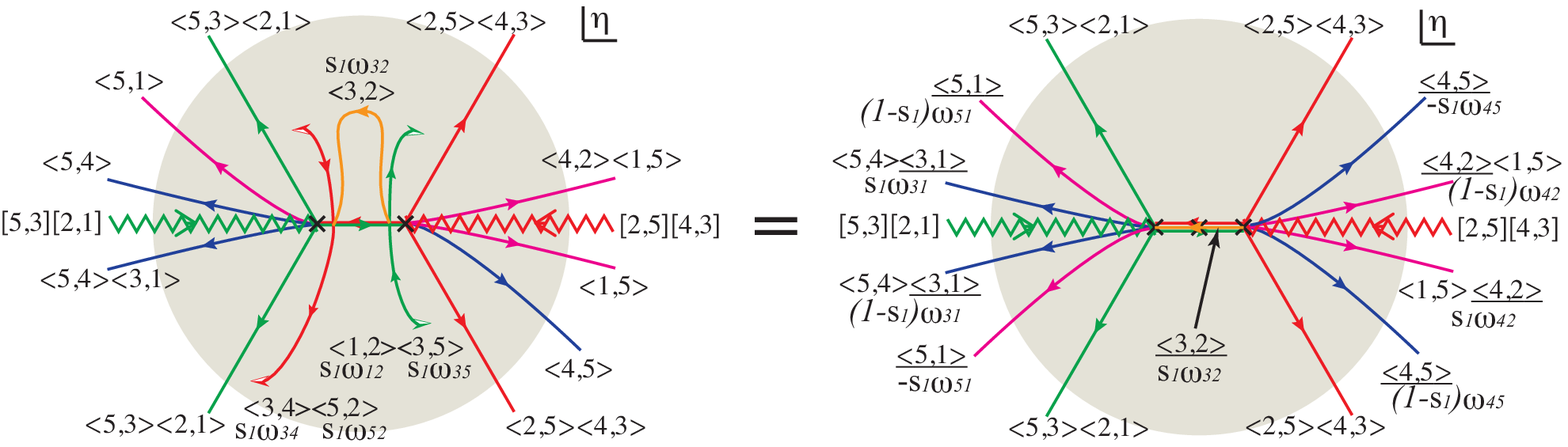}
\end{array}. \label{EquationFigure52systemDZofDKazakov5}
\end{align}
Here the underlined \underline{$<j,l>$} indicates that the weight is not the canonical value, and its value is explicitly written. If there is no underline, it means that their weights are the canonical ones and are not written in the spectral network. 
This solution also has the small instanton correction (related to $s_1$ and given by $\widetilde \varphi^{(3,2)}(\eta_*)$) to the free-energy. Importantly, this solution does not break the multi-cut BC. 

One then turns on the large instanton ($\alpha_2\neq 0$ in $(2,5)$-system). The corresponding instanton in $(5,2)$-system can be picked up by a vertical Stokes tail of $<4,1>$ (i.e.~is given by $\widetilde \varphi^{(4,1)}(\eta_*)$: See also Theorem \ref{TheoremUniformSignatureProperty}). Therefore, the corresponding proper network is obtained by weaving the network as 
\begin{align}
\begin{array}{c}
\includegraphics[scale=0.8]{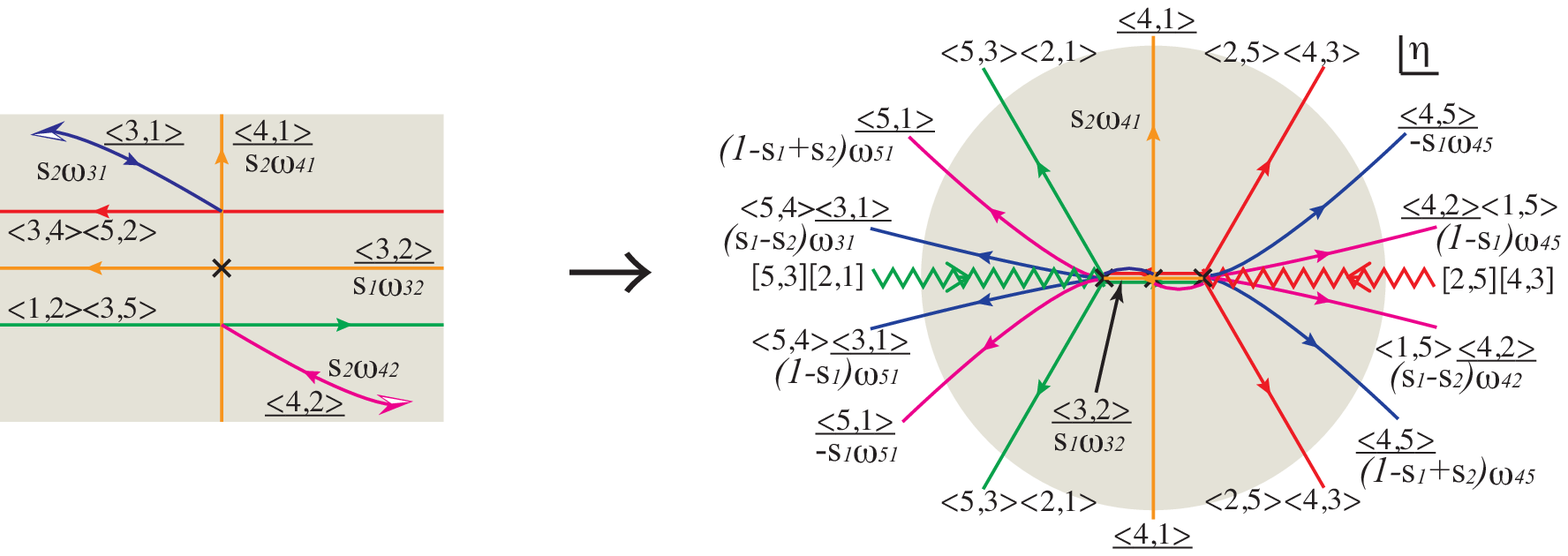}
\end{array}. \label{Figure52systemLargeDZofDKazakov5}
\end{align}
Here the behavior around the origin is shown in the left-hand-side. This solution has two-complex dimensional non-perturbative ambiguity (parametrized by $s_1$ and $s_2$) and possesses the large instanton correction (proportional to $s_2$). For further precise correspondence among those solutions (in $(2,5)$ and $(5,2)$ systems), it is again necessary to evaluate RH calculus. In any cases, however, {\em such a solution with the large instanton always breaks the multi-cut boundary condition in $(5,2)$-system}. As one can see in the profile (Fig.~\ref{FigureDualityDictionary23322552}), the multiplier corresponding to Stokes tail $<4,1>$ is forbidden by the multi-cut BC equation. Any solution with the large instanton should pass the saddle of $\widetilde \varphi^{(4,1)}(\eta)$ and provides the forbidden multiplier. 
We also note that, as one can also see in Proposition \ref{PropositionDualityDictionaryForInstantons}, this kind of phenomenon always occurs when $p=4m+1$. 

Instead of this large instanton $\widetilde \varphi^{(4,1)}(\eta_*)$, we can also add a horizontal Stokes tail $<1,4>$ which passes the saddle point of a small instanton $\widetilde \varphi^{(1,4)}(\eta_*)$: 
\begin{align}
\begin{array}{c}
\includegraphics[scale=0.8]{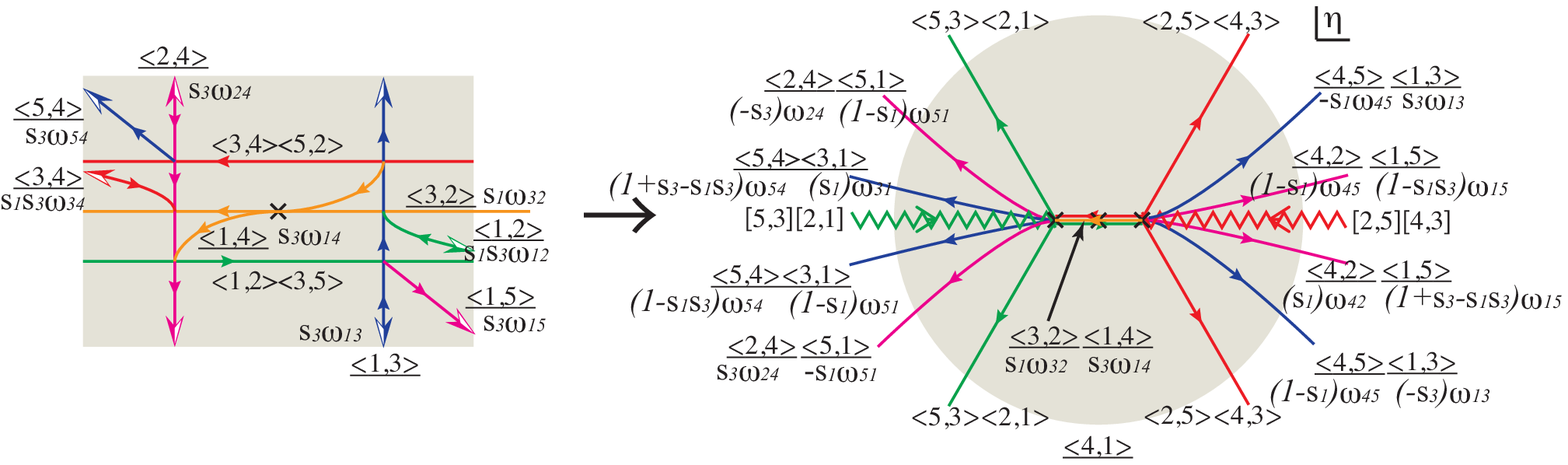}
\end{array}. \label{Figure52systemSmallDZofDKazakov5}
\end{align}
This solution also has two-complex-dimensional non-perturbative ambiguity (parametrized by $s_1$ and $s_3$) and possesses only small instantons (related to these two parameters). Importantly, this solution preserves the multi-cut boundary condition. However, the corresponding non-perturbative completions in $(2,5)$-system are given by the model with the up-side-down potential (which also has two-complex-dimensional non-perturbative ambiguity of $\alpha_1$ and $\alpha_3$): 
\begin{align}
\begin{array}{c}
\includegraphics[scale=0.6]{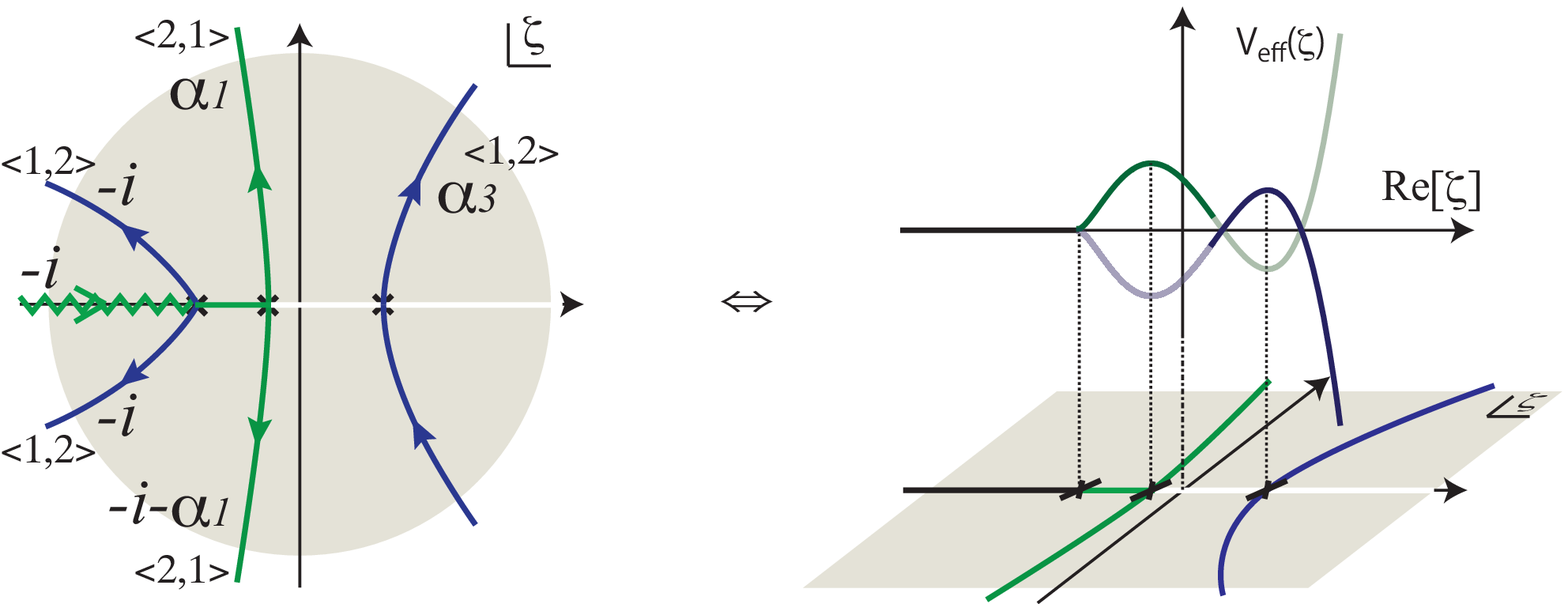}
\end{array}. \label{Equation25systemIrregularDZnetworkMatrixModel}
\end{align}
As is studied in \cite{CIY4}, such a solution with the up-side-down potential also breaks the multi-cut boundary condition in $(2,5)$-system. Therefore, as a natural consequence, this solution is forbidden from the viewpoint of $X$-system. 

\subsection{Resolvents and dynamics of eigenvalues \label{SubsectionResolventsAndDynamicsOfEigenvaluesDualityConstraints}}
In this subsection, we discuss an implication of {\em the breakdown of multi-cut boundary condition}. The actual geometry of eigenvalue cuts can be read from the spectral network obtained above (by using Definition \ref{DefinitionOfPhysicalCuts}), by using the scaling function of the orthogonal polynomial $\vec\psi_{\rm orth}(\zeta)$ (or $\vec \chi_{\rm orth}(\eta)$ for $Y$-system) which is given by Definition \ref{DefinitionMCBC}. 

\subsubsection{The $(5,2)$-system with multi-cut boundary condition \label{Section52systemWithMCBC}}

We first study $(5,2)$-system given by the proper network of Eq.~\eq{Figure52systemSmallDZofDKazakov5} (parametrized by $\alpha_3$ and $\alpha_5$), as an example which satisfies multi-cut boundary condition. We here consider the case of $(5,2)$-system, since $(2,5)$-system can be understood by a traditional and intuitive picture provided by one-matrix models.  

 As we discussed above, the dual side of this system, $(2,5)$-system with Eq.~\eq{Equation25systemIrregularDZnetworkMatrixModel}, does not satisfy multi-cut boundary condition and also there is no conventional matrix-model description. Despite of this fact, $(5,2)$-system of Eq.~\eq{Figure52systemSmallDZofDKazakov5} itself satisfies its multi-cut boundary condition and possesses a consistent picture described by the dynamics of $Y$-eigenvalues. 

\begin{figure}[htbp]
\begin{center}
\includegraphics[scale=0.7]{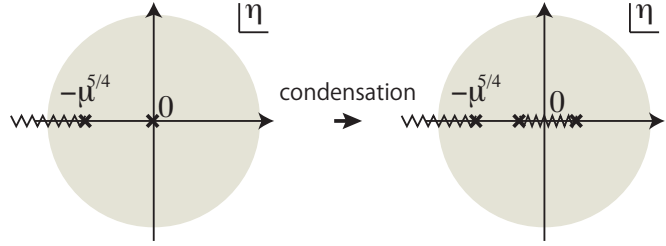}
\end{center}
\caption{\footnotesize Behavior of the resolvent cuts with and without condensations of instantons. } 
\label{FigureYLedgeSmallResolvents}
\end{figure}

The scaling function of the orthogonal polynomial $\vec\chi_{\rm orth}(\eta)$ is uniquely obtained and is given as the first component of the canonical solution $\widetilde \Psi_0(\eta)=\bigl(\widetilde \psi_0^{(1)}, \cdots ,\widetilde \psi_0^{(p)}\bigr)$. That is, 
\begin{align}
\vec \chi_{\rm orth}(\eta) = \widetilde \psi_0^{(1)}(\eta). 
\end{align}
In fact, the Stokes phenomenon of this function, $\vec \chi_{\rm orth}(\eta) = \widetilde \Psi_n(\eta) \vec v_n$ (of Eq.~\eq{ExpressionByCanonicalSolutionsInMCBC} and Eq.~\eq{ExpressionByCanonicalSolutionsInMCBC22}), is obtained as  
\begin{align}
\footnotesize
\left[
\begin{array}{c|ccccc}
n & v_n^{(1)} & v_n^{(2)} & v_n^{(3)} & v_n^{(4)} & v_n^{(5)} \\
\hline
13 & 0 & -\frac{1}{\sqrt{\omega }} & 0 & 0 & 0 \\
12 & 0 & -\frac{1}{\sqrt{\omega }} & 0 & 0 & 0 \\
11 & 0 & -\frac{1}{\sqrt{\omega }} & 0 & 0 & 0 \\
10 & 0 & -\frac{1}{\sqrt{\omega }} & 0 & 0 & 0 \\
9 & 0 & -\frac{1}{\sqrt{\omega }} & 0 & 0 & 0 \\
8 & 0 & -\frac{1}{\sqrt{\omega }} & 0 & 0 & 0 \\
7 & 1 & -\frac{1}{\sqrt{\omega }} & 0 & 0 & 0 \\
6 & 1 & -\frac{1}{\sqrt{\omega }} & -\frac{s_1}{\omega } & 0 & 0 \\
5 & 1 & -\frac{1}{\sqrt{\omega }} & -\frac{s_1}{\omega } & 0 &
\frac{s_1-1}{\omega ^2} \\
4 & 1 & -\frac{1}{\sqrt{\omega }} & 0 & 0 & \frac{s_1-1}{\omega ^2} \\
3 & 1 & -\frac{1}{\sqrt{\omega }} & 0 & 0 & 0 \\
2 & 1 & 0 & 0 & 0 & 0 \\
1 & 1 & 0 & 0 & 0 & 0 \\
0 & 1 & 0 & 0 & 0 & 0 \\
\end{array}
\right], 
\end{align}
and the asymptotic behavior of $\vec{\chi}_{\rm orth}(\eta)$ is the same as the cases of $s_3=0$ (i.e.~Eq.~\eq{EquationFigure52systemDZofDKazakov5}). 
Therefore, this system preserves the multi-cut boundary condition, and the single eigenvalue cut appears around $\eta\to\infty$ (as shown in the left-hand-side of Fig.~\ref{FigureYLedgeSmallResolvents})

As one can see in the spectral network (Eq.~\eq{Figure52systemSmallDZofDKazakov5}), there are (small) instantons related to $\widetilde \varphi^{(3,2)}(\eta_*)$ and $\widetilde \varphi^{(1,4)}(\eta_*)$. One can condense the instanton of $\widetilde \varphi^{(1,4)}(\eta_*)$ and the resulting geometry is shown as the right-hand-side of Fig.~\ref{FigureYLedgeSmallResolvents}. In order to obtain this eigenvalue cut from Definition \ref{DefinitionOfPhysicalCuts}, we should consider the Stokes lines defined as follows: 
\begin{align}
{\rm Re}\bigl[\widetilde \varphi^{(1,4)}(\eta)\bigr] = {\rm Re}\bigl[\widetilde \varphi^{(1,4)}(\eta_*)\bigr], \label{BranchCutStokesLinesEquation}
\end{align}
where $\eta_*$ is a saddle point of $\widetilde \varphi^{(1,4)}(\eta)$. Note that, since the instanton action of the saddle point is not vanishing,  this configuration of eigenvalues is obtained as a particular instanton sector related to the value of the instanton action, Eq.~\eq{BranchCutStokesLinesEquation}. Condensation of eigenvalues is given as follows: 
\begin{align}
\begin{array}{c}
\includegraphics[scale=0.7]{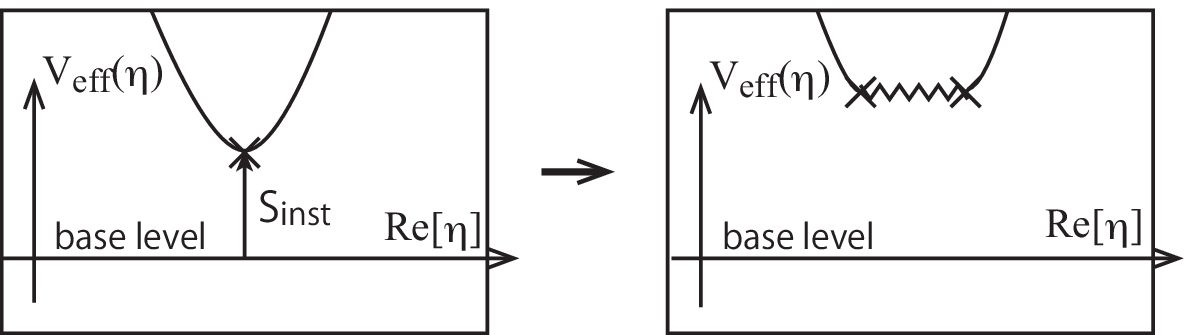}
\end{array}. 
\end{align}

\subsubsection{The $(2,5)$-system with the small instanton}

We next consider the completion of $(2,5)$-system given by the spectral network of Eq.~\eq{Equation25systemIrregularDZnetworkMatrixModel}, possessing two-complex-dimensional non-perturbative ambiguity (parametrized by $\alpha_1$ and $\alpha_3$). This system is the dual side of $(5,2)$-system defined with the spectral network Eq.~\eq{Figure52systemSmallDZofDKazakov5}, i.e.~the system which we have just discussed. 

\begin{figure}[htbp]
\begin{center}
\includegraphics[scale=0.7]{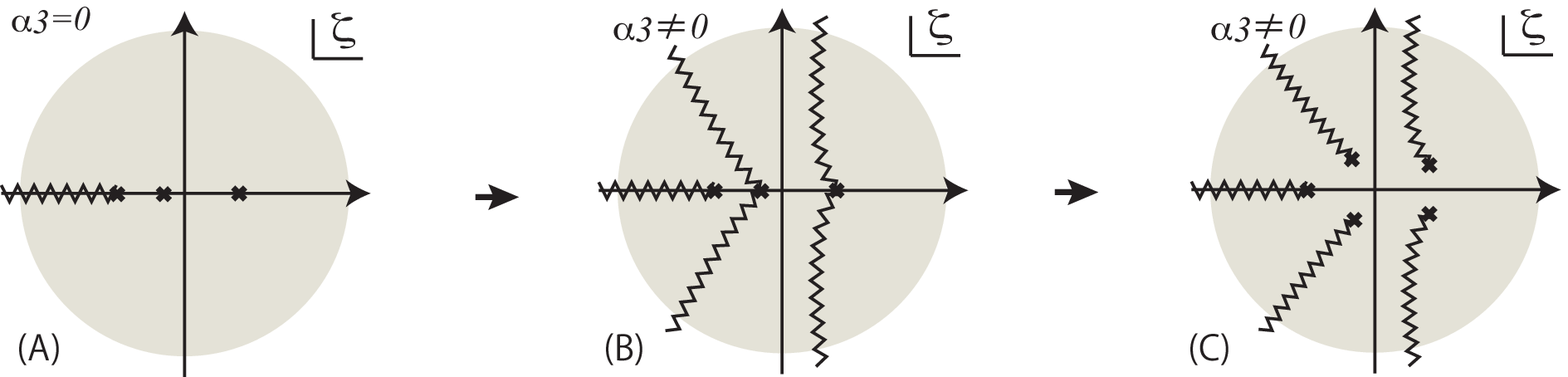}
\end{center}
\caption{\footnotesize Behavior of the resolvent cuts with perturbation of parameter $\alpha_6$. } 
\label{FigureYLedgeResolvents25}
\end{figure}

In this case, the scaling function of the orthogonal polynomial $\vec\psi_{\rm orth}(\zeta)$ is the second component of the canonical solution $\Psi_0(\zeta)=\bigl(\vec \psi_0^{(1)},\vec \psi_0^{(2)}\bigr)$. That is, 
\begin{align}
\vec \psi_{\rm orth}(\zeta) = \vec \psi_0^{(2)}(\zeta). 
\end{align}
In fact, the Stokes phenomenon of this function, $\vec \psi_{\rm orth}(\zeta) = \Psi_n(\zeta) \vec v_n$ (of Eq.~\eq{ExpressionByCanonicalSolutionsInMCBC} and Eq.~\eq{ExpressionByCanonicalSolutionsInMCBC22}), is obtained as  
\begin{align}
\footnotesize
\left[
\begin{array}{c|cc}
n & v_n^{(1)} & v_n^{(2)} \\
\hline
13 & i & 0 \\
11 & i & -\alpha _3 \\
9 & i \left(-\alpha _1 \alpha _3+\alpha _3+1\right) & -\alpha _3 \\
7 & i \left(-\alpha _1 \alpha _3+\alpha _3+1\right) & 1-\alpha _1 \alpha
_3 \\
5 & i \alpha _3 & 1-\alpha _1 \alpha _3 \\
3 & i \alpha _3 & 1 \\
1 & 0 & 1 \\
\end{array}
\right]\qquad \leftrightarrow \qquad 
\begin{array}{c}
\includegraphics[scale=0.7]{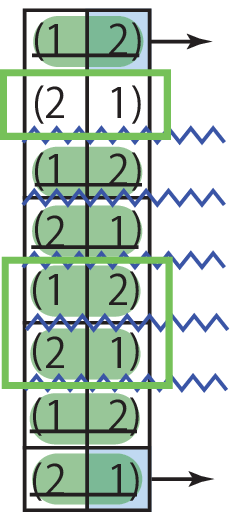}
\end{array}. 
\end{align}
Therefore, there are five eigenvalue cuts around $\zeta\to\infty$, along the angle of 
\begin{align}
\arg(\zeta) = \pi,\quad \frac{r\pm 2}{r}\pi,\quad \frac{r\pm 4}{r}\pi. 
\end{align}
These eigenvalue cuts are given as their Stokes lines and are drawn to be connected to saddle points. 
The geometry then can be seen as in Fig.~\ref{FigureYLedgeResolvents25}. It behaves as a first-order transition when $\alpha_3\neq 0$ (A $\to$ B in Fig.~\ref{FigureYLedgeResolvents25}). 

One can also consider condensations of instantons, similar to the cases of $(5,2)$-system. The condensation can be seen as the behavior of the resolvent space in Fig.~\ref{FigureYLedgeResolvents25} (B $\to$ C). 
If one writes it as the fermi-sea level, it is expressed as follows: 
\begin{align}
\begin{array}{c}
\includegraphics[scale=0.7]{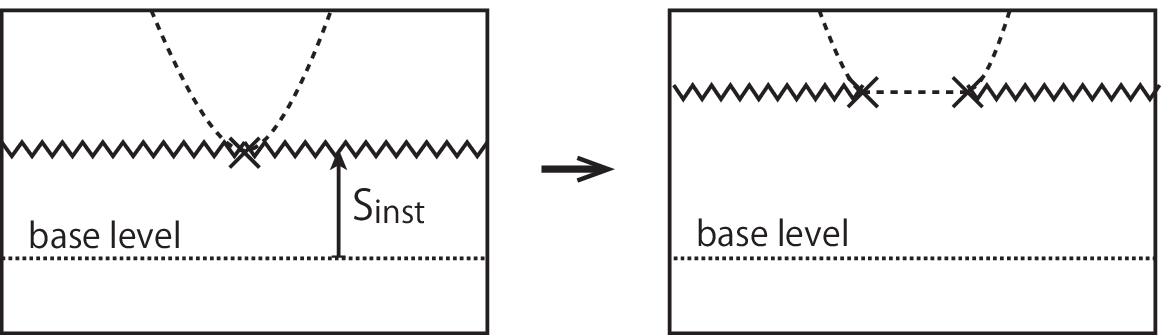}
\end{array}. \label{FigureEquationBoutroux25system}
\end{align}
That is, there is a (locally) gaussian potential around the saddle point as a local minimum of the effective potential. Intuitively, one may imagine that eigenvalues fill inside the gaussian potential, but the fact is that the eigenvalue cut of the resolvent goes to opposite directions for infinity. 

This kind of geometry does not appear in the usual one-matrix models. It is not surprising because this system is not realized in the conventional matrix models. Since one-matrix models are simple enough, one can easily identify which proper networks in $(2,5)$-system are described by matrix models. On the other hand, in the two-matrix models, it is not really clear {\em which proper networks are realized by matrix models and which are not?} In this sense, the multi-cut boundary condition and the eigenvalue geometry are one of the criterions to extract a necessary condition for being matrix models.

\subsubsection{The $(5,2)$-system with the large instanton}

\begin{figure}[htbp]
\begin{center}
\includegraphics[scale=0.7]{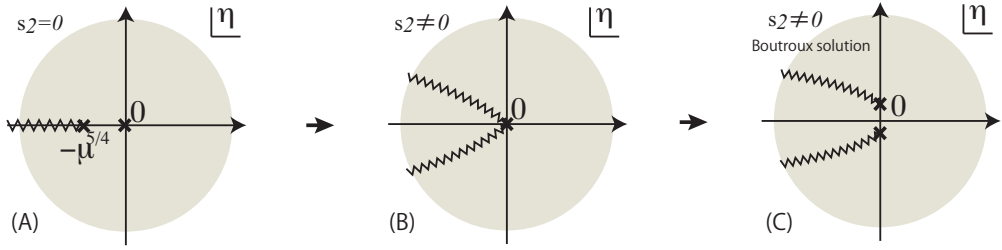}
\end{center}
\caption{\footnotesize Behavior of the resolvent cuts with perturbation of parameter $a_2$. } 
\label{FigureYLedgeResolvents}
\end{figure}

We next consider the completion of $(5,2)$-system given by the spectral network of Eq.~\eq{Figure52systemLargeDZofDKazakov5}, possessing two-complex-dimensional non-perturbative ambiguity (parametrized by $a_1$ and $a_2$). 
As is discussed in Section \ref{Section52systemWithMCBC}, the scaling function of the orthogonal polynomial $\vec\chi_{\rm orth}(\eta)$ is the first component of the canonical solution $\widetilde \Psi_0(\eta)=\bigl(\widetilde \psi_0^{(1)}, \cdots ,\widetilde \psi_0^{(p)}\bigr)$. That is, 
$\vec \chi_{\rm orth}(\eta) = \widetilde \psi_0^{(1)}(\eta)$. 
In fact, the Stokes phenomenon of this function, $\vec \chi_{\rm orth}(\eta) = \widetilde \Psi_n(\eta) \vec v_n$ (of Eq.~\eq{ExpressionByCanonicalSolutionsInMCBC} and Eq.~\eq{ExpressionByCanonicalSolutionsInMCBC22}), is obtained as  
\begin{align}
\footnotesize
\left[
\begin{array}{c|ccccc}
n & v_n^{(1)} & v_n^{(2)} & v_n^{(3)} & v_n^{(4)} & v_n^{(5)} \\
\hline
13 & 0 & -\frac{1}{\sqrt{\omega }} & 0 & 0 & 0 \\
12 & 0 & -\frac{1}{\sqrt{\omega }} & 0 & 0 & 0 \\
11 & 0 & -\frac{1}{\sqrt{\omega }} & 0 & 0 & 0 \\
10 & 0 & -\frac{1}{\sqrt{\omega }} & 0 & 0 & 0 \\
9 & 0 & -\frac{1}{\sqrt{\omega }} & 0 & 0 & 0 \\
8 & 0 & -\frac{1}{\sqrt{\omega }} & 0 & -\frac{s_2}{\omega ^{3/2}}
& 0 \\
7 & 1 & -\frac{1}{\sqrt{\omega }} & 0 & -\frac{s_2}{\omega ^{3/2}}
& 0 \\
6 & 1 & -\frac{1}{\sqrt{\omega }} & -\frac{s_1}{\omega } &
-\frac{s_2}{\omega ^{3/2}} & 0 \\
5 & 1 & -\frac{1}{\sqrt{\omega }} & \frac{s_2-s_1}{\omega }
& -\frac{s_2}{\omega ^{3/2}} & \frac{s_1-1}{\omega ^2} \\
4 & 1 & -\frac{1}{\sqrt{\omega }} & 0 & -\frac{s_2}{\omega ^{3/2}}
& \frac{s_1-s_2-1}{\omega ^2} \\
3 & 1 & -\frac{1}{\sqrt{\omega }} & 0 & -\frac{s_2}{\omega ^{3/2}}
& 0 \\
2 & 1 & 0 & 0 & -\frac{s_2}{\omega ^{3/2}} & 0 \\
1 & 1 & 0 & 0 & 0 & 0 \\
0 & 1 & 0 & 0 & 0 & 0 \\
\end{array}
\right]\qquad \leftrightarrow \qquad 
\begin{array}{c}
\includegraphics[scale=0.7]{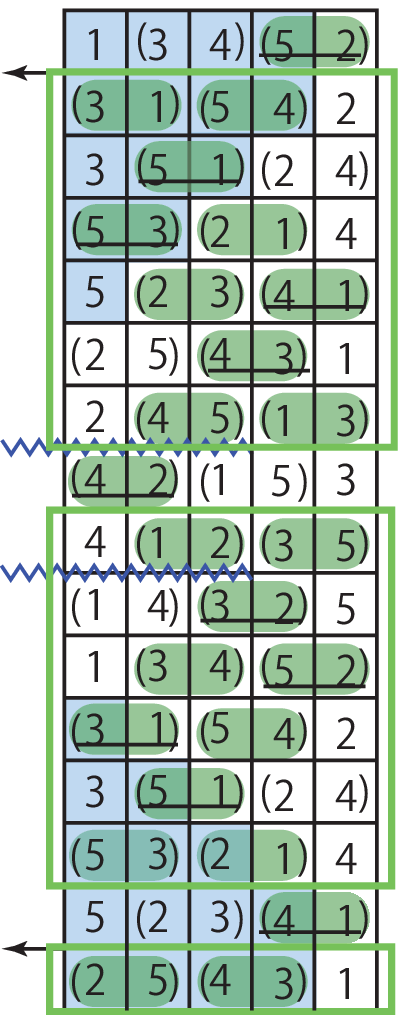}
\end{array}. 
\end{align}
Here the profile in the right-hand-side represents the leading behavior of $\vec \chi_{\rm orth}(\eta)$ around $\eta\to\infty$.%
\footnote{The underline of \underline{$(j|l)$} means that its Stokes multipliers are not the canonical value (as shown in the network, Eq.~\eq{Figure52systemLargeDZofDKazakov5}).} Therefore, there appear two eigenvalue cuts around $\eta\to\infty$ with angles of 
\begin{align}
\arg(\eta) = \pi \pm \frac{\pi}{r}, 
\end{align}
as an exchange of the dominant exponents $\widetilde \varphi^{(1)}\leftrightarrow \widetilde \varphi^{(4)}$ and $\widetilde \varphi^{(4)}\leftrightarrow \widetilde \varphi^{(2)}$. 
The geometry of eigenvalue cuts is then given by Stokes lines (not to confuse the branches, we use the coordinate $\widetilde \lambda$ and $\widetilde \tau$): 
\begin{align}
{\rm Re}[\widetilde \varphi^{(4,1)}(\widetilde \lambda)] = {\rm Re}[\widetilde \varphi^{(4,1)}(\widetilde \tau(1))],\qquad {\rm Re}[\widetilde \varphi^{(2,4)}(\widetilde \lambda)] = {\rm Re}[\widetilde \varphi^{(2,4)}(\widetilde \tau(2+1))],
\end{align}
where $\widetilde \tau(L)= \frac{L}{2p} \pi i$ is the saddle point, into which the Stokes lines can flow. By this, the behavior of eigenvalue geometry is shown in Fig.~\ref{FigureYLedgeResolvents}. The orthogonal polynomial shows a first-order phase transition when $s_2\neq 0$ (A $\to$ B in Fig.~\ref{FigureYLedgeResolvents}). 

Note that the Chebyshev solution is not the Boutroux solution in this spectral network, and there is a true vacuum $\widetilde \varphi_{\rm true}(\mu;\eta) \in\mathcal L_{\rm str_{\rm Chebyshev}}^{\hat {\mathcal K} \text{ of Eq.~\eq{Figure52systemLargeDZofDKazakov5} }}(\mu)$. The transition procedure is shown in Fig.~\ref{FigureYLedgeResolvents} (B $\to$ C). If one writes the transition as that of the fermi-sea level, it is expressed as follows: 
\begin{align}
\begin{array}{c}
\includegraphics[scale=0.7]{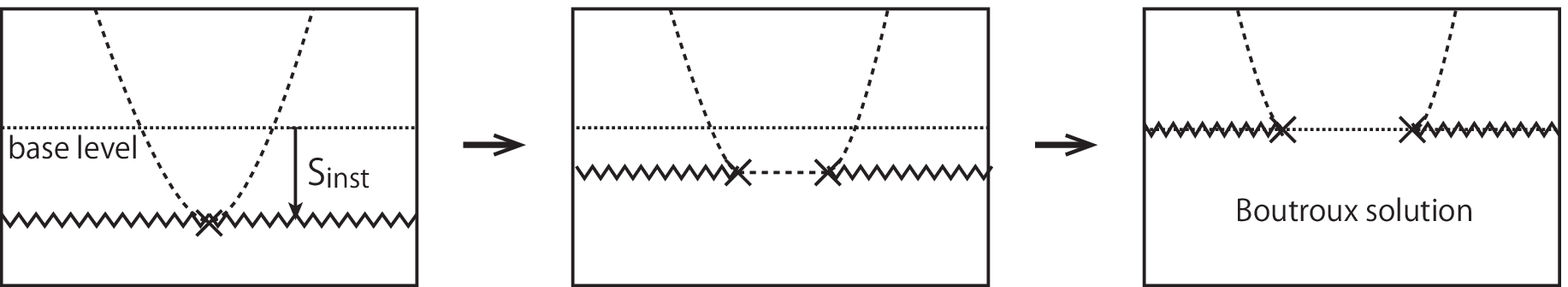}
\end{array}. 
\end{align}
This again is similar to the case of Eq.~\eq{FigureEquationBoutroux25system}. 
That is, there is a (locally) gaussian potential around the saddle point with a lower minimum than the base level, and the eigenvalue cut of the resolvent goes to opposite directions for infinity. 
This is the typical behavior which results from breakdown of multi-cut boundary condition. Therefore, this result indicates that this completion of $(5,2)$-system given by the spectral network Eq.~\eq{Figure52systemLargeDZofDKazakov5} is not realized by the conventional matrix models, or in other words, is not described by the conventional picture given by the dynamics of $Y$-eigenvalues.

\subsection{Duality constraints on string theory \label{SubsubsectionDualityConstraintsOnStringTheory}}

\begin{figure}[htbp]
\begin{center}
\includegraphics[scale=0.6]{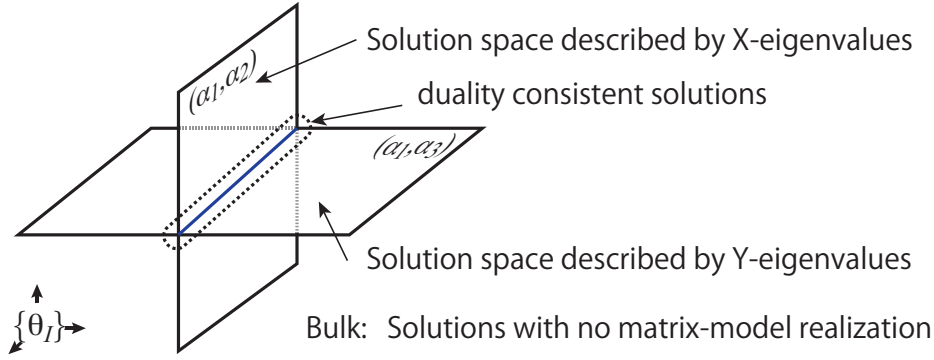}
\end{center}
\caption{\footnotesize The solution space of string equation and multi-cut BC for resolvents} 
\label{FigureDualConst}
\end{figure}

We now summarize the results of this section in Fig.~\ref{FigureDualConst}, where the solution space of string equation in $(2,5)\leftrightarrow (5,2)$-system is shown. The total dimension of the solution space is {\em four}, which is the order of the string equation. The dimension of the space of non-perturbative completions in $(2,5) \leftrightarrow (5,2)$ minimal string theory (based on Chebyshev solutions) is {\em three}, which are parametrized by $(\alpha_1,\alpha_2,\alpha_3)$ in $X$-system and by $(s_1,s_2,s_3)$ in $Y$-system. 

The non-trivial discrepancy appears when one considers whether the system is described by dynamics of $X$-eigenvalues and $Y$-eigenvalues. As we have discussed in this section, only the subspace can possess the conventional matrix-model descriptions: 
\begin{itemize}
\item The subspace parametrized by $(\alpha_1,\alpha_2) \simeq (s_1,s_2)$ is described by $X$-eigenvalues. 
\item The subspace parametrized by $(\alpha_1,\alpha_3) \simeq (s_1,s_3)$ is described by $Y$-eigenvalues. 
\end{itemize}
Therefore, only the further subspace (parametrized by $(\alpha_1)\simeq (s_1)$) is consistent with non-perturbative string duality. Otherwise, the system is described only either by $X$- or $Y$-eigenvalues. Therefore, non-perturbative string duality gives a constraint on the non-perturbative ambiguity and results in the system which only possesses one-dimensional non-perturbative ambiguity. This is the {\em duality constraint} caused by spectral duality in matrix models. 

We also note that the $(2,3) \leftrightarrow (3,2)$-system is also the same. In this case, the total solution space is two-dimensional; and the dimension of non-perturbative completions is one-dimension (parametrized by $\alpha_1$ of Eq.~\eq{FigureEquationDZnetwork23systemMatrixModels}). Although the discrepancy of the string duality does not appear in this case (due to the simplicity of the system), non-perturbative duality also results in the system which only possesses one-dimensional non-perturbative ambiguity. 

\subsubsection{(Non-)commutativity of two integrals and large $N$ limit}

If one admits our results on duality constraints, one should consider the implication of spectral duality in two-matrix integral of Fig.~\ref{FigureDualConstIntroduction}. That is, it implies that {\em the two matrix-integrals of $X$ and $Y$ does not commute in general.} 
As for commutativity of integrals, we should be concerned about Fubini's theorem (See e.g.~\cite{FunctionalAnalysisKF}). According to the theorem, non-commutativity appears when {\em the integrals are not absolutely integrable}, i.e.~the integral has oscillational (conditional) convergence. 

If one sees the system after double scaling limit, then the mean-field integral of the effective potentials passes along the part of oscillational convergence, which is the integral along eigenvalue branch cuts. Therefore, non-commutativity of integrals naturally takes place in this case. On the other hand, if one sees the system in finite $N$, since the potential is bounded from below in two-matrix integral, one naively expects that Fubini's theorem still holds, i.e.~the integral is commutative.

One can discuss this part in more detail. For example, in the following two-matrix models with $w(X,Y) = \frac{X^2}{2} + \frac{Y^2}{2} - a XY$, the integral order does not commute for $a>1$: 
\begin{align}
\mathcal Z = \int_{\mathcal C_X \times \mathcal C_Y} dX dY e^{-N \tr \bigl[\frac{X^2}{2} + \frac{Y^2}{2} -a XY \bigr]}. 
\end{align}
It is because this integral breaks the conditions of Fubini's theorem. One can also observe the following: The integration of $Y$-matrix does affect the leading behavior of the potential for $X$-matrix: 
\begin{align}
\mathcal Z = \int_{\mathcal C_X} dX \, e^{-N \tr \frac{X^2}{2}}\int_{\mathcal C_Y = {\mathfrak H}_N} dY \, e^{-N \tr \bigl[ \frac{Y^2}{2} -a XY\bigr]} = \int_{\mathcal C_X} dX \, e^{-N (1-a^2) \tr \frac{X^2}{2} }, 
\end{align}
and we should assign the contours of $X$ and $Y$ as 
\begin{align}
\mathcal C_X = i {\mathfrak H}_N,\qquad \mathcal C_Y = {\mathfrak H}_N \qquad \bigl(\text{first integrate $Y$}\bigr), 
\end{align}
according to the ordering of integrals. 
Here ${\mathfrak H}_N$ is the set of $N\times N$ Hermitian matrices, and $i{\mathfrak H}_N$ is that of $N\times N$ anti-Hermitian matrices:
\begin{align}
 i {\mathfrak H}_N \equiv \bigl\{X \in {{\mathfrak g}{\mathfrak l}}(N)\big|X^\dagger = - X\bigr\},\qquad {\mathfrak H}_N \equiv \bigl\{Y \in {{\mathfrak g}{\mathfrak l}}(N)\big|Y^\dagger = Y\bigr\}. 
\end{align}
The situation therefore becomes opposite when one first integrates $X$-matrix:
\begin{align}
\mathcal C_X = {\mathfrak H}_N,\qquad \mathcal C_Y = i {\mathfrak H}_N \qquad \bigl(\text{first integrate $X$}\bigr). 
\end{align}
However, such a phenomenon does not take place in the general cases. It is because the integration of one side does not affect the leading behavior of the potential for the other side. 

The difference of these two pictures are therefore double scaling limit. However, double scaling limit itself is essentially the procedure to scale up (or focus on) the local behavior of the system at the same time with large $N$ limit. Therefore, it is natural to conclude that {\em the non-commutativity appears as a result of large $N$ limit.} Therefore, we expect that this is the phenomenon caused by a large number of degrees of freedom inside the system. 

\subsubsection{Spectral networks and mean-field contours}
On the other hand, one can also see the non-commutativity of integrals from the relation between spectral networks and mean-field contours, discussed in Section \ref{SectionDiscussionEffectivePotentials}. If one focuses on one-matrix models, then the corresponding two-matrix models are simply given by 
the gaussian potential of $Y$, $V_2(Y) = \frac{Y^2}{2}$: 
\begin{align}
\mathcal Z &= \int_{\mathcal C_X} dX e^{-N \tr V(X)} = \int_{\mathcal C_X} dX e^{-N \tr V_1(X)} \int_{\mathcal C_Y=  {\mathfrak H}_N } dY e^{-N \tr \bigl[\frac{Y^2}{2} -XY\bigr]} \nn\\
&\simeq \int_{\mathcal C_x} dx e^{-N V_{\rm eff}(x)} 
\end{align}
In this case, $\mathcal C_Y$ is given by the set of $N \times N$ Hermitian matrices ${\mathfrak H}_N$: $\mathcal C_Y = {\mathfrak H}_N$, 
and the eigenvalues are along the real axes $\mathbb R$. Provided that the ordering of two integrals commutes, the mean-field integral should be expressed as an integral along real axes: 
\begin{align}
\mathcal Z  & = \int_{\mathcal C_Y= {\mathfrak H}_N } dY e^{-N \tr \frac{Y^2}{2}} \int_{\mathcal C_X} e^{-N \tr \bigl[V_1(X) - XY\bigr]} \nn\\
&\simeq  \int_{\mathbb R} dy\, e^{-N \widetilde V_{\rm eff}(y)}. \label{EqDualityConstraintEffectivePotentialForY}
\end{align}
However, if $X$-system has large (i.e.~ghost) instanton corrections, the spectral networks of $Y$-system should include vertical contours as in Eq.~\eq{Figure52systemLargeDZofDKazakov5} (due to the uniform signature property of Theorem \ref{TheoremUniformSignatureProperty}). Therefore, the ``equivalent'' mean-field integral in $Y$-system should include vertical contours, which do not originally exist in Eq.~\eq{EqDualityConstraintEffectivePotentialForY}: 
\begin{align}
\mathcal Z \simeq \int_{\mathcal C_x} dx\,  e^{-N V_{\rm eff}(x)} \simeq \int_{\mathbb R+ \text{vertical contours}} dy \, e^{-N \widetilde V_{\rm eff}(y)}\,\, \not \simeq \,\, \int_{\mathbb R} dy\, e^{-N \widetilde V_{\rm eff}(y)}. 
\end{align}
Therefore, the ordering of integrals does not commute in this case. In this sense, the key structure of duality constraint is the uniform signature property of Theorem \ref{TheoremUniformSignatureProperty}. That is, the directions of contour-integrals in $X$-systems and $Y$-systems (for the same instanton contributions) always differ by ninety degrees. 

In general, the solution space is parametrized by $2\mathfrak g (= (p-1)(q-1))$ D-instanton fugacity $\bigl\{ \theta_a\bigr\}_{a=1}^{2 \mathfrak g}$ which are those of D-instantons and of ghost D-instantons. Equivalently, the number of integration constants of string equation in the $(p,q)$-system is given by $2\mathfrak g$. Because of the uniform signature property of Theorem \ref{TheoremUniformSignatureProperty}, it is natural to conjecture the following: 
\begin{Conjecture}
Non-perturbative completions possessing non-perturbative spectral $p-q$ duality are give by a one-parameter family of the primitive solution. In particular, duality consistent $(p,q)$ minimal string theory is the true vacuum of the system. $\quad \square$
\end{Conjecture}
These conjectures should be checked with direct analysis in two-matrix models, or further by Riemann-Hilbert analysis. One may think that this conclusion is stronger than expected, because string theory should include many meta-stable vacua in the landscape. A possible reason would be because minimal string theory is too simple to have variety of the landscape. In the multi-cut matrix models \cite{MultiCut,fi1,irie2,CISY1,CIY1}, on the other hand, there can appear a number of dispersion-less vacua (obtained by DKK prescription) in a single string theory \cite{CISY1}. In this sense, it is also interesting to study duality constraints in these multi-cut systems.

\section{Conclusion and discussions \label{SectionDiscussion}}

In this paper, we have mainly studied two themes: 
\begin{itemize}
\item Non-perturbative completions of $(p,q)$ minimal string theory
\begin{itemize}
\item Classical monodromy matrices and spectral networks
\item Uniform signature property of instantons on the networks
\item Explicit solutions to local Riemann-Hilbert problems
\end{itemize}
\item Non-perturbative spectral $p-q$ duality and duality constraints
\begin{itemize}
\item Discrepancy in string duality as breakdown of multi-cut BC 
\end{itemize}
\end{itemize}
As a result of the breakdown of multi-cut BC, we have observed that string duality generally is broken in non-perturbative regimes. That is, {\em non-perturbative duality provides constraints rather than a dictionary of correspondence}: If one requires that string theory is described by matrix models associated with perturbative string theories in duality web, then it results in a constraint on non-perturbative ambiguity of string theory. 
As is discussed in Introduction, our quantitative results on the discrepancy should provide a missing piece in our understanding of non-perturbative string theory. Our proposal on the principle of non-perturbative string duality should be applied in general to various other systems in string theory, matrix model and integrable systems. 

\begin{itemize}
\item A simple extension of our analysis should be considered to various models which are governed by the usual topological recursions and spectral curves (c.f.~\cite{EynardOrantin}), therefore possibly also to knot theory (c.f.~\cite{KnotTheory}). It is because there is spectral duality in topological recursions, and duality constraints appear as its realization in matrix models. 
\item The necessity for our use of isomonodromy systems is due to complexity of effective potentials in the matrix-integral of two-matrix type (i.e.~including dual degree of freedom in the integral). In this sense, if the correspondence is explicitly given by that of two different matrix-integrals of one-matrix type,%
\footnote{We note that ``matrix-integrals of one-matrix type'' does not just mean one-matrix models nor a single kind of matrix. The point is that the effective potentials are easily obtained. For example, if spectral curve is given by second order in resolvent, the effective potential is obtained up to sign difference. In this case, the sign is usually easily chosen. Therefore, the matrix-integral would also include a number of matrices (or eigenvalues) of different kinds.  } 
then the application is even more straightforward (if there can be found a string duality connecting them): Just find contours which are consistent with the string duality. 
\item It is also interesting to see duality constraints caused by S-duality in matrix models (c.f.~\cite{EynardMarino}) which would be even closed within one-matrix models. 
\item For $\beta$-ensemble systems (c.f.~\cite{QuantumSpectralCurves}), it would not be straightforward if one adapts isomonodromy systems, since it is described by quantum isomonodromy systems. That is, it requires evaluation of effective potentials in the two-matrix-model description of $\beta$-ensemble which may not be an easy task. 
\item Also, instead of analytic study, one can also attempt to evaluate the duality constraints by use of numerical analysis which should provide another check of our argument on duality constraints. It should be also interesting to approach from some other developments of RHP for finite $N$ or orthogonal polynomials (c.f.~\cite{BleherIts,BertolaTovbisRHP}). 
\item Also it should be noted that double field theory \cite{DoubleFieldTheory} shares the same construction with two-matrix models. In this sense, it would be an example for us to apply the principle of non-perturbative string duality to critical string theory. In this sense, it is also interesting to study various other examples of duality constraints in matrix-type integrals obtained from relatively higher-dimensional field theory by localization technique \cite{Localization}. 
\item We also note that AGT correspondence \cite{AGT} (and also its extensions) also share the same structure. It claims a duality between four-dimensional quiver gauge theory and two-dimensional Liouville theory. If these theories are obtained by integrating four or two dimensional degree of freedom in six-dimensional theory, then the situation is the same as the two-matrix models and causes duality constraints on the system. In this sense, single-valued (also crossing-symmetric) correlators in Liouville theory (not only by conformal block) should be compared as non-perturbatively corresponding objects related to gauge theory. 
\item From more global viewpoints, there are studied a number of string dualities in string theory, mainly with use of D-branes (see e.g.~\cite{PolchinskiStringTheory}). However, the use of D-branes already means perturbative analysis. Therefore, duality constraints discussed in this paper will never be observed from that approach. It is essentially because these perturbative dualities observed by D-branes are duality of each perturbative saddle in the string-theory path integral. Only when one tries to extend the perturbative duality to the non-perturbative one acting on string-theory path integral itself, the duality constraints will appear. In this sense, we should reconsider all the string dualities from the non-perturbative point of views in order to extract their duality constraints on string theory. In our current understanding, the ways to extract the duality constraints in critical string theory are 1) analysis of Stokes phenomena/non-perturbative completions; 2) direct analysis of matrix model/gauge theory. 
\end{itemize}
These related study of our duality proposal should be interesting and should provide further understanding on non-perturbative string theory which would lead us to the unique non-perturbative string theory. 

\vspace{1cm}
\noindent
{\bf \large Note added in revision}
\vspace{0.2cm}

\noindent 
As new materials in the revision, we added solvability of RHP with the local RHP (Section \ref{SectionLocalRHP}), resolvents and dynamics of eigenvalues (Section \ref{SubsectionResolventsAndDynamicsOfEigenvaluesDualityConstraints}) and further discussion on non-commutativity of integrals (Section \ref{SubsubsectionDualityConstraintsOnStringTheory}). Explanations are clarified and expanded with additions of reviews (Section \ref{SubsubsectionAllOrderPerturbativeAnalysisAndSymplecticInvariants} and \ref{SubsubsectionEquivalenceWithInstantonsForAllOrderPT}), some detail proofs and detail calculations (e.g.~Appendix \ref{SectionIMSLaxBQ}). Some additional explanations are based on talks delivered by one of the authors (H.~Irie) in JPS meeting (on Sep.~21, 2013 in Japanese, around Fig.~\ref{FigureIntroductionDualityConstraintonMatrices}) and in YITP workshop 2013 (on Aug.~23, 2013 in Japanese, about Section \ref{SubsectionNonperturbativeAmbiguityOfMinimalStringTheory} and \ref{SubsectionComparisonOfInstantonsInDualityConstraints}).

\vspace{1cm}
\noindent
{\bf \large Acknowledgment}
\vspace{0.2cm}

\noindent
The authors would like to thank Masafumi Fukuma, Pei-Ming Ho, Kazuo Hosomichi, Goro Ishiki, Kohei Iwaki, Hikaru Kawai, Vladimir Kazakov, Ivan Kostov, Tsunehide Kuroki, Andrey Marshakov, Alexei Morozov, Toshio Nakatsu, Fumihiko Sugino and Mithat \"Unsal for useful discussions and/or comments. H.~Irie would like to thank people in Taiwan String Theory Focus Group, especially Pei-Ming Ho, for kind supports and hospitality during his stay in NCTS north at NTU, where some of the key progress has been done. The authors thank the Yukawa Institute for Theoretical Physics at Kyoto University, especially organizers of the YITP workshop on ``Field Theory and String Theory'' (YITP-W-13-12) for giving a chance to report some of the main results (on Aug.~23, 2013). Discussions during the workshop were also useful to complete this paper. C.-T.~Chan and C.-H.~Yeh also would like to thank YITP for hospitality during their stay which is useful to complete the paper. C.-T.~Chan and C.-H.~Yeh are supported in part by National Science Council (NSC) of Taiwan under the contract 
No.~99-2112-M-029-001-MY3 (C.-T.~Chan) and No.~102-2119-M-007-003 (C.-H.~Yeh). 
H.~Irie is supported in part by Japan Society for the Promotion of Science (JSPS) and Grant-in-Aid for JSPS Fellows No.~24-3610.


\appendix 

\section{Non-perturbative v.s.~Stokes ambiguity} \label{AppendixNonperturbativeVSStokesAmbiguity}
To avoid confusions, we also comment on Stokes phenomena appearing in analytic continuation of string coupling $g$ (or $t$). In general, if one performs asymptotic expansion in $g$ (or $t$), then we should understand that observables (say $u(g)$ for instance) are expanded in the following way:
\begin{align}
\left\{
\begin{array}{rl}
\text{stable: } & \ds 
u(g) \asymeq u^{[0]}(g) + O(g^{\infty}) \cr
\text{meta-stable: }& \ds 
u(g) \asymeq \sum_{n=0}^\infty \Bigl(\theta_A\,e^{\frac{A}{g}}\Bigr)^n\Bigl[u^{[n]}(g) + O(g^{\infty}) \Bigr]
\end{array}
\right. \quad \bigl(g\to +0 \in \mathbb R\bigr), \label{EqAsymExpTransFirstOrderRes}
\end{align}
where $\bigl\{u^{[n]}(g)\bigr\}_{n,m}$ are asymptotic series of multi-instanton sectors given by 
\begin{align}
u^{[n]}(g) = g^{n\gamma}\sum_{m=0}^\infty u^{[n]}_m \, g^m. 
\end{align}
The meaning of ``asymptotic'' is encoded in $O(g^\infty)$:  
\begin{align}
u(g) &\asymeq u^{[0]}(g) + O(g^{\infty}) \qquad \bigl(g\to +0\in \mathbb R\bigr)\nn\\
&\Leftrightarrow\quad 
\left\{
\begin{array}{l}
\text{For ${}^\forall M \in \mathbb N$, ${}^\exists \epsilon_M >0$ and ${}^\exists C_M >0$, } \cr
\ds \quad 
\text{s.t.}\quad 
\left|u(g)-\sum_{m=0}^M u_m^{[0]}\right|< C_M g^{M+1} \qquad 0 < g <  \epsilon_M 
\end{array}
\right.. \label{EqDefinitionAsymptoticExpansion}
\end{align}
In the cases of meta-stable expansion, this consideration should be applied to coefficients of the large-instanton expansion, keeping in our mind that we are expanding the observable around zero-instanton sector $u^{[0]}(g)$. In this sense, these coefficients $\bigl\{u_m^{[n]}\bigr\}_{n,m}$ are observable coefficients in $g\to +0\in \mathbb R$. 

On the other hand, invisible terms inside $O(g^{\infty})$ are exponentially small terms compared with the observable series $\bigl\{u^{[n]}(g)\bigr\}_{n,m}$. Since they are invisible, they are allowed to possess ``ambiguity''. Here, since this ``ambiguity'' is different from ``the non-perturbative ambiguity'' which we mainly discuss in this paper,%
\footnote{Non-perturbative ambiguity discussed in this paper is ambiguity arising in identifying a solution of Painlev\'e equation (or string equation) corresponding to string theory. In the trans-series analysis, such an ambiguity is caused by the fact that trans-series parameters (i.e.~D-instanton fugacity) are free parameters. } this ambiguity is temporarily called {\em Stokes ambiguity} to avoid confusion. If the invisible terms are also given by trans-series expansion: 
\begin{align}
u(g) \asymeq u^{[0]}(g) + \sum_{m=1}^\infty \bigl(\sigma_A e^{-\frac{A}{g}}\bigr)^m u^{[0;m]}(g) \qquad g\to +0 \in \mathbb R, \label{EqResTransGzeroP}
\end{align}
then $\sigma_A$ may possess Stokes ambiguity in the following way: The coefficient $\sigma_A$ can become observed if one analytically continues $u(g)$ till these exponentially small instanton terms become large enough to be observable. There are two ways of analytic continuation, $g \to |g|e^{\pm i\theta}$ (${}^\exists\theta >\frac{\pi}{2}$): 
\begin{align}
u(g) \asymeq u^{[0]}(g) + \sum_{m=1}^\infty \bigl(\sigma_A^{(\pm)} e^{-\frac{A}{g}}\bigr)^m u^{[0;m]}(g) \qquad g\to 0\times e^{\pm i\theta}\qquad \bigl({}^\exists \theta >\frac{\pi}{2}\bigr), 
\end{align}
and the coefficients $\sigma_A^{(\pm)}$ are generally different: 
\begin{align}
\sigma_A^{(+)} \neq \sigma_A^{(-)}, 
\end{align}
i.e.~Stokes phenomenon occurs. 
Therefore, $\sigma_A$ in Eq.~\eq{EqResTransGzeroP} is ambiguous.%
\footnote{It often happens that the imaginary part of ${\rm Im}[\sigma_A]$ become visible if the perturbative part $u^{[0]}(g)$ is real-valued series. In this case, this value has a physical meaning, like decay rate caused by a bounce solution (given by D-instanton). In this case, the jump should satisfies ${\rm Im}[\sigma_A^{(+)}- \sigma_A^{(-)}] = 0$. } 
That is, one cannot tell which values (i.e.~$\sigma_A^{(\pm)}$) the coefficient $\sigma_A$ should take in $g\to +0\in \mathbb R$. One should admit that this Stokes ambiguity is an inevitable nature of invisible terms and even is a result consistent with asymptotic expansion, defined by Eq.~\eq{EqDefinitionAsymptoticExpansion}. Note that Stokes ambiguity does not mean any problem of the system, rather it is a buffer mechanism in Stokes phenomena, which connects different asymptotic expansions in different directions. These behaviors can be rigorously captured by resurgent analysis and concretely analyzed (see e.g.~\cite{CiteStokesAmbiguity}%
\footnote{Note that ``non-perturbative ambiguity'' in \cite{CiteStokesAmbiguity} is our ``Stokes ambiguity''. }). 

Invisible terms in the meta-stable expansion are also considered as: 
\begin{align}
u(g) \asymeq \sum_{n=0}^\infty \Bigl(\theta_A\,e^{\frac{A}{g}}\Bigr)^n\Bigl[u^{[n]}(g) + \sum_{m=1}^\infty \bigl(\sigma_A e^{-\frac{A}{g}}\bigr)^m u^{[n;m]}(g) \Bigr], 
\end{align}
which are extensively discussed in \cite{GKIM,KMR,resurgentRef} (which are also called {\em new sectors}). If one naively sees, $\bigl\{u^{[n;n]}(g)\bigr\}_{n=1}^\infty$ sectors seem to be compatible with the zero-instanton sector $u^{[0]}(g)$. We however should not mix them with the zero-instanton sector $u^{[0]}(g)$, because they are already invisible in the asymptotic expansion, Eq.~\eq{EqAsymExpTransFirstOrderRes}, and $\sigma_A$ may even possess Stokes ambiguity.%
\footnote{These invisible sectors $\bigl\{ u^{[n;m]}(g) \bigr\}_{n=1,2,\cdots}^{m=1,2,\cdots}$ of $g\to +0\in\mathbb R$ are referred to as {\em new sectors}, since there is not found their worldsheet/matrix-model interpretation \cite{GKIM}. Naively one may think that it would be $n$-ghost and $m$-instanton sectors, but it does not have the expected property; Visible sectors, like $\bigl\{u^{[n]}(g)\bigr\}_{n=0}^\infty$ (or sometimes the pure-imaginary part of $\bigl\{u^{[0;m]}(g)\bigr\}_{m=1}^\infty$), however, can be understood as (ghost) D-instantons, though.  } 

A lesson of this section is following: In order to evaluate the asymptotic expansion, we first consider expansion, like Eq.~\eq{EqAsymExpTransFirstOrderRes}, and then later take into account its invisible sectors in discussion of Stokes phenomena in $g$ (or $t$). In the RHP, we actually do not need to care such invisible sectors, since we can evaluate the RHP for each value of $g$ (or $t$) which should give correct asymptotics caused by Stokes phenomena in $g$ (or $t$). This then automatically gives rise what the invisible sectors should be (See e.g.~\cite{ItsBook}).

\section{Duality at the minimal-string saddles \label{SectionDualityDKKprescription}}

In this subsection, we check how spectral $p-q$ duality in two-matrix models correctly acts on the minimal-string vacuum, i.e.~their saddle points of the string-theory path-integral. 

Non-perturbative realization of the $p-q$ duality (given by Eq.~\eq{EqPQdualPQLaxOperators}) explicitly maps the amplitudes of the one side, $\bigl\{u_n;v_m\bigr\}_{1\leq n\leq p}^{1\leq m\leq q}$ in $\bigl(\bP,\bQ\bigr)$ to those of the other side, $\bigl\{\widetilde u_n;\widetilde v_m\bigr\}_{1\leq n\leq q}^{1\leq m\leq p}$ in $\bigl(\widetilde \bP,\widetilde \bQ\bigr)$, without using any expansion. 
On the other hand, a perturbative saddle appears in the weak-coupling regime, $g\to 0$, and can be evaluated with the prescription proposed by Daul, Kazakov and Kostov \cite{DKK}, as a dispersion-less limit of the Lax operators, $\bigl(\bP^{\rm (cl)},\bQ^{\rm (cl)}\bigr)$ and $\bigl(\widetilde \bP^{\rm (cl)},\widetilde \bQ^{\rm (cl)}\bigr)$. 

Duality means that we look at the same system with different descriptions. 
If the duality consistently acts on the saddle, then we also should look at the same saddle point from different descriptions. This means that classical saddles of $\bigl(\bP^{\rm (cl)},\bQ^{\rm (cl)}\bigr)$ and $\bigl(\widetilde \bP^{\rm (cl)},\widetilde \bQ^{\rm (cl)}\bigr)$ should be the same saddle and be mapped with the relation, Eq.~\eq{EqPQdualPQLaxOperators}, to each other. Therefore, we check the following relation:
\begin{Proposition} 
In $(p,q)=(p,p\pm 1)$ unitary minimal string theory, the following relation holds: 
\begin{align}
\begin{CD}
\bigl(\bP,\bQ\bigr)_{p,q} @>{\text{$p-q$ dual}}>> \bigl(\widetilde \bP,\widetilde \bQ\bigr)_{q,p}  @= \bigl(\bP,\bQ\bigr)_{q,p} \\
@V{\text{DKK: }g\to 0}VV   @. @VV{\text{DKK: }g\to 0}V \\
\bigl(\bP^{\rm (cl)},\bQ^{\rm (cl)}\bigr)_{p,q} @>> {\text{$p-q$ dual}}> \bigl( \widetilde \bP^{\rm (cl)}, \widetilde\bQ^{\rm (cl)}\bigr)_{q,p}  @= \bigl( \bP^{\rm (cl)}, \bQ^{\rm (cl)}\bigr)_{q,p} 
\end{CD} \label{EqPerturbativeDualityCheckCD}
\end{align}
Here, notations are defined as follow:
\begin{itemize}
\item $\bigl(\bP,\bQ\bigr)_{p,q}$ is the Lax operators of Eq.~\eq{EqLaxOpPQ1}. 
\item $\bigl(\widetilde \bP,\widetilde \bQ\bigr)_{q,p}$ is the corresponding dual operators of Eq.~\eq{EqLaxOpPQ2}. According to Eq.~\eq{EqIdentificationOfPQandQPoperatorsInDual}, this operator is identified with $\bigl(\bP,\bQ\bigr)_{q,p}$. 
\item $\bigl(\bP^{\rm (cl)},\bQ^{\rm (cl)}\bigr)_{p,q}$ is a dispersion-less limit of the Lax operators obtained by the DKK procedure. 
\item $\bigl( \widetilde \bP^{\rm (cl)}, \widetilde\bQ^{\rm (cl)}\bigr)_{q,p}$ is a classical Lax operator obtained from $\bigl(\bP^{\rm (cl)},\bQ^{\rm (cl)}\bigr)_{p,q}$ by applying the relation, Eq.~\eq{EqPQdualPQLaxOperators}. 
\end{itemize}
$\square$
\end{Proposition}
\paragraph{The DKK prescription}
In the DKK prescription, the Lax operators $\bigl(\bP,\bQ\bigr)$ in the weak-coupling limit $g\to 0$ is evaluated by the dispersion-less limit: 
\begin{align}
g^{-1}\bigl[ \del, t\bigr] = 1 \quad \to\quad \bigl\{\del,t\bigr\}_{\rm PB} = 1,\qquad \bigl(\del = g \del_t,\, g\to 0\bigr). 
\end{align}
Here $\{\,,\,\}_{\rm PB}$ is the Poisson bracket between $t$ and $\del$. In particular, as is often observed, we assume that the coefficient of the most relevant operator, $t$, is negative:
\begin{align}
t<0 \qquad \bigl(\mu = - t >0\quad \text{if $\,q-p = \pm 1$}\bigr).  \label{EqSignOfTSection2}
\end{align}
The scaling (i.e.~perturbative) ansatz of the operators in its dispersion-less procedure is then given as 
\begin{align}
\bP(t;\del) \asymeq \bP^{\rm (cl)}(t;\del) + O(g),\qquad \bQ(t;\del) \asymeq \bQ^{\rm (cl)}(t;\del)+O(g),\qquad \bigl(g\to 0\bigr), 
\end{align}
with
\begin{align}
&\bP^{\rm (cl)}(t;\del) = (- t)^{\frac{p}{p+q-1}}\Bigl[\Pi_p(z)\Bigr],\qquad \bQ^{\rm (cl)}(t;\del) = \beta_{p,q}\,(- t)^{\frac{q}{p+q-1}}\Bigl[\Xi_q(z)\Bigr],\nn\\
& \qquad\qquad\bigl(z \equiv g(-t)^{\frac{-1}{p+q-1}} \del_t = -\kappa t \del_t,\,\kappa \equiv(- t)^{-\frac{p+q}{p+q-1}}g,\,g\to+ 0\bigr).
\label{EqClassicalLaxOperatorDKK}
\end{align}
With taking into account $\bigl[z,t^n\bigr] = -n\kappa t^n$, 
the Douglas equation becomes the DKK-EZJ equation \cite{DKK,EynardZinnJustin}:%
\footnote{Here sign of the equation is different from the usual equation, since we consider $t<0$. As is shown later, with this choice of sign, the system is consistent with the multi-cut boundary condition on Stokes phenomena. }
\begin{align}
q\,\Pi_p'(z)\, \Xi_q(z) - p \,\Xi_q'(z) \,\Pi_p(z) = -\frac{q+p-1}{\beta_{p,q}}. 
\end{align}
As a solution of this equation, one can find a solution written with Chebyshev polynomials of the first kind, which was first found in \cite{Kostov1}. For the two systems of $(p,q)$ and $(q,p)$ in its unitary series, one obtains
\begin{itemize}
\item If  $(p,q)=(p,p+1)$ then 
\begin{align}
\Pi_p(z) = T_p(z),\qquad \Xi_q(z) = T_q(z),\qquad \beta_{p,p+1} =  \frac{p+q-1}{pq}. \label{EqSolOfDKK1}
\end{align}
\item If $(p,q)=(q+1,q)$ then
\begin{align}
\Pi_p(z) = T_p(z),\qquad \Xi_q(z) =  T_q(z),\qquad \beta_{q+1,q} = -\frac{p+q-1}{pq}. \label{EqSolOfDKK2}
\end{align}
\end{itemize}

\paragraph{Check of duality}
The spectral $p-q$ dual operators $\bigl(\widetilde \bP^{\rm (cl)}(t;\del),\widetilde \bQ^{\rm (cl)}(t;\del)\bigr)$ of the classical Lax operators, Eq.~\eq{EqClassicalLaxOperatorDKK}, are then given with the same $\Pi_p(z)$ and $\Xi_q(z)$ as 
\begin{align}
\left\{
\begin{array}{l}
\widetilde \bP^{\rm (cl)}(t;\del) = (-1)^q \beta_{p,q}^{-1} \bigl[ \bQ^{\rm (cl)}(t;\del)\bigr]^{\rm T} = (- t)^{\frac{q}{p+q-1}}\Bigl[(-1)^q\Xi_q( - z)\Bigr],\cr 
\widetilde \bQ^{\rm (cl)}(t;\del) = (-1)^q\beta_{p,q}\bigl[ \bP^{\rm (cl)}(t;\del)\bigr]^{\rm T} = (-1)^r \beta_{p,q}(- t)^{\frac{p}{p+q-1}}\Bigl[(-1)^p\Pi_p(-z)\Bigr]. 
\end{array}
\right.. 
\end{align}
Since the solutions of the DKK equation, Eq.~\eq{EqSolOfDKK1} and Eq.~\eq{EqSolOfDKK2}, satisfy
\begin{align}
(-1)^p \Pi_p(-z) = \Pi_p(z),\qquad (-1)^q\Xi_q(-z) = \Xi_q(z), 
\end{align}
as a result of $ T_n(-z) = (-1)^nT_n(z)$, and 
\begin{align}
\beta_{p,p+1} = (-1)^r \beta_{p+1,p},\qquad (-1)^r = (-1)^{2p+1} = -1, 
\end{align}
one obtains 
\begin{align}
\widetilde \bP^{\rm (cl)}(t;\del) = (- t)^{\frac{q}{p+q-1}} T_q(z),\qquad 
\widetilde \bQ^{\rm (cl)}(t;\del) = \beta_{q,p}(- t)^{\frac{p}{p+q-1}} T_p(z). 
\end{align}
These are the solutions of the DKK equation in $(q,p)$ systems, and therefore it shows the relation, Eq.~\eq{EqPerturbativeDualityCheckCD}. 
$\quad \blacksquare$

It is also worth deriving the spectral curves $F(P,Q)=0$ by the DKK prescription, which appear as follows: 
\begin{align}
\lim_{g\to 0}F\Bigl(\frac{\zeta}{\sqrt{\mu}},\frac{g \beta^{-1}_{p,q}}{(\sqrt{\mu})^{q/p}} \del_\zeta\Bigr) \psi(t;\zeta) = \lim_{g\to 0}F\Bigl(\frac{g \beta^{-1}_{q,p}}{(\sqrt{\widetilde \mu})^{p/q}} \del_\eta, \frac{\eta}{\sqrt{\widetilde \mu}}\Bigr) \chi(t;\eta) = 0,
\end{align}
with the algebraic equation, 
\begin{align}
F(P,Q) = T_p(Q) - T_q(P)=0. 
\end{align}
In particular, the associated (dual) cosmological constants, $\mu$ and $\widetilde \mu$, have the following relation in the unitary cases: 
\begin{align}
\mu \equiv (-t)^{\frac{2p}{p+q-1}}>0,\qquad \widetilde \mu \equiv (-t)^{\frac{2q}{p+q-1}}>0,\qquad \bigl(\text{if $q-p = \pm 1$}\bigr). 
\end{align}
We note that the asymptotics of the BA functions is therefore generally given as
\begin{align}
\psi(t;\zeta) \asymeq \sum_{j=1}^p c_j(\zeta)\, e^{ \frac{1}{g}\varphi^{(j)}(t;\zeta)},\qquad \chi(t;\eta)\asymeq \sum_{l=1}^q d_l(\eta) \, e^{\frac{1}{g} \widetilde \varphi^{(l)}(t;\eta)}\qquad \bigl(g\to 0\bigr). \label{EqAsympOfBAFunctionAsDKK}
\end{align}
Here coefficients $\bigl\{c_j(\zeta)\bigr\}_{j=1}^p$ and $\bigl\{d_l(\eta)\bigr\}_{l=1}^q$ are sectionally holomorphic functions which possess some jumps causes by asymptotic expansion and Stokes phenomena. The functions $\bigl\{\varphi^{(j)}(\zeta)\bigr\}_{j=1}^p$ and $\bigl\{\widetilde \varphi^{(l)}(\zeta)\bigr\}_{l=1}^q$ are most important observables obtained from spectral curve and their properties are summarized in the next subsection.

\section{Isomonodromy Lax operators $(\mathcal B,\mathcal Q)$ \label{SectionIMSLaxBQ}}
The lax operators of $(\bP,\bQ)$ is now expressed as 
\begin{align}
\bP = 2^{p-1} \del^p + \sum_{n=1}^p u_n(t) \del^{p-n},\qquad \bQ= \beta\Bigl[2^{q-1} \del^q + \sum_{m=1}^q v_m(t) \del^{q-m}\Bigr], 
\end{align}
and here we show some details of how to obtain the isomonodromy Lax operators $(\mathcal B,\mathcal Q)$ from the BA system: 
\begin{align}
\zeta \psi (t;\zeta) = \bP(t;\del) \psi(t;\zeta),\qquad g \frac{\del \psi(t;\zeta)}{\del \zeta} = \bQ(t;\del) \psi(t;\zeta)\qquad \bigl(\del = g\del_t\bigr). \label{BAinAppendixBQ}
\end{align}
In this Appendix, $g=1$ for simplicity. 

\subsection{Fixing $\{v_n(t)\}_n$}
The Douglas equation $\bigl[\bP,\bQ\bigr]= 1$ is given as 
\begin{align}
\beta^{-1} &= \beta^{-1}\bigl[\bP,\bQ\bigr] = \bigl[2^{p-1} \del^{p} +  \sum_{n=2}^p u_n\, \del^{p-n}, 2^{q-1} \del^q + \sum_{n=2}^q v_n \, \del^{q-n}\bigr] \nn\\
&= 2^{p-1} \sum_{n=2}^q \bigl[\del^p,v_n\bigr] \del^{q-n} - 2^{q-1} \sum_{n=2}^p \bigl[\del^q,u_n\bigr] \del^{p-n} + \nn\\
&\quad + \sum_{n=2}^p \sum_{m=2}^q \Bigl(u_n\bigl[\del^{p-n},v_m\bigr] \del^{q-m} - v_m\bigl[\del^{q-m},u_n\bigr] \del^{p-n}\Bigr) \nn\\
&=  \sum_{n=3}^\infty \Bigl[\sum_{i=1}^{n-2} \Bigl(2^{p-1}\, \binom{p}{i}\, v_{n-i}^{(i)} - 2^{q-1} \, \binom{q}{i} \, u_{n-i}^{(i)}\Bigr)\Bigr] \del^{r-n} + \nn\\
&\quad + \sum_{n=5}^\infty \Bigl[ \sum_{m=2}^{n-3} \sum_{i=1}^{n-4} \Bigl( \binom{p-m}{i} u_m\, v_{n-m-i}^{(i)} - \binom{q-m}{i} v_m \, u_{n-m-i}^{(i)}\Bigr)\Bigr] \del^{r-n}
\end{align}
and therefore, the relation between $\{u_n\}_{n=2}^\infty$ and $\{v_n\}_{n=2}^\infty$ is following: 
\begin{align}
1.\quad & 2^{p-1} \binom{p}{1} v_2' - 2^{q-1} \binom{q}{1} u_2' = 0,  \nn\\
2.\quad & 2^{p-1} \binom{p}{1} v_3' + 2^{p-1} \binom{p}{2} v_2''  
- 2^{q-1} \binom{q}{1} u_3' - 2^{q-1} \binom{q}{2} u_2'' =0, \nn\\
3.\quad & \sum_{i=1}^{n-2} \Bigl(2^{p-1}\, \binom{p}{i}\, v_{n-i}^{(i)} - 2^{q-1} \, \binom{q}{i} \, u_{n-i}^{(i)}\Bigr) +  \nn\\
&\quad +\Bigl[ \sum_{m=2}^{n-3} \sum_{i=1}^{n-4} \Bigl( \binom{p-m}{i} u_m\, v_{n-m-i}^{(i)} - \binom{q-m}{i} v_m \, u_{n-m-i}^{(i)}\Bigr) =0, 
\end{align}
which result in 
\begin{align}
v_2(t)& =\Bigl(2^{q-1} \frac{q}{p}\Bigr)  \frac{u_2}{2^{p-1}} =  2^{q-p} \frac{q}{p} u_2(t),\nn\\
v_3(t) &= \Bigl(2^{q-1} \frac{q}{p}\Bigr) \Bigl( \frac{u_3}{2^{p-1} } + \frac{q-p}{2} \frac{u_2'}{2^{p-1}}\Bigr) =  2^{q-p} \frac{q}{p}\Bigl[u_3(t) + \frac{q-p}{2} u_2'(t)\Bigr], \nn\\
v_4(t)& = 2^{q-p} \frac{q}{p}\Bigl( u_4(t) + \frac{q-p}{2} u_3'(t) + \frac{(q-p)(p-2q+3)}{12} u_2''(t) + \frac{(q-p)}{2}\frac{u_2^2(t)}{p\, 2^{p-1}}\Bigr), \nn\\
&\cdots
\end{align}
In particular, the part which is linear in $u$: 
\begin{align}
v_n = v_n^{(linear)}[u] + O(u^2).
\end{align}
 are given as follows: 
\begin{Proposition} [The linear part of $v$ in $u$]
The linear part of $v_n$ is given as 
\begin{align}
&p \,2^{p-1}v_n^{(linear)} = q\, 2^{q-1} \sum_{m=0}^{n-2} B_m\, u_{n-m}^{(m)},\qquad B_n = \sum_{a=0}^{n-1} \mathcal S_a\bigl(A^{(p)}\bigr) \, \delta A_{n-a}, \nn\\
& A_n^{(p)} = \frac{1}{n+1} \binom{p-1}{n},\qquad  \delta A_{n} \equiv A_{n}^{(q)} - A_n^{(p)}\nn\\
& \mathcal S_n(x) = P_n(y),\qquad x_n = P_n(-y),
\end{align}
where $P_n(x)$ is a Schur polynomial: 
\begin{align}
\sum_{n=0}^\infty z^n P_n(x) = \exp\biggl[\sum_{n=1}^\infty z^n x_n\biggr]. 
\end{align}
In particular, 
\begin{align}
1.\quad& p\, 2^{p-1}\, v_2^{(linear)} = q\, 2^{q-1}\, \Bigl[u_2\Bigr] \nn\\
2.\quad&  p\, 2^{p-1}\, v_3^{(linear)} = q\, 2^{q-1}\, \Bigl[u_3 +\delta A_1\, u_2'\Bigr] \nn\\
3.\quad&  p\, 2^{p-1}\, v_4^{(linear)} = q\, 2^{q-1}\, \Bigl[u_4 +\delta A_1 \,u_3' +\Bigl( \delta A_2 + \bigl(-A_1^{(p)}\bigr) \delta A_1\Bigr)\, u_2''\Bigr] \nn\\
4.\quad&  p\, 2^{p-1}\, v_5^{(linear)} = q\, 2^{q-1}\, \Bigl[u_5 +\delta A_1 \,u_4' +\Bigl( \delta A_2 + \bigl(-A_1^{(p)}\bigr) \delta A_1\Bigr)\, u_3'' + \nn\\
&\qquad\qquad + \Bigl( \delta A_3 + \bigl(-A_1^{(p)}\bigr) \delta A_2 + \bigl(-A_2^{(p)} + (A_1^{(p)})^2 \bigr) \delta A_1\Bigr) \, u_2'''\Bigr] \nn\\
5.\quad&  p\, 2^{p-1}\, v_6^{(linear)} = q\, 2^{q-1}\, \Bigl[u_6 +\delta A_1 \,u_5' +\Bigl( \delta A_2 + \bigl(-A_1^{(p)}\bigr) \delta A_1\Bigr)\, u_4'' + \nn\\
&\qquad\qquad + \Bigl( \delta A_3 + \bigl(-A_1^{(p)}\bigr) \delta A_2 + \bigl(-A_2^{(p)} + (A_1^{(p)})^2 \bigr) \delta A_1\Bigr) \, u_3''' + \nn\\
&\qquad\qquad+ \Bigl( \delta A_4 + \bigl(-A_1^{(p)}\bigr) \delta A_3 + \bigl(-A_2^{(p)} + (A_1^{(p)})^2 \bigr) \delta A_2  \nn\\
&\qquad\qquad\qquad\qquad\qquad+ \bigl(-A_3^{(p)} + 2 A_2^{(p)} A_1^{(p)} - (A_1^{(p)})^3 \bigr) \delta A_1\Bigr) \, u_2''''
\Bigr]  \nn\\
&\cdots
\end{align}
$\square$
\end{Proposition}
The non-linear parts with non-derivative terms: 
\begin{align}
v_n = v_n^{(linear)}[u] + \widetilde  v_n[u] + O(\text{non-linear with derivatives}), 
\end{align}
are given as
\begin{align}
p 2^{p-1}\, \widetilde v_4 &= q 2^{q-1}\Bigl[ \frac{1}{p 2^{p-1} } (q-p) \frac{u_2^2}{2}\Bigr], \nn\\
p 2^{p-1}\, \widetilde v_5 &= q 2^{q-1}\Bigl[ \frac{1}{p 2^{p-1} } (q-p)\, u_2  u_3\Bigr], 
\end{align}
and generally summarized as 

\begin{Proposition} [Non-linear part of $v$ in $u$ (but no derivative in $t$)]
The non-linear parts of the above relations are given by 
\begin{align}
p 2^{p-1}\, \widetilde v_n = q2^{q-1}\Bigl[ \sum_{m=2}^\infty D_{m-1}  P_n^{(m)}(u)\Bigr]. 
\end{align}
Here $\bigl\{D_m\bigr\}_{m=1}^\infty$ are coefficients given by 
\begin{align}
D_m = \prod_{a=1}^m\Bigl( \frac{q-a p}{p\,2^{p-1}}\Bigr)
\end{align}
and $\bigl\{P^{(m)}_n(u) \bigr\}_{m=1}^n$ are the $n$-th order Schur polynomial of length $m$: 
\begin{align}
P_n^{(m)}(u) = \sum_{|Y|=n;\,l(Y)=m} \frac{u_{1}^{\nu_1} u_2^{\nu_2}\cdots}{\nu_1!\, \nu_2 ! \cdots},\qquad u= (u_1,u_2,\cdots), 
\end{align}
where Young diagram $Y$ is denoted as 
\begin{align}
Y = \bigl(\lambda_1,\lambda_2,\cdots,\lambda_i,\cdots\bigr),\qquad \lambda_1\geq \lambda_2 \geq \cdots \geq \lambda_i \geq \cdots \geq 0,
\end{align}
and $\nu_a= \lambda_a - \lambda_{a+1}$. Therefore, size and length of Young diagram, $|Y|$ and $l(Y)$, are given by 
\begin{align}
|Y| = \sum_{i=1}^\infty \lambda_i = \sum_{i=1}^\infty i \nu_i,\qquad l(Y) = \sum_{i=1}^\infty \nu_i = \lambda_1. 
\end{align}
Note that these polynomials satisfy
\begin{align}
P_n(u) = \sum_{m=1}^n P_n^{(m)}(u). 
\end{align}
$\square$
\end{Proposition}

\subsection{Isomonodromy $(\mathcal B,\mathcal Q)$}
\subsubsection{$\mathcal B$-operator}
Here we use the convention of 
\begin{align}
\zeta = 2^{p-1} \lambda^p,
\end{align}
and 
\begin{align}
\vec{\psi}_{\rm pre}(t;\zeta) \equiv 
\begin{pmatrix}
\psi \cr \psi' \cr \cdots \cr \psi^{(p-1)}
\end{pmatrix}, \qquad 
\vec{\psi} (t;\zeta) \equiv 
V^{-1}(\lambda)
\vec{\psi}_{\rm pre}(t;\zeta),\qquad 
V(\lambda) = 
\begin{pmatrix}
1  \cr
 &\lambda \cr
 & & \ddots \cr
 & & & \lambda^{p-1}
\end{pmatrix}, \label{DefinitionOfVmatrix}
\end{align}
and then the differential equation of $\bP$ operator (in Eq.~\eq{BAinAppendixBQ}) is represented as 
\begin{align}
\del \vec{\psi}_{\rm pre} = 
\begin{pmatrix}
0 & 1 \cr
\vdots & & \ddots \cr
0 & & & 1 & 0\cr
0 & & & 0& 1 \cr
\bigl[\lambda^p - \frac{u_p}{2^{p-1}}\bigr] & - \frac{u_{p-1}}{2^{p-1}} & \cdots &  - \frac{u_{2}}{2^{p-1}} & 0
\end{pmatrix} \vec{\psi}_{\rm pre} \equiv \mathcal B_{\rm pre}(t;\lambda) \vec{\psi}_{\rm pre}. 
\end{align}
This gives the following result: 
\begin{Proposition} [$\mathcal B$-operator]
The operator $\mathcal B$ is given as 
\begin{align}
\mathcal B(t;\lambda) = \Gamma \lambda - \frac{1}{2^{p-1}} \sum_{n=2}^p \frac{u_n(t)}{\lambda^{n-1}} E_{p,p-n+1},\qquad \del \vec{\psi}(t;\lambda) = \mathcal B(t;\lambda) \, \vec{\psi}(t;\lambda)
\end{align}
which in diagonal basis given as 
\begin{align}
\text{diagonal basis: }\qquad 
\mathcal B(t;\lambda) = \Omega^{-1} \lambda - \frac{1}{2^{p-1}} \sum_{n=2}^p \frac{u_n(t)}{\lambda^{n-1}} \xi_{p} \xi_{p-n+1}^\dagger,
\end{align}
where $\xi_a$ is a vector given as 
\begin{align}
\xi_{a} = \frac{1}{\sqrt{p}}
\begin{pmatrix}
1 \cr
\omega^{(a-1)} \cr
\omega^{2(a-1)} \cr
\hdots \cr
\omega^{(p-1)(a-1)}
\end{pmatrix},\qquad 
\xi_a^\dagger \xi_b =  \delta_{a-b,0},\qquad \Omega\, \xi_a = \xi_{a+1}, 
\end{align}
and $U=(\xi_1,\xi_2,\cdots,\xi_p)$. $\quad \square$
\end{Proposition}

\subsubsection{$\mathcal Q$-operator}
\begin{Lemma} \label{QoperatorLemmaAppendix}
The basic components of the operator $\mathcal Q$ is following: 
\begin{align}
\frac{\del \vec \psi}{\del \lambda}& = \mathcal Q(\lambda)\vec \psi = \biggl[p\beta_{p,q}(2\lambda)^{p-1} \widetilde {\mathcal Q}(\lambda) - \frac{{\rm P}_1}{\lambda}\biggr] \vec \psi,\qquad \frac{\del \vec \psi}{\del \zeta} =\Bigl[\beta_{p,q}\widetilde{ \mathcal Q}(\lambda) -\frac{{\rm P}_1}{p\zeta}\Bigr]\,\vec \psi,
\end{align}
and 
\begin{align}
\widetilde{\mathcal Q}(\lambda) &= \bQ(t;\mathcal B(\lambda)) + \sum_{i=1}^{p-1}\lambda^{-i}\,{\rm P}_i\Gamma^{-i}\,\frac{\del^i \bQ}{\del t^i}(t;\mathcal B(\lambda)), 
\end{align}
with 
\begin{align}
{\rm P}_i = &
\begin{pmatrix}
{}_0 C_i \cr
 & {}_1 C_i \cr
& & \ddots \cr
& & &{}_{p-1} C_{i}
\end{pmatrix},\qquad 
{\rm P}_i \Gamma^{-i}= 
\begin{pmatrix}
0 &\hdots &\hdots & \hdots & 0 & \hdots & 0\cr
\vdots & & & & \vdots & & \vdots  \cr
0 &\hdots &\hdots & \hdots & 0 & \hdots & 0\cr
{}_i C_i  & & & & \vdots & & \vdots \cr
 & {}_{i+1} C_i  & & & \vdots & & \vdots \cr
& & \ddots  & & \vdots & & \vdots \cr
& & &{}_{p-1} C_{i} & 0 & \hdots & 0
\end{pmatrix}
\end{align}
where ${}_n C_m$ denotes the binomial coefficient: ${}_n C_m = \binom{n}{m} = \frac{n(n-1)\cdots (n-m+1)}{m!}$ and satisfies ${}_n C_m = 0$ if $n<m$. Note that $\del_t^{n} \bQ(t;\mathcal B)$ denotes the partial derivative of $\bQ$ acting on explicit dependence on $t$. Also note the following notation:
\begin{align}
\bQ(t;\mathcal B(\lambda)) = 2^{q-1} \mathcal B^{[q]}(\lambda) + \sum_{n=2}^q v_n(t) \mathcal B^{[q-n]}(\lambda),\qquad \del^n \vec \psi = \mathcal B^{[n]}(\lambda)\vec\psi,
\end{align}
and 
\begin{align}
\mathcal B^{[n]}(\lambda) = \del \circ \mathcal B^{[n-1]}(\lambda) = \frac{\del \mathcal B^{[n-1]}(\lambda)}{\del t} + \mathcal B^{[n-1]}(\lambda)\,\mathcal B(\lambda),\qquad \mathcal B^{[1]}(\lambda) = \mathcal B(\lambda). 
\end{align}
$\square$
\end{Lemma}
Here note that $\mathcal B^{[n]}$ here is nothing to do with the $\mathcal B_n$ in Eq.~\eq{EqIsomonodromySystemGeneral}. 

Then the $\mathcal B^{[n]}$-operators are calculable in the following way: 
First of all, the operators $\mathcal B^{[n]}$ have the following expansion: 
\begin{align}
\mathcal B^{[n]} (\lambda) = \Gamma^n \lambda^n +  \lambda^{n-2} \sum_{a=0}^\infty \frac{\mathcal U_{n,a} \bigl[\{u_i(t)\}_{i=2}^p\bigr]}{\lambda^{a}}, 
\end{align}
and these coefficient functions satisfy: 
\begin{align}
\mathcal B^{[n]} &= \Gamma^n \lambda^n +  \lambda^{n-2} \sum_{a=0}^\infty \mathcal U_{n,a} \lambda^{-a} 
=\mathcal B^{[n-1]}{}'(\lambda) + \mathcal B^{[n-1]}(\lambda) \mathcal B(\lambda) \nn\\
&= \lambda^{n-3} \sum_{a=0}^\infty \mathcal U_{n-1,a}' \lambda^{-a}
+ \Bigl[ \Gamma^{n-1} \lambda^{n-1} +  \lambda^{n-3} \sum_{a=0}^\infty \mathcal U_{n-1,a} \lambda^{-a} \Bigr]\Bigl[\Gamma \lambda + \lambda^{-1} \sum_{a=0}^\infty \mathcal U_{a} \lambda^{-a} \Bigr]. 
\end{align}
That is, 
\begin{align}
\mathcal U_{n,a} &=  \Gamma^{n-1} \,\mathcal U_a + \mathcal U_{n-1,a}\,\Gamma + \mathcal U_{n-1,a-1}' + \sum_{b=0}^{a-2}\mathcal U_{n-1,b}\, \mathcal U_{a-b-2} \qquad\bigl(a=0,1,2,\cdots\bigr). 
\end{align}
The sequence of equations are given by 
\begin{align}
1.& \quad \mathcal U_{n,0} = \Gamma^{n-1} \,\mathcal U_0 + \mathcal U_{n-1,0}\,\Gamma \nn\\
2.& \quad \mathcal U_{n,1} = \Gamma^{n-1} \,\mathcal U_1 + \mathcal U_{n-1,1}\,\Gamma + \mathcal U_{n-1,0}' \nn\\
3.& \quad \mathcal U_{n,a} = \Gamma^{n-1} \,\mathcal U_a + \mathcal U_{n-1,a}\,\Gamma + \mathcal U_{n-1,a-1}' + \sum_{b=0}^{a-2}\mathcal U_{n-1,b}\, \mathcal U_{a-2-b}, \qquad \bigl(a = 2,3,\cdots\bigr). 
\end{align}
In particular, one can see the following: 
$\bigl\{\mathcal U_{n,a}\bigr\}_{n,a}$ are given as 
\begin{align}
1. \quad \mathcal U_{n,0} &= \sum_{a=0}^{n-1} \Gamma^a \,\mathcal U_0 \,\Gamma^{n-a-1} = -\frac{u_2}{2^{p-1}}\, \mathcal E_{n,p-1}
\nn\\
2. \quad \mathcal U_{n,1} &= \sum_{a=0}^{n-1} \Gamma^a \, \mathcal U_1 \, \Gamma^{n-a-1} + \sum_{m=1}^{n-1} \mathcal U_{m,0}'\, \Gamma^{n-1-m}= - \frac{u_3}{2^{p-1}} \, \mathcal E_{n,p-2}
-\frac{u_2'}{2^{p-1}} \mathcal E_{n-1,p-1}^{(1)}
, \nn\\ 
3. \quad \mathcal U_{n,c} &=\sum_{a=0}^{n-1} \Gamma^a \, \mathcal U_{c} \, \Gamma^{n-a-1} + \sum_{m=1}^{n-1}\biggl( \mathcal U_{m,c-1}' + \sum_{b=0}^{c-2} \mathcal U_{m,b}\,\mathcal U_{c-2-b}\biggr)\, \Gamma^{n-1-m} \nn\\
& \qquad\qquad\qquad\qquad\qquad\qquad\qquad\qquad\qquad\qquad \bigl(c=2,3,\cdots\bigr). \nn\\
4. \quad \mathcal U_{n,2} &= - \frac{u_4}{2^{p-1}} \,\mathcal E_{n,p-3}
-\frac{u_3'}{2^{p-1}} \mathcal E_{n-1,p-2}^{(1)}
 -\frac{u_2''}{2^{p-1}} \mathcal E_{n-2,p-1}^{(2)}
+ \frac{u_2^2}{2^{2(p-1)}} \mathcal E_{n-2,p-1}
 \nn\\
5. \quad \mathcal U_{n,3} &= - \frac{u_5}{2^{p-1}} 
\mathcal E_{n,p-4}
- \frac{u_4'}{2^{p-1}} 
\mathcal E_{n-1,p-3}^{(1)}
 -\frac{u_3''}{2^{p-1}} 
\mathcal E_{n-2,p-2}^{(2)}
 - \frac{u_2'''}{2^{p-1}} 
\mathcal E_{n-3,p-1}^{(3)}
+\nn\\ 
&\quad + \frac{(u_2^2)'}{2^{2(p-1)}} \mathcal E_{n-3,p-1}^{(1)}
 + \frac{u_2\, u_3}{2^{2(p-1)}} 
\mathcal E_{n-2,p-2}
+ \frac{u_3\, u_2}{2^{2(p-1)}} 
\mathcal E_{n-3,p-1}  +{}_2C_1 \frac{u_2'\, u_2}{2^{2(p-1)}} \, \mathcal E_{n-3,p-1}
\end{align}
Note that 
\begin{align}
\sum_{m=0}^{n}{}_{m-a} C_k = {}_{n-a+1} C_{k+1},
\end{align}
and 
\begin{align}
\mathcal E_{n,m}\equiv \sum_{a=0}^{n-1} \Gamma^a E_{p,m} \Gamma^{n-a-1},\qquad
\mathcal E_{n,m}^{(i)} \equiv \sum_{a=0}^{n-1} {}_{(n-1)-a+i} C_i \, \Gamma^a E_{p,m} \Gamma^{n-a-1},
\end{align}
which satisfies 
\begin{align}
&\mathcal E_{n+p,m} = \Gamma^{n+m-1} + \mathcal E_{n,m},\qquad \mathcal E_{p,m} = \Gamma^{m-1},\qquad \mathcal E_{0,m} = 0,\nn\\
&\mathcal E_{n,m}^{(0)} = \mathcal E_{n,m},\qquad \sum_{m=1}^{n} \mathcal E_{m-c,d}^{(i)} \Gamma^{n-1-m-c}= \mathcal E_{n-c,d}^{(i+1)},\nn\\
& \sum_{m=1}^{n-1} \mathcal E_{m-c,p-l}^{(i)}\, E_{p,p-l'} \Gamma^{n-m-1} = {}_{l+i}C_i\, \mathcal E_{n-c-l-1,p-l'}. 
\end{align}
The $\mathcal Q$ operator is expressed as 
\begin{align}
\mathcal Q(\lambda) = p \beta (2\lambda)^{p-1} \biggl[2^{q-1} \mathcal B^{[q]}(\lambda) + \sum_{n=2}^q v_n\, \mathcal B^{[q-n]}(\lambda) + \sum_{i=1}^{p-1} \lambda^{-i}\,{\rm P}_i\Gamma^{-i} \sum_{n=2}^{q} v_n^{(i)}\, \mathcal B^{[q-n]}(\lambda)\biggr] - \frac{{\rm P}_1}{\lambda}. 
\end{align}
Therefore, with the above coefficients, one can obtain the expression of $\mathcal Q$-operator: 
\begin{Proposition} [$\mathcal Q$-operator]
The asymptotics of $\mathcal Q$ is given as
\begin{align}
\mathcal Q(\lambda) =& p  \beta\, (2\lambda)^{p-1} \biggl[
2^{q-1} \Gamma^q \lambda^q + \sum_{n=2}^\infty \biggl( 2^{q-1} \mathcal U_{q,n-2} + \sum_{i=0}^{p-1} v_{n-i}^{(i)} \, \bigl({\rm P}_i \Gamma^{-i}\bigr) \Gamma^{q-n+i} \biggr)\lambda^{q-n} + \nn\\
&\qquad +\sum_{n=4}^\infty \biggl(\sum_{m=2}^{n-2} \sum_{i=0}^{p-1} v_{m}^{(i)} \, \bigl({\rm P}_i \Gamma^{-i}\bigr) \mathcal U_{q-m,n-m-i-2} \biggr) \lambda^{q-n}
\biggr] - \frac{{\rm P}_1}{\lambda} \nn\\
&\equiv \frac{p\beta }{2} \biggl[ \Gamma^q (2\lambda)^{r-1} + \sum_{n=2}^\infty 2^n \,\mathcal Q_n \, (2\lambda)^{r-n-1}\biggr] - \frac{{\rm P}_1}{\lambda},
\end{align}
The expansion coefficients of $\mathcal Q$ operator are given as 
\begin{align}
1.\quad&\mathcal Q_2 = \Bigl( \mathcal U_{q,0} + \frac{1}{2^{q-1}} v_2 \Gamma^{q-2}\Bigr)
=\frac{u_2}{2^{p-1}}\mathcal H_{u_2},\nn\\
2.\quad&\mathcal Q_3 = \Bigl( \mathcal U_{q,1} + \frac{1}{2^{q-1}} \big(v_3 + v_2'\, {\rm P}_1\bigr)\Gamma^{q-3} \Bigr) = \frac{u_3}{2^{p-1}} \mathcal H_{u_3} + \frac{u_2'}{2^{p-1}} \mathcal H_{u_2'}, \nn\\
3.\quad&\mathcal Q_4 = \Bigl( \mathcal U_{q,2} + \frac{1}{2^{q-1}} \bigl[\bigl(v_4 + v_3' \, {\rm P}_1 + v_2'' {\rm P}_2\bigr) \Gamma^{q-4} + v_2 \,\mathcal U_{q-2,0}\bigr]\Bigr) \nn\\
&\ \quad= \frac{u_4}{2^{p-1}} \mathcal H_{u_4} +  \frac{u_3'}{2^{p-1}} \mathcal H_{u_3'} +  \frac{u_2''}{2^{p-1}} \mathcal H_{u_2''} +  \frac{u_2^2}{2^{2(p-1)}} \mathcal H_{u_2^2}\nn\\
4.\quad&\mathcal Q_5 = \Bigl( \mathcal U_{q,3} + \frac{1}{2^{q-1}} \bigl[\bigl(v_5 + v_4' \, {\rm P}_1 + v_3'' {\rm P}_2+v_2'' {\rm P}_3\bigr) \Gamma^{q-5} + \nn\\
&\qquad\qquad \qquad\qquad + v_3 \,\mathcal U_{q-3,0}+ \bigl(v_2 \, \mathcal U_{q-2,1} + v_2' \,({\rm P}_1\Gamma^{-1})\,\mathcal U_{q-2,0}\bigr) \bigr]\Bigr) \nn\\
&\ \quad = \frac{u_5}{2^{p-1}} \mathcal H_{u_5}+\frac{u_4'}{2^{p-1}} \mathcal H_{u_4'} + \frac{u_3''}{2^{p-1}} \mathcal H_{u_3''} + \frac{u_2'''}{2^{p-1}} \mathcal H_{u_2'''} + \frac{u_3\,u_2}{2^{2(p-1)}} \mathcal H_{u_3\,u_2} +  \frac{u_2'\,u_2}{2^{2(p-1)}} \mathcal H_{u_2'\,u_2} \nn\\
5.\quad&\mathcal Q_n = \mathcal U_{q,n-2} + \frac{1}{2^{q-1}} \Bigl[ \sum_{i=0}^{p-1} v_{n-i}^{(i)} \, \bigl({\rm P}_i \Gamma^{-i}\bigr) \Gamma^{q-n+i} + \sum_{m=2}^{n-2} \sum_{i=0}^{p-1} v_{m}^{(i)} \, \bigl({\rm P}_i \Gamma^{-i}\bigr) \mathcal U_{q-m,n-m-i-2}\Bigr]
\end{align}
where $\mathcal H_{u_{n}^{(m)}}$ has its general form: 
\begin{align}
\mathcal H_{u_{n}^{(m)}} = \frac{q}{p}\Bigl[ \sum_{i=0}^{p-1} B_{m-i}{\rm P}_i\Bigr] \Gamma^{q-n-m} - \mathcal E_{q-m,p-n+1}^{(m)}
\end{align}
Here are also an example of $\mathcal H_{u_2^2}$: 
\begin{align}
\mathcal H_{u_2^2} = \Bigl(\frac{q(q-p)}{2p^2} \Gamma^{q-4} - \frac{q-p}{p} \mathcal E_{q-2,p-1}\Bigr)
\end{align}
$\square$
\end{Proposition}
Finally, we note the relation between the matrix-model basis and the diagonal basis: 
\begin{align}
(\mathcal B^{\rm (diagonal)},\mathcal Q^{\rm (diagonal)}) = U (\mathcal B^{\rm (matrix)},\mathcal Q^{\rm (matrix)}) U^\dagger. 
\end{align}
In the main text, only diagonal basis is shown.

\section{Monodromy calculus for the classical BA function\label{AppendixMonodromyCalculusOfCBAF}}

Here we summarize the calculation of monodromy in the classical BA function. 
First of all, the function $h^{(j|a)}(\lambda)$ is expressed as 
\begin{align}
h^{(j|a)}(\lambda) &=  \sqrt{\frac{2\cosh\bigl(\tau- 2\pi i \frac{j-1}{p}\bigr) - 2\cosh\bigl(\tau- 2\pi i \frac{a-1}{p}\bigr)}{\bigl(\omega^{-(j-1)}-\omega^{-(a-1)}\bigr)\bigl(2\cosh p \tau\bigr)^{1/p}}}\nn\\
&=  \sqrt{\frac{4 \sinh\bigl( \tau - \frac{\pi i n_a}{p}\bigr) \sinh \bigl( \frac{a-j}{p} \pi i\bigr)}{\bigl(\omega^{-(j-1)}-\omega^{-(a-1)}\bigr)\bigl(2\cosh p \tau\bigr)^{1/p}}}\nn\\
&=\sqrt{\dfrac{2\sinh(\tau-\frac{\pi i n_a}{p})}{\omega^{-\frac{n_a}{2}}\bigl(2\cosh p\tau\bigr)^{1/p}}},
\qquad \bigl(j+a = n_a+2 \bigr). 
\end{align}
One can check that $n_a \to n_a + p$ does not change the expression (i.e.~it is consistent with modulo $p$) and therefore one can choose $\{n_a\}_{a=1}^p$ as 
\begin{align}
\{n_a\}_{a=1}^p = \{0,1,2,\cdots,p-1\}. 
\end{align}
Around $\lambda \to \infty$, the function does not have branch cuts and behaves as 
\begin{align}
h^{(j|a)}(\lambda)  = 1 +O(\lambda^{-1})\qquad  (\lambda\to \infty), \label{hAsympInfty}
\end{align}
and therefore one can show that 
\begin{align}
\Delta^{(j)}(\lambda) = 1 + O(\lambda^{-1}),\qquad \bigl(\lambda\to \infty\bigr), 
\end{align}
therefore, 
\begin{align}
Z(\lambda)  &= \frac{1}{p} 
\begin{pmatrix}
\ds 
\sum_{n=1}^p \Bigl(\omega^{i-j}\Bigr)^{n-1}
\end{pmatrix}_{1\leq i,j\leq p}  + O(\lambda^{-1}) \nn\\
&= I_p  + O(\lambda^{-1}),\qquad \bigl(\lambda\to \infty\bigr). 
\end{align}

With keeping the asymptotic behavior (i.e.~the branch of square root) of Eq.~\eq{hAsympInfty}, we extract the behavior around the branch points $\lambda\to  \lambda_n = \omega^{\frac{n}{2}}$ or $\tau \to \tau_n = \dfrac{\pi i n}{p}$ ($n=0,1,\cdots,2p-1$). First of all, the monodromy relation of $\bigl\{z^{(j)}\bigr\}_{j=1}^p$ is given as 
\begin{align}
z^{(j)}(\lambda_n + e^{2\pi i} \delta \lambda) = z^{(l)}(\lambda_n+\delta \lambda),\qquad j+l \equiv n+2 \mod p. 
\end{align}
In fact, 
\begin{align}
z^{(j)}(\lambda)-z^{(l)}(\lambda) &= \cosh\bigl(\tau- 2\pi i \frac{j-1}{p}\bigr) - \cosh\bigl(\tau- 2\pi i \frac{l-1}{p}\bigr) \nn\\
& = 2i \sinh\bigl( \tau - \frac{\pi i n}{p}\bigr) \sin \Bigl( \frac{l-j}{p} \pi\Bigr) \nn\\
& =\Bigl[ 2i \sin \Bigl(\frac{l-j}{p} \pi \Bigr) \Bigr]\delta \tau+ O(\delta\tau^3)\qquad \bigl( \delta\tau \equiv \tau - \frac{\pi i n}{p}\bigr) \nn\\
&= \Bigl[ 2i \sin \Bigl(\frac{l-j}{p} \pi \Bigr) \Bigr] \sqrt{\frac{2(\lambda-\lambda_n)}{p\,\lambda_n}} \Bigl(1+ O(\lambda^{-1})\Bigr),
\end{align}
where 
\begin{align}
\lambda = &\frac{\Bigl[2\sqrt{\mu}\cosh p\bigl(\delta\tau + \frac{\pi i n}{p}\bigr) \Bigr]^{1/p}}{2} = \frac{(2\sqrt{\mu})^{1/p}}{2} e^{\frac{\pi i n}{p}}\Bigl[1 + \frac{1}{2p} (p\delta \tau)^2+O(\delta\tau^4)\Bigr].\nn\\
& \qquad \therefore  \quad \frac{\lambda-\lambda_n}{\lambda_n} =  \frac{p}{2} \delta \tau^2+O(\delta \tau^4).
\end{align}
Therefore, it is appropriate to approach the branch points from the direction of 
\begin{align}
\delta \tau >0 \qquad \Leftrightarrow \qquad 
 \frac{\lambda-\lambda_n}{\lambda_n}>0,
\end{align}
and one can easily see that 
\begin{align}
\Bigl[z^{(j)}-z^{(l)}\Bigr](\lambda_n+e^{2\pi i }\delta\lambda) = e^{\pi i } \,
\Bigl[z^{(j)}-z^{(l)}\Bigr](\lambda_n+\delta\lambda),
\end{align}
and 
\begin{align}
\sqrt{\Delta^{(j)}(\lambda_n+e^{2\pi i}\delta \lambda)} = c_{n,l,j}\, \sqrt{\Delta^{(l)}(\lambda_n+\delta \lambda)}.
\end{align}
Below, we calculate the coefficient $c_{n,l,j}$. 
In the same way as before, we can show that 
\begin{align}
h^{(j|a)}(\lambda) &= \sqrt{\dfrac{2\sinh\bigl(\delta \tau + \frac{n-n_a}{p} \pi i\bigr)}{\omega^{\frac{n-n_a}{2}}\bigl(2\cosh p\delta \tau\bigr)^{1/p}}}   \nn\\
&= \left\{
\begin{array}{cc}
\sqrt{\dfrac{2\sinh\bigl(\frac{n-n_a}{p} \pi i\bigr)}{2^{1/p} \,\omega^{\frac{n-n_a}{2}}}} \Bigl(1+ O(\lambda^{-1})\Bigr) & n-n_a \not\equiv 0 \mod p \cr
\ds
2^{\frac{p-1}{2p}} \sqrt[4]{\frac{2(\lambda-\lambda_n)}{p\,\omega^{n/2}}}\Bigl(1+ O(\lambda^{-1})\Bigr) & n-n_a \equiv 0 \mod p 
\end{array}
\right.,
\end{align}
which means that 
\begin{align}
\sqrt{\Delta^{(j)}(\lambda)} 
=2^{\frac{p-1}{2p}} \sqrt[4]{\frac{2(\lambda-\lambda_n)}{p\,\omega^{n/2}}} \Bigl(1+ O(\lambda^{-1})\Bigr)\times \prod_{1\leq a \,(\neq j,l)\leq p} 
\sqrt{\dfrac{2\sinh\bigl(\frac{l-a}{p} \pi i\bigr)}{2^{1/p} \,\omega^{\frac{l-a}{2}}}}
\end{align}
Therefore, 
\begin{align}
c_{n,l,j} &= i  \prod_{1\leq a \,(\neq j,l)\leq p} 
\sqrt{\dfrac{ \omega^{-\frac{l-a}{2}}\sinh\bigl(\frac{l-a}{p} \pi i\bigr)}{\omega^{-\frac{j-a}{2}}\sinh\bigl(\frac{j-a}{p} \pi i\bigr)}}
=  i  
\sqrt{\dfrac{\ds \prod_{1\leq a \,(\neq \eta)\leq p-1}  \omega^{-\frac{a}{2}}\sinh\bigl(\frac{a}{p} \pi i\bigr)}{\ds \prod_{1\leq a \,(\neq p - \eta)\leq p-1}\omega^{-\frac{a}{2}}\sinh\bigl(\frac{a}{p} \pi i\bigr)}} \nn\\
&= i  
\sqrt{\dfrac{ \omega^{-\frac{p-\eta}{2}}\sinh\bigl(\frac{p-\eta}{p} \pi i\bigr)}{\omega^{-\frac{\eta}{2}}\sinh\bigl(\frac{\eta}{p} \pi i\bigr)}} 
= i \sqrt{\frac{\omega^{-\frac{p-\eta}{2}} }{\omega^{-\frac{\eta}{2}} }} 
= \omega^{\frac{\eta}{2}} = \omega^{\frac{l-j}{2}}. 
\end{align}
with $\eta \equiv l-j>0$. Note that for $0<\theta < \pi$, 
\begin{align}
\sinh(\tau  + i \theta) \to e^{\frac{\pi i}{2}} |\sin(\theta)| \qquad (\tau \to 0\in \mathbb R). 
\end{align}
Here we assumed that $l-j=\eta>0$, but the opposite case is also similarly obtained and satisfies $c_{n,l,j} c_{n,j,l}=-1$. 
Therefore, the monodromy is finally given as 
\begin{align}
\left\{
\begin{array}{c}
\ds
\sqrt{\Delta^{(j)}(\lambda_n+e^{2\pi i}\delta\lambda)} = \omega^{\frac{l-j}{2}} \sqrt{\Delta^{(l)}(\lambda_n+\delta \lambda)} \cr
\ds
\sqrt{\Delta^{(l)}(\lambda_n+e^{2\pi i}\delta\lambda)} = \frac{-1}{\omega^{\frac{l-j}{2}}} \sqrt{\Delta^{(j)}(\lambda_n+\delta \lambda)}
\end{array}
\right.,
\qquad l>j, j+l \equiv n+2 \mod p.
\end{align}

\section{BA functions in $(5,2)$-system}
Here we show other solutions to the isomonodromy system than orthogonal polynomials: 
\begin{align}
\widetilde \Psi_0(\eta) = \bigl(\widetilde \psi_0^{(1)}(\eta),\cdots, \widetilde \psi_0^{(5)}(\eta)\bigr). 
\end{align}
\begin{align}
\widetilde\psi_0^{(2)}= 
\tiny
\left[
\begin{array}{c|ccccc}
n & v_n^{(1)} & v_n^{(2)} & v_n^{(3)} & v_n^{(4)} & v_n^{(5)} \\
13 & 0 & 0 & -\frac{1}{\sqrt{\omega }} & 0 & 0 \\
12 & 0 & 0 & -\frac{1}{\sqrt{\omega }} & 0 & \frac{s_1-1}{\omega
^{3/2}} \\
11 & 0 & 1 & -\frac{1}{\sqrt{\omega }} & 0 & \frac{s_1-1}{\omega
^{3/2}} \\
10 & 0 & 1 & -\frac{1}{\sqrt{\omega }} & \frac{s_1-s_2-1}{\omega } & \frac{s_1-1}{\omega ^{3/2}} \\
9 & \sqrt{\omega } & 1 & -\frac{1}{\sqrt{\omega }} & -\frac{s_2}{\omega } & \frac{s_1-1}{\omega ^{3/2}} \\
8 & \sqrt{\omega } & 1 & -\frac{1}{\sqrt{\omega }} & 0 & \frac{s_1-1}{\omega ^{3/2}} \\
7 & 0 & 1 & -\frac{s_1}{\sqrt{\omega }} & 0 & \frac{s_1-1}{\omega ^{3/2}} \\
6 & 0 & 1 & 0 & 0 & \frac{s_1-1}{\omega ^{3/2}} \\
5 & 0 & 1 & 0 & 0 & 0 \\
4 & 0 & 1 & 0 & 0 & 0 \\
3 & 0 & 1 & 0 & 0 & 0 \\
2 & 0 & 1 & 0 & 0 & 0 \\
1 & 0 & 1 & 0 & 0 & 0 \\
0 & 0 & 1 & 0 & 0 & 0 \\
\end{array}
\right]; \quad 
\widetilde\psi_0^{(3)} =
\tiny
\left[
\begin{array}{c|ccccc}
n & v_n^{(1)} & v_n^{(2)} & v_n^{(3)} & v_n^{(4)} & v_n^{(5)} \\
13 & 0 & 0 & 0 & -\frac{1}{\sqrt{\omega }} & 0 \\
12 & 0 & 0 & 0 & -\frac{1}{\sqrt{\omega }} & 0 \\
11 & 0 & 0 & 0 & -\frac{1}{\sqrt{\omega }} & 0 \\
10 & 0 & 0 & 0 & -\frac{1}{\sqrt{\omega }} & 0 \\
9 & 0 & 0 & 0 & -\frac{1}{\sqrt{\omega }} & 0 \\
8 & 0 & 0 & 0 & -\frac{1}{\sqrt{\omega }} & 0 \\
7 & 0 & 0 & 0 & -\frac{1}{\sqrt{\omega }} & 0 \\
6 & 0 & 0 & 0 & -\frac{1}{\sqrt{\omega }} & 0 \\
5 & 0 & 0 & 1 & -\frac{1}{\sqrt{\omega }} & 0 \\
4 & 0 & 0 & 1 & -\frac{1}{\sqrt{\omega }} & -\frac{1}{\omega } \\
3 & 0 & 0 & 1 & -\frac{1}{\sqrt{\omega }} & -\frac{1}{\omega } \\
2 & 0 & 0 & 1 & -\frac{1}{\sqrt{\omega }} & 0 \\
1 & 0 & 0 & 1 & -\frac{1}{\sqrt{\omega }} & 0 \\
0 & 0 & 0 & 1 & 0 & 0 \\
\end{array}
\right] \nn\\
\widetilde \psi_0^{(4)}=
\tiny 
\left[
\begin{array}{c|ccccc}
n & v_n^{(1)} & v_n^{(2)} & v_n^{(3)} & v_n^{(4)} & v_n^{(5)} \\
13 & 0 & 0 & 0 & 0 & -\frac{1}{\sqrt{\omega }} \\
12 & 0 & 0 & 0 & 0 & -\frac{1}{\sqrt{\omega }} \\
11 & 0 & 0 & 0 & 0 & -\frac{1}{\sqrt{\omega }} \\
10 & 0 & 0 & 0 & 0 & -\frac{1}{\sqrt{\omega }} \\
9 & 0 & 0 & 0 & 1 & -\frac{1}{\sqrt{\omega }} \\
8 & 0 & 0 & 0 & 1 & -\frac{1}{\sqrt{\omega }} \\
7 & 0 & 0 & \sqrt{\omega } & 1 & -\frac{1}{\sqrt{\omega }} \\
6 & 0 & 0 & \sqrt{\omega } & 1 & -\frac{1}{\sqrt{\omega }} \\
5 & 0 & 0 & 0 & 1 & -\frac{1}{\sqrt{\omega }} \\
4 & 0 & 0 & 0 & 1 & 0 \\
3 & 0 & 0 & 0 & 1 & 0 \\
2 & 0 & 0 & 0 & 1 & 0 \\
1 & 0 & 0 & 0 & 1 & 0 \\
0 & 0 & 0 & 0 & 1 & 0 \\
\end{array}
\right];\quad
\widetilde \psi_0^{(5)}=
\tiny 
\left[
\begin{array}{c|ccccc}
n & v_n^{(1)} & v_n^{(2)} & v_n^{(3)} & v_n^{(4)} & v_n^{(5)} \\
13 & \omega ^2 & 0 & 0 & 0 & s_1 \\
12 & \omega ^2 & \omega ^{3/2} & 0 & 0 & s_1 \\
11 & \omega ^2 & \omega ^{3/2} & 0 & \sqrt{\omega } \left(s_1-s_2\right) & s_1 \\
10 & \omega ^2 & \omega ^{3/2} & 0 & \sqrt{\omega } \left(s_1-s_2\right) & s_1 \\
9 & \omega ^2 & \omega ^{3/2} & 0 & -\sqrt{\omega } s_2 & s_1 \\
8 & \omega ^2 & \omega ^{3/2} & 0 & 0 & s_1 \\
7 & 0 & \omega ^{3/2} & -\omega s_1 & 0 & s_1 \\
6 & 0 & \omega ^{3/2} & 0 & 0 & s_1 \\
5 & 0 & \omega ^{3/2} & 0 & 0 & 1 \\
4 & 0 & \omega ^{3/2} & 0 & 0 & 1 \\
3 & 0 & \omega ^{3/2} & 0 & 0 & 1 \\
2 & 0 & \omega ^{3/2} & 0 & 0 & 1 \\
1 & 0 & \omega ^{3/2} & 0 & 0 & 1 \\
0 & 0 & 0 & 0 & 0 & 1 \\
\end{array}
\right]
\end{align}



\begin{thebibliography}{99}

\bibitem{PolchinskiStringTheory}
  J.~Polchinski,
  ``String theory. Vol. 1: An introduction to the bosonic string,''
  Cambridge, UK: Univ. Pr. (1998) 402 p; 
  ``String theory. Vol. 2: Superstring theory and beyond,''
  Cambridge, UK: Univ. Pr. (1998) 531 p



\bibitem{Polyakov}
  A.~M.~Polyakov,
   ``Quantum geometry of bosonic strings,''
  Phys.\ Lett.\ B {\bf 103} (1981) 207; 
   ``Quantum geometry of fermionic strings,''
  Phys.\ Lett.\ B {\bf 103} (1981) 211.




\bibitem{DSL}
  E.~Brezin and V.~A.~Kazakov,
   ``Exactly solvable field theories of closed strings,''
  Phys.\ Lett.\  B {\bf 236} (1990) 144;\\
  M.~R.~Douglas and S.~H.~Shenker,
  Nucl.\ Phys.\  B {\bf 335} (1990) 635;\\
  D.~J.~Gross and A.~A.~Migdal,
   ``Nonperturbative Two-Dimensional Quantum Gravity,''
  Phys.\ Rev.\ Lett.\  {\bf 64} (1990) 127.

\bibitem{BIPZ}
  E.~Brezin, C.~Itzykson, G.~Parisi and J.~B.~Zuber,
  ``Planar Diagrams,''
  Commun.\ Math.\ Phys.\  {\bf 59} (1978) 35.

\bibitem{KazakovSeries}
  V.~A.~Kazakov,
  ``The Appearance of Matter Fields from Quantum Fluctuations of 2D Gravity,''
  Mod.\ Phys.\ Lett.\  A {\bf 4} (1989) 2125.

\bibitem{Kostov1}
  I.~K.~Kostov,
  ``Strings embedded in Dynkin diagrams,''
  Cargese 1990, Proceedings, Random surfaces and quantum gravity, pp.135-149. 

\bibitem{Kostov2}
  I.~K.~Kostov,
  ``Loop amplitudes for nonrational string theories,''
  Phys.\ Lett.\  B {\bf 266} (1991) 317.

\bibitem{Kostov3}
  I.~K.~Kostov,
  ``Strings with discrete target space,''
  Nucl.\ Phys.\  B {\bf 376} (1992) 539
  [arXiv:hep-th/9112059].

\bibitem{BDSS}
  T.~Banks, M.~R.~Douglas, N.~Seiberg and S.~H.~Shenker,
  ``Microscopic and macroscopic loops in nonperturbative two-dimensional gravity,''
  Phys.\ Lett.\  B {\bf 238} (1990) 279.

\bibitem{BMPNonP}
  E.~Brezin, E.~Marinari and G.~Parisi,
  ``A Nonperturbative Ambiguity Free Solution Of A String Model,''
  Phys.\ Lett.\  B {\bf 242} (1990) 35.

\bibitem{TwoMatString}
  E.~Brezin, M.~R.~Douglas, V.~Kazakov and S.~H.~Shenker,
   ``The Ising model coupled to 2-d Gravity: A nonperturbative analysis,''
  Phys.\ Lett.\  B {\bf 237} (1990) 43;\\
  D.~J.~Gross and A.~A.~Migdal,
   ``Nonperturbative Solution of the Ising Model on a Random Surface,''
  Phys.\ Rev.\ Lett.\  {\bf 64} (1990) 717.

\bibitem{GrossMigdal2}
  D.~J.~Gross and A.~A.~Migdal,
   ``A nonperturbative treatment of two-dimensional quantum gravity,''
  Nucl.\ Phys.\  B {\bf 340} (1990) 333.

\bibitem{DouglasGeneralizedKdV}
  M.~R.~Douglas,
   ``Strings in less than one-dimension and the generalized KdV hierarchies,''
  Phys.\ Lett.\  B {\bf 238} (1990) 176.

\bibitem{TadaYamaguchiDouglas}
  T.~Tada and M.~Yamaguchi,
   ``$P$ and $Q$ operator analysis for two matrix model,''
  Phys.\ Lett.\  B {\bf 250} (1990) 38; \\
  M.~R.~Douglas,
   ``The Two matrix model,''
{\it  In *Cargese 1990, Proceedings, Random surfaces and quantum gravity* 77-83. (see HIGH ENERGY PHYSICS INDEX 30 (1992) No. 17911)}; \\
  T.~Tada,
   ``$(Q,P)$ Critical Point From Two Matrix Models,''
  Phys.\ Lett.\  B {\bf 259} (1991) 442.
  
\bibitem{Moore}
  G.~W.~Moore,
  ``Geometry Of The String Equations,''
  Commun.\ Math.\ Phys.\  {\bf 133} (1990) 261;
  ``Matrix Models Of 2-D Gravity And Isomonodromic Deformation,''
  Prog.\ Theor.\ Phys.\ Suppl.\  {\bf 102} (1990) 255.

\bibitem{FIK}
  A.~S.~Fokas, A.~R.~Its, A.~V.~Kitaev,
  ``The Isomonodromy approach to matrix models in 2-D quantum gravity,''
  Commun.\ Math.\ Phys.\  {\bf 147 } (1992)  395-430; 
  ``Discrete Painleve equations and their appearance in quantum gravity,''
  Commun.\ Math.\ Phys.\  {\bf 142 } (1991)  313-344.

\bibitem{GinspargZinnJustin}
  P.~H.~Ginsparg and J.~Zinn-Justin,
  ``Action principle and large order behavior of nonperturbative gravity,''
  in Proceedings, Random Surfaces and Quantum Gravity, Cargese, 1990, pp.\ 85-109. 

\bibitem{Shenker}
  S.~H.~Shenker,
  ``The Strength of nonperturbative effects in string theory,''
  in Proceedings, Random Surfaces and Quantum Gravity, Cargese, 1990, pp.\ 191-200. 

\bibitem{fkn}
  M.~Fukuma, H.~Kawai and R.~Nakayama,
  ``Continuum Schwinger-Dyson equations and universal structures 
   in two-dimensional quantum gravity,''
  Int.\ J.\ Mod.\ Phys.\ A {\bf 6} (1991) 1385;
  ``Infinite dimensional Grassmannian structure 
   of two-dimensional quantum gravity,''
  Commun.\ Math.\ Phys.\  {\bf 143} (1992) 371;

\bibitem{fkn3}
    M.~Fukuma, H.~Kawai and R.~Nakayama,
  ``Explicit solution for $p$--$q$ duality
   in two-dimensional quantum gravity,''
  Commun.\ Math.\ Phys.\  {\bf 148} (1992) 101.
  
\bibitem{DVV}
  R.~Dijkgraaf, H.~L.~Verlinde and E.~P.~Verlinde,
  ``Loop Equations And Virasoro Constraints In Nonperturbative 2-D Quantum
  Gravity,''
  Nucl.\ Phys.\  B {\bf 348} (1991) 435.

\bibitem{MultiCut}
  C.~Crnkovic and G.~W.~Moore,
   ``Multicritical multicut matrix models,''
  Phys.\ Lett.\  B {\bf 257} (1991) 322.

\bibitem{David0}  
F.~David,
  ``Phases of the large N matrix model and nonperturbative effects in 2-d gravity,''
  Nucl.\ Phys.\  B {\bf 348} (1991) 507. 

\bibitem{MSS}
  G.~W.~Moore, N.~Seiberg and M.~Staudacher,
  ``From loops to states in 2-D quantum gravity,''
  Nucl.\ Phys.\  B {\bf 362} (1991) 665.

\bibitem{David}
   F.~David,
  ``Nonperturbative effects in matrix models and vacua of two-dimensional
  gravity,''
  Phys.\ Lett.\  B {\bf 302} (1993) 403
  [arXiv:hep-th/9212106].

\bibitem{EynardZinnJustin}
  B.~Eynard and J.~Zinn-Justin,
  ``Large order behavior of 2-D gravity coupled to $d < 1$ matter,''
  Phys.\ Lett.\  B {\bf 302} (1993) 396
  [arXiv:hep-th/9301004].

\bibitem{DKK}
  J.~M.~Daul, V.~A.~Kazakov and I.~K.~Kostov,
   ``Rational theories of 2-D gravity from the two matrix model,''
  Nucl.\ Phys.\  B {\bf 409} (1993) 311
  [arXiv:hep-th/9303093].

\bibitem{fy1}
  M.~Fukuma and S.~Yahikozawa,
  ``Nonperturbative effects in noncritical strings with soliton 
  backgrounds,''
  Phys.\ Lett.\ B {\bf 396} (1997) 97
  [arXiv:hep-th/9609210]. 
  
 \bibitem{fy2}
   M.~Fukuma and S.~Yahikozawa,
  ``Combinatorics of solitons in noncritical string theory,''
  Phys.\ Lett.\ B {\bf 393} (1997) 316
  [arXiv:hep-th/9610199]. 

\bibitem{fy3}
  M.~Fukuma and S.~Yahikozawa,
  ``Comments on D-instantons in c $<$ 1 strings,''
  Phys.\ Lett.\  B {\bf 460} (1999) 71
  [arXiv:hep-th/9902169].

\bibitem{MultiCutUniversality}
  G.~Bonnet, F.~David and B.~Eynard,
  ``Breakdown of universality in multi-cut matrix models,''
  J.\ Phys.\ A  {\bf 33} (2000) 6739
  [arXiv:cond-mat/0003324].
  
\bibitem{McGreevyVerlinde}
  J.~McGreevy and H.~L.~Verlinde,
  ``Strings from tachyons: The c = 1 matrix reloaded,''
  JHEP {\bf 0312} (2003) 054
  [arXiv:hep-th/0304224].

\bibitem{Martinec}
  E.~J.~Martinec,
  ``The annular report on non-critical string theory,''
  arXiv:hep-th/0305148.

\bibitem{KMS}
  I.~R.~Klebanov, J.~M.~Maldacena and N.~Seiberg,
  ``D-brane decay in two-dimensional string theory,''
  JHEP {\bf 0307} (2003) 045
  [arXiv:hep-th/0305159].

\bibitem{AKK}
  S.~Y.~Alexandrov, V.~A.~Kazakov and D.~Kutasov,
  ``Non-perturbative effects in matrix models and D-branes,''
  JHEP {\bf 0309} (2003) 057
  [arXiv:hep-th/0306177].

\bibitem{KazakovKostov}
  V.~A.~Kazakov and I.~K.~Kostov,
  ``Instantons in non-critical strings from the two-matrix model,''
  arXiv:hep-th/0403152.
  
\bibitem{MMSS}
  J.~M.~Maldacena, G.~W.~Moore, N.~Seiberg and D.~Shih,
   ``Exact vs. semiclassical target space of the minimal string,''
  JHEP {\bf 0410} (2004) 020
  [arXiv:hep-th/0408039].

\bibitem{SeSh2}
  N.~Seiberg and D.~Shih,
   ``Flux vacua and branes of the minimal superstring,''
  JHEP {\bf 0501} (2005) 055
  [arXiv:hep-th/0412315].

\bibitem{HHIKKMT}
  M.~Hanada, M.~Hayakawa, N.~Ishibashi, H.~Kawai, T.~Kuroki, Y.~Matsuo and T.~Tada,
  ``Loops versus matrices: The nonperturbative aspects of noncritical string,''
  Prog.\ Theor.\ Phys.\  {\bf 112} (2004) 131
  [arXiv:hep-th/0405076].

\bibitem{SatoTsuchiya}
  A.~Sato and A.~Tsuchiya,
  ``ZZ brane amplitudes from matrix models,''
  JHEP {\bf 0502} (2005) 032
  [arXiv:hep-th/0412201]. 
 
 \bibitem{IshibashiKurokiYamaguchi}
  N.~Ishibashi and A.~Yamaguchi,
  ``On the chemical potential of D-instantons in c = 0 noncritical string
  theory,''
  JHEP {\bf 0506} (2005) 082
  [arXiv:hep-th/0503199];\\
  N.~Ishibashi, T.~Kuroki and A.~Yamaguchi,
  ``Universality of nonperturbative effects in c $<$ 1 noncritical string
  theory,''
  JHEP {\bf 0509} (2005) 043
  [arXiv:hep-th/0507263].
 
\bibitem{fis}
  M.~Fukuma, H.~Irie and S.~Seki,
  ``Comments on the D-instanton calculus in $(p,p+1)$ minimal string theory,''
  Nucl.\ Phys.\  B {\bf 728} (2005) 67
  [arXiv:hep-th/0505253].

\bibitem{fim}
  M.~Fukuma, H.~Irie and Y.~Matsuo,
   ``Notes on the algebraic curves in $(p,q)$ minimal string theory,''
  JHEP {\bf 0609} (2006) 075
  [arXiv:hep-th/0602274].

\bibitem{fi1}
  M.~Fukuma and H.~Irie,
   ``A string field theoretical description of $(p,q)$ minimal superstrings,''
  JHEP {\bf 0701} (2007) 037
  [arXiv:hep-th/0611045].

\bibitem{EynardOrantin}
  B.~Eynard,
  ``Topological expansion for the 1-hermitian matrix model correlation
  functions,''
  JHEP {\bf 0411} (2004) 031
  [arXiv:hep-th/0407261];\\
  B.~Eynard and N.~Orantin,
  ``Invariants of algebraic curves and topological expansion,''
  arXiv:math-ph/0702045;
  ``Geometrical interpretation of the topological recursion, and integrable
  string theories,''
  arXiv:0911.5096 [math-ph].
  
\bibitem{EynardOrantinXYsym}
  B.~Eynard and N.~Orantin,
  ``Topological expansion of mixed correlations in the hermitian 2 Matrix Model and x-y symmetry of the F(g) invariants,''
  arXiv:0705.0958 [math-ph]; 
  ``About the x-y symmetry of the $F_g$ algebraic invariants,''
  arXiv:1311.4993 [math-ph].

\bibitem{EynardMarino}
  B.~Eynard,
  ``Large N expansion of convergent matrix integrals, holomorphic anomalies,
  and background independence,''
  JHEP {\bf 0903} (2009) 003
  [arXiv:0802.1788 [math-ph]]; \\
  B.~Eynard and M.~Marino,
  ``A holomorphic and background independent partition function for matrix
  models and topological strings,''
  arXiv:0810.4273 [hep-th].

\bibitem{irie2}
  H.~Irie,
   ``Fractional supersymmetric Liouville theory and the multi-cut matrix models,''
  Nucl.\ Phys.\ B {\bf 819} (2009) 351 
  [arXiv:0902.1676 [hep-th]]. 

\bibitem{CISY1}
  C.~T.~Chan, H.~Irie, S.~Y.~Shih and C.~H.~Yeh,
  ``Macroscopic loop amplitudes in the multi-cut two-matrix models,''
  Nucl.\ Phys.\  B {\bf 828} (2010) 536
  [arXiv:0909.1197 [hep-th]].
  
\bibitem{CIY1}
  C.~T.~Chan, H.~Irie and C.~H.~Yeh,
  ``Fractional-superstring amplitudes, multi-cut matrix models and non-critical
  M theory,''
  Nucl.\ Phys.\  B {\bf 838} (2010) 75
  [arXiv:1003.1626 [hep-th]].

\bibitem{CIY2}
  C.~-T.~Chan, H.~Irie, C.~-H.~Yeh,
  ``Stokes Phenomena and Non-perturbative Completion in the Multi-cut Two-matrix Models,''
  arXiv:1011.5745v3 [hep-th] to be published in Nucl.\ Phys.\ B. 
 
\bibitem{CIY3}
  C.~-T.~Chan, H.~Irie and C.~-H.~Yeh,
  ``Stokes Phenomena and Quantum Integrability in Non-critical String/M Theory,''
  Nucl.\ Phys.\ B {\bf 855} (2012) 46
  [arXiv:1109.2598 [hep-th]].

\bibitem{CIY4}
  C.~-T.~Chan, H.~Irie and C.~-H.~Yeh,
  ``Analytic Study for the String Theory Landscapes via Matrix Models,''
  Phys.\ Rev.\ D {\bf 86} (2012) 126001
  [arXiv:1206.2351 [hep-th]].

  \bibitem{Ecalle}
  J.~Ecalle, 
  ``Introduction aux fonctions analysables et preuve constructive de la conjecture de Dulac,'' 
  Hermann, Paris, 1992. 


\bibitem{KPZ}
  V.~G.~Knizhnik, A.~M.~Polyakov and A.~B.~Zamolodchikov,
   ``Fractal structure of 2d-quantum gravity,''
  Mod.\ Phys.\ Lett.\ A {\bf 3} (1988) 819.

\bibitem{DDK}
  F.~David,
   ``Conformal field theories coupled to 2-D gravity in the conformal gauge,''
  Mod.\ Phys.\ Lett.\ A {\bf 3} (1988) 1651;\\
  J.~Distler and H.~Kawai,
   ``Conformal field theory and 2-D quantum gravity, 
    or who's afraid of Joseph Liouville?,''
  Nucl.\ Phys.\ B {\bf 321} (1989) 509.

\bibitem{DOZZ}
H.~Dorn and H.~J.~Otto,
``Two and three point functions in Liouville theory,''
Nucl.\ Phys.\ B {\bf 429} (1994) 375
[arXiv:hep-th/9403141]; \\
A.~B.~Zamolodchikov and Al.~B.~Zamolodchikov,
``Structure constants and conformal bootstrap in Liouville field theory,''
Nucl.\ Phys.\ B {\bf 477} (1996) 577
[arXiv:hep-th/9506136]. 

\bibitem{Teschner}
  J.~Teschner,
  ``On the Liouville three point function,''
  Phys.\ Lett.\  B {\bf 363} (1995) 65
  [arXiv:hep-th/9507109].

\bibitem{FZZT}
V.~Fateev, A.~B.~Zamolodchikov and Al.~B.~Zamolodchikov,
``Boundary Liouville field theory. I: 
Boundary state and boundary two-point function,''
arXiv:hep-th/0001012;\\
%
J.~Teschner,
``Remarks on Liouville theory with boundary,''
arXiv:hep-th/0009138.

\bibitem{ZZ}
  A.~B.~Zamolodchikov and Al.~B.~Zamolodchikov,
  ``Liouville field theory on a pseudosphere,''
  arXiv:hep-th/0101152.

\bibitem{SeSh}
  N.~Seiberg and D.~Shih,
   ``Branes, rings and matrix models in minimal (super)string theory,''
  JHEP {\bf 0402} (2004) 021
  [arXiv:hep-th/0312170].

\bibitem{KOPSS}
  D.~Kutasov, K.~Okuyama, J.~w.~Park, N.~Seiberg and D.~Shih,
  ``Annulus amplitudes and ZZ branes in minimal string theory,''
  JHEP {\bf 0408} (2004) 026
  [arXiv:hep-th/0406030].


\bibitem{FMS-CFT}
  P.~Di Francesco, P.~Mathieu and D.~Senechal,
  ``Conformal field theory,''
  New York, USA: Springer (1997) 890 p
  
\bibitem{Mehta}
  M.~L.~Mehta,
  ``A Method Of Integration Over Matrix Variables,''
  Commun.\ Math.\ Phys.\  {\bf 79}, 327 (1981).

\bibitem{DoubleFieldTheory}
  C.~Hull and B.~Zwiebach,
  ``Double Field Theory,''
  JHEP {\bf 0909} (2009) 099
  [arXiv:0904.4664 [hep-th]].
  
  
  


\bibitem{KharchevMarshakov}
  S.~Kharchev and A.~Marshakov,
  ``Topological versus nontopological theories and $p - q$ duality in $c \leq 1$ 2-d gravity models,''
  hep-th/9210072; 
  ``On $p - q$ duality and explicit solutions in $c \leq 1$ 2-d gravity models,''
  Int.\ J.\ Mod.\ Phys.\ A {\bf 10} (1995) 1219
  [hep-th/9303100].
  
  
\bibitem{BEH}
  M.~Bertola, B.~Eynard and J.~P.~Harnad,
  ``Duality, biorthogonal polynomials and multimatrix models,''
  Commun.\ Math.\ Phys.\  {\bf 229} (2002) 73
  [nlin/0108049 [nlin-si]]; 
  ``Differential systems for biorthogonal polynomials appearing in 2-matrix models and the associated Riemann-Hilbert problem,''
  Commun.\ Math.\ Phys.\  {\bf 243} (2003) 193
  [nlin/0208002 [nlin.SI]]; 
  ``Semiclassical orthogonal polynomials, matrix models and isomonodromic tau functions,''
  Commun.\ Math.\ Phys.\  {\bf 263} (2006) 401
  [nlin/0410043 [nlin.SI]].
  
\bibitem{KWduality}
  H.~A.~Kramers and G.~H.~Wannier,
  ``Statistics of the two-dimensional ferromagnet. Part 1.,''
  Phys.\ Rev.\  {\bf 60} (1941) 252.
  
\bibitem{AKOSY}
  T.~Asatani, T.~Kuroki, Y.~Okawa, F.~Sugino and T.~Yoneya,
  ``T duality transformation and universal structure of noncritical string field theory,''
  Phys.\ Rev.\ D {\bf 55} (1997) 5083
  [hep-th/9607218].
  
\bibitem{KurokiSugino}
  T.~Kuroki and F.~Sugino,
  ``T-duality of ZZ-brane,''
  arXiv:hep-th/0612042;
  ``T duality of the Zamolodchikov-Zamolodchikov brane,''
  Phys.\ Rev.\  D {\bf 75} (2007) 044008.

\bibitem{OkudaTakayanagi}
  T.~Okuda and T.~Takayanagi,
  ``Ghost D-branes,''
  JHEP {\bf 0603} (2006) 062
  [hep-th/0601024].

\bibitem{MarinoLecture}
  M.~Marino,
  ``Non-Perturbative effects in Matrix model, Chern-Simons Theory and Topological Strings,''
lectures in conference, {\em Topological Strings, Modularity and non-perturbative Physics}, June 21 - 31, 2010, at ESI in Vienna, Austria; 
  ``Lectures on non-perturbative effects in large N gauge theories, matrix models and strings,''
  arXiv:1206.6272 [hep-th].



\bibitem{ItsBook}
  A.~S.~Fokas, A.~R.~Its, A.~A.~Kapaev and V.~Y.~Novokshenov,
  ``Painlev\'e Transcendents: The Riemann-Hilbert Approach,''
Mathematical Surveys and Monographs, Vol.~128 (2006) 553 pp;\\
  A.~R.~Its and V.~Y.~Novokshenov, ``The Isomonodromic Deformation Method in the
Theory of Painlev\' e Equations,'' Springer-Verlag (1986). 


\bibitem{CiteOfStringLandscape}
  M.~R.~Douglas,
   ``The Statistics of string / M theory vacua,''  
JHEP {\bf 0305} (2003) 046  [hep-th/0303194]; 
  ``The string landscape and low energy supersymmetry,''  
arXiv:1204.6626 [hep-th] 
(for recent survay).  


\bibitem{AGT}
  L.~F.~Alday, D.~Gaiotto and Y.~Tachikawa,
  ``Liouville Correlation Functions from Four-dimensional Gauge Theories,''
  Lett.\ Math.\ Phys.\  {\bf 91} (2010) 167
  [arXiv:0906.3219 [hep-th]].
  
 
 \bibitem{JimboMiwaUeno}
  M.~Jimbo, T.~Miwa and K.~Ueno,
  ``Monodromy Preserving Deformations Of Linear Differential Equations With
  Rational Coefficients. I,''
  Physica D {\bf 2} (1981) 306-352; \\
  M.~Jimbo and T.~Miwa,
  ``Monodromy Preserving Deformations Of Linear Differential Equations With
  Rational Coefficients. II,''
  Physica D {\bf 2} (1981) 407-448; 
  ``Monodromy Preserving Deformations Of Linear Differential Equations With
  Rational Coefficients. III,''
  Physica D {\bf 4} (1981) 26-46;

\bibitem{KapaevPI}
A.~A.~Kapaev, 
``Quasi-linear Stokes phenomenon for the Painlev\'e first equation,''
J. Phys. A.: Math. Gen. 37 (2004) 11149-11167 [arXiv:nlin/0404026 [nlin.SI]]


  \bibitem{ExactWKB}
  T.~Kawai and Y.~Takei, 
  ``Algebraic Analysis of Singular Perturbation Theory,''
  Translations of Mathematical Monographs, Vol. 227, Amer. Math. Soc., 2005. 
  (Originally published in Japanese by Iwanami,. Tokyo in 1998.). 

\bibitem{DZmethod}
  P.~A.~Deift and X.~Zhou, 
  ``A steepest descent method for oscillatory Riemann-Hilbert problems. Asymptotics for the MKdV equation,''
  Ann.~of Math., {\bf 137} (1993) 295-368. 

\bibitem{ItsKapaev}
  A.~Its and A.~Kapaev, 
  ``The nonlinear steepest descent approach to the asymptotics of the second Painleve transcendent in the complex domain,''
  in MathPhys Odyssey 2001, Integrable Models and Beyond, eds. M.~Kashiwara and T.~Miwa, Prog. Math. Phys., 23, Birkhauser Boston (2002) 273-311. 

\bibitem{SpectralNetworks}
  D.~Gaiotto, G.~W.~Moore and A.~Neitzke,
  ``Spectral networks,''
  arXiv:1204.4824 [hep-th].
  
\bibitem{ODEIMCorrespondence}
  P.~Dorey, R.~Tateo,
  ``Anharmonic oscillators, the thermodynamic Bethe ansatz, and nonlinear integral equations,''
  J.\ Phys.\ A {\bf A32 } (1999)  L419-L425.
  [hep-th/9812211]; 
  ``On the relation between Stokes multipliers and the T - Q systems of conformal field theory,''
  Nucl.\ Phys.\  {\bf B563 } (1999)  573-602.
  [hep-th/9906219]; \\
  J.~Suzuki,
  ``Functional relations in Stokes multipliers and solvable models related to $U_q(A^{(1)}_n)$,''
  J.\ Phys.\ A {\bf A33 } (2000)  3507-3522.
  [hep-th/9910215].
  
\bibitem{RHcite}
  H.~Flaschka and A.~C.~Newell,
  ``Monodromy And Spectrum Preserving Deformations. 1,''
  Commun.\ Math.\ Phys.\  {\bf 76} (1980) 65. 
  
  
\bibitem{GeneralizedRecursions}
  V.~Bouchard, J.~Hutchinson, P.~Loliencar, M.~Meiers and M.~Rupert,
  ``A generalized topological recursion for arbitrary ramification,''
  arXiv:1208.6035 [math-ph]; \\
  V.~Bouchard and B.~Eynard,
  ``Think globally, compute locally,''
  JHEP {\bf 1302} (2013) 143
  [arXiv:1211.2302 [math-ph]].
  
  \bibitem{KressLIE}
  R.~Kress, 
  ``Linear Integral Equations,''
  Applied Mathematical Sciences, 
  Vol.~82 (2012) 316 pp. 
  
\bibitem{ShihAiry}
  A.~Hashimoto, M.-x.~Huang, A.~Klemm and D.~Shih,
  ``Open/closed string duality for topological gravity with matter,''
  JHEP {\bf 0505} (2005) 007
  [hep-th/0501141].
  

\bibitem{SecondPaper}
  C.~-T.~Chan, H.~Irie and C.~-H.~Yeh,
  ``Vacuum Connection Formula and Phase Structure in Minimal String Theory,'' in progress; ``Vacuum Connection Formula and Duality Bootstrap for Strings,'' presented by H.~Irie in KEK theory workshop 2014, Feb.~19, 2014. 

  
\bibitem{FunctionalAnalysisKF}
 A.~N.~Kolmogorov and S.~V.~Fomin, 
 ``Elements of the Theory of Functions and Functional Analysis,''
 Courier Dover Publications (1999) 257 pp. 
  
  


\bibitem{KnotTheory}
  R.~Dijkgraaf and H.~Fuji,
  ``The Volume Conjecture and Topological Strings,''
  Fortsch.\ Phys.\  {\bf 57} (2009) 825
  [arXiv:0903.2084 [hep-th]]; \\
  R.~Dijkgraaf, H.~Fuji and M.~Manabe,
  ``The Volume Conjecture, Perturbative Knot Invariants, and Recursion Relations for Topological Strings,''
  Nucl.\ Phys.\ B {\bf 849} (2011) 166
  [arXiv:1010.4542 [hep-th]]; \\
  G.~Borot and B.~Eynard,
  ``All-order asymptotics of hyperbolic knot invariants from non-perturbative topological recursion of A-polynomials,''
  arXiv:1205.2261 [math-ph].
  
\bibitem{QuantumSpectralCurves}
  B.~Eynard and O.~Marchal,
  ``Topological expansion of the Bethe ansatz, and non-commutative algebraic geometry,''
  JHEP {\bf 0903} (2009) 094
  [arXiv:0809.3367 [math-ph]]; \\
  L.~Chekhov, B.~Eynard and O.~Marchal,
  ``Topological expansion of the Bethe ansatz, and quantum algebraic geometry,''
  arXiv:0911.1664 [math-ph]; 
  ``Topological expansion of $\beta$-ensemble model and quantum algebraic geometry in the sectorwise approach,''
  Theor.\ Math.\ Phys.\  {\bf 166} (2011) 141
  [arXiv:1009.6007 [math-ph]].
  
\bibitem{BleherIts}
  P.~Bleher and A.~Its,
  ``Double scaling limit in the random matrix model: The Riemann-Hilbert approach,''
  Comm.~Pure Appl.~Math.~Vol. LVI, (2003) 0433?0516. 
  math-ph/0201003.
  
\bibitem{BertolaTovbisRHP}
M.~Bertola and A.~Tovbis,
``Universality in the profile of the semiclassical limit solutions to the focusing Nonlinear Schroedinger equation at the first breaking curve,''
Int.~Math.~Res.~Not., 2009. [arXiv:0909.3264 [nlin.SI]]; 
``Universality for the focusing nonlinear Schroedinger equation at the gradient catastrophe point: Rational breathers and poles of the tritronquee solution to Painleve I,''
Comm.~Pure Appl.~Math.~
Vol.~66, Issue 5 (2013) 678?752. [arXiv:1004.1828 [nlin.SI]]; 
``Asymptotics of orthogonal polynomials with complex varying quartic weight: global structure, critical point behaviour and the first Painleve' equation,'' arXiv:1108.0321 [nlin.SI]. 
  
  
\bibitem{Localization}
  V.~Pestun,
  ``Localization of gauge theory on a four-sphere and supersymmetric Wilson loops,''
  Commun.\ Math.\ Phys.\  {\bf 313} (2012) 71
  [arXiv:0712.2824 [hep-th]].

\bibitem{CiteStokesAmbiguity}
  I.~Aniceto and R.~Schiappa,
  ``Nonperturbative Ambiguities and the Reality of Resurgent Transseries,''
  arXiv:1308.1115 [hep-th]; \\
  G.~Basar, G.~V.~Dunne and M.~Unsal,
  ``Resurgence theory, ghost-instantons, and analytic continuation of path integrals,''
  arXiv:1308.1108 [hep-th].
  
\bibitem{GKIM}
  S.~Garoufalidis, A.~Its, A.~Kapaev and M.~Marino,
  ``Asymptotics of the instantons of Painleve I,''
  arXiv:1002.3634 [math.CA].
  
\bibitem{KMR}
  A.~Klemm, M.~Marino and M.~Rauch,
  ``Direct Integration and Non-Perturbative Effects in Matrix Models,''
  JHEP {\bf 1010} (2010) 004
  [arXiv:1002.3846 [hep-th]].

\bibitem{resurgentRef}
  I.~Aniceto, R.~Schiappa, M.~Vonk,
  ``The Resurgence of Instantons in String Theory,''
  [arXiv:1106.5922 [hep-th]]; \\
  R.~Schiappa and R.~Vaz,
  ``The Resurgence of Instantons: Multi-Cuts Stokes Phases and the Painleve II Equation,''
  arXiv:1302.5138 [hep-th].








\end{thebibliography}
\end{document}